\documentclass[prb, twocolumn, floatfix, superscriptaddress, showpacs, longbibliography]{revtex4-2}

\pdfoutput=1						

\usepackage[charter]{mathdesign}	
\usepackage{amsmath}				
\usepackage{mathrsfs}					

\usepackage{mathtools}
\usepackage[]{stackengine}

\newlength\correct
\usepackage[pdftex]{graphicx}
\usepackage{microtype}
\usepackage{bm}\let\vec\bm
\usepackage{gensymb}
\usepackage{mdframed}
\usepackage[caption=false]{subfig}

\usepackage[svgnames]{xcolor}
\usepackage[
	colorlinks=True,linkcolor=DarkRed,citecolor=ForestGreen,urlcolor=MediumBlue,
	pdfstartview=FitH,bookmarks=False,pdfpagemode=UseNone
]{hyperref}

\usepackage{feynmp-auto}
\usepackage{natbib}
\usepackage{hyperref}
\usepackage{float}
\usepackage[T1]{fontenc}
\usepackage{inputenc}
\usepackage{tcolorbox}
\usepackage{stackengine}
\newcommand\xrowht[2][0]{\addstackgap[.5\dimexpr#2\relax]{\vphantom{#1}}}

\makeatletter
\let\cat@comma@active\@empty
\makeatother

\begin{document}

\title{Kinetic theory of the non-local electrodynamic response in anisotropic metals: skin effect in 2D systems}

\author{Davide Valentinis}
\affiliation{Institut f\"{u}r Theorie der Kondensierten Materie, Karlsruher Institut
f\"{u}r Technologie, 76131 Karlsruhe, Germany}
\affiliation{Institut f\"{u}r Quantenmaterialien und Technologien, Karlsruher Institut
f\"{u}r Technologie, 76131 Karlsruhe, Germany}
\author{Graham Baker}
\affiliation{Department of Physics and Astronomy, University of British Columbia, Vancouver, BC V6T 1Z1, Canada}
\affiliation{Quantum Matter Institute, University of British Columbia, Vancouver, BC V6T 1Z4, Canada}
\author{Douglas A. Bonn}
\affiliation{Department of Physics and Astronomy, University of British Columbia, Vancouver, BC V6T 1Z1, Canada}
\affiliation{Quantum Matter Institute, University of British Columbia, Vancouver, BC V6T 1Z4, Canada}
\author{J\"{o}rg Schmalian}
\affiliation{Institut f\"{u}r Theorie der Kondensierten Materie, Karlsruher Institut
f\"{u}r Technologie, 76131 Karlsruhe, Germany}
\affiliation{Institut f\"{u}r Quantenmaterialien und Technologien, Karlsruher Institut
f\"{u}r Technologie, 76131 Karlsruhe, Germany}
\date{\today}

\begin{abstract}
The electrodynamic response of ultra-pure materials at low temperatures becomes spatially non-local. This non-locality gives rise to phenomena such as hydrodynamic flow in transport and the anomalous skin effect in optics. In systems characterized by an anisotropic electronic dispersion, the non-local dynamics becomes dependent on the relative orientation of the sample with respect to the applied field, in ways that go beyond the usual, homogeneous response. Such orientational dependence should manifest itself not only in transport experiments, as recently observed, but also in optical spectroscopy. 
In this paper we develop a kinetic theory for the distribution function and the transverse conductivity of two- and three-dimensional Fermi systems with anisotropic electronic dispersion. By expanding the collision integral into the eigenbasis of a collision operator, we include momentum-relaxing scattering as well as momentum-conserving collisions. We examine the isotropic 2D case as a reference, as well as anisotropic hexagonal and square Fermi-surface shapes. We apply our theory to the quantitative calculation of the skin depth and the surface impedance, in all regimes of skin effect. We find qualitative differences between the frequency dependence of the impedance in isotropic and anisotropic systems. Such differences are shown to persist even for more complex 2D Fermi surfaces, including the ``supercircle'' geometry and an experimental parametrization for PdCoO$_2$, which deviate from an ideal polygonal shape. We study the orientational dependence of skin effect due to Fermi-surface anisotropy, thus providing guidance for the experimental study of non-local optical effects. 

\end{abstract}

\maketitle

\section{Introduction}\label{Intro}

\begin{figure}[H] \centering
\includegraphics[width=0.75\columnwidth]{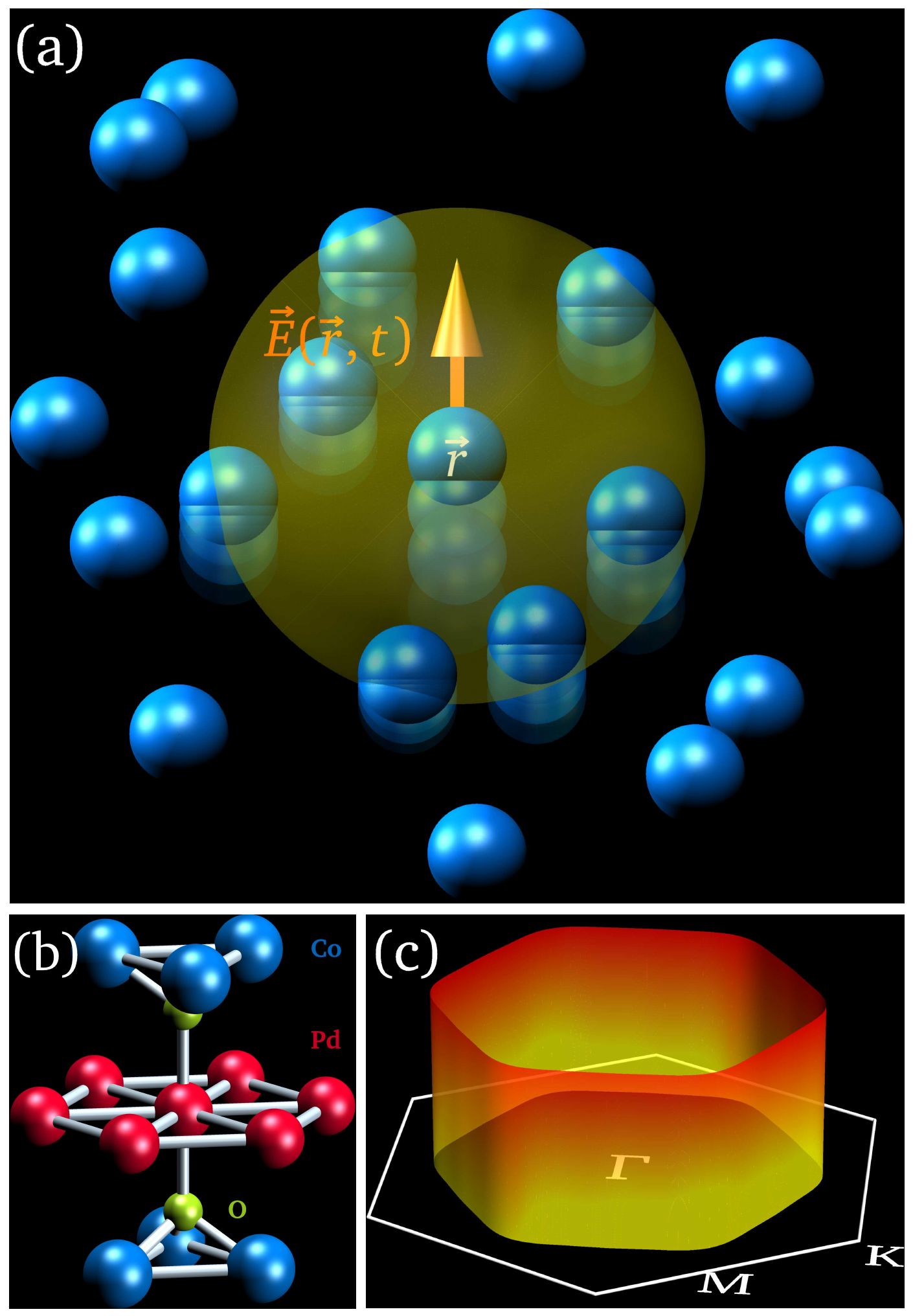}
\caption{\label{fig:Coherence} (a) Visual representation of the spatially non-local dielectric response in three dimensions: the application of an electric field $\vec{E}(\vec{r},t)$ at coordinate $\vec{r}$ excites a response of electrons (blue spheres) in an extended coherence volume (yellow-shaded sphere). (b) Schematics of part of the unit cell of PdCoO$_2$, showing a conducting Pd layer with hexagonal in-plane coordination and out-of-plane bonds with CoO$_2$ tetrahedra. (c) The anisotropic, hexagonal Fermi surface of PdCoO$_2$ from a tight-binding fit of de Haas–van Alphen magnetotransport data \cite{Hicks-2012}.
}
\end{figure}

The symmetry and non-locality of many-body correlations are responsible for a variety of quantum phenomena observed in solid-state electronic systems. The non-equilibrium response of strongly correlated metals to external probes, like electromagnetic fields or thermal gradients, is in general spatially non-local: the external perturbation, applied at one spatial coordinate $\vec{r}$, elicits an electronic reaction in an extended region of space $\left\{\vec{r}' \right\}\neq \vec{r}$ surrounding the probe; see Fig.\@ \ref{fig:Coherence}(a). The radius of such region, depicted for an isotropic material by the yellow sphere in Fig.\@ \ref{fig:Coherence}(a), is often determined by the carrier mean-free path $l_{\rm mr}=v_F/\gamma_{\rm mr}$, where $v_F$ is the Fermi velocity and $\gamma_{\rm mr}$ is the momentum-relaxation rate. In Fourier space the non-locality entails a dependence of the response functions on wave vector $\vec{q}$.

Striking consequences of non-local behavior are electronic viscosity and elasticity. In the regime where momentum-conserving electron collisions at rate $\gamma_{\rm mc}$ are the dominant source of scattering, electrons collectively respond as a macroscopic viscous substance \cite{Landau-1987fm} described by the laws of hydrodynamics \cite{Gurzhi-1959,Gurzhi-1968,Fritz-2008,Muller-2009, Andreev-2011, Briskot-2015, Levitov-2016, Narozhny-2017,Guo-2018,Gorbar-2018,Link-2018b,Ledwith-2019, Gorbar-2019,Kiselev-2019,Narozhny-2019,Levchenko-2020,DeLuca-2021,Baghramyan-2021,Belitz-2022}. Evidence of such phenomena is retrieved in transport experiments, where crossovers between hydrodynamic and ballistic flow of electron fluids have been reported in systems as diverse as (Al,Ga)As \cite{Molenkamp-1994,deJong-1995}, graphene \cite{Crossno-2016,Bandurin-2016,Krishna-Kumar-2017, Sulpizio-2019,Ku-2020}, PdCoO$_2$ \cite{Moll-2016}, WP$_2$ \cite{Gooth-2018,Jaoui-2018}, Sb \cite{Jaoui-2021}, WTe$_2$ \cite{Vool-2021}. 
Likewise, for finite-frequency and low-momentum excitations in the collisionless regime $\gamma_{\rm mr} \ll \gamma_{\rm mc}\rightarrow 0^+$, dissipationless correlations due to interactions between electrons are encoded by elastic moduli \cite{Landau-1986el} \cite{Tokatly-1999,Conti-1999,Tokatly-2000,Vignale-2005,FZVM-2014,Link-2015,Link-2018a,Alvarez-2020,Dufty-2020,Valentinis-2021a,Valentinis-2021b,Belitz-2022}. 
This visco-elastic regime can be detected through optical spectroscopy, as testified by the analysis of plasmon resonances in In:CdO nanofilms \cite{DeCeglia-2018} and coaxial nanoapertures \cite{Yoo-2019}. 

Besides recently engineered applications, non-locality is firmly established as the origin of anomalous skin effect, whereby an oscillating electromagnetic field at frequency $\omega$, applied at a vacuum-metal interface, generates currents at anomalously long depths into the metal \cite{Reuter-1948,Chambers-1952, Casimir-1967a,Casimir-1967b,Casimir-1967c,Wooten-1972opt,Sondheimer-2001,Dressel-2001}.
In turn, the generated bulk currents significantly enlarge the penetration depth (skin depth) of radiation inside the metal, with respect to the usual exponential damping due to relaxational dynamics and ensuing Ohmic conduction.
The anomalous response regime manifests itself for $\omega \ll v_F q$, where $v_F$ is the velocity of charged carriers (i.e.\@, the Fermi velocity), and it is due to ballistic conduction: the motion of electrons in the metal is essentially unimpeded by any form of scattering, i.e.\@, $\omega \gg \left\{ \gamma_{\rm mr}, \gamma_{\rm mc}\right\}$. Moreover, if momentum-conserving collisions are sufficiently strong compared to relaxation, i.e.\@, $\gamma_{\rm mr} \ll \gamma_{\rm mc}$, electron hydrodynamics shows in a finite-frequency window between the normal and anomalous regime, thus producing a viscous form of skin effect \cite{FZVM-2014,Levchenko-2020,Valentinis-2021a,Matus-2021}. All aforementioned regimes can be distinguished by their distinct dependence of the skin depth on $\omega$, $\gamma_{\rm mc}$, and $\gamma_{\rm mr}$, as we will appreciate in the following. 
Thus, the skin effect, observed through surface impedance measurements \cite{Pippard-1947a,Pippard-1947b,Pippard-1947c}, represents an ideal probe of the non-local transverse current response in multiple regimes of momentum and frequency.

The above picture of non-locality is convoluted in the solid state by the interactions of electrons with their chemical surroundings: the crystalline lattice and impurities break translational and rotational invariance. In particular, strong electron-lattice interactions mold the electronic bandstructure, producing anisotropic Fermi surfaces with reduced symmetry in accordance with the crystalline potential. For instance, the hexagonal coordination of Pd atoms in the conducting planes of the delafossite compound PdCoO$_2$, represented by the red spheres in Fig.\@ \ref{fig:Coherence}(b), produces an essentially 2D Fermi surface with in-plane section shaped as a hexagon with rounded corners \cite{Hicks-2012} \footnote{Still, the relatively slight warping of the Fermi surface of PdCoO$_2$, in the direction orthogonal to the Pd conducting planes, significantly impacts the skin effect in experiments \cite{Baker-2022_preprint}. Therefore, from the point of view of optical spectroscopy, PdCoO$_2$ behaves as a highly anisotropic 3D material. In this paper, we focus on the ideal case of 2D anisotropic systems, in which electrodynamics and anisotropy are entirely confined to a 2D conductive plane.}. The reduced symmetry of the electronic dispersion may also be accompanied by a variation of the momentum-relaxation rates along different spatial coordinates. The ensuing anisotropic conductivity tensor generates a sizable anisotropy in the electrodynamic properties of novel materials, as observed in quasi-2D layered systems like delafossites \cite{Kim-2009,Noh-2009,Ong-2010,Hicks-2012,Takatsu-2013,Nandi-2018,Mackenzie-2017} and van der Waals compounds \cite{Geim-2013,Basov-2016,Ajayan-2016,Ma-2018,Ruta-2020,Toshiya-2021}. 
Notice that, if the local ($q=0$) conductivity tensor is not rotational-invariant, the electrodynamic anisotropy manifests itself already in the local limit. However, even in systems possessing isotropic conductivities at $q=0$, anisotropic transport and optical effects may emerge in the finite-$q$ response, due to the lack of Fermi-surface rotational invariance. These effects, which are symmetry-forbidden within local response, entail a number of novel phenomena \cite{Varnavides-2020}, such as the appearance of rotational \cite{Cook-2019} and Hall \cite{Toshio-2020} components of the viscosity tensor in hydrodynamic regime. Different dissipative components, including the rotational viscosity, may then be selectively studied in systems with dihedral symmetry, by measuring the local heat generated at the center of a square sample by electric currents injected in a 8-contact geometry on the sample perimeter \cite{Cook-2021}. 
Therefore, non-locality can unveil novel electrodynamic regimes due to anisotropic Fermi surfaces, which would be invisible in the $q=0$ response. 

\begin{figure}[ht] \centering
\includegraphics[width=\columnwidth]{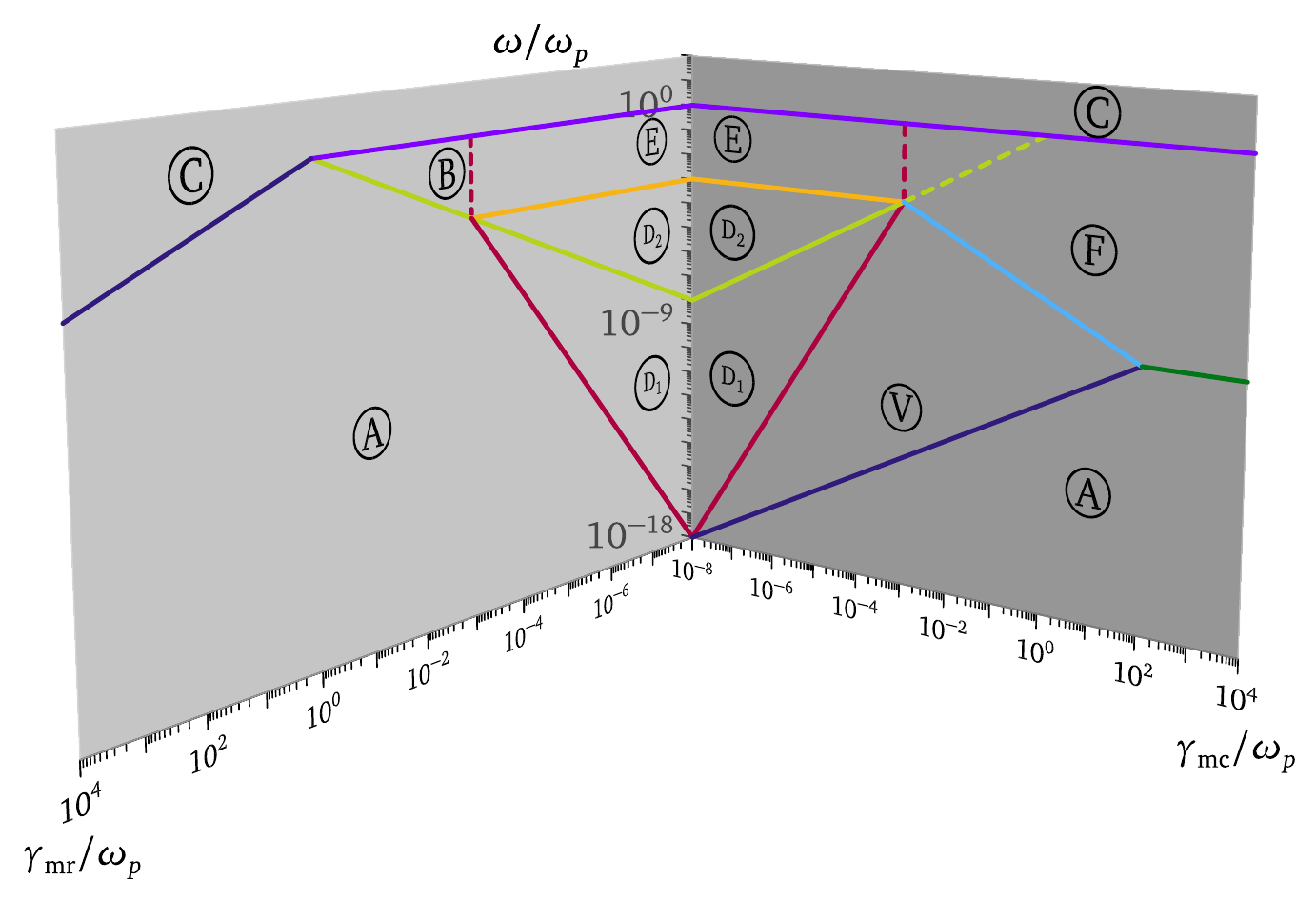}
\caption{\label{fig:3D_skin_effect} Summary of skin effect regimes as a function of relaxation rate $\gamma_{\rm mr}$, collision rate $\gamma_{\rm mc}$, and frequency $\omega$, normalized to the plasma frequency $\omega_p$, assuming a Fermi velocity $v_F/c=0.001$. This skin effect ``phase diagram'' applies to both isotropic and anisotropic systems.}
\end{figure}

Various phenomenological and microscopic theories have been developed for modeling the non-equilibrium momentum-dependent electronic response. Such valuable efforts are often grounded on the Boltzmann kinetic equation, which determines the response in terms of a local-equilibrium distribution function. The latter is influenced by external driving forces and interactions, as well as by a collision integral that models the specific scattering channels (conserving or relaxing momentum) available for the given electron liquid.
However, in three- (3D) and two-dimensional (2D) materials alike, most earlier treatments have focused on isotropic systems with quadratic electronic dispersion $\varepsilon_{\vec{k}}\propto k^2$, where the shape of the Fermi surface is spherical and circular respectively. There, the local velocity field $v_{\vec{k}} \propto \nabla_{\vec{k}} \varepsilon_{\vec{k}}$ and the associated current density $J_{\vec{k}} \propto v_{\vec{k}}$ are proportional to the crystal momentum $\hbar k$. This assumption underlies the seminal work of Reuter and Sondheimer \cite{Reuter-1948,Sondheimer-2001}, which elucidates the crossover at finite excitation frequency $\omega$ between normal and anomalous skin effect, within a Boltzmann treatment that includes relaxation but not momentum-conserving collisions. At finite $\gamma_{\rm mc}$, isotropy allows one to relate the charge and current susceptibilities with the dependence of the electronic response on exchanged momentum $\vec{q}$, encoded, e.g.\@, by the viscosity tensor \cite{Bradlyn-2012}. 
The proportionality between current and crystal momentum is invalidated in anisotropic systems, having Fermi surfaces that explicitly break rotational invariance. Hence, the ensuing electrodynamic properties can be very different from the isotropic case, as they depend on the orientation of the Fermi surface with respect to the direction of the driving field. Such dependence was shown quantitatively by Sondheimer \cite{Sondheimer-1954}, generalizing his surface impedance calculations to single- and two-band systems with spheroidal Fermi surfaces: in these examples, the finite-$q$ optical properties show a different anisotropy than the one at $q=0$, but the conductivity already breaks rotational invariance in the local limit. The sensitivity of optics to Fermi-surface shape was also qualitatively discussed by Pippard \cite{Pippard-1954}, for generic dispersion anisotropy in 2D and 3D by the means of the ineffectiveness concept, which we will discuss in Sec.\@ \ref{Disc}.
These works considered relaxation but not momentum-conserving collisions, which further enrich the theoretical analysis with various crossovers between hydrodynamic, ballistic and anomalous response regimes as a function of $\omega$, $\gamma_{\rm mr}$, and $\gamma_{\rm mc}$. All crossovers are qualitatively described by specific ratios between characteristic length scales for skin effect, as summarized in Sec.\@ \ref{Summ}, Fig.\@ \ref{fig:3D_skin_effect}, and Table \ref{tab:skin_lengths}. 
Such complexity invariably complicates the analysis of experimental data, and calls for a theory that is general enough to capture and distinguish all essential manifestations of electrodynamic non-locality, while being adaptable enough to allow for quantitative comparisons among different static and dynamic probes. 

In this paper, we develop a kinetic theory for the distribution function and the conductivity tensor $\sigma_{\alpha \beta}(\vec{q},\omega)$ of a 2D or 3D Fermi system endowed with an arbitrary electronic structure $\varepsilon_{\vec{k}}$. Technically, we recast the Boltzmann equation into an infinite system of algebraic equations in the collision operator formalism. In this approach, both the local-equilibrium distribution function and the collision integral are expanded in terms of the eigenfunctions (modes) of the collision operator. These modes are constrained by the conservation of charge, and include slow momentum relaxation occurring at rate $\gamma_{\rm mr}$, as well as momentum-conserving collisions at rate $\gamma_{\rm mc} \geq \gamma_{\rm mr}$. Here we make the central assumption of our approach that all momentum-conserving processes are governed by the same rate. The two scattering rates  $\gamma_{\rm mr}$ and $\gamma_{\rm mc}$ are not tied to a specific scattering process, i.e. the detailed nature of these rates is not important for our considerations. 
The advantage of our formulation is that the conductivity can be evaluated independently of the explicit form of the collision operator eigenmodes, for an arbitrarily shaped Fermi surface. 
Under the minimal assumption that two mirror symmetry planes exist in $\vec{k}$ space, we concentrate on the response to transverse electric fields $\vec{E}(\vec{q},\omega)=E_y(q_x,\omega) \hat{u}_y \perp \vec{q}=q_x \hat{u}_x$, in a 2D reference system with unit vectors $\hat{u}_x$ and $\hat{u}_y$. We obtain a closed expression for the 2D transverse conductivity $\sigma_{yy}(q_x,\omega)$ as a function of frequency $\omega$, scattering rates $\gamma_{\rm mr}$ and $\gamma_{\rm mc}$, and angle $\theta$ describing the angular dependence of the anisotropic Fermi-surface velocity $\vec{v}_{\vec{k}}$. 
Our formalism retrieves the known results for circular Fermi surfaces in the hydrodynamic limit, and generalizes them to arbitrary frequencies and scattering rates. In addition, we discuss results for an hexagonal shape, inspired by systems like ultrapure PdCoO$_2$ \cite{Hicks-2012,Takatsu-2013} and transition-metal dichalcogenides like NbSe$_2$ \cite{Rossnagel-2001,Inosov-2008, Borisenko-2009}, and for a square shape, relevant for, e.g.\@, PrTe$_3$ \cite{Lee-2016}, GdTe$_3$ \cite{Liu-2020}. For both hexagonal and square shapes, the dependence of the conductivity on momentum and frequency qualitatively changes as a function of Fermi-surface orientation with respect to the applied field. In particular, we show that whenever large Fermi-surface segments with velocity parallel to the applied field exist, the dependence of $\sigma_{yy}(q_x,\omega)$ on $q_x$ vanishes, i.e.\@, the contribution of these segments to the conductivity is completely local. This insensitivity to $q_x$ qualitatively changes the character of conduction in anomalous regime, and in the case of a square Fermi surface -- see Fig.\@ \ref{fig:Square}(b) -- non-locality is completely suppressed and the material responds like a Ohmic (Drude) conductor.
We apply our kinetic theory for $\sigma_{yy}(q_x,\omega)$ to the calculation of the skin depth and of the associated surface impedance $Z(\omega)$, assuming specular scattering at the vacuum-metal boundary.
Our results for $Z(\omega)$ as a function of $\omega$, $\gamma_{\rm mr}$, and $\gamma_{\rm mc}$, are summarized in the qualitative skin effect ``phase diagrams'' of Figs.\@ \ref{fig:3D_skin_effect_iso}, \ref{fig:3D_skin_effect_misalign}, \ref{fig:3D_skin_effect_align}, and \ref{fig:3D_skin_effect_square}, which are discussed in Sec.\@ \ref{Summ}. These figures compare all impedance regimes existing for an isotropic dispersion with the ones resulting from different orientations of hexagonal and square Fermi surfaces. It is seen that the boundaries between different skin effect regimes, as well as the dependence of the surface impedance on frequency and scattering rates, significantly depend on Fermi-surface geometry and orientation. Such results are confirmed both numerically and analytically by the explicit computations of $|Z(\omega)|$ shown in Figs.\@ \ref{fig:omegacross_iso}, \ref{fig:omegacross_hex}, and \ref{fig:omegacross_square}. In particular, the effect of flat Fermi-surface portions with velocity parallel to the field qualitatively affects $|Z(\omega)|$ in anomalous regime: for $\omega < \left\{\gamma_{\rm mc},\gamma_{\rm mr}\right\}$ we have $|Z(\omega)| \propto \omega^{1/2}$, which would be normally associated with normal skin effect in isotropic systems, while for $\omega > \left\{\gamma_{\rm mc},\gamma_{\rm mr}\right\}$ the metal in anomalous regime behaves as a perfect conductor characterized by $\sigma_{yy}(\omega)\sim i/\omega$. 

\begin{table*}[t]
	\centering
		\begin{tabular}{| c | c | c |} 
 \hline \hline
  Symbol & Definition & Description \\
	\hline \hline
$\lambda_L$ & $c/\omega_p$ & London penetration depth \\
 \hline
 $l_\omega$ & $v_F/\omega$ & Distance traveled during one field oscillation \\ 
 \hline
$l_{\rm mr}$ & $v_F/\gamma_{\rm mr}$ & Relaxational mean-free path \\ 
\hline
$l_{\rm mc}$ & $v_F/\gamma_{\rm mc}$ & Collision length \\
\hline
$l_G$ & $\sqrt{l_{\rm mc} l_{\rm mr}}$ & Gurzhi length \\
\hline
$\delta_v$ & $\left[(\lambda_L)^2 l_\omega l_{\rm mc}\right]^{1/4}$ & Skin depth in viscous regime \\ 
 \hline
\end{tabular} \caption{\label{tab:skin_lengths} Characteristic length scales which determine the crossovers between different regimes of skin effect, as a function of momentum-relaxing rate $\gamma_{\rm mr}$, momentum-conserving collision rate $\gamma_{\rm mc}$, and radiation frequency $\omega$, for an electron liquid with Fermi velocity $v_F$.}
\end{table*}
In addition, in the absence of scattering we present analytical expressions for the skin depth and the surface impedance for specular boundary conditions in hexagonal and square geometries, valid at all ratios $\omega/(v_F q_x)$.

Hence, we construct a compact formalism that covers all forms of normal, anomalous, and viscous skin effects, and that makes direct contact to the basic electrodynamic quantity $Z(\omega)$. It also allows to make contact to Pippard’s effective carrier number $n^\ast$ in the non-local regime, as discussed in Sec.\@ \ref{Gen_ineffectiveness}. 

This paper is organized as follows. Sec.\@ \ref{Summ} contains a summary of main results and a qualitative discussion of all regimes of skin effect encountered in subsequent sections, for both isotropic and anisotropic 2D systems. In Sec.\@ \ref{Kin} we present and solve the kinetic equation for the distribution function of a metal with arbitrary Fermi surface in 2D and 3D, using the collision-operator formalism constructed in Sec.\@ \ref{Oper}. Sec.\@ \ref{T_cond} hosts the calculation of the transverse conductivity, which is specialized to the 2D isotropic case and to hexagonal and square geometries in Sec.\@ \ref{Cond_poly}, at arbitrary Fermi-surface orientations with respect to the transverse electric field. The skin depth in the anisotropic case is discussed in Sec.\@ \ref{Skin}, and applied to the hexagonal and square cases. Using the results in the preceding sections, the numerical and analytical results for the anisotropic surface impedance are collected and analyzed in Sec.\@ \ref{Z_spec}. Sec.\@ \ref{Disc} presents a generalization of Pippard's ``ineffectiveness concept'' for anisotropic dispersions, discusses the possible impact of short-ranged Landau interactions and long-ranged Coulomb forces on the results of this paper, as well as the effect of assuming different boundary conditions at the vacuum-metal surface. Our conclusions and perspectives for future work are summarized in Sec.\@ \ref{Concl}.

\section{Summary of main results}\label{Summ}

The fundamental theoretical result of this paper is the transverse conductivity for an arbitrary 2D or 3D Fermi-surface geometry, that possesses mirror symmetry planes in reciprocal space. Assuming an electric field $E_\beta(q_\alpha,\omega)$ and momentum $q_\alpha$ pointing in the spatial directions $\beta$ and $\alpha$ respectively (e.g.\@ $\left\{\alpha,\beta\right\}=\left\{x,y\right\}$ in 2D), the non-local transverse conductivity can be written as
\begin{equation}\label{eq:lonloc_sigma_T}
\sigma_{\beta \beta}(q_\alpha,\omega)= \epsilon_0 \Omega_p^2 \frac{G_0(q_\alpha,\omega)}{1-c_\beta^2 \delta \gamma G_0(q_\alpha,\omega)},
\end{equation}
where 
\begin{equation}\label{eq:delta_gamma}
\delta\gamma=\gamma_{\rm mc}-\gamma_{\rm mr}
\end{equation}
is the difference between momentum-conserving collision rate $\gamma_{\rm mc}$ and relaxation rate $\gamma_{\rm mr}$. $c_\beta$ is a geometry-dependent numerical constant, independent from $q_\alpha$ and $\omega$, and $G_0\left(q_\alpha,\omega\right)$ is an average of the angular variation of the Fermi velocity $\vec{v}_{\vec{k}}$ on the Fermi surface; see Eq.\@ (\ref{eq:G_0_separ}). $\epsilon_0$ is the vacuum dielectric constant, and $\Omega_p$ is a characteristic frequency for the anisotropic system, which is defined in Eq.\@ (\ref{eq:Omega_p_separ}) below. For the polygonal geometries that we analyze, assuming an isotropic Fermi velocity modulus $v_F$, $\Omega_p=2 \omega_p$ where $\omega_p=\sqrt{(n e^2)/(m \epsilon_0)}$ is the electronic plasma frequency of the isotropic electron gas, where $n=\mathscr{N}/\mathscr{A}$ is the electron density, i.e.\@, the number of electrons $\mathscr{N}$ per unit area $\mathscr{A}$, $e$ is the electron charge, $m$ is the free electron mass.
The conductivity (\ref{eq:lonloc_sigma_T}) allows us to treat isotropic and anisotropic systems on equal footing. This property is crucial for the application of our kinetic theory to the skin effect phenomenon, where qualitative criteria to distinguish all electrodynamic regimes can be traced on the basis of ratios among characteristic length scales, which are qualitatively independent from Fermi-surface geometry. Such length scales are listed in Table \ref{tab:skin_lengths}. Among these, the most fundamental quantity is the frequency-dependent skin depth
\begin{equation}\label{eq:skin_depth_gen}
\delta_s(\omega)\propto \omega^{\eta-1},
\end{equation}
which defines the damping constant of electromagnetic fields inside the metal and has a scaling exponent $\eta$ in frequency. Throughout the $\left(\gamma_{\rm mr},\gamma_{\rm mc},\omega\right)$ parameter space in Fig.\@ \ref{fig:3D_skin_effect}, the skin depth (\ref{eq:skin_depth_gen}) can be qualitatively linked with the experimentally observable surface impedance $Z(\omega)$ as: 
\begin{equation}\label{eq:skin_depth_Z_gen}
Z(\omega) \propto e^{-i\frac{\pi}{2} \eta} \omega^{\eta},
\end{equation}
as explicit computations in Secs.\@ \ref{Z_spec} and \ref{Gen_ineffectiveness} show. 
We will focus on the quantities $\delta_s(\omega)$ and $\left|Z(\omega)\right|$ to characterize skin effect throughout this paper. 

Additional length scales employed in the optics literature are: the London penetration depth $\lambda_L=c/\omega_p$, which is the skin depth associated with a perfect conductor; the distance $l_\omega$ traveled by electrons during one oscillation cycle of radiation; the mean-free path $l_{\rm mr}$ connected to momentum relaxation. These length scales are useful to identify the skin effect ``phase diagram'' in the $\left(\gamma_{\rm mr},\omega\right)$ plane assuming $\gamma_{\rm mr}=\gamma_{\rm mc}$ (see the left-hand panel of Fig.\@ \ref{fig:3D_skin_effect}). In addition, new regimes arise when $\gamma_{\rm mc}>\gamma_{\rm mr}$ (cf.\@ the right-hand panel of Fig.\@ \ref{fig:3D_skin_effect}). Correspondingly, new relevant length scales emerge: the collision length $l_{\rm mc}$, linked with momentum-conserving scattering; the Gurzhi length $l_G$, which governs the hydrodynamic skin effect, whereby electrons respond to radiation like a viscous fluid \cite{Gurzhi-1959,Gurzhi-1968,Levchenko-2020}; the associated skin depth $\delta_v$ in viscous regime \cite{Levchenko-2020,Valentinis-2021a}. Hence, Fig.\@ \ref{fig:3D_skin_effect} emphasizes that $\gamma_{\rm mc}$ offers a new ``axis'' in the skin-effect ``phase diagram'': momentum-conserving collisions represent an additional and experimentally controllable parameter, to explore the full parameter space for skin effect. 
The qualitative criteria to characterize the crossovers between adjacent regions of the ``phase diagram'' are collected in Tables \ref{tab:qual_cross_gammamr} and \ref{tab:qual_cross_gammamc}, and produce the colored lines in Fig.\@ \ref{fig:3D_skin_effect}. 
\begin{table*}[t]
	\centering 
		\begin{tabular}{| c | c | c |} 
\hline \hline
  Boundary & Criterion by length scales & Criterion by $\omega$, $\gamma_{\rm mr}$ \\
	\hline \hline \xrowht{15pt}
{\Large \textcircled{\normalsize A}}--{\Large \textcircled{\normalsize B}}, {\Large \textcircled{\small D$_1$}}--{\Large \textcircled{\small D$_2$}} & $l_\omega=l_{\rm mr}$ & $\omega=\gamma_{\rm mr}$ \\
 \hline \xrowht{15pt}
{\Large \textcircled{\normalsize A}}--{\Large \textcircled{\small D$_1$}} & $\delta_s(\omega)=l_{\rm mr}$ & $\omega=(\lambda_L/v_F)^2(\gamma_{\rm mr})^3$ \\
 \hline \xrowht{15pt}
{\Large \textcircled{\normalsize B}}--{\Large \textcircled{\normalsize E}} & $\delta_s(\omega)=l_{\rm mr}$ & $\gamma_{\rm mr}=v_F/\lambda_L$ \\ 
\hline \xrowht{15pt}
{\Large \textcircled{\small D$_2$}}--{\Large \textcircled{\normalsize E}} & $\delta_s(\omega)=l_\omega$ & $\omega=v_F/\lambda_L$ \\
\hline \xrowht{15pt}
{\Large \textcircled{\normalsize B}}--{\Large \textcircled{\normalsize C}}, {\Large \textcircled{\normalsize E}}--{\Large \textcircled{\normalsize C}} & $l_\omega=\lambda_L v_F/c$ & $\omega=\omega_p$ \\
\hline \xrowht{15pt}
{\Large \textcircled{\normalsize A}}--{\Large \textcircled{\normalsize C}} & $l_\omega=l_{\rm mr} (\omega/\omega_p)^2$ & $\omega=\omega_p^2/\gamma_{\rm mr}$ \\
 \hline 
\end{tabular} \caption{\label{tab:qual_cross_gammamr} Criteria for all possible crossovers between different regimes of skin effect in the $\left(\gamma_{\rm mr}, \omega\right)$ plane of Fig.\@ \ref{fig:3D_skin_effect}. $\delta_s(\omega)$ is the skin depth in the analyzed regime. }
\end{table*} 
The classification in the $\left(\gamma_{\rm mr},\omega\right)$ plane is well established for isotropic materials \cite{Casimir-1967a,Casimir-1967b,Wooten-1972opt,Dressel-2001,Tanner-2019opt}. The traditional nomenclature of known regimes in optics reads: {\Large \textcircled{\normalsize A}} for normal or classical skin effect, where $l_{\rm mr}\ll \delta_s(\omega)$ and $l_{\rm mr}\ll l_\omega$ and the electrodynamics in the skin layer is local and diffusive in nature; {\Large \textcircled{\normalsize B}} for the relaxational regime, where $l_\omega \ll l_{\rm mr}\ll \delta_s(\omega)$, where many oscillations of the electric field occur between relaxation events, which are still as strong as to keep the electrodynamics local and diffusive; {\Large \textcircled{\normalsize E}} for the ``extreme anomalous'' skin effect, or ``anomalous reflection'', where $l_\omega \ll \delta_s(\omega)\ll l_{\rm mr}$: here electrons experience many radiation cycles in the skin layer, and relaxation is weak enough for the conductivity to become spatially non-local. In practice, the regimes {\Large \textcircled{\normalsize B}} and {\Large \textcircled{\normalsize E}} differ only slightly, as the metal behaves in both cases as a perfect conductor with skin depth $\delta_s \equiv \lambda_L$. The regions {\Large \textcircled{\small D$_1$}} and {\Large \textcircled{\small D$_2$}}, where $\delta_s(\omega)\ll l_{\rm mr}$ and $l_\omega \ll l_{\rm mr}$ and the conduction is essentially ballistic and non-local, are known together as anomalous skin effect. They are collectively labeled {\Large \textcircled{\normalsize D}} in isotropic systems; this is because the scaling $\eta=2/3$ is identical in regions {\Large \textcircled{\small D$_1$}} and {\Large \textcircled{\small D$_2$}} if the Fermi surface is spherically symmetrical \cite{Reuter-1948,Sondheimer-2001,Pippard-1947b}. We will shortly appreciate that, in anisotropic systems, regions {\Large \textcircled{\small D$_1$}} and {\Large \textcircled{\small D$_2$}} can exhibit different scaling exponents $\eta$. For this reason, in the following we refer to cases {\Large \textcircled{\small D$_1$}} and {\Large \textcircled{\small D$_2$}} as ``anomalous-1' and ``anomalous-2'', respectively. The regime {\Large \textcircled{\normalsize C}} is transparent: it occurs above the plasma edge $\omega \geq \omega_p$, and the radiation absorption by the metal is frequency-independent. 
\begin{table*}[t]
	\centering 
		\begin{tabular}{| c | c | c |} 
\hline \hline
  Boundary & Criterion by length scales & Criterion by $\omega$, $\gamma_{\rm mc}$ \\
	\hline \hline \xrowht{15pt}
{\Large \textcircled{\small D$_1$}}--{\Large \textcircled{\small D$_2$}} & $l_\omega=l_{\rm mc}$ & $\omega=\gamma_{\rm mc}$ \\
 \hline \xrowht{15pt}
{\Large \textcircled{\small D$_1$}}--{\Large \textcircled{\normalsize V}} & $\delta_s(\omega)=l_{\rm mc}$ & $\omega=(\lambda_L/v_F)^2\gamma_{\rm mc}^3$ \\
 \hline \xrowht{15pt}
{\Large \textcircled{\small D$_2$}}--{\Large \textcircled{\normalsize E}} & $\delta_s(\omega)=l_\omega$ & $\omega=v_F/\lambda_L$ \\
\hline \xrowht{15pt}
{\Large \textcircled{\normalsize E}}--{\Large \textcircled{\normalsize F}} & $\delta_s(\omega)=l_{\rm mc}$ & $\gamma_{\rm mc}=v_F/\lambda_L$ \\ 
\hline \xrowht{15pt}
{\Large \textcircled{\normalsize A}}--{\Large \textcircled{\normalsize V}} & $\delta_s(\omega)=l_G$ & $\omega=(\lambda_L/v_F)^2\gamma_{\rm mr}^2\gamma_{\rm mc}$ \\
\hline \xrowht{15pt}
{\Large \textcircled{\normalsize V}}--{\Large \textcircled{\normalsize F}} & $\delta_v=\lambda_L$ & $\omega=(v_F/\lambda_L)^2\gamma_{\rm mc}^{-1}$ \\
\hline \xrowht{15pt}
{\Large \textcircled{\normalsize E}}--{\Large \textcircled{\normalsize C}}, {\Large \textcircled{\normalsize F}}--{\Large \textcircled{\normalsize C}} & $l_\omega=\lambda_L v_F/c$ & $\omega=\omega_p$ \\
 \hline 
\end{tabular} \caption{\label{tab:qual_cross_gammamc} Criteria for all possible crossovers between different regimes of skin effect in the $\left(\gamma_{\rm mc}, \omega\right)$ plane of Fig.\@ \ref{fig:3D_skin_effect}. $\delta_s(\omega)$ denotes the skin depth in the analyzed regime. }
\end{table*}
Notice that the criteria involving $\delta_s(\omega)$ actually refer to the skin depth in the regime where the crossover occurs. For instance, to estimate the boundary between {\Large \textcircled{\normalsize A}} and {\Large \textcircled{\small D$_1$}} we need to employ the skin depth for normal skin effect, which is $\delta_s(\omega) \sim \lambda_L (\gamma_{\rm mr}/\omega)^{1/2}$ (cfr.\@ Eq.\@ (\ref{eq:skin_depth_normal})), while the boundaries {\Large \textcircled{\normalsize B}}--{\Large \textcircled{\normalsize E}} and {\Large \textcircled{\small D$_2$}}--{\Large \textcircled{\normalsize E}} require $\delta_s(\omega)\equiv \lambda_L$ in perfect-conductor regime. 

In the $\left(\gamma_{\rm mc},\omega\right)$ plane for $\gamma_{\rm mc}>\gamma_{\rm mr}$, hydrodynamic behavior occurs when the Gurzhi length becomes larger than the skin depth in normal regime \cite{Levchenko-2020,Valentinis-2021a}: $l_G>\delta_s(\omega)$. In viscous regime, the skin depth becomes $\delta_s(\omega) \equiv \delta_v$. Still, one needs $\delta_v>\lambda_L$, or equivalently $l_G>\lambda_L$, for hydrodynamics to be enabled, with $\delta_v=\lambda_L$ marking the crossover to the regime {\Large \textcircled{\normalsize F}}. In the latter region {\Large \textcircled{\normalsize F}} the metal also responds as a perfect conductor, similarly to the cases {\Large \textcircled{\normalsize B}} and {\Large \textcircled{\normalsize E}}. 
A notable feature is that, increasing frequency at fixed and small $\gamma_{\rm mc}$, the value of $\omega$ at which hydrodynamics breaks down actually depends on Fermi-surface geometry, as we will see below. Such feature also has consequences in the $\left(\gamma_{\rm mr},\omega\right)$ plane.  

\begin{figure}[ht] \centering
\includegraphics[width=\columnwidth]{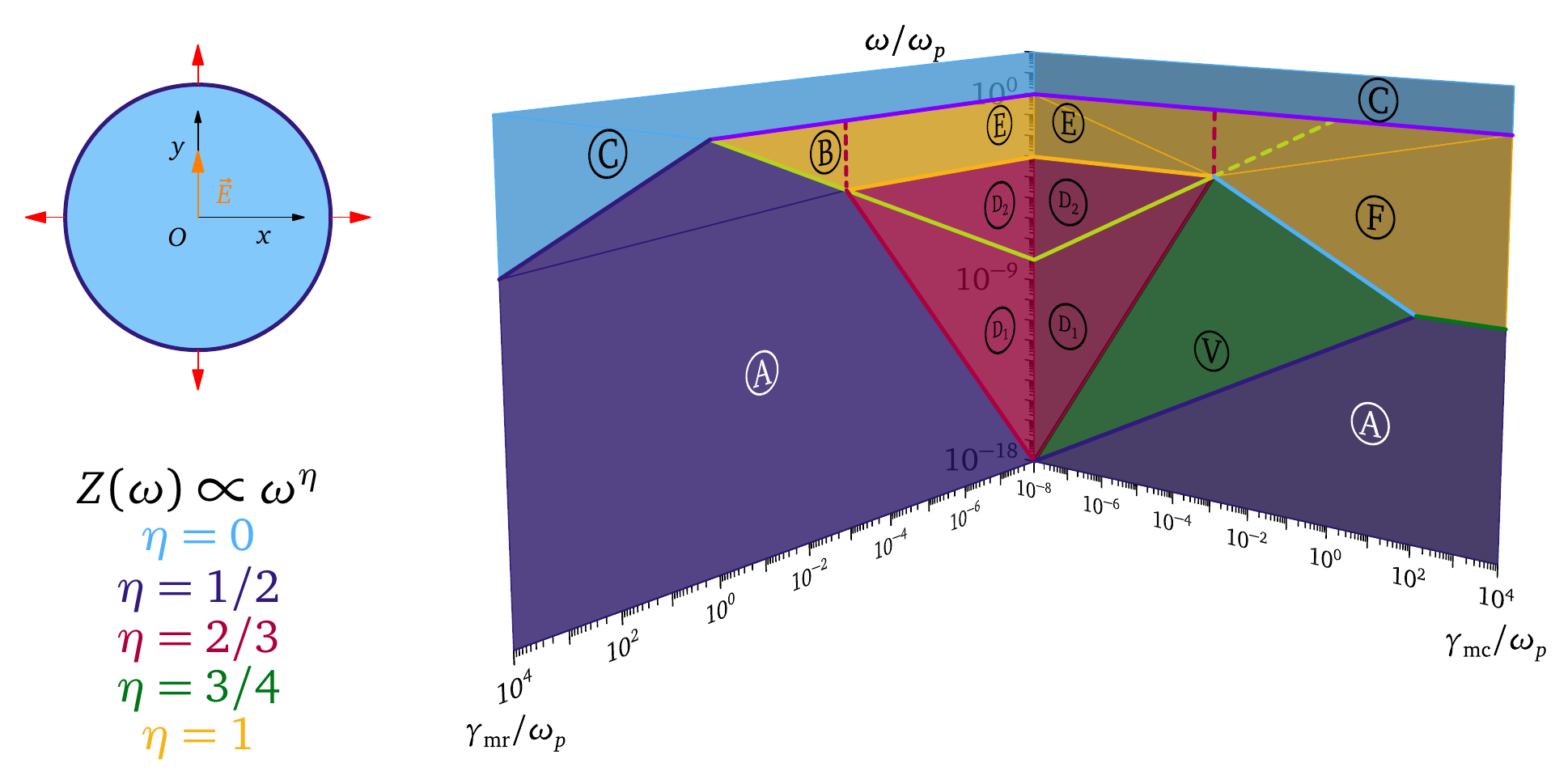}
\caption{\label{fig:3D_skin_effect_iso}Summary of skin effect regimes as a function of relaxation rate $\gamma_{\rm mr}$, momentum-conserving collision rate $\gamma_{\rm mc}$, and frequency $\omega$, normalized to the plasma frequency $\omega_p$, for an isotropic 2D (circular) Fermi surface. The boundaries are calculated assuming a Fermi velocity $v_F/c=0.001$.
}
\end{figure}

In all cases, the boundaries between confining regions in Fig.\@ \ref{fig:3D_skin_effect} are to be considered order-of-magnitude estimates of crossovers, as they neglect numerical prefactors: they amount to ratios
\begin{equation}
l_1=\alpha_g l_2
\end{equation}
between given length scales $l_1$ and $l_2$ with $\alpha_g\equiv 1$. In reality, the numerical coefficient $\alpha_g$ does depend on Fermi-surface geometry and orientation, and may contain higher-order corrections in $\gamma_{\rm mc}$ and $\delta \gamma$. These details are analyzed in Appendix \ref{App:Z_analys}, where we derive explicit expressions for the impedance in all relevant regimes, for a circular, hexagonal, and square Fermi surface, as well as more precise crossover boundaries which include the geometry- and orientation-dependent coefficient $\alpha_g$. These precise estimations yield the dashed lines in Figs.\@ \ref{fig:omegacross_iso}, \ref{fig:omegacross_hex}, and \ref{fig:omegacross_square}: there, we see that deviations of $\alpha_g$ from unity are especially important where two crossover boundaries are close to each other. 

Fig.\@ \ref{fig:3D_skin_effect_iso} summarizes the frequency dependence of the skin depth and surface impedance, for a material endowed with a circular Fermi surface and a Fermi velocity $v_F=0.001c$ with $c$ speed of light in vacuum. We retrieve all the properties and regimes of skin effect known for isotropic systems: normal skin effect {\Large \textcircled{\normalsize A}} at low frequencies, exhibiting $\eta=1/2$; anomalous regimes {\Large \textcircled{\small D$_1$}} and {\Large \textcircled{\small D$_2$}}, where $\eta=2/3$ as previously mentioned; the viscous regime {\Large \textcircled{\normalsize V}}, existing exclusively for $\gamma_{\rm mc}>\gamma_{\rm mr}$ and where $\eta=3/4$; the ``perfect-conductor'' regimes {\Large \textcircled{\normalsize B}}, {\Large \textcircled{\normalsize E}}, and {\Large \textcircled{\normalsize F}}, where the skin depth is $\omega$-independent and so $\eta=1$. We notice that the diagram \ref{fig:3D_skin_effect_iso} is qualitatively similar for isotropic dispersions in 2D and 3D. 

\begin{figure}[ht] \centering
\includegraphics[width=\columnwidth]{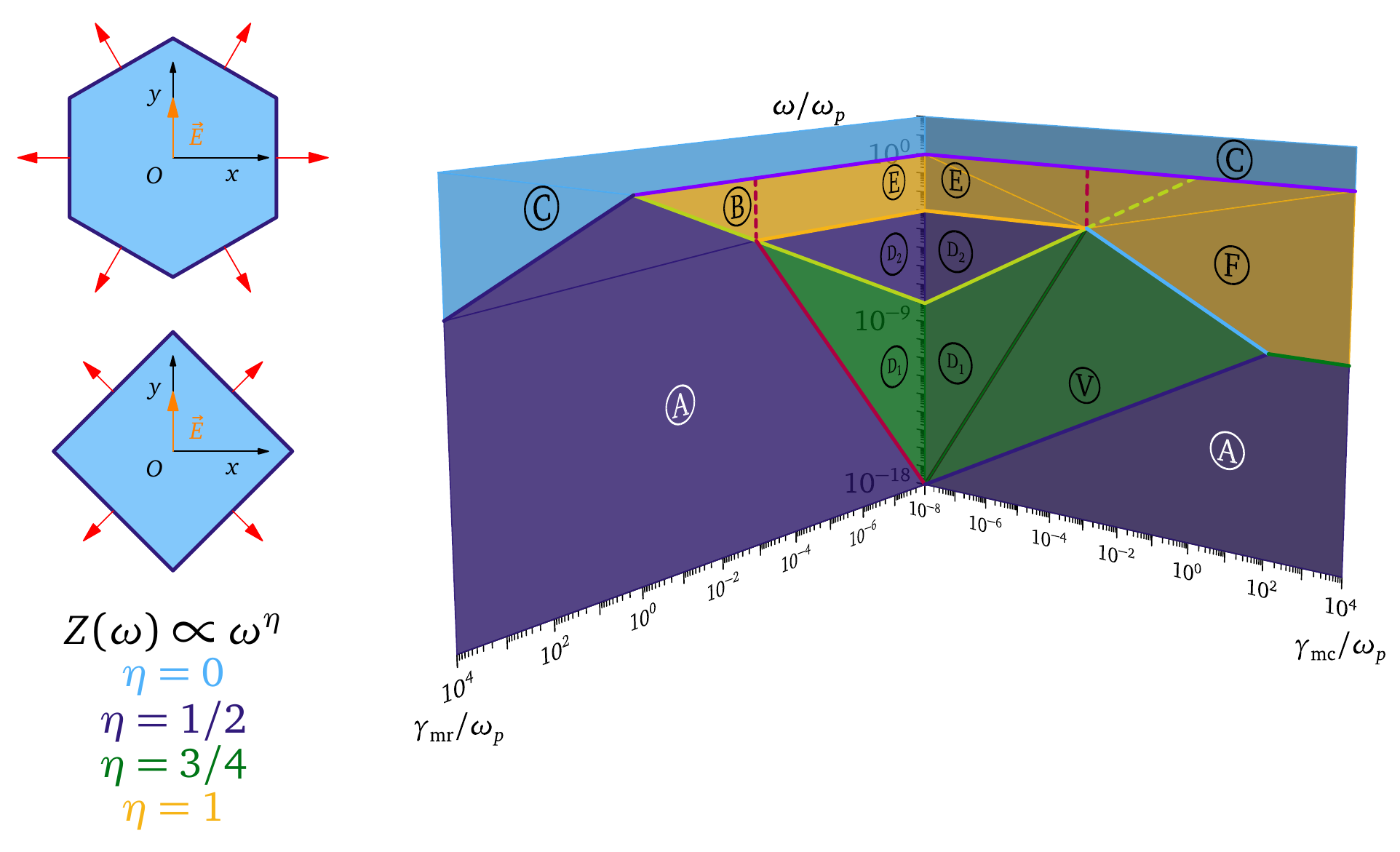}
\caption{\label{fig:3D_skin_effect_misalign}Summary of skin effect regimes as a function of relaxation rate $\gamma_{\rm mr}$, momentum-conserving collision rate $\gamma_{\rm mc}$, and frequency $\omega$, normalized to the plasma frequency $\omega_p$, for the hegagonal and square Fermi surface, oriented as depicted on the left of the figure. The two polygonal geometries share a similar ``phase diagram'' of the frequency scaling exponent $\eta$. The boundaries are evaluated assuming a Fermi velocity $v_F/c=0.001$.
}
\end{figure}
Fermi-surface anisotropy significantly alters the exponent $\eta$ in many regions of the ``phase diagram''. We can appreciate such modifications in Fig.\@ \ref{fig:3D_skin_effect_misalign}, which pertains to two polygonal geometries with similar behavior: the hexagonal Fermi surface with two faces parallel to the electric field, and the ``diamond-shaped'' Fermi surface, as depicted on the left of Fig.\@ \ref{fig:3D_skin_effect_misalign}. Here and for all polygonal geometries discussed in this section, we assume for simplicity that the Fermi velocity modulus $v_F$ is isotropic in the 2D plane, although the conductivity (\ref{eq:sigma_yy_separ}) is also capable of treating cases where $v_F$ depends on orientation. The most striking changes with respect to the isotropic case of Fig.\@ \ref{fig:3D_skin_effect_iso} are that the anomalous-1 and anomalous-2 regimes have different scaling exponents, $\eta=3/4$ and $\eta=1/2$, respectively. This means that the viscous regime extends throughout regions {\Large \textcircled{\normalsize V}} and {\Large \textcircled{\small D$_1$}} for these orientations and for $\gamma_{\rm mc}\geq\gamma_{\rm mr}$. The viscous physics at play is the same, as one can confirm through the analysis of the impedance for the present orientations: we have 
\begin{equation}\label{eq:Z_qual_hydro}
\left|Z(\omega)\right| \propto \sqrt{\frac{v_F}{c}} \frac{\omega^{\frac{3}{4}}}{\gamma_{\rm mc}^{\frac{1}{4}}}
\end{equation}
in both areas {\Large \textcircled{\normalsize V}} and {\Large \textcircled{\small D$_1$}} (see Eqs.\@ (\ref{eq:Z_hex_par_hydro}) and (\ref{eq:Z_square_dia_hydro})), which indicates that the two regions are indistinguishable. 
Equally outstanding is the fact that the region {\Large \textcircled{\small D$_1$}} in the $\left(\gamma_{\rm mr},\omega\right)$ for $\gamma_{\rm mc}=\gamma_{\rm mr}$ is characterized by viscous skin effect, analogously to the regime {\Large \textcircled{\normalsize V}} that occurs for $\gamma_{\rm mc}>\gamma_{\rm mr}$. This phenomenon appears because hydrodynamics persists in region {\Large \textcircled{\small D$_1$}}. In turn, this is compatible with the hydrodynamic condition $l_G>\delta_s(\omega)$, that is satisfied even for $\gamma_{\rm mc}=\gamma_{\rm mr}$ and it corresponds to $\delta_s(\omega)\equiv \delta_v <l_{\rm mr}$, i.e.\@, to the crossover {\Large \textcircled{\normalsize A}} $\rightarrow$ {\Large \textcircled{\small D$_1$}} for equal scattering rates. Hence, if viscous skin effect is at play in region {\Large \textcircled{\small D$_1$}} for $\gamma_{\rm mc}>\gamma_{\rm mr}$, as for the present orientations, it also appears for $\gamma_{\rm mc}=\gamma_{\rm mr}$ in {\Large \textcircled{\small D$_1$}}. We expect such feature to bear significant consequences for DC transport experiments as well, for the anisotropic configurations in Fig.\@ \ref{fig:3D_skin_effect_misalign}. Indeed, Fermi-surface anisotropy has been shown to significantly affect the DC conductance and the ballistic-to-hydrodynamic crossover in narrow channels \cite{Cook-2019,Cook-2021}, leading to nonmonotonic dependence of the conductance with channel width and temperature, and to the appearance of a new anisotropy-related component of the viscosity tensor.

Furthermore, notice that in anomalous (ballistic) regime {\Large \textcircled{\small D$_2$}}, the isotropic case shows $\eta=2/3$, while $\eta=1/2$ in the presence of anisotropy. This change is surprising, as a scaling $\eta=1/2$ is usually associated with normal (diffusive) skin effect in isotropic materials \cite{Reuter-1948,Chambers-1952, Casimir-1967a,Casimir-1967b,Casimir-1967c,Wooten-1972opt,Sondheimer-2001,Dressel-2001}. 

However, the regions {\Large \textcircled{\normalsize A}} and {\Large \textcircled{\small D$_2$}} are physically distinct: in the former regime, one has
\begin{equation}\label{eq:Z_qual_q0} 
\left|Z(\omega)\right| \propto \frac{\sqrt{\omega \gamma_{\rm mr}}}{\omega_p},
\end{equation}
which depends on relaxation and signals diffusive dynamics (see Eqs.\@ (\ref{eq:Z_hex_lowq})), while in the latter regime, we obtain 
\begin{equation}\label{eq:Z_qual_anom}
\left|Z(\omega)\right| \propto \sqrt{\frac{v_F}{c}} \sqrt{\frac{\omega}{\omega_p}},
\end{equation}
which is independent from $\gamma_{\rm mc}$ and $\gamma_{\rm mr}$, and thus indicates ballistic behavior (see Eqs.\@ (\ref{eq:Z_hex_par_anom}) and (\ref{eq:Z_square_dia_anom})). 
Hence, even though they share the same frequency scaling $\eta=1/2$, the regions {\Large \textcircled{\normalsize A}} and {\Large \textcircled{\small D$_2$}} host normal and anomalous skin effect, respectively, and can be distinguished on the basis of the dependence on the scattering rates.
\begin{figure}[ht] \centering
\includegraphics[width=\columnwidth]{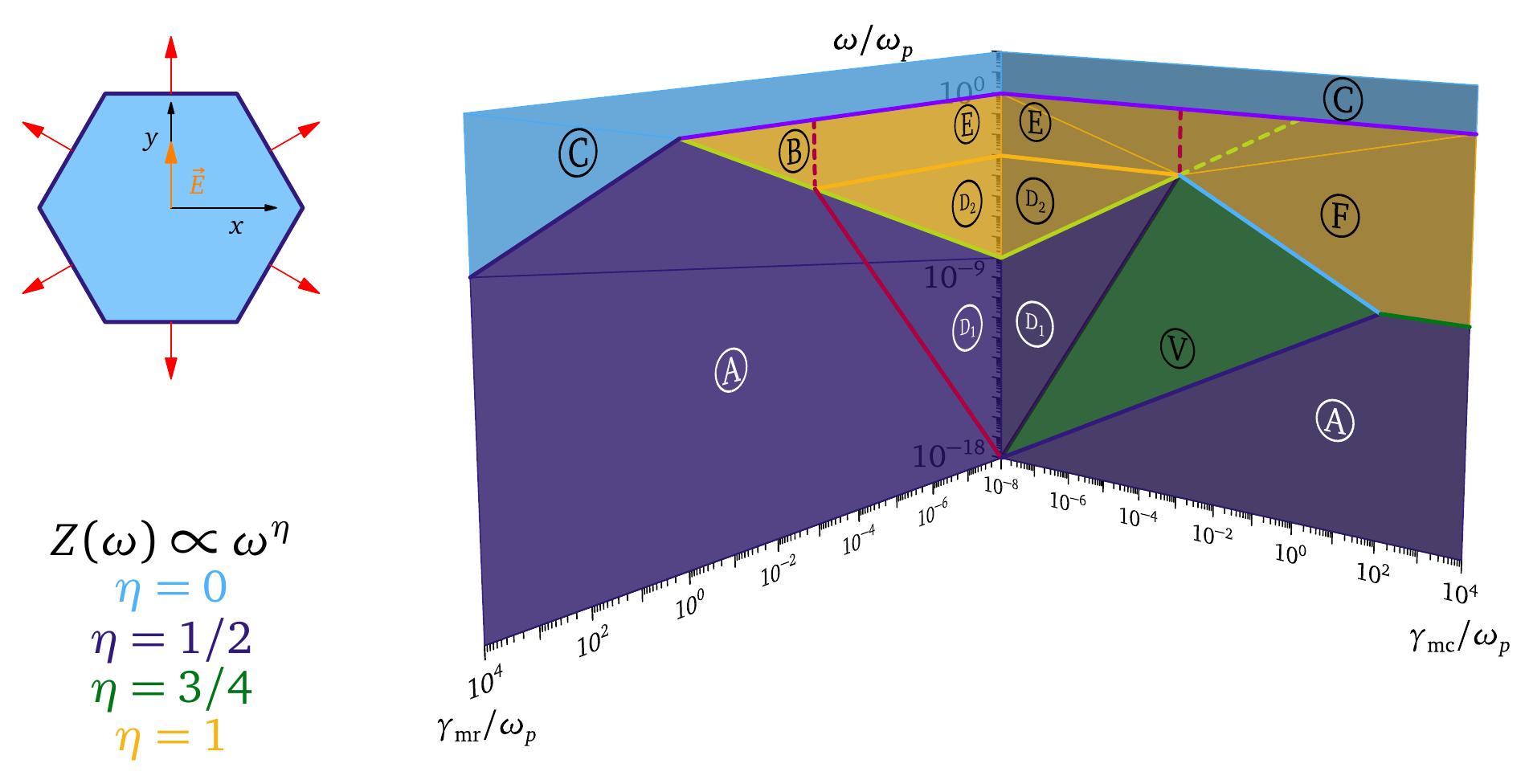} 
\caption{\label{fig:3D_skin_effect_align} Summary of skin effect regimes as a function of relaxation rate $\gamma_{\rm mr}$, momentum-conserving collision rate $\gamma_{\rm mc}$, and frequency $\omega$, normalized to the plasma frequency $\omega_p$, for the hegagonal Fermi surface oriented as depicted on the left of the figure, and rotated with respect to Fig.\@ \ref{fig:3D_skin_effect_misalign} by an angle $\phi=\pi/2$. All boundaries are calculated assuming a Fermi velocity $v_F/c=0.001$.
} 
\end{figure}

To highlight the dependence of skin effect on the orientation of the anisotropic Fermi surface with respect to the applied field, we now rotate the hexagonal geometry of Fig.\@ \ref{fig:3D_skin_effect_misalign} by $\phi=\pi/2$. This produces the Fermi surface displayed on the left of Fig.\@ \ref{fig:3D_skin_effect_align}. There are evident differences with respect to the orientations of Fig.\@ \ref{fig:3D_skin_effect_misalign}: the perfect-conductor regime with $\eta=1$ extends down to the regime {\Large \textcircled{\small D$_2$}}, while the region {\Large \textcircled{\small D$_1$}} has $\eta=1/2$ as for the normal regime {\Large \textcircled{\normalsize A}}. Nevertheless, the physics is not identical: for $\gamma_{\rm mc}\geq \gamma_{\rm mr}$ in the normal regime, the impedance still follows Eq.\@ (\ref{eq:Z_qual_q0}) as for $\phi=0$, while in the anomalous region $\left|Z(\omega)\right|=\sqrt{(\gamma_{\rm mc}+2 \gamma_{\rm mr}) \omega)/2}$ (see Eq.\@ (\ref{eq:Z_hex_pi2_anom})). 
In the $\left(\gamma_{\rm mr},\omega\right)$ plane for $\gamma_{\rm mc}=\gamma_{\rm mr}$, there is no viscous region, and the regime {\Large \textcircled{\small D$_2$}} shows $\eta=1$ (perfect conductivity): from the point of view of the scaling exponent $\eta$, the ``phase diagram'' is identical to the one of the Drude model of free electrons with relaxation rate $\gamma_{\rm mr}$, where the electrodynamics is completely local and diffusive. The regions {\Large \textcircled{\normalsize A}} and {\Large \textcircled{\small D$_1$}} have $|Z(\omega)|\propto \sqrt{\gamma_{\rm mr} \omega}$ and $|Z(\omega)|\propto \sqrt{3}/2 \sqrt{\gamma_{\rm mr} \omega}$, respectively, so they differ only in the numerical factor $\sqrt{3/2}$, as deduced from Eqs.\@ (\ref{eq:Z_hex_lowq}) and (\ref{eq:Z_hex_pi2_anom}). 
In essence, Fig.\@ \ref{fig:3D_skin_effect_align} shows that Fermi-surface anisotropy is able to completely suppress the non-locality (dependence on $q_x$) of the conductivity in the $\left(\gamma_{\rm mr},\omega\right)$ plane, which would yield anomalous (ballistic) and viscous skin effect. 
\begin{figure}[ht] \centering
\includegraphics[width=\columnwidth]{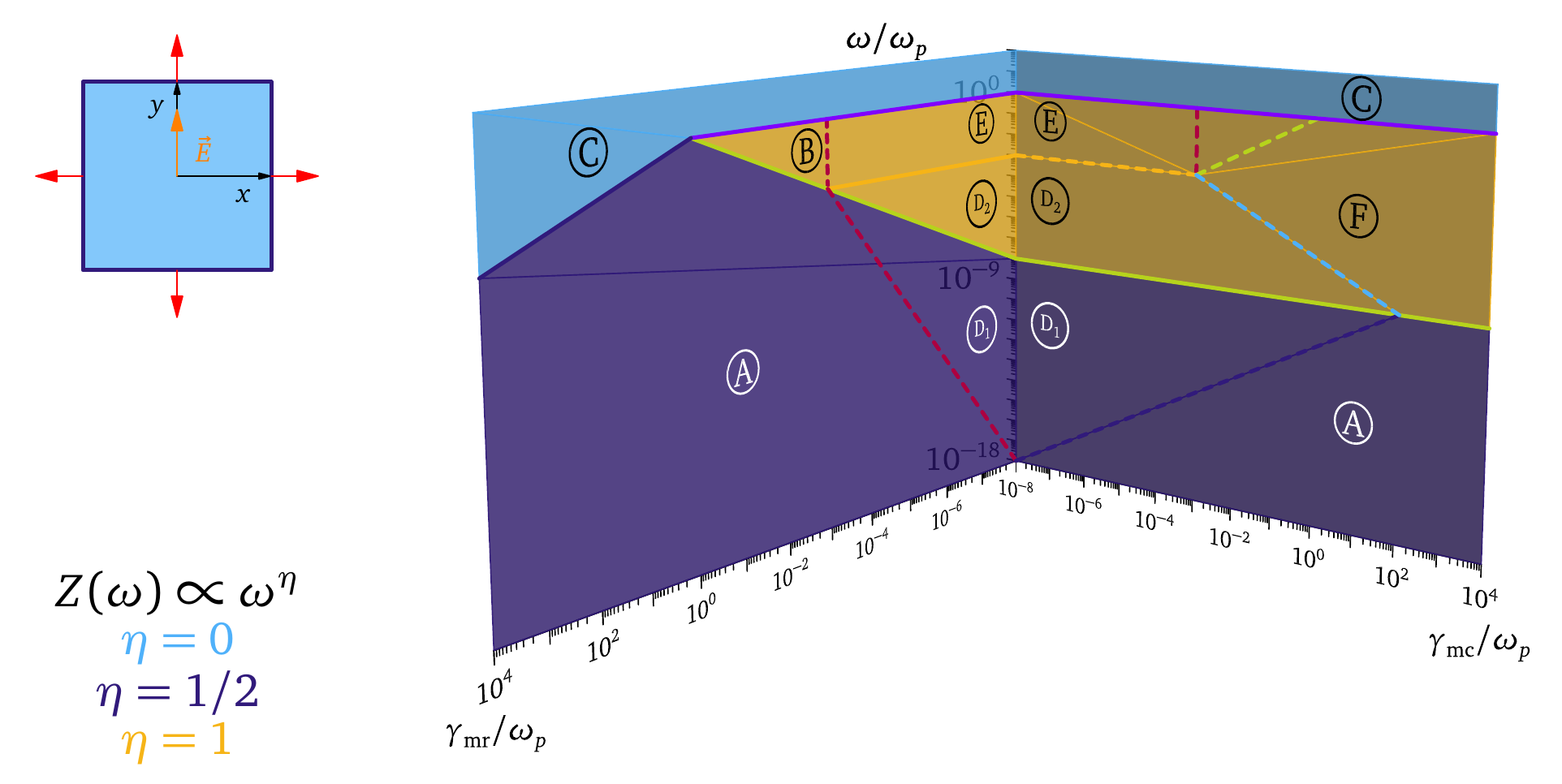} 
\caption{\label{fig:3D_skin_effect_square} Summary of skin effect regimes as a function of relaxation rate $\gamma_{\rm mr}$, momentum-conserving collision rate $\gamma_{\rm mc}$, and frequency $\omega$, normalized to the plasma frequency $\omega_p$, for the square Fermi surface oriented as depicted on the left of the figure, and rotated with respect to Fig.\@ \ref{fig:3D_skin_effect_misalign} by an angle $\phi=\pi/4$. All boundaries are evaluated assuming a Fermi velocity $v_F/c=0.001$.
} 
\end{figure}

Such suppression of non-locality is total for the special case of a square Fermi surface with two sides parallel to the electric field, as depicted on the left of Fig.\@ \ref{fig:3D_skin_effect_square}. This geometry is obtained by rotating the ``diamond-shaped'' case of Fig.\@ \ref{fig:3D_skin_effect_misalign} by $\phi=\pi/4$. Here we find the striking result that ballistic and viscous skin effects are completely absent from the ``phase diagram'', which reduces identically to the one of the Drude model with relaxation rate $\gamma_{\rm mr}$. Thus, only normal and perfect-conductor regimes of skin effect persist for this orientation: the material behaves as a local conductor even at finite momentum. 

The aforementioned analysis of the conductivity (see Sec.\@ \ref{Cond_poly}), and of the skin depth (see Sec.\@ \ref{Skin}), suggests that large Fermi-surface segments with velocity $\vec{v}_F$ parallel to the applied field tend to suppress the non-local character of conduction, i.e.\@, to inhibit anomalous skin effect. In fact, if $\vec{v}_F$ is locally parallel to the field, there is no component of $\vec{v}_F$ along the direction of momentum $\vec{q}=q_x\hat{u}_x$, so non-local currents depending on $q_x$ cannot be generated. The exemplary case is in Fig.\@ \ref{fig:3D_skin_effect_square}: there, two Fermi-surface sides have $\vec{v}_F \parallel \hat{u}_y$. Hence, the diagram is identical to the Drude model because the conductivity for such square geometry is momentum-independent: see Eq.\@ (\ref{eq:sigma_yy_sq_pi4_gamma}). 
We can intuitively view this feature from the alternative viewpoint of Pippard's \emph{ineffectiveness concept}, which postulates that only the fraction of electrons with velocity nearly parallel to $\vec{E}$ can contribute to screening of electromagnetic fields when $\delta_s(\omega) \ll l_{\rm mr}$: the screening is reduced by all other electrons being ineffective, and this determines the long penetration depth $\delta_s(\omega)$, the spatial non-locality, and ballistic character of conduction typical of anomalous skin effect. However, if $\vec{v}_F \parallel \hat{u}_y$, all electrons on that Fermi-surface segment are effective, and the ineffectiveness concept does not apply. These metallic electrons respond like in a perfect conductor, in the absence of scattering, and generate the smallest possible skin depth $\delta_s(\omega)\approx \lambda_L$; see, e.g.\@, the skin depths (\ref{eq:aniso_flat3}) and $\lambda_L$ at large momentum for the geometries in Figs.\@ \ref{fig:3D_skin_effect_align} and \ref{fig:3D_skin_effect_square}.
Notice that the aforementioned geometric effect is different from the familiar behavior in regions {\Large \textcircled{\normalsize B}} and {\Large \textcircled{\normalsize E}}: these perfect-conductor regions occur because $\omega \gg \left\{\gamma_{\rm mr}, \gamma_{\rm mc}\right\}$, and also exist in the isotropic case (cf.\@ Fig.\@ \ref{fig:3D_skin_effect_iso}). On the contrary, the perfect-conductor behavior in the high-momentum regime {\Large \textcircled{\small D$_2$}} of Fig.\@ \ref{fig:3D_skin_effect_align} is due to the effective electrons in anisotropic systems. For the isotropic case, the anomalous regime is characterized by $\sigma_{yy}(q_x,\omega) \propto q_x^{-1}$, which has the special property that the separating line between regions {\Large \textcircled{\small D$_1$}} and {\Large \textcircled{\small D$_2$}} is irrelevant; however, on anisotropic Fermi-surface segments with velocity parallel to $\vec{E}$ the dependence on $q_x$ is removed, so that those electrons are again affected by the relative size of $\omega$, $\gamma_{\rm mr}$, and $\gamma_{\rm mc}$, and regions {\Large \textcircled{\small D$_1$}} and {\Large \textcircled{\small D$_2$}} are distinct. If scattering is dominant over frequency, we have $\eta=1/2$, as seen through Eqs.\@ (\ref{eq:aniso_flat3_gammamc}) for the skin depth (region {\Large \textcircled{\small D$_1$}}). In the opposite high-frequency limit, $\eta=1$ (region {\Large \textcircled{\small D$_2$}}). The different character of conduction in the scattering-less regime is further analyzed in Appendix \ref{App_Z_hex_analysis_noscat} and \ref{Square_Z_analys_noscat}. 
The ineffectiveness concept also yields a simple geometrical interpretation of the factor-$\sqrt{3/2}$ difference between regions {\Large \textcircled{\small D$_1$}} and {\Large \textcircled{\normalsize A}} in Fig.\@ \ref{fig:3D_skin_effect_align}: in this orientation, two sides of the hexagon have Fermi velocity parallel to $\vec{E}$, therefore they are acting locally (similarly to normal skin effect), while the Fermi velocity on the remaining four sides is almost perpendicular to $\vec{E}$, so that those sides negligibly contribute to skin effect; therefore, the response in region {\Large \textcircled{\small D$_1$}} is local, as if there were a $\vec{q}$-independent conductivity, analogously to region {\Large \textcircled{\normalsize A}}, but in {\Large \textcircled{\small D$_1$}} the density (or the effective plasma frequency) of electrons that respond to the field is reduced compared to {\Large \textcircled{\normalsize A}}, due to only two hexagon sides being effective.  

To rationalize the frequency behavior of the skin depth and the impedance in all non-local regimes, we generalize the ineffectiveness concept to momentum-dependent conductivities and anisotropic geometries (see Sec.\@ \ref{Gen_ineffectiveness}), thus showing that if the conductivity scales as $\sigma_{yy}(q_x,\omega)\propto \omega^\eta/q_x^\zeta$, then the skin depth behaves as $\delta_s(\omega)\propto \omega^{-(\eta+1)/(\zeta+2)}$. Through Eq.\@ (\ref{eq:skin_depth_Z_gen}), this scaling is seen to be completely consistent with the impedance ``phase diagrams'' in Figs.\@ \ref{fig:3D_skin_effect_iso}, \ref{fig:3D_skin_effect_misalign}, and \ref{fig:3D_skin_effect_align}. 

The equivalence of the scaling exponent $\eta$ in physically distinct regions of the ``phase diagrams'' \ref{fig:3D_skin_effect_iso}, \ref{fig:3D_skin_effect_misalign}, and \ref{fig:3D_skin_effect_align}, demonstrates that the frequency dependence of the impedance is not sufficient, by itself, to uniquely characterize some skin effect regimes in anisotropic systems. In these situations, one can refer to the analytical dependence of the impedance on $\gamma_{\rm mr}$ and $\gamma_{\rm mc}$, to evaluate whether the ``$\omega$-degenerate'' regions share the same physics or are distinguishable on the basis of the scattering rates; see Appendix \ref{App:Z_analys}. An example of this method is the aforementioned difference between Eqs.\@ (\ref{eq:Z_qual_q0}) and (\ref{eq:Z_qual_anom}), which allows us to discern normal skin effect in region {\Large \textcircled{\normalsize A}} from the anomalous character of region {\Large \textcircled{\small D$_2$}} in Fig.\@ \ref{fig:3D_skin_effect_misalign}. Conversely, the regimes {\Large \textcircled{\normalsize V}} and {\Large \textcircled{\small D$_1$}} are seen to correspond to the same viscous behavior, based on Eq.\@ (\ref{eq:Z_qual_hydro}).

To practically investigate how anisotropic skin effect depends separately on $\gamma_{\rm mr}$ and $\gamma_{\rm mc}$, one can refer to the $\gamma_{\rm mc}$-dependent saturation of the surface resistance $\mathrm{Re}Z(\omega)$ in the relaxationless limit $\gamma_{\rm mr} \rightarrow 0$: this is a well-established method for the experimental observation of anomalous skin effect in isotropic materials \cite{Reuter-1948,Sondheimer-2001,Pippard-1947b}, and can be readily extended to anisotropic configurations, as shown in Figs.\@ \ref{fig:Zs_phi090_taumr_Hex} and \ref{fig:Zs_phi045_taumr_Square}. The relaxation rate $\gamma_{\rm mr}$ may be tuned in practice by controlling the amount of disorder in the crystal, while both $\gamma_{\rm mr}$ and $\gamma_{\rm mc}$ can be varied with temperature. 

Furthermore, to assess the robustness of our results for the anisotropic ``phase diagrams'' \ref{fig:3D_skin_effect_misalign} and \ref{fig:3D_skin_effect_align} with respect to Fermi-surface nonidealities, we consider more complex shapes which are qualitatively similar to the ideal square and hexagon: these are the ``supercircle'' -- see Sec.\@ \ref{Supercircle} -- which allows us to controllably introduce a finite curvature of individual Fermi-surface segments and/or rounded corners, and a 2D parametrization of the approximately hexagonal Fermi surface of PdCoO$_2$ stemming from quantum oscillations experiments \cite{Hicks-2012} -- see Sec.\@ \ref{PdCoO2_2D} -- to make contact with a more realistic case of anisotropic system. In all cases, we find that the effect of anisotropy is quantitatively modified by the non-polygonal shape, but the orientational dependence of skin effect and the alterations in the exponent $\eta$ with respect to the isotropic case are qualitatively robust features of our kinetic theory.

In the next section, we begin constructing the kinetic theory that underlies all results summarized so far.

\section{Kinetic theory for arbitrary Fermi surface}\label{Kin}

The Boltzmann equation with collision term is given as
\begin{multline}\label{eq:BE}
\frac{\partial f_{\vec{k}}\left(\vec{r},t\right)}{\partial t}-e\vec{E}\left(\vec{r},t\right)\cdot\mathbf{\nabla}_{\vec{k}}f_{\vec{k}}\left(\vec{r},t\right)+\vec{v}_{\vec{k}}\cdot\mathbf{\nabla}_{\vec{r}}f_{\vec{k}}\left(\vec{r},t\right) \\ =-\left({\cal C} f\right)_{\vec{k}},
\end{multline}
where $\left({\cal C} f\right)_{\vec{k}}$ is the collision integral. Let $\varepsilon_{\vec{k}}$ be  the single-particle dispersion relation  with velocity $\vec{v}_{\vec{k}}=\nabla_{\vec{k}} \varepsilon_{\vec{k}}$. 
We expand the distribution function around its equilibrium value $f_{\vec{k}}=f_{\vec{k}}^{\left(0\right)}+\delta f_{\vec{k}}$. To leading order in  $\delta f_{\vec{k}}$, we obtain after a Fourier transformation from the space of coordinates and time $\left\{\vec{r},t\right\}$ to momenta and frequencies $\left\{\vec{q},\omega\right\}$:
\begin{multline}\label{eq:BElin}
-i\omega\delta f_{\vec{k}}\left(\vec{q},\omega\right)-e\vec{E}\left(\vec{q},\omega\right)\cdot\vec{v}_{\vec{k}}\frac{\partial f^{\left(0\right)}\left(\varepsilon_{\vec{k}}\right)}{\partial\varepsilon_{\vec{k}}} \\ +i\vec{q}\cdot\vec{v}_{\vec{k}}\delta f_{\vec{k}}\left(\vec{q},\omega\right)=-\left({\cal C} f\right)_{\vec{k}}.
\end{multline}
Eq.\@ (\ref{eq:BElin}) determines the deviation $\delta f_{\vec{k}}\left(\vec{q},\omega\right)$ of the distribution function from the equilibrium solution $f_{\vec{k}}^{(0)}\equiv f^{\left(0\right)}\left(\varepsilon_{\vec{k}}\right)$ in the presence of the driving field $\vec{E}\left(\vec{q},\omega\right)$ and of collisions parametrized by the integral $\left({\cal C} f\right)_{\vec{k}}$. 

\subsection{Operator formalism and inner product}\label{Oper}

In order to efficiently analyze the linearized Boltzmann equation (\ref{eq:BElin})  for arbitrary $\varepsilon_{\vec{k}}$, it is convenient to expand both $\delta f_{\vec{k}}$ and the collision term $\left({\cal C} f\right)_{\vec{k}}$ in terms of a complete set of eigenfunctions, which define the collision operator $\hat{C}$ as follows. 
First, we the write our distribution function in the form
\begin{equation}\label{eq:f_psi}
\delta f_{\vec{k}}\left(\vec{q},\omega\right)=-k_B T\frac{\partial f^{\left(0\right)}\left(\varepsilon_{\vec{k}}\right)}{\partial\varepsilon_{\vec{k}}}\psi_{\vec{k}}\left(\vec{q},\omega\right).
\end{equation}
The linearization of the collision integral can be expressed as
\begin{equation}\label{eq:C_lin}
\left({\cal C} f\right)_{\vec{k}} =\int_{k'}\frac{\delta \left({\cal C} f\right)_{\vec{k}}}{\delta\psi_{\vec{k}'}(\vec{q},\omega)}\psi_{\vec{k}'}(\vec{q},\omega).
\end{equation}
With this formulation, it follows that
\begin{multline}\label{eq:Boltz_psi}
\left(-i\omega+i\vec{q}\cdot\vec{v}_{\vec{k}}\right)\psi_{\vec{k}}\left(\vec{q},\omega\right)+\hat{C}\psi_{\vec{k}}(\vec{q},\omega) \\ =-\frac{e}{k_B T}\vec{E}\left(\vec{q},\omega\right)\cdot\vec{v}_{\vec{k}}
\end{multline}
with collision operator 
\begin{equation}\label{eq:Coll_op_psi}
\hat{C}\psi_{\vec{k}}=\left[-k_B T\frac{\partial f^{\left(0\right)}\left(\varepsilon_{\vec{k}}\right)}{\partial\varepsilon_{\vec{k}}}\right]^{-1}\int_{k'}\frac{\delta({\cal C} \delta f)_{\vec{k}}}{\delta\psi_{\vec{k}'}\left(\vec{q},\omega\right)}\psi_{\vec{k}'}\left(\vec{q},\omega\right).
\end{equation}
Let $\psi_{\vec{k}}$ be an element of a function space with inner product \cite{Arnold-2000,Ziman-1960el,Fritz-2008,Cook-2019} 
\begin{equation}\label{eq:scalar_prod}
\left\langle \phi\mid\psi\right\rangle =\int_{\vec{k}}w_{\vec{k}}\phi_{\vec{k}}^{*}\psi_{\vec{k}},
\end{equation}
where we use the weight function $w_{\vec{k}}=-k_B T \partial f^{\left(0\right)}\left(\varepsilon_{\vec{k}}\right)/\partial\varepsilon_{\vec{k}}=f^{\left(0\right)}\left(\varepsilon_{\vec{k}}\right)\left[1-f^{\left(0\right)}\left(\varepsilon_{\vec{k}}\right)\right]>0$.  This definition obeys all the usual properties of a scalar product. Notice that all functions, and henceforth the scalar product, depend parametrically on frequency $\omega$ and momentum $\vec{q}$.

For the matrix elements of the collision operator, it holds in particular that 
\begin{eqnarray}\label{eq:C_matrix_el}
\left\langle \phi\left|\hat{C}\right|\psi\right\rangle &=& \int_{\vec{k}}w_{\vec{k}}\phi_{\vec{k}}^{*}\hat{C}\psi_{\vec{k}} \nonumber \\ &=&\int_{\vec{k}\boldsymbol{k}'}\phi_{\vec{k}}^{*}\frac{\delta({{\cal C} \delta f}_{\vec{k}})}{\delta\psi_{\vec{k}'}\left(\vec{q},\omega\right)}\psi_{\vec{k}'}\left(\vec{q},\omega\right).
\end{eqnarray}
The quantity (\ref{eq:C_matrix_el}) is associated with production of entropy $S$ in the system. The density of the entropy production rate $Q\left(\vec{r},t \right)=\partial s\left(\vec{r},t \right)/\partial t$ with entropy density $s\left(\vec{r},t \right)$  can be written as
\begin{eqnarray}
Q\left(\vec{r},t \right) &=& -k_{B}\int_{\vec{k}}{{\cal C} \delta f}_{\vec{k}}\ln\left[\frac{1}{f_{\vec{k}}\left(\vec{r},t\right)}-1\right] \nonumber \\
  &\approx&  -k_B  \left\langle \psi\left|\hat{C}\right|\psi\right\rangle,
\end{eqnarray}
where we expanded the distribution function around equilibrium in the second step.
Hence, it follows that the entropy production vanishes in the local equilibrium, and that it is equal to $\left\langle \psi\left|\hat{C}\right|\psi\right\rangle$ per unit volume. To ensure that $Q\left(\vec{r},t \right)>0$, the operator $\hat{C}$ must be positive definite, the reason for the variational formulation of the Boltzmann approach.
In particular, it follows that 
\begin{equation}
{{\cal C} \delta f}_{\vec{k}}=\int_{\vec{k}'} \frac{\delta{{\cal C} \delta f}_{\vec{k}}}{\delta \psi_{\vec{k}'}}\psi_{\vec{k}'} =\frac{1}{2}\frac{\delta Q\left[\psi\right]}{\delta\psi_{\vec{k}}},
\end{equation}
 a property that can be used to show that 
$\left\langle \psi\mid\hat{C}\phi\right\rangle =\left\langle \hat{C}\psi\mid\phi\right\rangle $, i.e.\@ the operator $\hat{C}$ is Hermitian under the above scalar product. Therefore, the eigenvalues of $\hat{C}$ are real, and its eigenfunctions form an orthonormal set of basis functions. 

Let the eigenfunctions of the collision operator be $\chi_{\vec{k},m}$ with eigenvalues $\gamma_{m}$ \footnote{A straightforward possible generalization to the present treatment allows the eigenvalues $\gamma_{\vec{k},m}$ to depend on $\vec{k}$ \cite{Baker-2022_preprint}, which implies wave-vector dependent momentum-relaxing and momentum-conserving scattering rates. In this work, we consider $\gamma_m$ to be independent of $\vec{k}$.}:
\begin{equation}\label{eq:eigen_C}
\hat{C}\chi_{\vec{k},m}=\gamma_{m}\chi_{\vec{k},m},
\end{equation}
\begin{equation}\label{eq:chi_complete}
\sum_{m} \left| \chi_{\vec{k}, m} \right\rangle  \left\langle \chi_{\vec{k}, m'} \right|=1, 
\end{equation}
and 
\begin{equation}\label{eq:chi_orthonorm}
\left\langle \chi_{\vec{k}, m} \right|\left. \chi_{\vec{k}, m'} \right\rangle=\delta_{m m'}.
\end{equation} 
We can now expand our distribution function 
\begin{equation}\label{eq:psi_chi_ansatz}
\psi_{\vec{k}}=\sum_{m}a_{m}\chi_{\vec{k},m}
\end{equation}
with coefficients 
\begin{equation}
a_{m}=\int_{\vec{k}}w_{\vec{k}}\chi_{\vec{k},m}^{*}\psi_{\vec{k}}.
\end{equation}
We insert the ansatz (\ref{eq:psi_chi_ansatz}) into the Boltzmann equation (\ref{eq:Boltz_psi}) and expand the source term 
\begin{equation}
S_{\vec{k}}\equiv-\frac{e}{k_B T}\vec{E}\left(\vec{q},\omega\right)\cdot\vec{v}_{\vec{k}}=\sum_{m}s_{m}\chi_{\vec{k},m},
\end{equation}
such that 
\begin{equation}\label{eq:s_m_chik}
s_{m}=-\frac{e}{k_B T}\vec{E}\left(\vec{q},\omega\right)\int_{\vec{k}}w_{\vec{k}}\chi_{\vec{k},m}^{*}\vec{v}_{\vec{k}}.
\end{equation}
Now the Boltzmann equation takes the form 
\begin{equation}\label{eq:Boltz_chik}
\left(\gamma_{m}-i\omega\right)a_{m}+\sum_{m'}i\vec{q}\cdot\left\langle m\left|\vec{v}\right|m'\right\rangle a_{m'}=s_{m},
\end{equation}
where the matrix elements of the velocity operator are
\begin{equation}
\left\langle m\left|\vec{v}\right|m'\right\rangle =\int_{\vec{k}}w_{\vec{k}}\chi_{\vec{k},m}^{*}\vec{v}_{\vec{k}}\chi_{\vec{k},m'}.
\end{equation}
Hence, at $\vec{q}=\vec{0}$ the distribution function $a_m=s_m/(\gamma_m-i \omega)$ gives rise to a Drude response while non-local events at finite-$\vec{q}$ couple the distinct eigenmodes. Once this coupled system of equations is solved, one obtains expressions for the distribution function $\delta f$ in terms of the eigenvalues of the collision operator. Such formulation will be useful for the evaluation of the conductivity tensor. Recently, a full solution of this problem was achieved for graphene in the limit of weak electron-electron interaction \cite{Kiselev-2019b,Kiselev-2020}.

\subsection{Conservation laws}

Before proceeding with our analysis, it is convenient to express conservation laws for the system in terms of the eigenfunctions $\chi_{\vec{k}, m}$. The associated continuity equations are the result of the conservation of electric charge, energy, and possibly momentum. They enter the analysis as zero modes  $\hat{C}\chi_{\vec{k},a}=0$ of the collision operator.  
If we multiply the Boltzmann equation (\ref{eq:Boltz_psi}) with no applied sources (i.e.\@, $E(\vec{q},\omega)=0$) by such a zero mode $w_{\vec{k}}\chi_{\vec{k},a}^{*}$, we find that  
\begin{equation}
\rho_{a}\left(\vec{q},\omega\right)=\int_{\vec{k}}w_{\vec{k}}\chi_{\vec{k},a}^{*}\psi_{\vec{k}}\left(\vec{q},\omega\right)
\end{equation}
is a conserved quantity with current 
\begin{equation}
j_{a,\alpha}\left(\vec{q},\omega\right)=\int_{\vec{k}}w_{\vec{k}}\chi_{\vec{k},a}^{*}v_{\vec{k},\alpha}\psi_{\vec{k}}\left(\vec{q},\omega\right)
\end{equation}
and obeys the continuity equation 
\begin{equation}
-i\omega\rho_{a}\left(\vec{q},\omega\right)+i\vec{q}\cdot\boldsymbol{j}_{a}\left(\vec{q},\omega\right)=0.
\end{equation}
Charge conservation corresponds to $\chi_{\vec{k},\rho}=1$, while energy conservation gives $\chi_{\vec{k},\varepsilon}=\varepsilon_{\vec{k}}$, and momentum conservation would demand $\chi_{\vec{k},a}=v_{\vec{k},a}$. Notice, these modes are not normalized to unity with our scalar product (\ref{eq:scalar_prod}).

\subsection{Distribution function}\label{Distr}

Equipped with the constraints given by the continuity equations, we proceed to solve the infinite set of linear equations (\ref{eq:Boltz_chik}) to find $\delta f_{\vec{k}}$. 
The spirit of our analysis is similar to the one used in Ref.\@ \onlinecite{Callaway-1959}, yet we also allow for momentum relaxation.
As we will see below, our results will depend on two aspects of the electronic  structure $\varepsilon_{\vec{k}}$: the electronic plasma frequency $\Omega_p$, to be defined in Eq.\@ (\ref{eq:Omega_p_separ}) below, and the Fermi velocity $\vec{v}_{\vec{k}}$, as it depends on wave vector $\vec{k}$. 

Let us consider the Boltzmann equation (\ref{eq:Boltz_psi}) with the collision operator (\ref{eq:Coll_op_psi}) and the scalar product (\ref{eq:scalar_prod}) for an arbitrary Fermi surface. In the eigenvalue equation (\ref{eq:eigen_C}) for the collision operator we impose charge conservation, which demands $\gamma_0=0$, and momentum relaxation with scattering rate $\gamma_{\rm mr}$. The other eigenfunctions are assumed to be equal, i.e.\@, they all correspond to the eigenvalue $\gamma_{\rm mc}$. Hence, we choose the complete set of basis states
\begin{subequations}\label{eq:chi_states}
\begin{equation}\label{eq:chi_0}
\chi_{\vec{k},0}=c_0 \, : \, \gamma_0=0,
\end{equation}
\begin{equation}\label{eq:chi_i}
\chi_{\vec{k},i}=c_0 c_i \tilde{v}_{\vec{k},i} \, : \, \gamma_i\equiv \gamma_{\rm mr},
\end{equation}
\begin{equation}\label{eq:chi_mg2}
\chi_{\vec{k},m} \, : \, \gamma_m\equiv \gamma_{\rm mc}, \, \forall m\neq\left\{0,i\right\},
\end{equation}
\end{subequations}
where $i=\left\{x,y\right\}$ for $d=2$ while $i=\left\{x,y,z\right\}$ for $d=3$, and
\begin{equation}\label{eq:vtilde}
\vec{\tilde{v}}_{\vec{k}}=\frac{\vec{v}_{\vec{k}}}{\left\langle v_{\vec{k}} \right\rangle_0}
\end{equation}
is the velocity vector for state $\vec{k}$ normalized to the average velocity $\left\langle v_{\vec{k}} \right\rangle_0$ at the Fermi wave vector $k=k_F$ (which is a function of electron density, $k_F=k_F(n)$). 

The states (\ref{eq:chi_mg2}) are assumed to be orthonormal to the states (\ref{eq:chi_0}) and (\ref{eq:chi_i}), and to each other, according to Eq.\@ (\ref{eq:chi_orthonorm}). While in the hydrodynamic limit it holds $\gamma_{\rm mc}\gg\gamma_{\rm mr}$, our results are valid for all $\gamma_{\rm mc}\geq\gamma_{\rm mr}$ and for all values of the scattering rates in comparison to typical frequencies and momenta. Of course the rates have to be small compared to the Fermi energy to justify the Boltzmann approach. We further use the notation (\ref{eq:delta_gamma}) for the difference between the two rates.

In performing the scalar product (\ref{eq:scalar_prod}), we conveniently choose to integrate over surfaces $S(\epsilon)$ of constant energy $\epsilon$ \cite{Dressel-2001} assuming twofold spin degeneracy, as
\begin{equation}\label{eq:int_k_CE}
\int_{\vec{k}} \equiv \int \frac{2 d \vec{k}}{(2 \pi)^d}=\frac{2}{(2 \pi)^d \hbar}\int d \epsilon \int_{S(\epsilon)} \frac{d S}{ v_{\vec{k}}}.
\end{equation}
From now on, we assume that the reciprocal-space integral (\ref{eq:int_k_CE}) is confined to states on the Fermi surface $S_F$, $S(\epsilon) \approx S_F$, for sufficiently low temperature $T \ll T_F$ with $T_F$ Fermi temperature of the electron ensemble. This way, $-\partial f_{FD}(\epsilon)/\partial \epsilon \approx \delta(\epsilon)$. Thus, the average velocity $\left\langle v_{\vec{k}} \right\rangle_0$ in Eq.\@ (\ref{eq:vtilde}) results
\begin{equation}\label{eq:v_av_F}
\left\langle v_{\vec{k}} \right\rangle_0=\frac{\int_{S_F} \frac{d S}{v_{\vec{k}_F}} v_{\vec{k}_F}}{\int_{S_F} \frac{d S}{v_{\vec{k}_F}}} \\ =\frac{S_F}{\int_{S_F} \frac{d S}{v_{\vec{k}_F}}},
\end{equation}
where we have defined the Fermi-surface area for arbitrary, possibly anisotropic, shape
\begin{equation}\label{eq:S_F}
S_F=\int_{S_F} d S.
\end{equation}
For an isotropic free-electron gas, Eq.\@ (\ref{eq:S_F}) gives $S= 2 \pi k_F$ and $S= 4 \pi (k_F)^2$ in 2D and 3D, respectively.
Notice that with Eq.\@ (\ref{eq:v_av_F}) we take into account the possibility that the Fermi velocity itself is anisotropic in space, i.e., that it varies in magnitude along the Fermi surface. We will return on this point in Sec.\@ \ref{v_F_anisotropy}, where we will suitably parametrize the possible Fermi-velocity anisotropy. 
 
We assume that the distribution function only depends on the direction $\hat{k}=\vec{k}/k$: $\psi_{\vec{k}}\equiv \psi_{\hat{k}}$. 
The normalization of the eigenfunction $\chi_{\vec{k},0}$ requires 
\begin{equation}\label{eq:c_0_vF_var}
c_0=\left[\frac{2 k_B T}{(2 \pi)^d \hbar} \int_{S_F} \frac{d S}{v_{\vec{k}_F}}\right]^{-\frac{1}{2}}.
\end{equation}
Notice that the integral in Eq.\@ (\ref{eq:c_0_vF_var}) is connected to the spinful Fermi-level density of states $N_0^{el}(0)$ per unit volume:
\begin{multline}\label{eq:DOS_kF}
N_0^{el}(0)=2\frac{1}{\mathscr{V}}\delta(\epsilon-E_F)=\frac{1}{\mathscr{V}}\int \frac{2 d \vec{k}}{(2 \pi)^d} \delta(\epsilon-E_F) \\ \equiv \frac{2}{(2 \pi)^d} \int_{S_F} \frac{d S}{v_{\vec{k}_F}},
\end{multline}
where $\mathscr{V}$ is the system volume.
In the same way, the normalization coefficient $c_i$ results 
\begin{equation}\label{eq:c_1_vF_var}
c_i=\left[\frac{ \int_{S_F} \frac{d S}{v_{\vec{k}_F}} (\tilde{v}_{\vec{k},i})^2}{\int_{S_F} \frac{d S}{v_{\vec{k}_F}}}\right]^{-\frac{1}{2}}.
\end{equation}
Eqs.\@ (\ref{eq:c_0_vF_var}) and (\ref{eq:c_1_vF_var}) are proved in Appendix \ref{App:coll_op_states}.

Within the above assumptions, we obtain for the collision operator
\begin{multline}\label{eq:C_op_scal_expl}
\hat{C}\psi_{\vec{k}} = \hat{C}\psi_{\hat{k}}= \gamma_{\rm mc}\psi_{\hat{k}}-\gamma_{\rm mc}n_0(\vec{q},\omega) \\  -\delta \gamma \sum_i \left[ c_i^2 \tilde{v}_{\vec{k},i} p_i(\vec{q},\omega)\right]
\end{multline}
with particle density
\begin{equation}\label{eq:n_0}
n_0(\vec{q},\omega)=\frac{\int_{S_F} \frac{d S}{v_{\vec{k}_F}} \psi_{\hat{k}}(\vec{q},\omega)}{\int_{S_F} \frac{d S}{v_{\vec{k}_F}}}
\end{equation}
and momentum density
\begin{equation}\label{eq:p_i_d}
p_i(\vec{q},\omega)=\frac{\int_{S_F} \frac{d S}{v_{\vec{k}_F}} \tilde{v}_{\vec{k},i} \psi_{\hat{k}}(\vec{q},\omega)}{\int_{S_F} \frac{d S}{v_{\vec{k}_F}}}.
\end{equation}
along the spatial directions $i$, as shown in Appendix \ref{App:coll_op_states}. 

This finally yields for the distribution function 
\begin{multline}\label{eq:distr_S}
\left|\psi_{\hat{k}} \right\rangle=\frac{1}{-i \omega+\gamma_{\rm mc}+i \vec{q}\cdot \vec{v}_{\vec{k}}}\left\{-\frac{e}{k_B T} \vec{E}(\vec{q},\omega) \cdot \vec{v}_{\vec{k}} \right. \\ \left.+\gamma_{\rm mc}  n_0(\vec{q},\omega)+\delta \gamma \sum_i \left[ c_i^2 \tilde{v}_{\vec{k},i} p_i(\vec{q},\omega)\right]\right\}.
\end{multline}
Eq.\@ (\ref{eq:distr_S}) is the main result of this section: it relates the distribution function (\ref{eq:f_psi}) with the conserved density (\ref{eq:n_0}), the momenta (\ref{eq:p_i_d}), as well as with the scattering rates $\gamma_{\rm mr}$ and $\gamma_{\rm mc}$, as a function of the anisotropic Fermi-surface velocity $\vec{v}_{\vec{k}}$. The explicit form of $\psi_{\hat{k}}$ naturally depends on the relative orientation of $\vec{E}$, $\vec{q}$, and $\vec{v}_{\hat{k}}$. In the next section, we specialize Eq.\@ (\ref{eq:distr_S}) to purely transverse excitations exerted by an electric field $\vec{E} \perp \vec{q}$. 

\subsection{Transverse conductivity}\label{T_cond}

Let us consider an electric field $\vec{E}(\vec{q},\omega) \equiv E_\beta(\vec{q},\omega) \hat{u}_\beta$ along the $\beta$ axis, and focus on momenta $\vec{q}=q_\alpha \hat{u}_\alpha$ pointing along the $\alpha \neq \beta$ direction: this setup will give rise to a transverse conductivity. Then, it follows
\begin{multline}\label{eq:distr_S_T}
\left|\psi_{\hat{k}} \right\rangle=\frac{1}{-i \omega+\gamma_{\rm mc}+i q_\alpha v_{\vec{k}_F, \alpha}}\left\{-\frac{e}{k_B T} E_\beta(q_\alpha,\omega)  v_{\vec{k}_F, \beta}  \right. \\ \left. +\gamma_{\rm mc} n_0(q_\alpha,\omega)+\delta \gamma \sum_i \left[ c_i^2 \tilde{v}_{\vec{k},i} p_i(q_\alpha,\omega)\right]\right\}. 
\end{multline}
Notice that we can arrange the equations for the moments of $\left|\psi_{\hat{k}} \right\rangle$ in a matrix form, by conveniently defining the average
\begin{equation}
\left \langle \left \langle A \right \rangle \right \rangle=\left[ \int_{S_F} \frac{d S}{v_{\vec{k}_F}}\right]^{-1} \int_{S_F} \frac{d S}{v_{\vec{k}_F}} \frac{A}{\gamma_{\rm mc}-i \omega +i q_\alpha v_{\vec{k}_F,\alpha}}
\end{equation}
for the quantity $A$. These equations for the moments are obtained by multiplying Eq.\@ (\ref{eq:distr_S_T}) by $\left[d S/ v_{\vec{k}_F} \right]/\left[\int_{S_F} d S/ v_{\vec{k}_F} \right]$ and $\left[d S/ v_{\vec{k}_F} \tilde{v}_{\vec{k},i} \right]/\left[\int_{S_F} d S/ v_{\vec{k}_F} \right]$, $i=\left\{\alpha,\beta\right\}$, and integrating over the Fermi surface $S_F$. 
\begin{widetext}
For $d=2$, where $\left\{\alpha,\beta\right\}=\left\{x,y\right\}$, we obtain 
\begin{equation}\label{eq:moments_psik_2D}
  \begin{bmatrix}
    \gamma_{\rm mc}\left \langle \left \langle 1 \right \rangle \right \rangle-1  &  c_\alpha^2 \delta \gamma \left \langle \left \langle \tilde{v}_{\vec{k},\alpha} \right \rangle \right \rangle & c_\beta^2 \delta \gamma \left \langle \left \langle  \tilde{v}_{\vec{k},\beta} \right \rangle \right \rangle \\
		  \gamma_{\rm mc}\left \langle \left \langle \tilde{v}_{\vec{k},\alpha} \right \rangle \right \rangle  &  c_\alpha^2 \delta \gamma \left \langle \left \langle \tilde{v}_{\vec{k},\alpha}^2 \right \rangle \right \rangle-1 & c_\beta^2 \delta \gamma \left \langle \left \langle \tilde{v}_{\vec{k},\alpha} \tilde{v}_{\vec{k},\beta} \right \rangle \right \rangle \\
			 \gamma_{\rm mc}\left \langle \left \langle \tilde{v}_{\vec{k},\beta} \right \rangle \right \rangle  &  c_\alpha^2 \delta \gamma \left \langle \left \langle \tilde{v}_{\vec{k},\alpha} \tilde{v}_{\vec{k},\beta} \right \rangle \right \rangle & c_\beta^2 \delta \gamma \left \langle \left \langle \tilde{v}_{\vec{k},\beta}^2 \right \rangle \right \rangle-1 
  \end{bmatrix}
	  \begin{bmatrix}
    n_0(q_\alpha,\omega) \\ p_\alpha(q_\alpha,\omega) \\ p_\beta(q_\alpha,\omega) 
  \end{bmatrix}=
	\frac{e}{k_B T} E_\beta
	  \begin{bmatrix}
    \left \langle \left \langle v_{\vec{k}_F,\beta} \right \rangle \right \rangle \\ \left \langle \left \langle v_{\vec{k}_F,\beta} \tilde{v}_{\vec{k},\alpha} \right \rangle \right \rangle \\ \left \langle \left \langle v_{\vec{k}_F,\beta} \tilde{v}_{\vec{k},\beta} \right \rangle \right \rangle
  \end{bmatrix}.
	\end{equation}
Eq.\@ (\ref{eq:moments_psik_2D}) is simplified if we assume that there are two mirror planes in reciprocal space of momenta $\hbar \vec{k}$, in the $xy$ plane formed by $\vec{q}$ and $\vec{E}$. This means that, if we take a unit vector $\hat{u}(k_x,k_y)$ such that $\left|\hat{u}(k_x,k_y)\right|=1$, 
\begin{equation}\label{eq:mirror_planes}
\hat{u}(k_x,k_y)=\hat{u}(k_x,-k_y)=\hat{u}(-k_x,k_y).
\end{equation}
The condition (\ref{eq:mirror_planes}) ensures a good integration property for a function $a(\tilde{v}_{\vec{k},x},\tilde{v}_{\vec{k},y})$ which is odd under reflection about the $x$ or $y$ axis: in particular, a generic function $\mathscr{F}(\tilde{v}_{\vec{k},x})$ multiplied by $a(\tilde{v}_{\vec{k},x},\tilde{v}_{\vec{k},y})=-a(-\tilde{v}_{\vec{k},x},\tilde{v}_{\vec{k},y})$ produces a null result upon integration over the Fermi surface:
\begin{equation}\label{eq:int_null_SF_symm}
\left[\frac{\int_{S_F} d S}{ v_{\vec{k}_F}} \right]^{-1} \int_{S_F} \frac{d S}{v_{\vec{k}_F}} a(\tilde{v}_{\vec{k},x},\tilde{v}_{\vec{k},y}) \mathscr{F}(\tilde{v}_{\vec{k},x})=0.
\end{equation} 
An example is $a(\tilde{v}_{\vec{k},x},\tilde{v}_{\vec{k},y})=\tilde{v}_{\vec{k},y}$. 
Within the hypothesis (\ref{eq:mirror_planes}), the matrix equation (\ref{eq:moments_psik_2D}) simplifies to 
\begin{equation}\label{eq:moments_psik_2D_mirror}
  \begin{bmatrix}
    \gamma_{\rm mc}\left \langle \left \langle 1 \right \rangle \right \rangle-1  &  c_\alpha^2 \delta \gamma \left \langle \left \langle \tilde{v}_{\vec{k},\alpha} \right \rangle \right \rangle & 0 \\
		  \gamma_{\rm mc}\left \langle \left \langle \tilde{v}_{\vec{k},\alpha} \right \rangle \right \rangle  &  c_\alpha^2 \delta \gamma \left \langle \left \langle \tilde{v}_{\vec{k},\alpha}^2 \right \rangle \right \rangle-1 & 0 \\
			 0  &  0 & c_\beta^2 \delta \gamma \left \langle \left \langle \tilde{v}_{\vec{k},\beta}^2 \right \rangle \right \rangle-1 
  \end{bmatrix}
	  \begin{bmatrix}
    n_0(q_\alpha,\omega) \\ p_\alpha(q_\alpha,\omega) \\ p_\beta(q_\alpha,\omega) 
  \end{bmatrix}=
	\frac{e}{k_B T} E_\beta 
	  \begin{bmatrix}
    0 \\ 0 \\ \left \langle \left \langle v_{\vec{k}_F,\beta} \tilde{v}_{\vec{k},\beta} \right \rangle \right \rangle
  \end{bmatrix}.
	\end{equation}		
For $d=3$, where $\left\{\alpha,\beta,\gamma\right\}=\left\{x,y,z\right\}$, we obtain 
\begin{multline}\label{eq:moments_psik_3D}
  \begin{bmatrix}
    \gamma_{\rm mc}\left \langle \left \langle 1 \right \rangle \right \rangle-1  &  c_\alpha^2 \delta \gamma \left \langle \left \langle \tilde{v}_{\vec{k},\alpha} \right \rangle \right \rangle & c_\beta^2 \delta \gamma \left \langle \left \langle  \tilde{v}_{\vec{k},\beta} \right \rangle \right \rangle & c_\gamma^2 \delta \gamma \left \langle \left \langle  \tilde{v}_{\vec{k},\gamma} \right \rangle \right \rangle  \\
		  \gamma_{\rm mc}\left \langle \left \langle \tilde{v}_{\vec{k},\alpha} \right \rangle \right \rangle  &  c_\alpha^2 \delta \gamma \left \langle \left \langle \tilde{v}_{\vec{k},\alpha}^2 \right \rangle \right \rangle-1 & c_\beta^2 \delta \gamma \left \langle \left \langle \tilde{v}_{\vec{k},\alpha}  \tilde{v}_{\vec{k},\beta} \right \rangle \right \rangle & c_\gamma^2 \delta \gamma \left \langle \left \langle \tilde{v}_{\vec{k},\alpha}  \tilde{v}_{\vec{k},\gamma} \right \rangle \right \rangle \\
			 \gamma_{\rm mc}\left \langle \left \langle \tilde{v}_{\vec{k},\beta} \right \rangle \right \rangle  &  c_\alpha^2 \delta \gamma \left \langle \left \langle \tilde{v}_{\vec{k},\alpha} \tilde{v}_{\vec{k},\beta} \right \rangle \right \rangle & c_\beta^2 \delta \gamma \left \langle \left \langle \tilde{v}_{\vec{k},\beta}^2 \right \rangle \right \rangle-1 & c_\gamma^2 \delta \gamma \left \langle \left \langle \tilde{v}_{\vec{k},\beta}  \tilde{v}_{\vec{k},\gamma} \right \rangle \right \rangle \\
\gamma_{\rm mc}\left \langle \left \langle \tilde{v}_{\vec{k},\gamma} \right \rangle \right \rangle  &  c_\alpha^2 \delta \gamma \left \langle \left \langle \tilde{v}_{\vec{k},\alpha} \tilde{v}_{\vec{k},\gamma} \right \rangle \right \rangle & c_\beta^2 \delta \gamma \left \langle \left \langle \tilde{v}_{\vec{k},\beta} \tilde{v}_{\vec{k},\gamma} \right \rangle \right \rangle & c_\gamma^2 \delta \gamma \left \langle \left \langle \tilde{v}_{\vec{k},\gamma}^2 \right \rangle \right \rangle-1
  \end{bmatrix}
	  \begin{bmatrix}
    n_0(q_\alpha,\omega) \\ p_\alpha(q_\alpha,\omega) \\ p_\beta(q_\alpha,\omega) \\ p_\gamma(q_\alpha,\omega)
  \end{bmatrix} \\ =
	\frac{e}{k_B T} E_\beta
	  \begin{bmatrix}
    \left \langle \left \langle v_{\vec{k}_F,\beta} \right \rangle \right \rangle \\ \left \langle \left \langle v_{\vec{k}_F,\beta} \tilde{v}_{\vec{k},\alpha} \right \rangle \right \rangle \\ \left \langle \left \langle v_{\vec{k}_F,\beta} \tilde{v}_{\vec{k},\beta} \right \rangle \right \rangle \\ \left \langle \left \langle v_{\vec{k}_F,\beta} \tilde{v}_{\vec{k},\gamma} \right \rangle \right \rangle
  \end{bmatrix}.
	\end{multline}
Similarly to the 2D case of Eq.\@ (\ref{eq:moments_psik_2D_mirror}), Eq.\@ (\ref{eq:moments_psik_3D}) simplifies if we assume that there are three mirror planes in reciprocal space of momenta $\hbar \vec{k}$. Taking a unit vector $\hat{u}(k_x,k_y,k_z)$ such that $\left|\hat{u}(k_x,k_y,k_z)\right|=1$, we have
\begin{equation}\label{eq:mirror_planes_3D}
\hat{u}(k_x,k_y,k_z)=\hat{u}(k_x,-k_y,k_z)=\hat{u}(-k_x,k_y,k_z)=\hat{u}(k_x,k_y,-k_z).
\end{equation}
Assuming Eq.\@ (\ref{eq:mirror_planes}) to hold, the matrix equation (\ref{eq:moments_psik_3D}) becomes
\begin{multline}\label{eq:moments_psik_3D_mirror}
  \begin{bmatrix}
    \gamma_{\rm mc}\left \langle \left \langle 1 \right \rangle \right \rangle-1  &  c_\alpha^2 \delta \gamma \left \langle \left \langle \tilde{v}_{\vec{k},\alpha} \right \rangle \right \rangle & 0 & 0  \\
		  \gamma_{\rm mc}\left \langle \left \langle \tilde{v}_{\vec{k},\alpha} \right \rangle \right \rangle  &  c_\alpha^2 \delta \gamma \left \langle \left \langle \tilde{v}_{\vec{k},\alpha}^2 \right \rangle \right \rangle-1 & 0 & 0 \\
			 0  &  0 & c_\beta^2 \delta \gamma \left \langle \left \langle \tilde{v}_{\vec{k},\beta}^2 \right \rangle \right \rangle-1 & 0 \\
0  &  0 &0 & c_\gamma^2 \delta \gamma \left \langle \left \langle \tilde{v}_{\vec{k},\gamma}^2 \right \rangle \right \rangle-1
  \end{bmatrix}
	  \begin{bmatrix}
    n_0(q_\alpha,\omega) \\ p_\alpha(q_\alpha,\omega) \\ p_\beta(q_\alpha,\omega) \\ p_\gamma(q_\alpha,\omega)
  \end{bmatrix} \\ =
	\frac{e}{k_B T} E_\beta v_F
	  \begin{bmatrix}
    0 \\ 0 \\ \left \langle \left \langle v_{\vec{k}_F,\beta}\tilde{v}_{\vec{k},\beta} \right \rangle \right \rangle \\ 0
  \end{bmatrix}.
	\end{multline} 
\end{widetext}
In the following we will work within the mirror-planes assumptions (\ref{eq:mirror_planes}) and (\ref{eq:mirror_planes_3D}), which lead to Eqs.\@ (\ref{eq:moments_psik_2D_mirror}) and (\ref{eq:moments_psik_3D_mirror}) in 2D and 3D respectively.

Since the electric field $E_\beta(\vec{q},\omega)$ only couples to momentum in the same direction $\hat{u}_\beta$, we have to ensure that $\psi_{\hat{k}}$ vanishes for $E_\beta(\vec{q},\omega)=0$. This is accomplished if $n_0(\vec{q},\omega)=0$ and $p_i(\vec{q},\omega)=0$ for $i \neq \beta$. 
Then, it follows for the momentum density 
\begin{equation}\label{eq:py_expl_vanish}
p_\beta(q_\alpha,\omega)=-\frac{e}{k_B T} \frac{E_\beta(q_\alpha,\omega) \left\langle \left\langle v_{\vec{k}_F,\beta} \tilde{v}_{\vec{k},\beta} \right\rangle \right \rangle}{1-\delta \gamma c_\beta^2 \left\langle \left\langle \tilde{v}_{\vec{k},\beta}^2 \right\rangle \right \rangle }. 
\end{equation}
Therefore, Eq.\@ (\ref{eq:py_expl_vanish}) specifies the relation between the induced momentum density and the applied transverse electric field.
For the distribution function (\ref{eq:distr_S_T}), this means
\begin{multline}\label{eq:distr_S_T_transv}
\left|\psi_{\hat{k}} \right\rangle=-\frac{e}{k_B T} \frac{E_\beta}{-i \omega +\gamma_{\rm mc} +i  q_\alpha \hat{v}_{\vec{k}_F,\alpha}} \\ \times \left[ v_{\vec{k}_F,\beta} + \frac{c_\beta^2 \delta \gamma \tilde{v}_{\vec{k},\beta} \left\langle \left\langle \tilde{v}_{\vec{k},\beta} v_{\vec{k}_F,\beta} \right\rangle \right \rangle}{1-c_\beta^2 \delta \gamma \left\langle \left\langle \tilde{v}_{\vec{k},\beta}^2 \right\rangle \right \rangle}\right].
\end{multline}
Since the electrical current density is 
\begin{multline}\label{eq:J_def}
\vec{J}(\vec{q},\omega)=-e \int_{\vec{k}} \vec{v}_{\vec{k}} \delta f_{\vec{k}}(\vec{q},\omega)\\ =-e \int_{\vec{k}} w_{\vec{k}} \vec{v}_{\vec{k}}  \left|\psi_{\vec{k}}(\vec{q},\omega) \right\rangle,
\end{multline}
the transverse conductivity results
\begin{multline}\label{eq:sigma_yy_gen}
\sigma_{\beta \beta}(q_\alpha,\omega)=\frac{J_y(q_\alpha,\omega)}{E_y(q_\alpha,\omega)}\\ =\frac{2 e^2 v_F}{(2 \pi)^d \hbar}\int_{S_F} \frac{d S}{v_{\vec{k}_F}}\frac{1}{-i \omega+\gamma_{\rm mc} +i v_{\vec{k}_F,\alpha} q_\alpha} \\ \times \left[(v_{\vec{k}_F,\beta})^2+\frac{c_\beta^2 \delta \gamma v_{\vec{k}_F,\beta} \tilde{v}_{\vec{k},\beta} \left\langle\left\langle \tilde{v}_{\vec{k},\beta} v_{\vec{k}_F,\beta} \right\rangle\right\rangle}{1-c_\beta^2 \delta \gamma \left\langle\left\langle (\tilde{v}_{\vec{k},\beta})^2 \right\rangle\right\rangle}\right] \\=\frac{2 e^2}{(2 \pi)^d \hbar} \left[\int_{S_F} \frac{ d S}{v_{\vec{k}_F}}\right] \\ \times \left[\left\langle\left\langle(v_{\vec{k}_F,\beta})^2\right\rangle\right\rangle +\frac{c_\beta^2 \delta \gamma (\left\langle\left\langle \tilde{v}_{\vec{k},\beta} v_{\vec{k}_F,\beta} \right\rangle\right\rangle)^2}{1-c_\beta^2 \delta \gamma \left\langle\left\langle (\tilde{v}_{\vec{k},\beta})^2 \right\rangle\right\rangle}\right].
\end{multline} 
Thus, we have derived a closed expression for the transverse conductivity of an arbitrary Fermi surface, that covers the ballistic, Ohmic, and viscous regimes. Besides the premise of quasiparticle transport in the Boltzmann regime, the only major assumption is that all momentum-conserving processes are governed by the same relaxation rate $\gamma_{\rm mc}$.
Notice that Eq.\@ (\ref{eq:sigma_yy}) holds whenever the symmetry planes (\ref{eq:mirror_planes}) or (\ref{eq:mirror_planes_3D}) exist for the dispersion relation, which is the case when the Fermi surface is approximated by, e.g.\@, a regular polygon in 2D or solid in 3D. This approximation can easily be relaxed by referring to the general expressions (\ref{eq:moments_psik_2D}) and (\ref{eq:moments_psik_3D}). 

\subsection{Parametrization of the anisotropic Fermi-surface velocity}\label{v_F_anisotropy}

The transverse conductivity (\ref{eq:sigma_yy_gen}) may be recast in a more compact form, if we select a convenient parametrization of the velocity $\vec{v}_{\vec{k}_F}$. We employ
\begin{equation}\label{eq:vF_anis}
\vec{v}_{\vec{k}_F}=\left\langle v_{\vec{k}_F}\right\rangle_0 \nu_{\vec{k}} \hat{n}_{\vec{k}} \equiv v_F \nu_{\vec{k}} \hat{n}_{\vec{k}},
\end{equation}
where $v_F=\left\langle v_{\vec{k}_F}\right\rangle_0$, in accordance with Eq.\@ (\ref{eq:v_av_F}), $\hat{n}_{\vec{k}}$ is the unit vector ($\left|\hat{n}_{\vec{k}} \right|=1$)in the direction of $\vec{k}$, while $\nu_{\vec{k}}$ encodes the orientational variation of the Fermi-surface velocity and is unitary upon averaging:
\begin{equation}\label{eq:nu_k}
\frac{\int_{S_F} \frac{d S}{v_{\vec{k}_F}} \nu_{\vec{k}}}{\int_{S_F} \frac{d S}{v_{\vec{k}_F}}}=1.
\end{equation}
Eq.\@ (\ref{eq:vF_anis}) also implies that $\vec{\tilde{v}}_{\vec{k}}=\vec{v}_{\vec{k}_F}/v_F\equiv \nu_{\vec{k}} \hat{n}_{\vec{k}}$. Notice that $v_F$ in Eq.\@ (\ref{eq:vF_anis}) is a free parameter: this means that, by fixing a value for $v_F$, we impose a value for the average (\ref{eq:v_av_F}) of the Fermi velocity over all orientations. 
With the parametrization (\ref{eq:vF_anis}), Eq.\@ (\ref{eq:n_0}) simplifies to 
\begin{equation}\label{eq:n_0_separ}
n_0(\vec{q},\omega)=\frac{\int_{S_F} \frac{d S}{\nu_{\vec{k}}} \psi_{\hat{k}}(\vec{q},\omega)}{\int_{S_F} \frac{d S}{\nu_{\vec{k}}}}. 
\end{equation}
with normalization coefficient (\ref{eq:c_0_vF_var}) translating as
\begin{equation}\label{eq:c_0_vF_var_separ}
c_0=\left[\frac{2 k_B T}{(2 \pi)^d \hbar v_F} \int_{S_F} \frac{d S}{\nu_{\vec{k}}}\right]^{-\frac{1}{2}}.
\end{equation}
In the same way, using Eq.\@ (\ref{eq:vF_anis}) the momentum density components (\ref{eq:p_i_d}) become 
\begin{equation}\label{eq:p_i_d_separ}
p_i(\vec{q},\omega)=\frac{\int_{S_F} \frac{d S}{\nu_{\vec{k}}} \nu_{\vec{k}} n_{\vec{k},i} \psi_{\hat{k}}(\vec{q},\omega)}{\int_{S_F} \frac{d S}{\nu_{\vec{k}}}}.
\end{equation}
with coefficients
\begin{equation}\label{eq:c_1_vF_var_separ}
c_i=\left[\frac{ \int_{S_F} \frac{d S}{\nu_{\vec{k}}} (\nu_{\vec{k}} n_{\vec{k},i})^2}{\int_{S_F} \frac{d S}{\nu_{\vec{k}}}}\right]^{-\frac{1}{2}}.
\end{equation}
This way, the transverse conductivity (\ref{eq:sigma_yy}) results  
\begin{multline}\label{eq:sigma_yy_separ}
\sigma_{\beta \beta}(q_\alpha,\omega)=\frac{2 e^2 v_F}{(2 \pi)^d \hbar}\int_{S_F} \frac{d S}{\nu_{\vec{k}}} \frac{\left\langle\left\langle (\nu_{\vec{k}} n_{\vec{k},\beta})^2 \right\rangle\right\rangle}{1-c_\beta^2 \delta \gamma \left\langle\left\langle (\nu_{\vec{k}} n_{\vec{k},\beta})^2 \right\rangle\right\rangle} \\ \equiv \epsilon_0 \Omega_p^2 \frac{G_0(q_\alpha,\omega)}{1-c_\beta^2 \delta \gamma G_0(q_\alpha,\omega)},
\end{multline}
where we define 
\begin{multline}\label{eq:G_0_separ}
G_0(q,\omega)=\left[\int_{S_F} \frac{d S}{\nu_{\vec{k}}}\right]^{-1} \int_{S_F} dS \frac{ \nu_{\vec{k}} ( n_{\vec{k},\beta})^2}{\gamma_{\rm mc} -i \omega +i v_F \nu_{\vec{k}} n_{\vec{k},\alpha} q } \\ \equiv \left\langle\left\langle (\nu_{\vec{k}} n_{\vec{k},\beta})^2 \right\rangle\right\rangle.
\end{multline}
We also define the plasma frequency
\begin{equation}\label{eq:Omega_p_separ}
\Omega_p^2 =\sum_i \omega_{p,ii}^2 \equiv \frac{2 e^2 v_F}{(2 \pi)^d \hbar \epsilon_0} \int_{S_F} \frac{d S}{\nu_{\vec{k}}},
\end{equation}
with individual contributions
\begin{equation}\label{eq:omega_p_ii_separ}
\epsilon_0 \omega_{p,ii}^2=\frac{e^2 v_F^2}{c_0^2 k_B T} \int_{S_F} \frac{d S}{\nu_{\vec{k}}} n_{\vec{k},i}^2=\epsilon_0\Omega_p^2 \int_{S_F} \frac{d S}{\nu_{\vec{k}}} n_{\vec{k},i}^2.
\end{equation}
The anisotropy taken into account by the conductivity (\ref{eq:sigma_yy_separ}) is twofold: the Fermi-surface shape is allowed to be anisotropic, and in addition the Fermi velocity (\ref{eq:vF_anis}) may vary along different orientations due to the factor $\nu_{\vec{k}}$; both anisotropies are only constrained by the assumed existence of two or three symmetry planes in reciprocal space, in accordance with Eqs.\@ (\ref{eq:mirror_planes}) and (\ref{eq:mirror_planes_3D}). 
In the next section and in the rest of this paper, for simplicity we will work under the assumption that the Fermi velocity is isotropic in space, while we will retain the anisotropy in the Fermi-surface shape. 

\subsection{Transverse conductivity for constant Fermi-velocity modulus}

In the limit where all states at the Fermi surface share the same velocity modulus, we have
\begin{equation}\label{eq:v_k_F}
\vec{v}_{\vec{k}_F} \approx \vec{v}_F=v_F \hat{n}_{\vec{k}},
\end{equation}
that is, $\nu_{\vec{k}} \equiv 1 \forall \vec{k}$, and Eqs.\@ (\ref{eq:n_0_separ})-(\ref{eq:omega_p_ii_separ}) considerably simplify. Specifically, we obtain for the coefficient (\ref{eq:c_0_vF_var_separ}) that
\begin{equation}\label{eq:c_0_intS}
c_0=\left[ \frac{2 k_B T}{(2 \pi)^d \hbar v_F} S_F\right]^{-1/2},
\end{equation}
while Eq.\@ (\ref{eq:c_1_vF_var_separ}) translates as 
\begin{equation}\label{eq:c_i_intS_iso}
c_i=\left[ \int_{S_F} \frac{d S}{S_F} n_{\vec{k},i}^2\right]^{-1/2}.
\end{equation}
The particle density (\ref{eq:n_0_separ}) and the momentum density (\ref{eq:p_i_d_separ}) become
\begin{equation}\label{eq:n_0_iso}
n_0(\vec{q},\omega)=\int_{S_F} \frac{d S}{S_F} \psi_{\hat{k}}(\vec{q},\omega)
\end{equation}
and
\begin{equation}\label{eq:p_i_d_iso}
p_i(\vec{q},\omega)=\int_{S_F} \frac{d S}{S_F} n_{\vec{k},i} \psi_{\hat{k}}(\vec{q},\omega),
\end{equation}
respectively. Finally, the transverse conductivity (\ref{eq:sigma_yy_separ}) reduces to
\begin{equation}\label{eq:sigma_yy}
\sigma_{\beta \beta}(q_\alpha,\omega) = \epsilon_0 \Omega_p^2 \frac{G_0(q_\alpha,\omega)}{1-c_\beta^2 \delta \gamma G_0(q_\alpha,\omega)},
\end{equation}
where from Eq.\@ (\ref{eq:G_0_separ}) we have
\begin{equation}\label{eq:G_0_iso}
G_0(q,\omega)=\int_{S_F} \frac{d S}{S_F} \frac{n_{\vec{k},\beta}^{2}}{-i \omega +\gamma_{\rm mc}+i v_F q n_{\vec{k},\alpha}}. 
\end{equation}
The plasma frequency (\ref{eq:Omega_p_separ}) for a constant Fermi-velocity modulus translates as 
\begin{equation}\label{eq:omega_p_S_iso}
\Omega_p^2=\sum_i \omega_{p,ii}^2 \equiv \frac{2e^2 S_F v_F}{(2 \pi)^d \hbar \epsilon_0}
\end{equation}
where
\begin{multline}\label{eq:omega_p_ii_iso}
\epsilon_0 \omega_{p,ii}^2=\frac{e^2 (v_F)^2}{c_0^2 k_B T} \int_{S_F} \frac{d S}{S_F} n_{\vec{k},i}^2 \\ \equiv \frac{2 e^2 S_F v_F}{(2 \pi)^d \hbar} \int_{S_F} \frac{d S}{S_F} n_{\vec{k},i}^2. 
\end{multline}
Moreover, notice that the DC conductivity in the limit $q_\alpha \rightarrow 0$, $\omega\rightarrow 0$ implied by Eq.\@ (\ref{eq:sigma_yy}) depends on the relaxation rate $\gamma_{\rm mr}$ but not on the momentum-conserving rate $\gamma_{\rm mc}$: 
\begin{equation}\label{eq:sigma_yy_DC}
\sigma_{\beta \beta}(0,0)=\frac{\epsilon_0 \omega_{p,\beta \beta}^2}{\gamma_{\rm mr}}.
\end{equation} 
Eq.\@ (\ref{eq:sigma_yy}) can, among others, be used to analyze the various versions of the skin effect, as we will appreciate in Secs.\@ \ref{Skin} and \ref{Z_spec}. In the following section, we first check that the conductivity (\ref{eq:sigma_yy}) reduces to the known expression for isotropic 2D systems, and then we specialize to the anisotropic examples of hexagonal and square Fermi-surface shapes. More complicated Fermi-surface geometries are discussed in Sec.\@ \ref{Rounded} and compared with the polygonal symmetries. Selected solutions for 3D systems are left for further work. 

\section{Transverse conductivity for 2D isotropic and polygonal geometries}\label{Cond_poly}

In applying the theory of Sec.\@ \ref{T_cond} to  cases of 2D circular and polygonal Fermi surfaces, we have to span a substantially large parameter space as a function of $\gamma_{\rm mr}$, $\gamma_{\rm mc}$, $v_F$, $q$, and $\omega$. This space can be conveniently constrained by considering how the ratio $\gamma_{\rm mr}/\gamma_{\rm mc}$ evolves in practice as a function of temperature. At room temperature, where phase-space constraints become less crucial and anharmonic phonon effects suppress phonon drag behavior one often expects $\gamma_{\rm mr} \approx \gamma_{\rm mc}$. On the other hand  $\gamma_{\rm mc} \gg \gamma_{\rm mr}$ is expected for ultra-clean systems at cryogenic temperatures. These two separate cases will be considered in the following analysis.
For definiteness, we fix the spatial coordinates $\alpha=x$ and $\beta=y$ in the what follows.
The expression for $\Omega_p$ and $c_y$ is common to all Fermi-surface shapes and orientation here considered. This is because, from the respective definitions (\ref{eq:omega_p_ii_iso}) and (\ref{eq:c_i_intS_iso}), we have 
\begin{equation}\label{eq:c_y_Omegap}
\omega_{p,ii}^2=\frac{\Omega_p^2}{c_i^2},
\end{equation}
and this quantity is set by the symmetry of the crystal system, analogously to the local ($q_x=0$) conductivity tensor \cite{Tanner-2019opt}. In all example Fermi surfaces discussed herein, the local conductivity tensor is isotropic and so by symmetry
\begin{equation}\label{eq:Omegap_c_i_expl}
c_{y}^{2}=\Omega_{p}^{2}/\omega_{p}^{2}=2.
\end{equation}
In Eq.\@ (\ref{eq:Omegap_c_i_expl}), we have defined the plasma frequency of the isotropic electron gas \cite{Ashcroft-1976, Dressel-2001}
\begin{equation}\label{eq:omega_p_iso}
\omega_p=\sqrt{\frac{n e^2}{m \epsilon_0}}.
\end{equation} 
Furthermore, another simplification occurs in the Fermi-surface integrations for all polygonal geometries, within the assumption that the scattering rates $\gamma_{\rm mr}$ and $\gamma_{\rm mc}$, and the Fermi-velocity modulus $v_F$, do not depend on orientation: in this case, since the regular polygon of $N_S$ sides can be decomposed into $N_S$ regular triangles, and the normal vectors $n_{\vec{k},i}$ do not vary within each individual side $i=\left\{1,\cdots N_S\right\}$, the integrals $\int_{S_F} F(n_{\vec{k}}) d S/S_F=\left[\sum_{i=1}^{N_S} F(n_{\vec{k},i})\right]/N_S$ for any function $F(n_{\vec{k}})$. In other words, the Fermi-surface integrations can be taken as discrete sums which depend on the values $n_{\vec{k},i}$ of the normal vector assumed on each polygon side. We will see concrete examples of this simplification in Secs.\@ \ref{Hex_FS} and \ref{Square_FS}, but first we analyze the simplest case of a circular Fermi surface.

\subsection{The isotropic limit}\label{Iso_circ}

Let us check the expression (\ref{eq:sigma_yy}) for a circular Fermi surface, in 2D. In this case, it is convenient to parametrize the Fermi-surface integrations with the angle $\theta \in \left[0,2 \pi\right]$: 
\begin{subequations}
\begin{equation}
n_{\vec{k},x} \equiv n_{\theta,x}=\cos \theta,
\end{equation}
\begin{equation}
n_{\vec{k},y} \equiv n_{\theta,y}=\sin \theta.
\end{equation}
\end{subequations} 
The Fermi-surface integrations reduce to 
\begin{equation}\label{eq:S_F_int_theta}
\int_{S_F} \frac{d S}{S_F}=\int_0^{2 \pi} \frac{k_F d\theta}{2 \pi k_F}=\int_0^{2 \pi}\frac{d \theta}{2 \pi}. 
\end{equation}
The distribution function $\left|\psi_{\hat{k}} \right\rangle \equiv \left|\psi_{\theta} \right\rangle$ follows: 
\begin{multline}\label{eq:psi_theta_circ}
\left|\psi_{\theta} \right\rangle=-\frac{e}{k_B T} \frac{E_y(q_x,\omega) v_F \sin \theta}{-i \omega +\gamma_{\rm mc} +i v_F q_x \cos\theta} \\ \times \frac{1}{1-2 \delta \gamma G_0(q_x,\omega)},
\end{multline}
where we have
\begin{multline}\label{eq:B}
G_0(q,\omega)\equiv\left\langle \left\langle \hat{v}_{\vec{k},y}^2 \right\rangle \right \rangle \\ = \int_0^{2 \pi} \frac{d \theta}{2 \pi}\frac{(\sin\theta)^2}{-i\omega+\gamma_{\rm mc}+i v_F q \cos \theta} \\ =\frac{-\gamma_{\rm mc} + i \omega +\sqrt{(\gamma_{\rm mc}- i \omega)^2 +(v_F q)^2}}{(v_F q )^2}. 
\end{multline}
in accordance with Eq.\@ (\ref{eq:G_0_iso}).
The full distribution function $\delta f_{\vec{k}} (\vec{q},\omega)$ is then \begin{widetext}
\begin{equation}\label{eq:deltaf_iso}
\delta f_{\vec{k}}=-k_B T \frac{\partial f^{0}(\epsilon_{\vec{k}})}{\partial \epsilon_{\vec{k}}} \left|\psi_{\theta} \right\rangle = e E_y \frac{\partial f^{0}(\epsilon_{\vec{k}})}{\partial \epsilon_{\vec{k}}} \frac{ v_F \sin \theta}{ -i \omega +\gamma_{\rm mc} + i v_F q_x \cos \theta -2 \delta \gamma (-i \omega +\gamma_{\rm mc} + i v_F q_x \cos \theta) G_0(q_x,\omega)}.
\end{equation}
\end{widetext}
Using Eqs.\@ (\ref{eq:Omegap_c_i_expl}), (\ref{eq:psi_theta_circ}), (\ref{eq:sigma_yy}) and (\ref{eq:B}), we obtain the transverse conductivity 
\begin{multline}\label{eq:sigmayy_iso}
\sigma_{yy}\left(q_x,\omega\right)=2 \epsilon_0 \omega_p^2 \\ \times \frac{1}{-i \omega+2 \gamma_{\rm mr}-\gamma_{\rm mc}+\sqrt{(v_F q_x)^2-(\omega+i \gamma_{\rm mc})^2}},
\end{multline}
which simplifies in the limit $\delta\gamma=0$ to 
\begin{equation}\label{eq:sigmayy_iso_gamma}
\sigma_{yy}\left(q_x,\omega\right)=2 \epsilon_0 \omega_p^2 \int\frac{d\theta}{2\pi}\frac{\sin^{2}\theta}{-i\omega+\gamma+iv_{F}q_x\cos\theta},
\end{equation}
consistently with the literature \cite{Cook-2019,Khoo-2021}. Equivalently, Eqs.\@ (\ref{eq:deltaf_iso}) and (\ref{eq:sigmayy_iso}) can be derived by recognizing that the eigenfunctions $\chi_{\vec{k},m}$ for isotropic Fermi surfaces are the angular momentum states $\chi_{\vec{k},m}=c_0 e^{i m \theta}$ \cite{Cook-2019}; see the Supplemental Material for the alternative derivation. 

In particular, in the limit of large $\gamma_{\rm mc}$ we retrieve the hydrodynamic result \cite{Cook-2019,Levchenko-2020}
\begin{equation}
\sigma_{yy}\left(q_x,\omega\right)=\frac{\epsilon_0 (\omega_{p})^2}{-i\omega+\gamma_{\rm mr}+\left(v_{F}q_x\right)^{2}/\left(4\gamma_{\rm mc}\right)}.
\end{equation}

\subsection{Hexagonal Fermi surface}\label{Hex_FS}

\begin{figure}[ht]
\includegraphics[width=0.8\columnwidth]{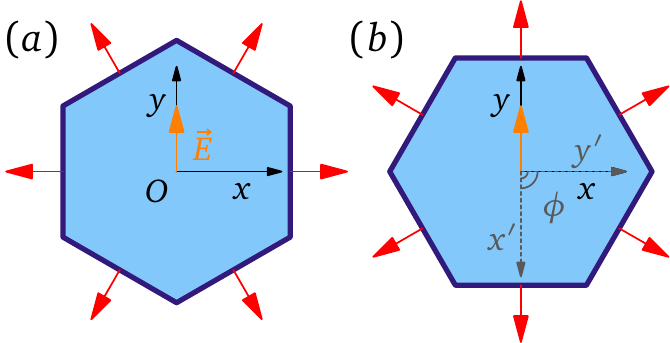}
\caption{\label{fig:Hexagon} Schematic representation of the hexagonal Fermi-surface geometry with electric field $\vec{E}=E_y \hat{u}_y$ aligned with the $y$ axis. Red arrows show the local Fermi velocity vectors $\vec{v}_F=v_F \vec{n}\left(\theta\right)$. (a) ``Parallel'' configuration with two hexagon faces aligned with the $y$ axis. (b) Configuration obtained by rotating the hexagon in panel (a) by an angle $\phi=\pi/2$, formed by the axes $\vec{x}'=\left\{x',y'\right\}$ with respect to the frame $\vec{x}=\left\{x,y\right\}$ in which the field $\vec{E}=E_y \hat{u}_y$. 
}
\end{figure}

We consider a perfectly hexagonal Fermi surface. This seems to be a good approximation for the 2D conducting Pd planes of PdCoO$_{2}$, while it neglects the slight Fermi-surface warping in the direction orthogonal to the planes \cite{Baker-2022_preprint}; see Fig.\@ \ref{fig:Coherence}(c). The in-plane approximately hexagonal shape is supported by the careful analysis of Ref.\@ \onlinecite{Bachmann-2021}, which shows that the directional distribution of Fermi-velocity unit vectors is strongly peaked in six directions. Moreover, bandstructure calculations fitted to angle-resolved photoemission data supports the view that the Fermi velocity is essentially constant within given sectors of the hexagon \cite{Nandi-2018}. The unphysical consequences of the discontinuities in the velocity derivative at the hexagon edges may be removed by rounding the edges themselves with smooth arcs, with a however small radius of curvature \cite{Cook-2019}, but we will relegate this refinement to section \ref{PdCoO2_2D}. 

\subsubsection{Reference geometry: parallel configuration}\label{Hex_par}

It is again convenient to parametrize the Fermi-surface integrations in terms of an angle $\theta \in \left[0,2 \pi\right]$. The normal vector $\vec{n}_{\vec{k}}=n_{\vec{k}}(\theta)$ is normalized as $\hat{n}_{\vec{k}}=\vec{n}_{\vec{k}}/n_{\vec{k}}$ (for a general definition, see App.\@ \ref{Curvature_2D}), and its Cartesian components are 
\begin{subequations}
\begin{equation}
n_{\vec{k},x} \equiv n_{x,\theta},
\end{equation}
\begin{equation}
n_{\vec{k},y} \equiv n_{y,\theta}.
\end{equation}
\end{subequations}
The vectors $n_{x,\theta}$ and $n_{y,\theta}$, pointing towards the Fermi velocity, assume different lengths depending on the orientation of the hexagon. From Eq.\@ (\ref{eq:mirror_planes}), we then have
\begin{equation}\label{eq:u_kx_ky_theta}
\hat{n}_{\vec{k}}=\hat{n}(\theta)=n_{x,\theta} \hat{u}_x+n_{y,\theta} \hat{u}_y.
\end{equation}
For a crystal that is perfectly aligned with the surface --- i.e.\@, $k_y$ and $k_{x}$ are oriented parallel and perpendicular to the surface respectively --- the unit vector $\hat{n}\left(\theta\right)$ takes the values given in Tab.\@ \ref{tab:v_hex_sample1}.
\begin{table}
	\centering
		\begin{tabular}{| c | c | c | c |} 
 \hline
 $\theta$ & $\left[-\frac{\pi}{6}, \frac{\pi}{6}\right]$ & $\left[\frac{\pi}{6}, \frac{\pi}{2}\right]$ & $\left[\frac{\pi}{2}, \frac{5\pi}{6}\right]$ \\ \hline 
$\hat{n}\left(\theta\right)$ & $\left(1, 0\right)$ & $\left(\frac{1}{2}, \frac{\sqrt{3}}{2}\right)$ & $\left(-\frac{1}{2}, \frac{\sqrt{3}}{2}\right)$ \\ \hline 
$\theta$ & $\left[\frac{5\pi}{6}, \frac{7\pi}{6}\right]$ & $\left[\frac{7\pi}{6}, \frac{3\pi}{2}\right]$ & $\left[\frac{3\pi}{2}, \frac{11\pi}{6}\right]$ \\ \hline 
$\hat{n}\left(\theta\right)$ & $\left(-1, 0\right)$ & $\left(-\frac{1}{2}, -\frac{\sqrt{3}}{2}\right)$ & $\left(\frac{1}{2}, -\frac{\sqrt{3}}{2}\right)$ \\
 \hline
\end{tabular} \caption{\label{tab:v_hex_sample1} Piecewise-constant orientation of the Fermi velocity for a hexagonal Fermi surface with two faces aligned with the direction of the applied electric field; see Fig.\@ \ref{fig:Hexagon}(a). $\hat{n}\left(\theta\right)$ is the unit vector in the velocity direction (locally orthogonal to the Fermi surface). }
\end{table}
This allows us to perform the momentum integrations $\int_{S_F} d S/S_F$ in Eq.\@ (\ref{eq:G_0_iso}) for a function $F(v_{\vec{k}})=F\left(v_{x,\theta},v_{y,\theta}\right)$ as discrete sums over the $N_S=6$ sides of the hexagon, as mentioned earlier in Sec.\@ \ref{Cond_poly}. It works as follows: 
\begin{multline}\label{eq:int_sum_hex}
\int_{S_F}\frac{d S}{S_F} F\left(v_{x,\theta},v_{y,\theta}\right)=\frac{\int_{S_F} n_{\vec{k}}(\theta) d \theta F\left(v_{x,\theta},v_{y,\theta}\right)}{\int_{S_F}n_{\vec{k}}(\theta) d \theta} \\ \equiv \frac{\sum_{i=1}^6 A_t F\left(v_{F}n_{x,i},v_{F}n_{y,i}\right)}{6 A_t} \\ =
\frac{1}{6}\sum_{i=1}^{6}F\left(v_{F}n_{x,i},v_{F}n_{y,i}\right),
\end{multline}
where $\hat{n}\left(\theta\right)\equiv \hat{n}_i=\left(n_{x,i}, n_{y,i}\right)$ are the six unit vectors given in Tab.\@ \ref{tab:v_hex_sample1}. $A_t=\int_{\theta_{i,a}}^{\theta_{i,b}} d\theta n_{\vec{k}}(\theta)$ is the integral of the line element $n_{\vec{k}}$ along each side, corresponding to angles $\theta \in \left[\theta_{i,a}, \theta_{i,b}\right]$; this integral is the same for each side $i=\left\{1,\cdots 6\right\}$ for regular polygons.

Next, we have to compute Eq.\@ (\ref{eq:G_0_iso}). We find
\begin{equation}\label{eq:G_0_hexpar}
G_0(q_x,\omega)=-\frac{2}{(\gamma_{\rm mc}-i \omega) \left(\lambda^2-4\right)},
\end{equation}
where
\begin{equation}\label{eq:lambda_var}
\lambda=\frac{v_F q_x}{\omega +i \gamma_{\rm mc}}.
\end{equation}
Using Eqs.\@ (\ref{eq:G_0_hexpar}) and (\ref{eq:sigma_yy}), we finally have the transverse conductivity
\begin{multline}\label{eq:sigma_yy_hexpar_expl}
\frac{\sigma_{yy}(q_x,\omega)}{\epsilon_0 (\omega_p)^2}=-\frac{4}{(\gamma_{\rm mc}-i \omega) \left(\lambda^2-4\right)} \\ \times \left[1+\frac{4 \delta \gamma}{(\gamma_{\rm mc}-i \omega) \left(\lambda^2-4\right)}\right]^{-1}.
\end{multline} 
Eq.\@ (\ref{eq:sigma_yy_hexpar_expl}) considerably simplifies for $\delta \gamma=0$. In this case, we have
\begin{equation}\label{eq:sigmayy_hex_par_gamma}
\frac{\sigma_{yy}(q_x,\omega)}{\epsilon_0 \omega_p^2}=\frac{4}{(\gamma-i \omega) \left(4-\lambda^2\right)}
\end{equation}
with $\gamma=\gamma_{\rm mc}=\gamma_{\rm mr}$. In particular, in the absence of scattering rates, that is for $\gamma_{\rm mr}=\gamma_{\rm mc}=0$, Eq.\@ (\ref{eq:sigmayy_hex_par_gamma}) translates as the simple result
\begin{equation}\label{eq:sigmayy_hex_par_gamma0}
\frac{\sigma_{yy}(q_x,\omega)}{\epsilon_0 \omega_p^2}=\frac{4 i \omega}{4 \omega^2-(v_F q_x)^2}, 
\end{equation}
which displays a pole at $\omega=\pm v_F q_x /2$. 

Now consider the general case $\gamma_{\rm mr} \neq \gamma_{\rm mc}$. 
In the high-momentum regime $q_x \rightarrow +\infty$, the expansion at leading order of Eq.\@ (\ref{eq:sigma_yy_hexpar_expl}) produces
\begin{equation}\label{eq:sigma_yy_phi0_highq} 
\frac{\sigma_{yy}(q_x,\omega)}{\epsilon_0 \omega_p^2}=4\frac{\gamma_{\rm mc}-i \omega}{(v_F q_x)^2}+o\left[(v_F q_x)^{-4}\right].
\end{equation} 
In the limit $q_x \rightarrow 0$, i.e.\@, for long-wavelength excitations, we have
\begin{multline}\label{eq:sigma_yy_phi0_lowq}
\frac{\sigma_{yy}(q_x,\omega)}{\epsilon_0 (\omega_p)^2}=\frac{1}{\gamma_{\rm mr}-i \omega}+\frac{(v_F q_x)^2}{4 (\gamma_{\rm mc}-i \omega)(\omega+i \gamma_{\rm mr})^2}\\ +o\left[(v_F q_x)^4\right],
\end{multline}
so at zero momentum the conductivity is governed by $\gamma_{\rm mr}$ but not $\gamma_{\rm mc}$. 
Finally, expanding Eq.\@ (\ref{eq:sigma_yy_hexpar_expl}) to linear order in $1/\gamma_{\rm mc} \rightarrow 0^+$, we obtain the conductivity in hydrodynamic regime dominated by momentum-conserving collisions: 
\begin{multline}\label{eq:sigma_yy_phi0_hydro}
\frac{\sigma_{yy}(q_x,\omega)}{\epsilon_0 \omega_p^2}=\frac{1}{\gamma_{\rm mr}-i \omega}-\frac{1}{\gamma_{\rm mc}}\frac{(v_F q_x)^2}{4 (\gamma_{\rm mr}-i \omega)^2} \\+o\left[(\gamma_{\rm mc})^{-2}\right]. 
\end{multline} 

The present configuration, with two hexagonal faces aligned with the $y$ axis, serves as a reference for the analysis of the conductivity for arbitrary Fermi-surface orientation with respect to the applied field: to perform such generalization, it is convenient to introduce rotated Fermi-surface coordinates with respect to the field axis, as described in the next section.

\subsubsection{Arbitrary crystal orientation}\label{Hex_phi}

Let us now analyze arbitrary crystal orientations. Let $\vec{x}=\left(x,y\right)$ refer to the coordinates along and perpendicular to the applied field $\vec{E}(\vec{q},\omega)\equiv E_y(q_x,\omega) \hat{u}_y$. We then call $\vec{x}'=\left(x',y'\right)$ the coordinates that are aligned with the crystalline axes, where $\boldsymbol{x}'=R^{-1}\boldsymbol{x}$ and the 2D rotation matrix $R$ is 
\begin{equation}\label{eq:rotation}
R=\begin{bmatrix}
\cos\phi & -\sin\phi\\
\sin\phi & \cos\phi
\end{bmatrix}
\end{equation}
if the axes $\vec{x}'$ are rotated by an angle $\phi$ with respect to the axes $\vec{x}$; see Fig.\@ \ref{fig:Hexagon}(b). 
This implies that the velocities in the frame aligned with the surface are $\vec{v}_{\vec{k}_F}=R\vec{v}'_{\vec{k}}=v_{F}R\hat{n}\left(\theta\right)$. It follows that the velocities can be written as 
\begin{multline}\label{eq:v_theta_phi}
v_{\vec{k}_F}\equiv\vec{v}\left(\theta,\phi\right)=v_{F}\vec{m}\left(\theta,\phi\right)\\=m_x\left(\theta,\phi\right)\hat{u}_x+m_y\left(\theta,\phi\right)\hat{u}_y
\end{multline}
with rotated unit vector.
The components are given as: 
\begin{eqnarray}\label{eq:v_vect_rot}
m_{x}\left(\theta,\phi\right) & = & \cos\left(\phi\right)n_{x,\theta}-\sin\left(\phi\right)n_{y,\theta},\nonumber \\
m_{y}\left(\theta,\phi\right) & = & \sin\left(\phi\right)n_{x,\theta}+\cos\left(\phi\right)n_{y,\theta}.
\end{eqnarray}
Inserting the parametrization (\ref{eq:v_theta_phi}) into Eq.\@ (\ref{eq:G_0_iso}), and performing the discrete summation according to Eq.\@ (\ref{eq:int_sum_hex}), we obtain 
\begin{multline}\label{eq:G0_phi}
G_0(q_x,\omega,\phi)=\int_{S_F} \frac{d S}{S_F} \frac{\left[m_y(\theta,\phi)\right]^2}{-i \omega +\gamma_{\rm mc} + i v_F q m_x(\theta,\phi)} \\=\frac{1}{\gamma_{\rm mc}-i \omega} \frac{-16+20\lambda^{2}+\lambda^{4}\left[\cos\left(6\phi\right)-5\right]}{-32+48\lambda^{2}-18\lambda^{4}+\lambda^{6}\left[1+\cos\left(6\phi\right)\right]},
\end{multline}
where we have used Eq.\@ (\ref{eq:lambda_var}). 
The transverse conductivity then follows from Eq.\@ (\ref{eq:sigma_yy}), and formally it is
\begin{equation}\label{eq:sigma_yy_hexphi}
\frac{\sigma_{yy}(q_x,\omega)}{2 \epsilon_0 (\omega_p)^2}= \frac{G_0(q_x,\omega,\phi)}{1-2 \delta \gamma G_0(q_x,\omega,\phi)}. 
\end{equation}
 Setting $\phi=0$, we recover of course Eq.\@ (\ref{eq:sigma_yy_hexpar_expl}) for $G_i\left(q_x,\omega,0\right)$. Also, notice that the conductivity has a 6-fold periodicity in rotation angle $\phi$ due to the D$_{3d}$ symmetry. 
The conductivity (\ref{eq:sigma_yy_hexphi}) significantly depends on the orientation angle $\phi$: in particular, we see from Eq.\@ (\ref{eq:G0_phi}) that $G_0\left(q_x,\omega,\phi\right)\propto\lambda^{-2}$ for $\lambda \gg 1$, which is the high-momentum regime where anomalous skin effect takes place; see Secs.\@ \ref{Skin} and \ref{Z_spec}. However, for $\phi=\pi/2$ two corners of the hexagon intersect the $x$ axis and $1+\cos\left(6\phi\right)=0$. Then we obtain $G_0\left(q_x,\omega,\pi/2\right)=\left[3(\gamma_{\rm mc}-i \omega)\right]^{-1}+{o}\left(\lambda^{-2}\right)$ for $\lambda \gg 1$. Thus, we do expect a qualitative difference between the orientations $\phi=0$ and $\phi=\pi/2$ in the regime of anomalous skin effect.  

For $\phi=\pi/2$ and $\gamma_{\rm mr}=\gamma_{\rm mc}=\gamma$, we obtain for the conductivity 
\begin{equation}\label{eq:sigma_yy_hex_90_gamma}
\frac{\sigma_{yy}\left(q_{x},\omega,\pi/2 \right)}{\epsilon_0 \omega_{p}^{2}}= \frac{4 (\gamma-i \omega)^2+2(v_F q_x)^2}{(\gamma-i \omega)\left[4(\gamma-i \omega)^2+3 (v_F q_x)^2\right]},
\end{equation}
which is manifestly different from the $\phi=0$ case (\ref{eq:sigmayy_hex_par_gamma}). In particular, for $\gamma=0$ 
\begin{equation}\label{eq:sigma_yy_hex_90_0}
\frac{\sigma_{yy}\left(q_{x},\omega,\pi/2 \right)}{\epsilon_0 \omega_{p}^{2}}= \frac{2 i}{\omega} \frac{2 \omega^2-(v_F q_x)^2}{4 \omega^2-3 (v_F q_x)^2},
\end{equation}
that has a pole for $\omega=\pm \sqrt{3/4} v_F q_x $. 

We now turn to the general case $\delta \gamma \neq 0$. Remarkably, the differences between different orientations of the hexagonal Fermi surface actually vanish both in the low-momentum regime $q_x\rightarrow 0^+$ to order $q_x^2$ and in the hydrodynamic limit $\gamma_{\rm mc}\rightarrow +\infty$. In fact, expanding Eq.\@ (\ref{eq:sigma_yy_hexphi}) for a generic angle $\phi$ to order $q_x^2$, we exactly obtain Eq.\@ (\ref{eq:sigma_yy_phi0_lowq}), as in the $\phi=0$ case. Similarly, expanding Eq.\@ (\ref{eq:sigma_yy_hexphi}) to linear order in $1/\gamma_{\rm mc} \rightarrow 0^+$, we obtain Eq.\@ (\ref{eq:sigma_yy_phi0_hydro}) for any $\phi$. Hence, the conductivity does not depend on orientation in the low-momentum and hydrodynamic regimes, for a hexagonal Fermi surface. In fact, the independence from $\phi$ in hydrodynamic regime can be also inferred from the structure of the Navier-Stokes equations with D$_{3d}$ symmetry \cite{Cook-2019}; see Sec.\@ III in the Supplemental Material.  

The conductivity does show orientational dependence in the high-momentum regime $q_x\rightarrow +\infty$: the leading-order expansion of Eq.\@ (\ref{eq:sigma_yy_hexphi}) for generic angle $\phi \neq \alpha \pi/6$, $\alpha \in \mathbb{Z}$ gives 
\begin{multline}\label{eq:sigma_yy_hex_phi_highq} 
\frac{\sigma_{yy}\left(q_{x},\omega,\phi \right)}{\epsilon_0 \omega_{p}^{2}}=\frac{(\gamma_{\rm mc}-i \omega)\left[5-\cos(6 \phi)\right]}{\left[v_F q_x \cos(3 \phi)\right]^2} \\ +o\left[(v_F q_x)^{-4}\right],
\end{multline} 
which correctly reduces to Eq.\@ (\ref{eq:sigma_yy_phi0_highq}) for $\phi=0$. However, Eq.\@ (\ref{eq:sigma_yy_hex_phi_highq}) diverges for $\phi=\pi/2$, which is an indication of the special nature of conduction and skin effect in this configuration, as we will analyze in Sec.\@ \ref{Skin}. To avoid this apparent divergence, we set $\phi=\pi/2$ in Eq.\@ (\ref{eq:sigma_yy_hexphi}) before performing the expansion for $q_x\rightarrow +\infty$, with the leading-order result
\begin{multline}\label{eq:sigma_yy_hex_90_highq}
\frac{\sigma_{yy}\left(q_{x},\omega,\pi/2 \right)}{2\epsilon_0 \omega_{p}^{2}}=\frac{2}{\gamma_{\rm mc}+2 \gamma_{\rm mr}-i \omega} \\ +\frac{(4 \gamma_{\rm mc}-i \omega)^3}{(\gamma_{\rm mc}+2 \gamma_{\rm mr}-3 i \omega)^2 (v_F q)^2} +o\left[(v_F q_x)^{-4}\right].
\end{multline}
In the large-momentum regime, there is a qualitative difference in the conductivity, depending on the presence or absence of scattering: the limits $\omega \rightarrow 0$ and $\gamma \rightarrow 0$ do not commute. This is most evident in the limit $\delta \gamma =0$ for Eq.\@ (\ref{eq:sigma_yy_hex_90_highq}), which yields
\begin{equation}\label{eq:sigma_yy_hex_90_highq_gamma}
\frac{\sigma_{yy}\left(q_{x},\omega,\pi/2 \right)}{\epsilon_0 \omega_{p}^{2}}=\frac{2}{3(\gamma-i \omega)}. 
\end{equation}
Then, we see a qualitative difference in the presence or absence of scattering: for $\gamma \ll \omega$, the conductivity (\ref{eq:sigma_yy_hex_90_highq_gamma}) is inversely proportional to $\omega$ for $\gamma_{\rm mr}=\gamma_{\rm mc}=0$; on the contrary, if $\gamma \gg \omega$, the conductivity is approximately frequency-independent. A similar dichotomy appears for the more general Eq.\@ (\ref{eq:sigma_yy_hex_90_highq}) with arbitrary $\gamma_{\rm mr}\leq \gamma_{\rm mc}$.
This difference will be at the origin of a qualitative distinction for the skin depths in anomalous regime, in the absence or presence of scattering; see Sec.\@ \ref{Skin_hex}.

To investigate how the anisotropic electrodynamics depends on the specific Fermi-surface shape, in the next section we analyze another exemplary case, namely the square point group $D_4$. 

\subsection{Square Fermi surface}\label{Square_FS}

To get a better intuition for the origin of the rather distinct behavior of the two orientations of the sample edge relative to the crystalline axis, we consider a square Fermi surface, as shown in Fig.\@ \ref{fig:Square}. The procedure is identical to the one for the hexagonal Fermi surface: we assume that the Fermi velocity is constant in the four segments. 

\subsubsection{Reference geometry: diamond-shaped configuration}\label{Square_dia}

\begin{figure}[ht]
\includegraphics[width=0.8\columnwidth]{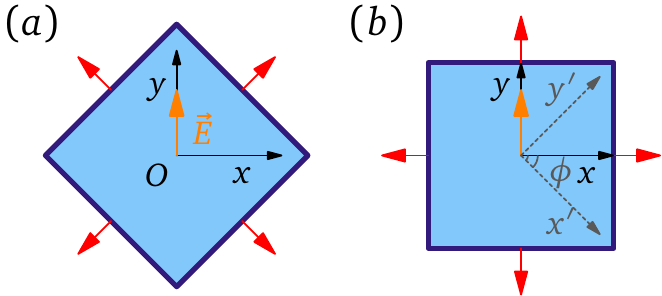}
\caption{\label{fig:Square} Schematic representation of the square Fermi-surface geometry with electric field $\vec{E}=E_y \hat{u}_y$ aligned with the $y$ axis of the frame $\vec{x}=\left\{x,y\right\}$ . Red arrows show the local Fermi velocity vectors $\vec{v}_F=v_F \hat{n}\left(\theta\right)$. (a) ``Diamond-shaped'' configuration with vertices intersecting the $x$ and $y$ axes. (b) Configuration obtained by rotating the square in panel (a) by an angle $\phi=\pi/4$, formed by the axes $\vec{x}'=\left\{x',y'\right\}$ with respect to the frame $\vec{x}$.
}
\end{figure}

In analyzing the diamond-shaped case, we again use the parametrization (\ref{eq:u_kx_ky_theta}) in terms of $\theta$. 
With the piecewise approximation for the Fermi velocity, the angular integrations over the Fermi surface in Eq.\@ (\ref{eq:G_0_iso}) reduce to the discrete sum
\begin{equation}\label{eq:int_sum_square}
\int_{S_F}\frac{d S}{S_F}F\left(v_{x}\left(\theta\right),v_{y}\left(\theta\right)\right)=\frac{1}{4}\sum_{i=1}^{4}F\left(v_{F}n_{x,i},v_{F}n_{y,i}\right),
\end{equation}
where $\hat{n}$ again satisfies Eq.\@ (\ref{eq:u_kx_ky_theta}). Eq.\@ (\ref{eq:int_sum_square}) for the square Fermi surface is analogous to its counterpart (\ref{eq:int_sum_hex}) for the hexagonal geometry. 
For the geometry in which each pair of vertices of the square intersect the $x$ and $y$ axis, as in Fig.\@ \ref{fig:Square}(a), the unit vectors $\hat{n}(\theta)$ are collected in Tab. \ref{tab:v_dia_sample1}. 
\begin{table}
	\centering
		\begin{tabular}{| c | c | c | c | c |} 
 \hline
 $\theta$ & $\left[0, \frac{\pi}{2}\right]$ & $\left[\frac{\pi}{2}, \pi\right]$ & $\left[\pi, \frac{3\pi}{2}\right]$ & $\left[\frac{3\pi}{2}, 2 \pi \right]$ \\ \hline 
$\hat{n}\left(\theta\right)$ & $\frac{1}{\sqrt{2}}\left(1, 1\right)$ & $\frac{1}{\sqrt{2}}\left(-1, 1\right)$ & $\frac{1}{\sqrt{2}}\left(-1, -1\right)$ & $\frac{1}{\sqrt{2}}\left(1, -1\right)$ \\ \hline 
\end{tabular} \caption{\label{tab:v_dia_sample1} Piecewise-constant orientation of the Fermi velocity for a square Fermi surface with two vertices intersecting the $x$ axis, and two intersecting the $y$ axis; see Fig.\@ \ref{fig:Square}(a). $\hat{n}\left(\theta\right)$ is the unit vector in the velocity direction (locally orthogonal to the Fermi surface).}
\end{table}

The angular sum in Eq.\@ (\ref{eq:G_0_iso}) leads to 
\begin{equation}\label{eq:G_0_dia}
G_0(q_x,\omega)=-\frac{2}{(\gamma_{\rm mc}-i \omega) \left(2-\lambda^2\right)}. 
\end{equation}
Eqs.\@ (\ref{eq:G_0_dia}) and (\ref{eq:sigma_yy}) yield the transverse conductivity
\begin{multline}\label{eq:sigma_yy_dia_expl}
\frac{\sigma_{yy}(q_x,\omega)}{\epsilon_0 (\omega_p)^2}=\frac{2}{(\gamma_{\rm mc}-i \omega) (2 -\lambda^2) } \\ \times \left\{1-\frac{2 \delta \gamma}{(\gamma_{\rm mc}-i \omega) (2-\lambda^2)}\right\}^{-1}.
\end{multline}
Let us first analyze the conductivity for equal momentum-relaxing and momentum-conserving scattering rates. For $\delta \gamma=0$, Eq.\@  (\ref{eq:sigma_yy_dia_expl}) simplifies to  
\begin{equation}\label{eq:sigmayy_square_dia_gamma}
\frac{\sigma_{yy}(q_x,\omega)}{\epsilon_0 \omega_p^2}=\frac{2}{(\gamma-i \omega) \left[2-\left(\frac{ v_F q_x}{\omega+i \gamma}\right)^2\right]},
\end{equation}
which in the scattering-less limit $\gamma_{\rm mr}=\gamma_{\rm mc}=0$ becomes simply
\begin{equation}\label{eq:sigmayy_square_par_gamma0}
\frac{\sigma_{yy}(q_x,\omega)}{\epsilon_0 \omega_p^2}=\frac{2 i}{\omega \left[2-\left(\frac{v_F q_x}{\omega}\right)^2\right]}
\end{equation}
with a pole at $\omega=\pm v_F q_x /\sqrt{2}$. 

Let us now focus on $\gamma_{\rm mr} \neq \gamma_{\rm mc}$. In the low-momentum regime $q_x\rightarrow 0^+$, an expansion of the conductivity (\ref{eq:sigma_yy_dia_expl}) to order $q_x^2$ gives
\begin{multline}\label{eq:sigmayy_square_lowq}
\frac{\sigma_{yy}(q_x,\omega)}{\epsilon_0 \omega_p^2}=\frac{1}{\gamma_{\rm mr}-i \omega}+\frac{(v_F q)^2}{2 (\gamma_{\rm mc}-i \omega) (i \gamma_{\rm mr}+\omega)^2} \\ +o[(v_F q_x)^4].
\end{multline}
The extreme hydrodynamic limit is obtained by expanding Eq.\@ (\ref{eq:sigma_yy_dia_expl}) to order $1/\gamma_{\rm mc}$:
\begin{multline}\label{eq:sigmayy_square_hydro}
\frac{\sigma_{yy}(q_x,\omega)}{\epsilon_0 \omega_p^2}=\frac{1}{\gamma_{\rm mr}-i \omega} -\frac{(v_F q)^2}{2 \gamma_{\rm mc} (\gamma_{\rm mr}-i \omega)^2}\\ +o[(\gamma_{\rm mc})^{-2}]. 
\end{multline}
In the high-momentum regime $q_x\rightarrow +\infty$, the conductivity (\ref{eq:sigma_yy_dia_expl}) gives at leading order
\begin{equation}\label{eq:sigmayy_square_highq}
\frac{\sigma_{yy}(q_x,\omega)}{\epsilon_0 \omega_p^2}=2\frac{\gamma_{\rm mc}-i \omega}{(v_F q_x)^2}+o[(v_F q_x)^{-4}],
\end{equation}
which differs only by a factor of $2$ from the hexagonal result (\ref{eq:sigma_yy_phi0_highq}). 

In the next section we generalize the above results to an arbitrary orientation of the square Fermi surface with respect to the applied field, in the same way as done in Sec.\@ \ref{Hex_phi} for the hexagonal Fermi surface. This analysis will allow us to draw general criteria to link the form of the transverse conductivity for polygonal geometries to the given Fermi-surface orientation. 

\subsubsection{arbitrary crystal orientation}\label{Square_phi}

For an arbitrary orientation we rotate the velocities by the angle $\phi$, in accordance with Eq.\@ (\ref{eq:v_theta_phi}), analogously to the case of a hexagonal Fermi surface in Sec.\@ \ref{Hex_phi}. Then, the velocity vectors in the frame aligned with the applied field are given by Eq.\@ (\ref{eq:v_vect_rot}), and the conductivity follows from Eq.\@ (\ref{eq:sigma_yy_hexphi}). The only difference with respect to the hexagonal case is that the expression for $G_0(q_x,\omega,\phi)$, with $\phi \in \left[0, 2 \pi\right]$ and $\gamma_{\rm mr}\neq \gamma_{\rm mc}$, is modified for a square Fermi surface:
\begin{equation}\label{eq:G0_phi_sq}
G_0(q_x,\omega,\phi)=\frac{1}{\gamma_{\rm mc}-i \omega} \frac{4 -\lambda^2\left[-3+\cos\left(4\phi\right)\right]}{8(1-\lambda^2)+\lambda^{4}\left[1+\cos\left(4\phi\right)\right]},
\end{equation}
where we have used the variable (\ref{eq:lambda_var}). Setting $\phi=0$ in Eqs.\@ (\ref{eq:G0_phi_sq}) we retrieve Eq.\@ (\ref{eq:sigma_yy_dia_expl}).

There is a qualitative difference between square and hexagonal conductivities in the $q_x \rightarrow 0^+$ limit. In fact, in the low-momentum regime and the extreme hydrodynamic limit, the conductivity for a square Fermi surface displays significant orientational dependence. This is best seen by expanding the conductivity (\ref{eq:sigma_yy_hexphi}) for generic $\phi$ to order $q_x^2$: 
\begin{multline}\label{eq:sigmayy_square_phi_lowq}
\frac{\sigma_{yy}(q_x,\omega)}{\epsilon_0 \omega_p^2}=\frac{1}{\gamma_{\rm mr}-i \omega}+\frac{\left[v_F q \cos(2 \phi)\right]^2}{2 (\gamma_{\rm mc}-i \omega)(i \gamma_{\rm mr}+\omega)^2}\\ +o\left[(v_F q)^4\right].
\end{multline}
In particular, an expansion of Eq.\@ (\ref{eq:sigma_yy_hexphi}) to order $1/\gamma_{\rm mc}$ (hydrodynamic regime) yields
\begin{multline}\label{eq:sigmayy_square_phi_hydro}
\frac{\sigma_{yy}(q_x,\omega)}{\epsilon_0 \omega_p^2}=\frac{1}{\gamma_{\rm mr}-i \omega}+\frac{\left[v_F q \cos(2 \phi)\right]^2}{2 \gamma_{\rm mc}(i \gamma_{\rm mr}+\omega)^2}\\ +o\left[(v_F q)^4\right].
\end{multline}
The dependence on $\phi$ shown by Eqs.\@ (\ref{eq:sigmayy_square_phi_lowq}) and (\ref{eq:sigmayy_square_phi_hydro}) is in stark contrast with the respective counterparts (\ref{eq:sigma_yy_phi0_lowq}) and (\ref{eq:sigma_yy_phi0_hydro}) for a hexagonal Fermi surface, which do not depend on orientation. Hence, we expect an angle-dependent electrodynamics in hydrodynamic regime for a square-shaped Fermi surface, but not for a hexagonal one; see also Sec.\@ III in the Supplemental Material.

In the high-momentum regime, Eq.\@ (\ref{eq:sigma_yy_hexphi}) becomes
\begin{multline}\label{eq:sigmayy_square_phi_highq}
\frac{\sigma_{yy}(q_x,\omega)}{\epsilon_0 \omega_p^2}=\frac{(\gamma_{\rm mc}-i \omega)\left[3-\cos(4 \phi)\right]}{\left[v_F q_x \cos(2 \phi)\right]^2} \\+o\left[(v_F q_x)^{-4}\right], 
\end{multline}
for generic angle $\phi \neq \alpha \pi/4$, $\alpha \in \mathbb{Z}$, which correctly gives Eq.\@ (\ref{eq:sigmayy_square_highq}) for $\phi=0$. 
The expression (\ref{eq:sigmayy_square_phi_highq}) diverges for $\phi=\pi/4$, that is for two square faces parallel to the applied field, in a similar way as what found for Eq.\@ (\ref{eq:sigma_yy_hex_phi_highq}), $\phi=\pi/2$ in the hexagonal case.

The conductivity for $\phi=\pi/4$, which corresponds to Fig.\@ \ref{fig:Square}(b), displays a remarkable property: it is \emph{identically} equivalent to the one of the Drude model, for a local conductor with relaxation rate $\gamma_{\rm mr}$:
\begin{equation}\label{eq:sigma_yy_sq_pi4_gamma}
\frac{\sigma_{yy}\left(q_{x},\omega\right)}{\epsilon_0 \omega_p^2}\equiv\frac{1}{\gamma_{\rm mr}-i \omega},
\end{equation}
independently from $q_x$ and $\gamma_{\rm mc}$. This result is of course dramatically different from the $\phi=0$ case of Eq.\@ (\ref{eq:sigma_yy_dia_expl}): for $\phi=\pi/4$, the non-local character of the transverse conductivity completely disappears. 

The appearance of perfect conductivities for specific orientations is actually linked with large Fermi-surface segments parallel to the electric field for polygonal geometries, as detailed in the following section. 

\subsection{Comparison between polygonal geometries}\label{Cond_poly_sigma}

The expressions for the conductivity found in Secs.\@ \ref{Hex_FS}-\ref{Square_phi} give us insight into how sensitive different anisotropic Fermi-surface geometries are to the orientation of the applied electric field. 
Both in the regime $q\rightarrow 0^+$ and $\gamma_{\rm mc}\rightarrow +\infty$ (hydrodynamic limit), the leading-order expansions of the conductivity do not depend on orientation for a hexagonal Fermi surface; see Eqs.\@ (\ref{eq:sigma_yy_phi0_lowq}) and (\ref{eq:sigma_yy_phi0_hydro}). Hence, we expect no variation of the electrodynamic properties with angle $\phi$ in this case. On the contrary, for a square geometry Eqs.\@ (\ref{eq:sigmayy_square_phi_lowq}) and (\ref{eq:sigmayy_square_phi_hydro}) predict a significant angular dependence of the low-momentum and hydrodynamic transverse conductivity. 

Let us now turn to the high-momentum regime characteristic of anomalous skin effect. Comparing Figs.\@ \ref{fig:Hexagon}(a) and \ref{fig:Square}(b), one finds that the common aspect is the propagation of electrons parallel to the surface, on a large portion of the Fermi surface. These flat segments, with velocity parallel to the current, give rise to a momentum-independent conductivity. To see this, consider the expression (\ref{eq:lonloc_sigma_T}) for the 2D conductivity with $\gamma_{\rm mr}=\gamma_{\rm mc}=0$, $\alpha=x$ and $\beta=y$: 
\begin{equation}\label{eq:sigma_yy_gamma0_gen}
\sigma_{yy}\left(q_{x},\omega\right)=\frac{e^{2}}{-i\omega}\int_{\vec{k}}\frac{v_{\vec{k},y}^{2}}{1-\frac{q_{x}}{\omega}v_{\vec{k},x}}.
\end{equation}
For a piecewise-constant Fermi surfaces, the momentum integral in Eq.\@ (\ref{eq:sigma_yy_gamma0_gen}) depends on $r=v_F q_x/\omega$ and it usually decays as $r^{-2}$, while for a spherical Fermi surface it decays like $r^{-1}$; see Eq.\@ (\ref{eq:sigmayy_iso_gamma}). These different power laws determine the difference in the $\omega$-evolution of the conductivities for polygonal and circular Fermi surfaces. However, if there is a flat piece of the Fermi surface with velocity $\vec{v}_{\vec{k}}\parallel\hat{u}_{y}$ parallel to the applied field, then $\vec{v}_{\vec{k},x}=0$ and the dependence on $q_x$ disappears from the conductivity (\ref{eq:sigma_yy_gamma0_gen}) even at finite momentum, as exemplified by Eq.\@ (\ref{eq:sigma_yy_sq_pi4_gamma}) for a square Fermi surface. 
Ultimately, this is due to the fact that we are looking at a transverse conductivity with momentum (decay mode) direction $\vec{q}\parallel \hat{u}_{x}$,  orthogonal to the direction of the current $\vec{J}\parallel \hat{u}_{y}$. 

The presence of Fermi-surface segments with velocity parallel to the current influences all the electrodynamic properties of the anisotropic system through the conductivity. To investigate this impact, in the next sections we apply our kinetic theory for the electrodynamic response to the calculation of the skin depth and the surface impedance. 

\section{Polaritons and skin depth}\label{Skin}

The dependence of the anisotropic conductivity on momentum and on orientation qualitatively has an immediate measurable consequence on the damping of electromagnetic fields inside the metal, i.e.\@, skin effect. To investigate the spatial profile of the electric field $E_y(q_x,\omega)$, we start from the electromagnetic wave equation \cite{Reuter-1948,Dressel-2001,Tanner-2019opt}
\begin{equation}\label{eq:wave}
\nabla^2 \vec{E}(\vec{r},\omega)+\frac{\omega^2}{c^2} \vec{E}(\vec{r},\omega) =-\mu_0 i \omega \vec{J}(\vec{r},\omega), 
\end{equation}
stemming from Maxwell's equations for a medium with no bound charges or magnetization. Eq.\@ (\ref{eq:wave}) is supplemented by the non-local relation between the free current density and the electric field, i.e.\@, the generalized Ohm's law
\begin{equation}\label{eq:Ohm_gen}
J_\alpha(\vec{r},\omega)=\int d \vec{r}' \sigma_{\alpha \beta}(\vec{r}-\vec{r}',\omega) E_\beta(\vec{r}',\omega),
\end{equation}
which provides a definition of the rank-2 conductivity tensor $\sigma_{\alpha \beta}(\vec{r}-\vec{r}',\omega)$. 
Consider a geometry where the crystal is cut along the $y z$ plane, i.e.\@, vacuum occupies $x < 0$ while our material takes the other half space $x > 0$. We again choose a polarization of the electric field along the y direction, $\vec{E}(x,\omega) \equiv E(x,\omega) \hat{u}_y$, which varies in space along the x direction. Notice that these coordinates do not have to be aligned with any of the crystalline axes. Hence, the wave equation (\ref{eq:wave}) and the linear-response relation (\ref{eq:Ohm_gen}) yield
\begin{multline}\label{eq:wave_y_compl}
\frac{\partial^2 E_y(x,\omega)}{\partial x^2}+\frac{\omega^2}{c^2} E_y(x,\omega) \\  =-\mu_0 i \omega \int d x' \sigma_{yy}(x-x',\omega) E_y(x',\omega). 
\end{multline}
It is convenient to Fourier-transform the non-local wave equation (\ref{eq:wave_y_compl}) to reciprocal space of momenta $q_x$, assuming the oscillatory evolution $E_y(x,\omega)=E_0 e^{i (q_x x -\omega t)}$, where $E_0=E_y(0,0)$. The current density also follows the same oscillatory pattern. Then, we have
\begin{multline}\label{eq:wave_y_compl_q}
-q_x^2 E_y(q_x,\omega)+\frac{\omega^2}{c^2} E_y(q_x,\omega) \\ =-\mu_0 i \omega \sigma_{yy}(q_x,\omega) E_y(q_x,\omega). 
\end{multline}
Using the relation between the dielectric function $\epsilon_{yy}(q_x,\omega)$ and the conductivity, $\sigma_{yy}(q_x,\omega)=-i \epsilon_0 \omega \left[\epsilon_{yy}(q_x,\omega)-1\right]$, we can implicitly solve Eq.\@ (\ref{eq:wave_y_compl_q}) for \emph{complex-valued} momentum $q_x \in \mathbb{C}$, and obtain
\begin{equation}\label{eq:qx_sq_gen}
q_x^2=\frac{\omega^2}{c^2} \left[ 1+\frac{i}{ \epsilon_0 \omega} \sigma_{yy}(q_x,\omega)\right] \equiv \frac{\omega^2}{c^2}\epsilon_{yy}(q_x,\omega). 
\end{equation}
In essence, Eq.\@ (\ref{eq:qx_sq_gen}) determines the self-consistent electromagnetic field inside the metal, i.e.\@ the polaritons. In local metals, where $\sigma_{yy}(q_x,\omega)$ is approximated by its value at zero-momentum $\sigma_{yy}(0,\omega)\equiv \sigma_{yy}(\omega)$, there is only one polariton mode satisfying Eq.\@ (\ref{eq:qx_sq_gen}): this is the standard case, whereby Fresnel's laws of refraction completely determine reflection and transmission at a boundary. However, when non-locality in the electrodynamic response is non-negligible, there can be multiple polariton branches which satisfy the self-consistent relation (\ref{eq:qx_sq_gen}). For instance, the transverse dielectric function of isotropic viscous charged fluids \cite{FZVM-2014,Levchenko-2020}, and of Fermi liquids at leading order in the expansion at low momenta $\omega \gg v_F q$ \cite{Valentinis-2021a,Valentinis-2021b}, give rise to two degenerate polaritons for each $\omega$. 

At frequencies $\omega \lessapprox \omega_p$ we can neglect the displacement-current term in Eq.\@ (\ref{eq:wave_y_compl_q}) (second term on the left-hand side) \cite{Dressel-2001,Valentinis-2021a}, so that the polariton momentum is
\begin{equation}\label{eq:qx_sq_lowom}
q_x^2=i \mu_0 \omega \sigma_{yy}(q_x,\omega). 
\end{equation}
If we explicitly write the dispersion and attenuation of the electric field as $E_y(x,\omega)\equiv E_0 e^{i (\mathrm{Re}\left\{ q_x\right\} x-\omega t)} e^{-x/\delta_s}$ for a given polariton branch stemming from Eqs.\@ (\ref{eq:qx_sq_gen}) or (\ref{eq:qx_sq_lowom}), we can identify a skin depth 
\begin{equation}\label{eq:skin_depth_mode}
\delta_s=\delta_s(\omega)=\frac{1}{\mathrm{Im}\left\{q_x (\omega)\right\}}
\end{equation}
for the given polariton mode \cite{Dressel-2001}. When $N_p$ branches $E_{y,\alpha}(x,\omega)$, $\alpha=\left\{1, \cdots N_p\right\}$ are present, the electric field that propagates inside the metal is a coherent superposition of all $N_p$ modes. The resulting spatial profile $E_y(x,\omega)$ may not be a simple damped exponential as a function of $x$, and may even exhibit interference patterns when the individual intensities $\left|E_{y,\alpha}(x,\omega)\right|^2$ of some modes are comparable \cite{FZVM-2014,Valentinis-2021a}. Nevertheless, we can always identify a skin depth for each mode $\alpha$, and the damping of $E_y(x,\omega)$ at the largest depths is determined by the mode which possesses the largest skin depth $\delta_{s,\alpha}$. 

Using the definition (\ref{eq:skin_depth_mode}), in the next sections we proceed to extract a value for the skin depth, at low and high frequencies compared to $v_F q_x$, for the hexagonal- and square-shaped Fermi surfaces studied in Sec.\@ (\ref{Cond_poly}). We will work in the absence of scattering for simplicity, to illustrate the connections among the conductivity, the skin depth and the impedance. The generalization to finite $\gamma_{\rm mc}$ and $\gamma_{\rm mr}$ is straightforward, using the results of Secs.\@ \ref{Hex_FS} and \ref{Square_FS}, and only the results for $\delta_s(\omega)$ that qualitatively affect the dependence on $\omega$ will be explicitly given. In general, the dominance of momentum-relaxing scattering $\gamma_{\rm mr} >0$ or of collisions $\gamma_{\rm mc}>0$ enables two additional regimes, those of normal (Ohmic) skin effect and hydrodynamic skin effect, where $\delta_s(\omega)\propto \omega^{-1/2}$ and $\delta_s(\omega) \propto \omega^{-1/4}$ \cite{Levchenko-2020,Valentinis-2021a}, respectively. A more comprehensive analysis with scattering is provided in terms of the surface impedance in Appendix \ref{App:Z_analys}. 
In Sec.\@ \ref{Z_spec}, we will see that the extracted skin depths determine the asymptotic value of the surface impedance for polygonal geometries in the scattering-less limit. 

\subsection{Hexagonal Fermi surface}\label{Skin_hex}

Let us begin with the hexagonal Fermi surface in the ``parallel'' configuration of Fig.\@ \ref{fig:Hexagon}(a). Then, the angle $\phi=0$ and, for $\gamma_{\rm mr}=\gamma_{\rm mc}=0$, we can employ Eq.\@ (\ref{eq:qx_sq_lowom}) together with the conductivity (\ref{eq:sigmayy_hex_par_gamma0}) to determine the polariton dispersion: 
\begin{equation}\label{eq:qx_hex_phi0}
q_{x}^{2}=-\left(\frac{\omega_p}{c}\right)^2 \frac{4 \omega^2}{4 \omega^2-(v_F q_x)^2}.
\end{equation}
In the limit $\omega\gg v_{F} \left|q_x\right|$, Eq.\@ (\ref{eq:qx_hex_phi0}) yields $q_x \approx i/\lambda_L$, where $\lambda_L=c/\omega_p$ is the London penetration depth. Then, the skin depth from Eq.\@ (\ref{eq:skin_depth_mode}) is frequency-independent:
\begin{equation}\label{eq:aniso_flat}
\delta_s\equiv\lambda_L=\frac{c}{\omega_p}. 
\end{equation}
In the opposite regime $\omega\ll v_{F} \left|q_x\right|$, Eq.\@ (\ref{eq:qx_hex_phi0}) gives $q_x=(1 \pm i) \sqrt{\omega/(v_F \lambda_L)}$ so that the skin depth 
\begin{equation}\label{eq:aniso_flat2}
\delta_s(\omega)=\sqrt{\frac{v_F \lambda_L}{\omega}}\propto \omega^{-\frac{1}{2}}. 
\end{equation}
The result (\ref{eq:aniso_flat2}) is surprising: it implies that the anomalous skin depth $\delta_s(\omega) \propto \omega^{-1/3}$, valid for isotropic systems, changes into $\delta_s(\omega) \propto \omega^{-1/2}$ for the anisotropic hexagonal case, a power law usually associated with normal skin effect; see, e.g.\@, Refs.\@ \onlinecite{Reuter-1948,Sondheimer-2001,Dressel-2001,Tanner-2019opt}, and the Supplemental Material. This behavior corresponds to region {\Large \textcircled{\small D$_2$}} in Fig.\@ \ref{fig:3D_skin_effect_misalign}. However, if scattering is dominant over frequency, we enter the viscous regime of region {\Large \textcircled{\small D$_1$}} and the skin depth is changed qualitatively. For instance, using the limit of Eq.\@ (\ref{eq:sigma_yy_phi0_highq}) for $\gamma_{\rm mc} \gg \omega$, we obtain
\begin{equation}\label{eq:skin_depth_visc_hex0}
\delta_s(\omega)=\frac{\sqrt{\lambda_L v_F}}{\sqrt{2} (\omega \gamma_{\rm mc})^{1/4}} \frac{1}{\cos(\pi/8)} \propto \omega^{-1/4}. 
\end{equation}
This behavior persists for $\gamma_{\rm mc}=\gamma_{\rm mr}$. 

For the crystal orientation $\phi=\pi/2$ of Fig.\@ \ref{fig:Hexagon}(b), the conductivity is given by Eq.\@ (\ref{eq:sigma_yy_hex_90_0}), therefore the polariton dispersion (\ref{eq:qx_sq_lowom}) stems from
\begin{equation}\label{eq:qx_hex_phi90}
q_{x}^{2}=-2\left(\frac{\omega_p}{c}\right)^2 \frac{2 \omega^2-(v_F q_x)^2}{4 \omega^2-3(v_F q_x)^2}.
\end{equation}
The limit $\omega\gg v_{F} \left|q_x\right|$ still yields $q_x \approx i/\lambda_L$ for Eq.\@ (\ref{eq:qx_hex_phi90}), so that the skin depth is (\ref{eq:aniso_flat}) for both $\phi=0$ and $\phi=\pi/2$ in such regime. The difference between the two orientations emerges for $\omega\ll v_{F} \left|q_x\right|$, as  Eq.\@ (\ref{eq:qx_hex_phi90}) implies $q_x=i\sqrt{2/3}/\lambda_L$ and hence the skin depth is $\omega$-independent:
\begin{equation}\label{eq:aniso_flat3}
\delta_s=\sqrt{\frac{3}{2}} \lambda_L.  
\end{equation}
The comparison between the results (\ref{eq:aniso_flat2}) and (\ref{eq:aniso_flat3}) indicates a significant orientational dependence of anomalous skin effect in the hexagonal system, which deviates from the standard isotropic case $\delta_s(\omega) \propto \omega^{-1/3}$. Notice that, in the presence of strong scattering $\omega \ll \left\{\gamma_{\rm mr},\gamma_{\rm mc}\right\}$, the skin depth in anomalous regime $\omega\ll v_{F} \left|q_x\right|$ from the conductivity (\ref{eq:sigma_yy_hex_90_highq}) at leading order changes to 
\begin{equation}\label{eq:aniso_flat3_gammamc}
\delta_s(\omega)= \lambda_L \sqrt{\frac{\gamma_{\rm mc}+2 \gamma_{\rm mr}}{\omega}}. 
\end{equation} 
Hence, scattering changes the anomalous skin depth from the constant value (\ref{eq:aniso_flat3}) to $\delta_s(\omega)\propto \omega^{-1/2}$ for $\phi=\pi/2$. This is a consequence of the non-commuting limits $\omega \ll \gamma_{\rm mc}$ and $\omega \gg \gamma_{\rm mc}$ in the conductivity for this geometry, as analyzed in Sec.\@ \ref{Hex_phi}; see Eq.\@ (\ref{eq:sigma_yy_hex_90_highq}). 
The angular sensitivity of the conductivity persists in the square-shaped case, as shown below. 

\subsection{Square Fermi surface}\label{Skin_square}

We now analyze the square Fermi surface in the $\phi=0$ configuration of Fig.\@ \ref{fig:Square}(a) and $\gamma_{\rm mr}=\gamma_{\rm mc}=0$, to compare with the hexagonal shape. The conductivity follows Eq.\@ (\ref{eq:sigmayy_square_par_gamma0}), and the complex momentum (\ref{eq:qx_sq_lowom}) results from
\begin{equation}\label{eq:qx_square_phi0}
(q_{x})^{2}=-2\left(\frac{\omega_p}{c}\right)^2 \frac{1}{2-(v_F q_x/\omega)^2}.
\end{equation} 
In the regime $\omega\gg v_{F} \left|q_x\right|$ we have $q_x \approx i/\lambda_L$ for Eq.\@ (\ref{eq:qx_hex_phi90}) as for the other geometries, giving the London skin depth (\ref{eq:aniso_flat}). In the opposite limit $\omega\ll v_{F} \left|q_x\right|$ there is a decaying solution $q_x= i 2^{1/4} \sqrt{\omega/(\lambda_L v_F)}$, which gives the skin depth 
\begin{equation}\label{eq:aniso_flat4}
\delta_s(\omega)=2^{-\frac{1}{4}} \sqrt{\frac{v_F \lambda_L}{\omega}} \propto \omega^{-\frac{1}{2}}.  
\end{equation} 
As for the hexagonal shape with $\phi=0$, the square Fermi surface gives $\delta_s(\omega)\propto \omega^{-1/2}$ in anomalous regime (region {\Large \textcircled{\small D$_2$}} in Fig.\@ \ref{fig:3D_skin_effect_misalign}), in contrast with the isotropic case (see Fig.\@ \ref{fig:3D_skin_effect_iso}). When scattering is dominant over frequency, we find ourselves in the viscous regime, and $\delta_s(\omega)\propto \omega^{-1/4}$, analogously to the hexagonal case (see region {\Large \textcircled{\small D$_1$}} in Fig.\@ \ref{fig:3D_skin_effect_misalign}); this can be obtained from the high-momentum conductivity (\ref{eq:sigmayy_square_highq}) in the limit $\gamma_{\rm mc} \gg \omega$.

The configuration shown in Fig.\@ \ref{fig:Square}(b), with $\phi=\pi/4$, gives equally surprising results, as the conductivity (\ref{eq:sigma_yy_sq_pi4_gamma}) is momentum-independent. Thus, for any $\omega/(v_F \left|q_x\right|)$ ratio we obtain the London skin depth $\delta_s\equiv \lambda_L$ for $\gamma_{\rm mr}=\gamma_{\rm mc}=0$. On the other hand, for $\gamma_{\rm mr} \gg \omega$, we obtain $\delta_s(\omega)=\lambda_L \sqrt{\gamma_{\rm mr}/\omega} \propto \omega^{-1/2}$. Thus, momentum-relaxing scattering changes the skin depth for the square Fermi surface with $\phi=\pi/4$, in the same way as in the Drude model. 

\subsection{London penetration depth and flat Fermi-surface segments}\label{Cond_poly_skin}

The polygonal examples analyzed in Secs.\@ \ref{Skin_hex} and \ref{Skin_square} help us deducing a general property of the skin depth with piece-constant Fermi surfaces: large portions of Fermi surface parallel to the surface generate the smallest possible skin depth $\lambda_L$, which is usually associated with the penetration depth of electromagnetic fields into a superconductor. With the benefit of hindsight, we have already seen through Eq.\@ (\ref{eq:sigma_yy_gamma0_gen}) that Fermi-surface segments with velocity parallel to the applied field $E_y(q_x,\omega) \hat{u}_y$ possess $v_{\vec{k},x}=0$ and generate a momentum-independent conductivity. This is the origin of the equivalence $\delta_s \equiv \lambda_L$ found in the hexagonal- and square-shaped examples. The limit $\omega \gg v_F \left|q_x \right|$ is equivalent for all geometries, as a perfect conductivity is found for all shapes. However, even in the regime $\omega \ll v_F \left|q_x \right|$ where anomalous skin effect takes place, the portions $\vec{v}_F \parallel \hat{u}_y$ modify the skin depth with respect to the isotropic result $\delta_s(\omega) \sim \omega^{-1/3}$: in the hexagonal case with $\phi=0$, the segments are not exactly parallel to the field and cannot completely suppress the $q_x$ dependence of the conductivity, but the power $\delta_s(\omega) \sim \omega^{-1/2}$ is modified by anisotropy; the square geometry with $\phi=\pi/4$ is even more extreme, as in this case the segments of $\vec{v}_F$ parallel to the field completely suppress non-locality. 

The local contribution to the conductivity, for electrons propagating parallel to the field, also has the consequence that the two anomalous regions ({\Large \textcircled{\small D$_1$}} and {\Large \textcircled{\small D$_2$}} in Fig.\@ \ref{fig:3D_skin_effect}) acquire two distinct scaling exponents for the frequency dependence of the skin depth. This is because the locally responding electrons are sensitive to the relative ratios of $\omega$, $\gamma_{\rm mr}$, and $\gamma_{\rm mc}$. Such effect is in contrast with the indistinguishability of regions {\Large \textcircled{\small D$_1$}} and {\Large \textcircled{\small D$_2$}} in the isotropic case, where the conductivity depends on $q_x$ in anomalous regime (see Fig.\@ \ref{fig:3D_skin_effect_iso}). 

We can anticipate that the orientational dependence of the skin depth will affect the surface impedance, due to the connection between the two quantities. 
We substantiate this statement with the numerical and analytical calculations presented in the next section. 

\section{Surface impedance for specular surface scattering}\label{Z_spec}

The surface impedance is defined as the ratio between the electric field impinging at the surface of the sample, located at $x=0$, and the generated current density inside the metal \cite{Reuter-1948,Dressel-2001}:
\begin{equation}\label{eq:Z_def}
Z(\omega)=\frac{E_y(0,\omega)}{\int_0^{+\infty} dx J_y(x,\omega)}=i \omega \mu_0 \frac{E_y(0,\omega)}{\left.\frac{\partial E_y(x,\omega)}{\partial x}\right|_{x=0^+}}.
\end{equation}
The second step in Eq.\@ (\ref{eq:Z_def}) can be obtained using Eq.\@ (\ref{eq:Ohm_gen}) combined with the wave equation (\ref{eq:wave_y_compl}), neglecting the displacement current term in the latter, and integrating the current density $J_y(x,\omega)$ over $x$ under the assumption that $\lim_{\left|x\right|\rightarrow +\infty }\partial E_y(x,\omega)/\partial x=0$.
The explicit calculation of the surface impedance requires modeling the electron dynamics at the interface between the external medium (here assumed to be vacuum for simplicity) and the interior of the metal. In our setup, we assume specular boundary conditions at the surface $x=0$. In such configuration, if we consider the vacuum-packed half of space $x<0$ filled with another piece of the same metal, the electrons in each half will have the same history as if the reflection were specular: the only information required is $E(0,\omega)$ and the gradient of this specular field at $x=0$, since $E_y(x,\omega)$ forms a cusp at $x=0$ due to being damped both for $x<0$ and $x>0$ \cite{Pitaevskii-2012kin,Dressel-2001,Kittel-1996}. Hence, $\left.\partial E_y(x,\omega)/\partial x \right|_{x=0^+}=-\left.\partial E_y(x,\omega)/\partial x \right|_{x=0^-}$. The latter condition can be incorporated into the wave equation (\ref{eq:wave_y_compl}) by adding a delta-function boundary term: 
\begin{multline}\label{eq:wave_y_compl_spec}
\frac{\partial^2 E_y(x,\omega)}{\partial x^2}+\frac{\omega^2}{c^2} E_y(x,\omega) \\ =-\mu_0 i \omega \int d x' \sigma_{yy}(x-x',\omega) E_y(x',\omega) \\ +2 \left.\frac{\partial E_y(x,\omega)}{\partial x}\right|_{x=0^+} \delta(z). 
\end{multline} 
Fourier-transforming Eq.\@ (\ref{eq:wave_y_compl_spec}) to momentum space, we have
\begin{multline}\label{eq:E_q_sigma}
E_y(q_x,\omega)=2 \left. \frac{\partial E_y(x,\omega)}{\partial x} \right|_{x=0^+} \\ \times \frac{1}{\omega^2/c^2+\mu_0 i \omega \sigma_{yy}(q_x,\omega) - q_x^2}. 
\end{multline}
Notice that Eq.\@ (\ref{eq:E_q_sigma}) allows us to express both the electric field and the conductivity $\sigma_{yy}(q_x,\omega)$ in momentum space even in the presence of a surface, which breaks translational invariance, thanks to the specular boundary conditions. 
An inverse Fourier transform of the electric field $E_y(x,\omega)=(2\pi)^{-1} \int_{-\infty}^{+\infty} d q E_y(q_x,\omega) e^{-i q_x x} $ at the interface $x=0$ yields 
\begin{multline}\label{eq:E0_sigma}
E_y(0,\omega)=\frac{1}{\pi} \left. \frac{\partial E_y(x,\omega)}{\partial x} \right|_{x=0^+} \times \\  \int_{-\infty}^{+\infty} d q_x \frac{1}{\omega^2/c^2+\mu_0 i \omega \sigma_{yy}(q_x,\omega) - q_x^2}. 
\end{multline}
Finally, inserting Eq.\@ (\ref{eq:E0_sigma}) into the definition (\ref{eq:Z_def}) gives the surface impedance in terms of the conductivity for specular boundary conditions: 
\begin{multline}\label{eq:Z_spec_displ} 
Z(\omega)=i \frac{\omega \mu_0}{\pi} \int_{-\infty}^{+\infty} d q_x \frac{1}{\omega^2/c^2+\mu_0 i \omega \sigma_{yy}(q_x,\omega) - (q_x)^2} \\ = i \frac{Z_0}{\pi} \frac{\omega}{\omega_p} \int_{-\infty}^{+\infty} d z \frac{1}{\omega^2/\omega_p^2+S_{yy}(z,\omega)-z^2}, 
\end{multline} 
where $Z_0=\mu_0 c$ is the vacuum surface impedance, and we defined
\begin{equation}\label{eq:S_nonloc}
S_{yy}(q_x,\omega)=\frac{\mu_0 c^2 i \omega \sigma_{yy}(q_x,\omega)}{\omega_p^2}=\frac{ i \omega}{\epsilon_0 \omega_p^2} \sigma_{yy}(q_x,\omega)
\end{equation}
The displacement-current term $\omega^2/c^2$ in Eq.\@ (\ref{eq:Z_spec_displ}), contained in the integrand at the denominator, is essentially responsible for the development of the transparent regime (region {\Large \textcircled{\normalsize C}} in Fig.\@ \ref{fig:3D_skin_effect}), and it is negligible for frequencies below the plasma edge \cite{Casimir-1967a,Casimir-1967b,Dressel-2001,Wooten-1972opt}. Hence, if we analyze the impedance for for $\omega< \omega_p$, we can neglect the displacement-current term in Eq.\@ (\ref{eq:Z_spec_displ}), with the result 
\begin{equation}\label{eq:Z_spec}
Z(\omega)= i \frac{Z_0}{\pi} \frac{\omega}{\omega_p} \int_{-\infty}^{+\infty} d z \frac{1}{S_{yy}(z,\omega)-z^2}. 
\end{equation} 
In the following, we will employ the full Eq.\@ (\ref{eq:Z_spec_displ}) for numerical computations of the impedance ``phase diagrams'', while we will revert to Eq.\@  (\ref{eq:Z_spec}) without the displacement current, to derive analytical results for the boundaries between different skin effect regimes. In any case, the conductivity that enters into the momentum-space integral for the impedance is given by Eq.\@ (\ref{eq:lonloc_sigma_T}) in general. The 2D isotropic limit, where the conductivity follows Eq.\@ (\ref{eq:sigmayy_iso}), is analyzed for reference in Appendix \ref{App_Z_iso_analysis}. 
We checked that our results for the impedance modulus $|Z(\omega)|$ are qualitatively robust with respect to the choice of interface boundary conditions: the difference between specular and diffusive surface scattering is negligible for all Fermi-surface geometries here considered. Boundary conditions do significantly influence the surface resistance (real part of the impedance, $\mathrm{Re} Z(\omega)$) in the relaxation regime {\Large \textcircled{\normalsize B}} and in the extreme anomalous regime {\Large \textcircled{\normalsize E}} indicated in Fig.\@ \ref{fig:3D_skin_effect}. We postpone the detailed analysis of this sensitivity to different boundary conditions to a subsequent work; see also the discussion in Sec.\@ \ref{Bound_cond_disc}. 

In what follows, we will discuss the dependence of the surface impedance, stemming from Eqs.\@ (\ref{eq:Z_spec}) and (\ref{eq:S_nonloc}), on frequency, scattering rates and orientation angle $\theta$, for the anisotropic 2D geometries of hexagonal and square Fermi surfaces, at fixed Fermi velocity $v_F$.
As we will shortly appreciate, all numerical and analytical results for the impedance confirm the qualitative ``phase diagrams'' sketched in Figs.\@ \ref{fig:3D_skin_effect_iso}, \ref{fig:3D_skin_effect_misalign}, \ref{fig:3D_skin_effect_align}, and \ref{fig:3D_skin_effect_square}, and agree with the discussion anticipated in Sec.\@ \ref{Summ}.
In particular, let us stress that the anomalous regime is the most sensitive to anisotropy: while all other forms of skin effect depend quantitatively but not qualitatively on Fermi-surface geometry, the impedance in anomalous regime changes from $\left|Z(\omega)\right|\propto \omega^{2/3}$ in isotropic systems \cite{Reuter-1948, Sondheimer-2001}, to different power laws. For example, for the geometries of Figs.\@ \ref{fig:3D_skin_effect_align} and \ref{fig:3D_skin_effect_square}, $\left|Z(\omega)\right|\propto \omega^{1/2}$ for the hexagonal and square geometries in the presence of scattering. This remarkable difference highlights that the $\sqrt{\omega}$ behavior of the impedance, usually associated with normal skin effect, is also encountered in anomalous regime in anisotropic systems. Such peculiar power-law further changes in the scattering-less limit to a perfect-conductor frequency-independent behavior. 
To identify the exponent $\eta$ such that $\left|Z(\omega)\right| \propto \omega^\eta$, we will follow Eq.\@ (\ref{eq:skin_depth_Z_gen}), which shows that $\eta$ can be equivalently extracted from $\mathrm{Arg}Z(\omega)/(-\pi/2)$; see also Sec.\@ \ref{Gen_ineffectiveness}. This way, in the following we will span the skin-effect parameter space as a function of $\omega$, $\gamma_{\rm mc}$, and $\gamma_{\rm mr}$. 

\subsection{Hexagonal Fermi surface}\label{Z_hex_spec}

To obtain the surface impedance for an hexagonal Fermi surface, we use Eqs.\@ (\ref{eq:Z_spec}) and (\ref{eq:S_nonloc}) in conjunction with the conductivity (\ref{eq:sigma_yy_hexpar_expl}) or (\ref{eq:sigma_yy_hexphi}), for the ``parallel'' configuration $\phi=0$ and for generic rotation angle $\phi$ respectively. 

\begin{figure}[ht] \centering
\includegraphics[width=0.95\columnwidth]{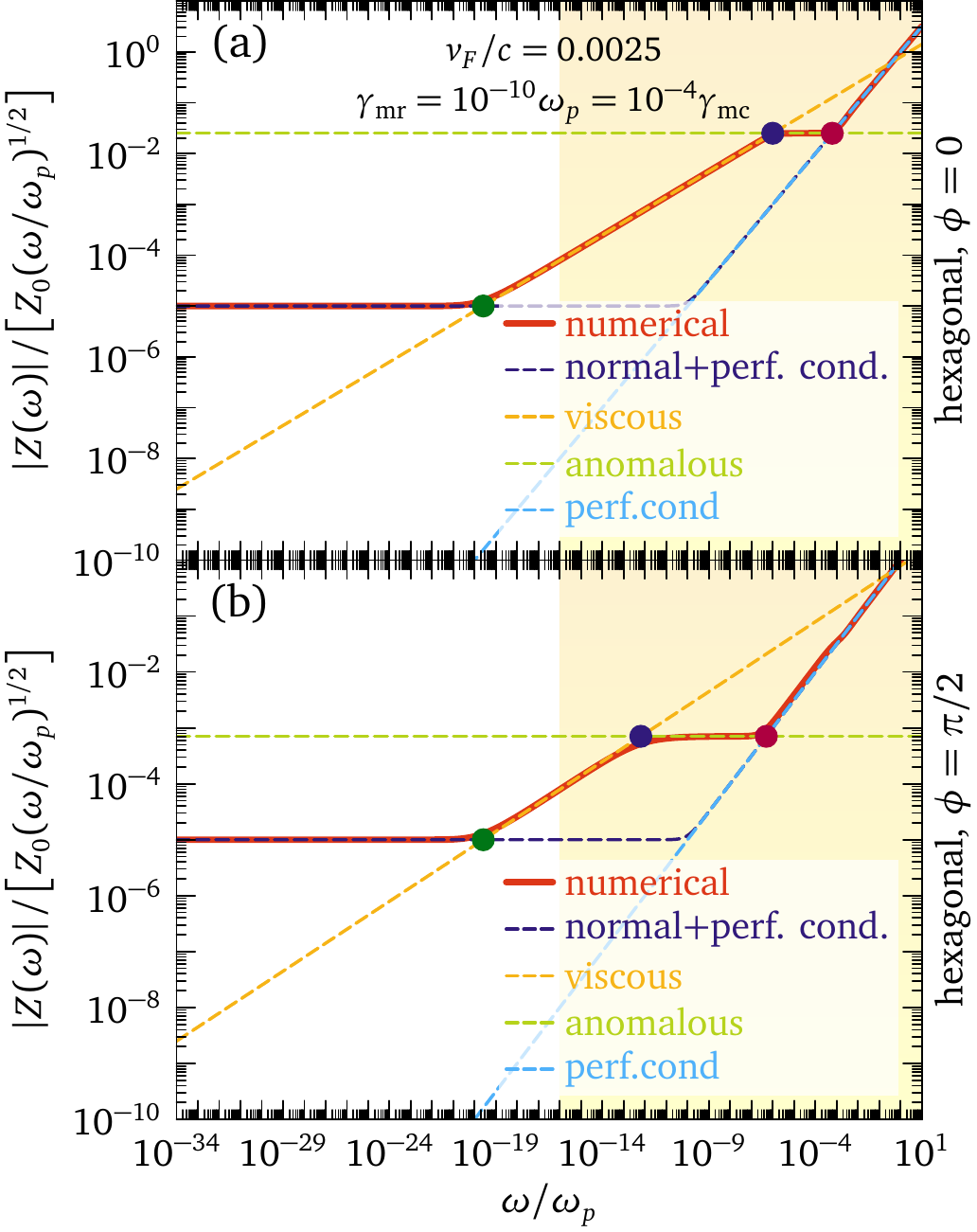} 
\caption{\label{fig:AbsZ_hex_crossovers} Absolute value of the surface impedance $\left|Z(\omega)\right|$, divided by the vacuum impedance $Z_0$ and by $(\omega/\omega_p)^{1/2}$, as a function of $\omega/\omega_p$, for a hexagonal Fermi surface. We employ the parameters $v_F/c=0.0025$, and $\gamma_{\rm mr}=10^{-4}\gamma_{\rm mc}=10^{-6}\omega_p$. Dashed curves show analytical results valid in each regime, derived in Appendix \ref{App_Z_hex_analysis}. The yellow-shaded area gives a qualitative estimation of the parameter space accessible to experiments \cite{Dressel-2001}. (a) ``Parallel'' configuration with rotation angle $\phi=0$. (b) Configuration with rotation angle $\phi=\pi/2$.
}
\end{figure}

Numerical results for the impedance modulus $\left|Z(\omega)\right|$, divided by the vacuum impedance $Z_0$ and by $(\omega/\omega_p)^{1/2}$ for visual clarity, are displayed as a function of $\omega/\omega_p$ by the solid red curves in Fig.\@ \ref{fig:AbsZ_hex_crossovers}. We use the parameters $v_F/c=0.0025$ (corresponding to the Fermi velocity of PdCoO$_2$ \cite{Nandi-2018,Bachmann-2021}), $\gamma_{\rm mr}=10^{-10} \omega_p$, and $\gamma_{\rm mc}=10^{-6}\omega_p$. 
Figs.\@ \ref{fig:AbsZ_hex_crossovers}(a) and \ref{fig:AbsZ_hex_crossovers}(b) show the results for $\phi=0$ and $\phi=\pi/2$, respectively. 

\begin{figure*}[ht] \centering
\includegraphics[width=0.8\textwidth]{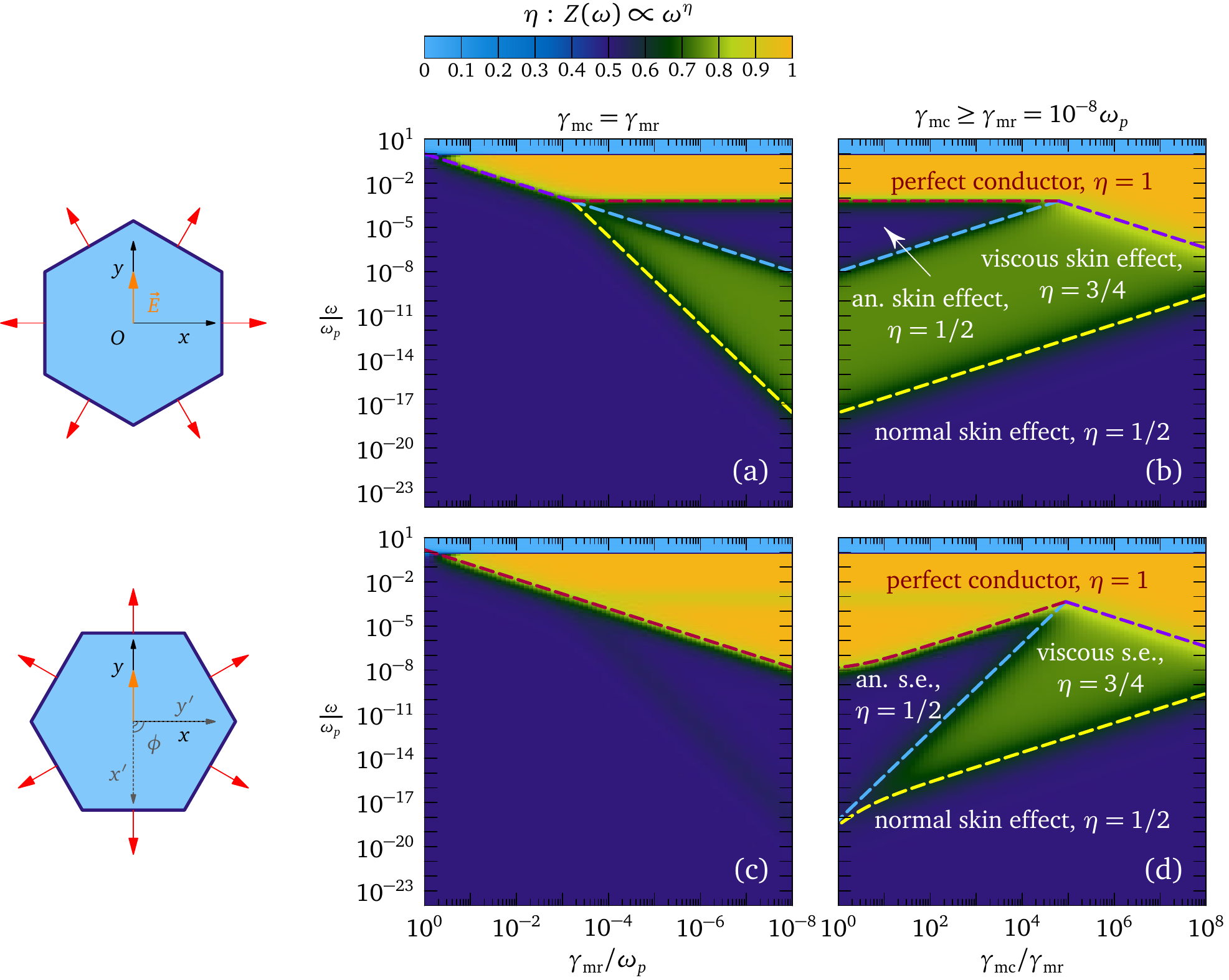}
\caption{\label{fig:omegacross_hex}  Orientational dependence of skin effect regimes for a hexagonal Fermi surface, as measured by the surface impedance $Z(\omega)$ as a function of relaxation rate $\gamma_{\rm mr}$, momentum-conserving collision rate $\gamma_{\rm mr}$, and frequency $\omega/\omega_p$, where $\omega_p$ is the plasma frequency (\ref{eq:omega_p_iso}). The corresponding orientation is sketched on the left-hand side of each plot, together with the applied electric field $\vec{E}=E_y \hat{u}_y$ aligned with the $y$ axis. Red arrows depict the local Fermi velocity vectors. (a) ``Parallel'' configuration in the $\left(\gamma_{\rm mr},\omega \right)$ plane, for $\gamma_{\rm mr}=\gamma_{\rm mc}$. (b) ``Parallel'' configuration in the $\left(\gamma_{\rm mc},\omega \right)$ plane, for fixed $\gamma_{\rm mr}=10^{-8}\omega_p$. (c) Fermi surface rotated by $\phi=\pi/2$ with respect to panels (a) and (b), in the $\left(\gamma_{\rm mr},\omega \right)$ plane, for $\gamma_{\rm mr}=\gamma_{\rm mc}$. (d) Fermi surface rotated by $\phi=\pi/2$ in the $\left(\gamma_{\rm mc},\omega \right)$ plane, for fixed $\gamma_{\rm mr}=10^{-8}\omega_p$. The color palette is the density plot of $\mathrm{Arg}Z(\omega)/(-\pi/2)$, giving the exponent $\eta$ of $Z(\omega)\propto \omega^\eta$. Dashed lines are the analytical crossover boundaries derived in Appendix \ref{Hex_crossover_omega}. 
}
\end{figure*} 
Analytical expressions, collected in Appendix \ref{App_Z_hex_analysis}, are available in each regime of skin effect, and are displayed as dashed lines in Fig.\@ \ref{fig:AbsZ_hex_crossovers}. Specifically: dashed blue lines show normal skin effect, given by Eq.\@ (\ref{eq:Z_hex_lowq}) for $\phi=0$ and $\phi=\pi/2$; dashed orange lines display viscous skin effect, given by Eq.\@ (\ref{eq:Z_hex_par_hydro}) for $\phi=0$; dashed green lines show anomalous skin effect, according to Eq.\@ (\ref{eq:Z_hex_par_highq}) (Eq.\@ (\ref{eq:Z_hex_pi2_highq})) for $\phi=0$ ($\phi=\pi/2$); light-blue dashed lines showing the perfect-conductivity result, common to both orientations. In Fig.\@ \ref{fig:AbsZ_hex_crossovers}(b), we also barely distinguish an intermediate frequency range between the anomalous and perfect-conductivity regimes,  where the scattering rates are negligible in comparison with frequency, but still $\omega <v_F \left|q_x\right|$. In this frequency window the impedance is $Z(\omega)=-i \mu_0 \sqrt{3/2} \lambda_L \omega$, in accordance with Eq.\@ (\ref{eq:Z_hex_gamma0_90_lowom}). 

The intersection points of the analytical results (dashed lines in Fig.\@ \ref{fig:AbsZ_hex_crossovers}) provide a natural criterion to identify the crossover frequency between different skin effect regimes. Such intersections are highlighted by the green, blue, and red dots in Fig.\@ \ref{fig:AbsZ_hex_crossovers}. The resulting respective crossover frequencies $\omega_{n v}$, $\omega_{v a}$ and $\omega_{a p}$ between normal/viscous, viscous/anomalous, and anomalous/perfect-conductor skin effect, are functions of $\gamma_{\rm mr}$, $\gamma_{\rm mc}$ and $v_F$ in general, and their analytical expressions are reported in Appendix \ref{App_Z_hex_analysis}, together with a viscous/perfect conductor boundary $\omega_{v p}$ not visible in Fig.\@ \ref{fig:AbsZ_hex_crossovers}. These functions generate the dashed curves in the ``phase diagrams'' presented in Fig.\@ \ref{fig:omegacross_hex}(a)-(d). 

The density plots of $\mathrm{Arg}Z(\omega)/(-\pi/2)$ in Fig.\@ \ref{fig:omegacross_hex} stem from Eq.\@ (\ref{eq:Z_spec_displ}) (including the displacement current), and hence they include the transparent regime (light blue-shaded area) found for $\omega \geq \omega_p$; see also region {\Large \textcircled{\normalsize C}} in Fig.\@ \ref{fig:3D_skin_effect}. 
Notice how the extension of each skin effect region in Fig.\@ \ref{fig:omegacross_hex} depends on orientation, and excellently agrees with the qualitative discussion of Figs.\@ \ref{fig:3D_skin_effect_misalign} and \ref{fig:3D_skin_effect_align} in Sec.\@ \ref{Summ}. In particular, for $\phi=0$ the viscous regime extends throughout regions {\Large \textcircled{\small D$_1$}} and {\Large \textcircled{\normalsize V}}, and it is present even for $\gamma_{\rm mr}=\gamma_{\rm mc}$; cfr.\@ Figs.\@ \ref{fig:omegacross_hex}(a),(b) and \ref{fig:3D_skin_effect_misalign}.
For $\phi=\pi/2$ and $\gamma_{\rm mr}=\gamma_{\rm mc}$, the viscous regime (green-shaded region) shrinks and there is a direct crossover from normal to anomalous skin effect, which share the same power law $\left|Z(\omega)\right|\propto \omega^{1/2}$ but with different numerical prefactors. This is fully consistent with the difference between regions {\Large \textcircled{\normalsize A}} and {\Large \textcircled{\small D$_1$}} in Fig.\@ \ref{fig:3D_skin_effect_align}. 

In the absence of scattering rates, a qualitative difference in anomalous regime appears, as shown in Appendix \ref{App_Z_hex_analysis_noscat}: while for $\phi=0$ the impedance still follows $\left|Z(\omega)\right|\propto \omega^{1/2}$, for $\phi=\pi/2$ the impedance is purely imaginary and $\left|Z(\omega)\right|\propto \sqrt{3/2} \lambda_L \omega$. This aspect explains the distinction between regions {\Large \textcircled{\small D$_1$}} and {\Large \textcircled{\small D$_2$}} in Fig.\@ \ref{fig:3D_skin_effect_align}. Hence, the anomalous impedance is very sensitive to the presence or absence of scattering, either momentum-conserving or relaxing. We emphasize that this contrasts with the isotropic case: for circular, spherical, or spheroidal Fermi surfaces, the regions {\Large \textcircled{\small D$_1$}} and {\Large \textcircled{\small D$_2$}} are indistinguishable \cite{Dressel-2001,Tanner-2019opt}; see Fig.\@ \ref{fig:3D_skin_effect_iso}.

\begin{figure}[ht]
\includegraphics[width=0.7\columnwidth]{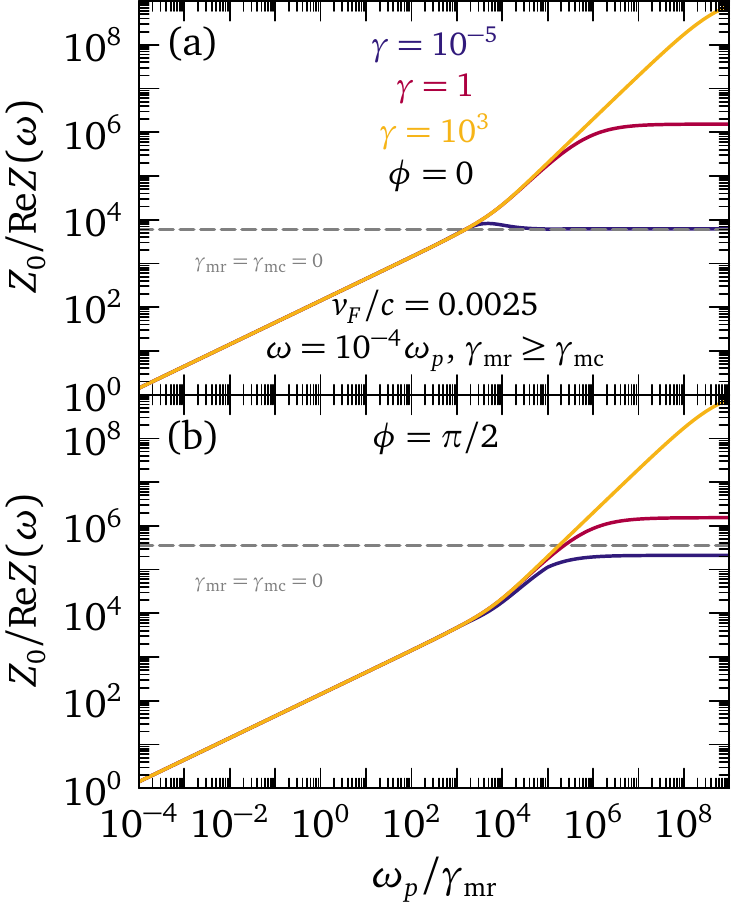} 
\caption{\label{fig:Zs_phi090_taumr_Hex} Inverse surface resistance $1/\mathrm{Re}Z(\omega)$ normalized to the vacuum-impedance $Z_0$, as a function of inverse momentum-relaxation rate $\omega_p/\gamma_{\rm mr}$, for a hexagonal Fermi surface with Fermi-velocity modulus $v_F=2.5 \times 10^{-3} c$, $\omega/\omega_p=10^{-4}$, and $\gamma_{\rm mc}=\mathrm{max}\left\{\gamma_{\rm mr},\bar{\gamma}\right\}$, with $\bar{\gamma}=\left\{10^{-5},1,10^3\right\}$ for the blue, red, and gold curves respectively. Specular boundary conditions are assumed, according to Eq.\@ (\ref{eq:Z_spec}). The dashed gray lines represent the limit $\gamma_{\rm mr}=\gamma_{\rm mc}=0$. (a) Results from Eq.\@ (\ref{eq:sigma_yy_hexpar_expl}), for tilting angle $\phi=0$ with respect to the surface; see Fig.\@ \ref{fig:Hexagon}(a). The scattering-less limit is given by Eq.\@ (\ref{eq:Z_noscat_0}). (b) Results from Eqs.\@ (\ref{eq:G0_phi}) and (\ref{eq:sigma_yy_hexphi}), for $\phi=\pi/2$; see Fig.\@ \ref{fig:Hexagon}(b). The scattering-less limit stems from Eq.\@ (\ref{eq:Z_noscat_90}).
}
\end{figure}

Another way to recognize the occurrence of anomalous forms of skin effect is to plot the inverse surface resistance $1/\mathrm{Re}Z(\omega)$ as a function of inverse momentum-relaxation rate $\omega_p/\gamma_{\rm mr}$, at fixed frequency $\omega$, as traditionally performed in the analysis of skin effect after Pippard \cite{Pippard-1947a,Pippard-1947b,Reuter-1948,Sondheimer-2001}. Since $\gamma_{\rm mr}$ is an increasing function of temperature $T$ (e.g.\@, $\gamma_{\rm mr} \propto T^2$ for Fermi-liquid electron-electron interactions), these graphs can be interpreted as the evolution of the impedance with decreasing $T$. 
Figs.\@ \ref{fig:Zs_phi090_taumr_Hex}(a),(b) show our numerical results for $1/\mathrm{Re}Z(\omega)$ in the hexagonal Fermi-surface geometry, for $\phi=0$ and $\phi=\pi/2$ respectively, which correspond to the geometries shown in Fig.\@ \ref{fig:Hexagon}(a),(b). 
Since we expect $\gamma_{\rm mr}\approx \gamma_{\rm mc}$ at high $T$, while $\gamma_{\rm mr}\ll \gamma_{\rm mc}$ at low $T$, we have taken $\gamma_{\rm mc}=\mathrm{max}\left\{\gamma_{\rm mr},\bar{\gamma}\right\}$, with $\bar{\gamma}=\left\{10^{-5},1,10^3\right\}$ for the blue, red, and gold curves in each panel respectively.
We see the characteristic saturation of the inverse surface resistance, the value of which parametrically depends on $\gamma_{\rm mc}$ and on Fermi-surface orientation parametrized by the angle $\phi$. In Reuter and Sondheimer theory of isotropic systems, such saturation is associated with anomalous skin effect. As we include momentum-conserving collisions as well, Fig.\@ \ref{fig:Zs_phi090_taumr_Hex} shows that the asymptotic saturation value depends on $\gamma_{\rm mc}$, which corresponds to either viscous or anomalous skin effect, depending on whether $\omega \ll \gamma_{\rm mc}$ or $\omega \gg \gamma_{\rm mc}$ respectively.  The specific saturation value for a given $\gamma_{\rm mc}$ is influenced by Fermi-surface orientation, as shown by comparing curves of the same color in Figs.\@ \ref{fig:Zs_phi090_taumr_Hex}(a),(b). 

Hence, the relaxationless saturation of the surface resistance in hydrodynamic regime $\omega \ll \gamma_{\rm mc}$ constitutes evidence for viscous skin effect, governed by momentum-conserving collisions, and quantitatively influenced by the orientation of the anisotropic Fermi surface. Analogous conclusions hold for a square Fermi-surface geometry, as detailed in the next section. 

\subsection{Square Fermi surface}\label{Z_sq_spec}

For a square Fermi surface, the surface impedance follows from Eqs.\@ (\ref{eq:Z_spec}) and (\ref{eq:S_nonloc}), and from the conductivity (\ref{eq:sigma_yy_dia_expl}) or (\ref{eq:sigma_yy_hexphi}), for the ``diamond-shaped'' configuration $\phi=0$ and for generic rotation angle $\phi$ respectively. 

\begin{figure}[ht] \centering
\includegraphics[width=0.95\columnwidth]{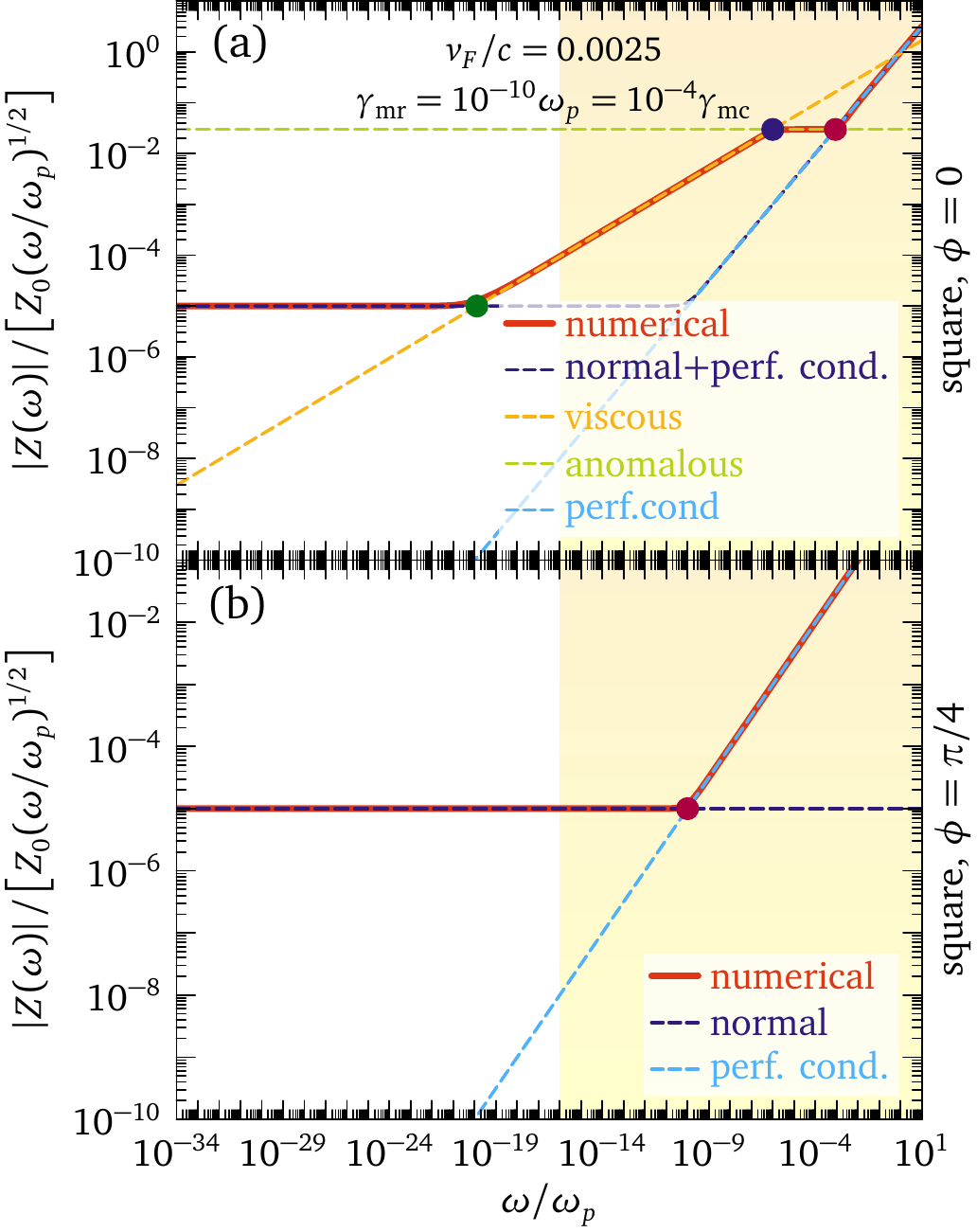} 
\caption{\label{fig:AbsZ_sq_crossovers} Absolute value of the surface impedance $\left|Z(\omega)\right|$, divided by the vacuum impedance $Z_0$ and by $(\omega/\omega_p)^{1/2}$, as a function of $\omega/\omega_p$, for a square Fermi surface. We use the parameters $v_F/c=0.0025$, and $\gamma_{\rm mr}=10^{-4}\gamma_{\rm mc}=10^{-6}\omega_p$. Dashed curves show analytical results valid in each regime, derived in Appendix \ref{Square_Z_analys}. The yellow-shaded area represents a qualitative estimation of the parameter space accessible to experiments \cite{Dressel-2001}. (a) ``Diamond-shaped'' configuration with rotation angle $\phi=0$. (b) Configuration with rotation angle $\phi=\pi/4$. 
}
\end{figure}

Numerical calculations for the impedance modulus $\left|Z(\omega)\right|$, divided by the vacuum impedance $Z_0$ and by $(\omega/\omega_p)^{1/2}$, are displayed as a function of $\omega/\omega_p$ by the solid red curves in Fig.\@ \ref{fig:AbsZ_sq_crossovers}. All parameters are the same as the ones employed for Fig.\@ \ref{fig:AbsZ_hex_crossovers}. Figs.\@ \ref{fig:AbsZ_sq_crossovers}(a) and \ref{fig:AbsZ_sq_crossovers}(b) display the calculations for $\phi=0$ and $\phi=\pi/4$, respectively, and the two configurations are visually represented in Fig.\@ \ref{fig:Square}(a),(b). 
\begin{figure*}[ht] \centering
\includegraphics[width=0.8\textwidth]{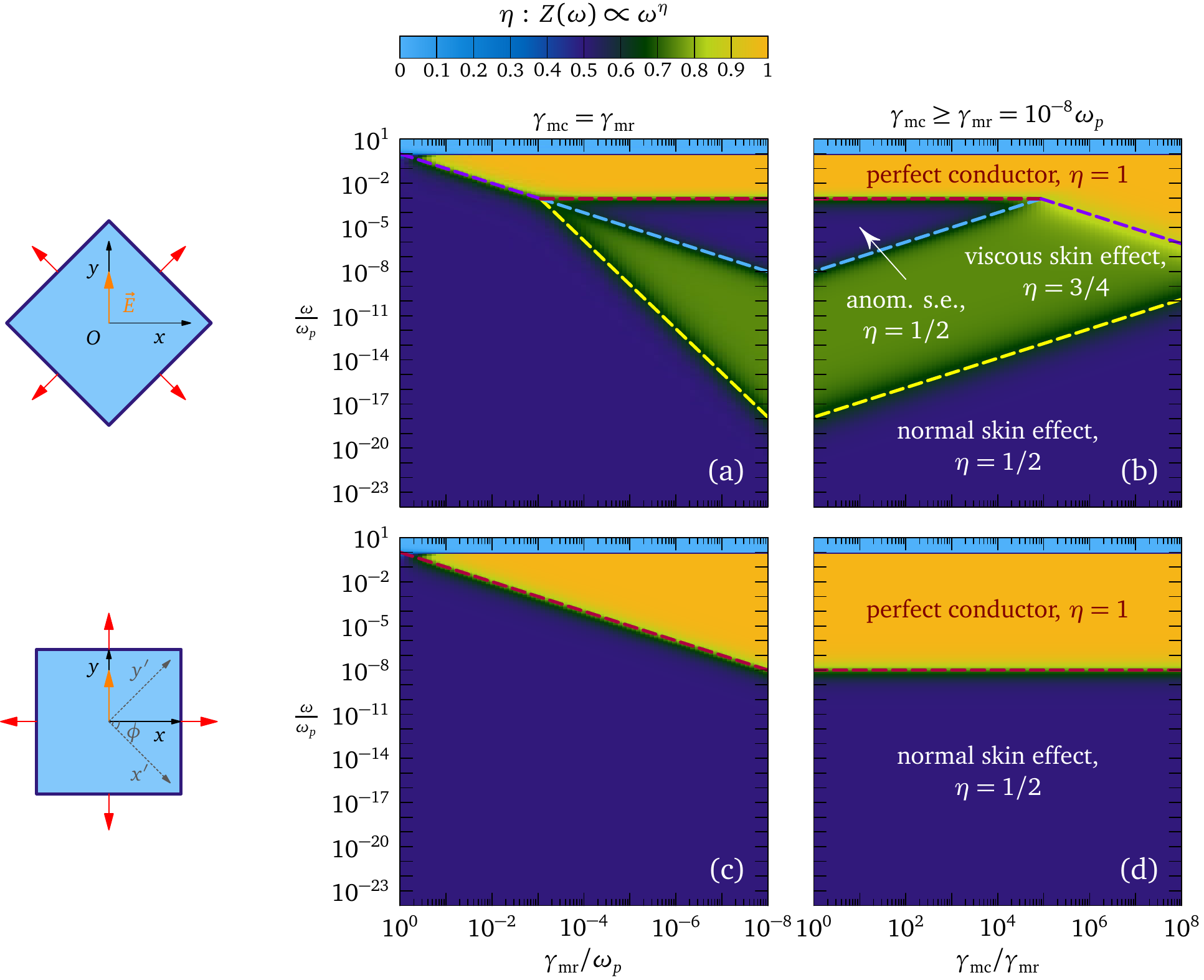}
\caption{\label{fig:omegacross_square} Orientational dependence of skin effect regimes for a square Fermi surface, as measured by the surface impedance $Z(\omega)$ as a function of relaxation rate $\gamma_{\rm mr}$, momentum-conserving collision rate $\gamma_{\rm mr}$, and frequency $\omega/\omega_p$, where $\omega_p$ is the plasma frequency (\ref{eq:omega_p_iso}). The corresponding orientation is sketched on the left-hand side of each plot, together with the applied electric field $\vec{E}=E_y \hat{u}_y$ aligned with the $y$ axis. Red arrows depict the local Fermi velocity vectors. (a) ``Diamond-shaped'' configuration in the $\left(\gamma_{\rm mr},\omega \right)$ plane, for $\gamma_{\rm mr}=\gamma_{\rm mc}$. (b) ``Diamond-shaped'' configuration in the $\left(\gamma_{\rm mc},\omega \right)$ plane, for fixed $\gamma_{\rm mr}=10^{-8}\omega_p$. (c) Fermi surface rotated by $\phi=\pi/4$ with respect to panels (a) and (b), in the $\left(\gamma_{\rm mr},\omega \right)$ plane, for $\gamma_{\rm mr}=\gamma_{\rm mc}$. (d) Fermi surface rotated by $\phi=\pi/4$ in the $\left(\gamma_{\rm mc},\omega \right)$ plane, for fixed $\gamma_{\rm mr}=10^{-8}\omega_p$. The color palette is the density plot of $\mathrm{Arg}Z(\omega)/(-\pi/2)$, giving the exponent $\eta$ of $Z(\omega)\propto \omega^\eta$. Dashed lines are the analytical crossover boundaries derived in Appendix \ref{Sq_crossover_omega}. 
} 
\end{figure*}
Appendix \ref{Square_Z_analys} reports analytical expressions in each regime of skin effect, which are displayed by dashed lines in Fig.\@ \ref{fig:AbsZ_sq_crossovers}: dashed blue lines show normal skin effect, given by Eq.\@ (\ref{eq:Z_hex_lowq}) for $\phi=0$ and $\phi=\pi/4$; the dashed orange line displays viscous skin effect, given by Eq.\@ (\ref{eq:Z_square_dia_hydro}) for $\phi=0$; the dashed green line shows anomalous skin effect, according to Eq.\@ (\ref{eq:Z_square_dia_anom}) for $\phi=0$; light-blue dashed lines showing the perfect-conductivity result, which is equal for both orientations. 

Analogously to the hexagonal case, we define the crossover frequencies between all skin effect regimes, as the intersection points between different analytical limits (dashed lines in Fig.\@ \ref{fig:AbsZ_sq_crossovers}). These intersections are marked by the green, blue, and red dots in Fig.\@ \ref{fig:AbsZ_sq_crossovers}, and correspond to the crossover frequencies $\omega_{n v}$, $\omega_{v a}$ and $\omega_{a p}$ between normal/viscous, viscous/anomalous, and anomalous/perfect-conductor skin effect. Together with a viscous/perfect conductor crossover $\omega_{v p}$, not visible in Fig.\@ \ref{fig:AbsZ_sq_crossovers}, the functions $\omega_{n v}$, $\omega_{v a}$ and $\omega_{a p}$ produce the dashed curves in the impedance ``phase diagrams'' of Fig.\@ \ref{fig:omegacross_square}. 

As for Fig.\@ \ref{fig:omegacross_hex}, the density plots of $\mathrm{Arg}Z(\omega)/(-\pi/2)$ in Fig.\@ \ref{fig:omegacross_square} stem from Eq.\@ (\ref{eq:Z_spec_displ}) and include the effect of the displacement current. 
The qualitative behavior of $\left|Z(\omega)\right|$ in each regime is similar for Figs.\@ \ref{fig:omegacross_hex}(a),(b) and \ref{fig:omegacross_square}(a),(b). However, Figs.\@ \ref{fig:omegacross_hex}(c),(d) and \ref{fig:omegacross_square}(c),(d) are radically different, as the former includes effects of non-locality due to the momentum-dependent conductivity, while the latter is equivalent to the ``phase diagram'' of the local Drude model. The lack of non-locality is due to the large Fermi-surface portions with velocity parallel to the surface and the electric field, as described in Sec.\@ \ref{Summ}. Overall, the boundaries between different regimes in Figs.\@ \ref{fig:omegacross_hex} and \ref{fig:omegacross_square} are consistent with the qualitative ``phase diagrams'' in Figs.\@ \ref{fig:3D_skin_effect_misalign}, \ref{fig:3D_skin_effect_align}, and \ref{fig:3D_skin_effect_square}.

\begin{figure}[ht]
\includegraphics[width=0.7\columnwidth]{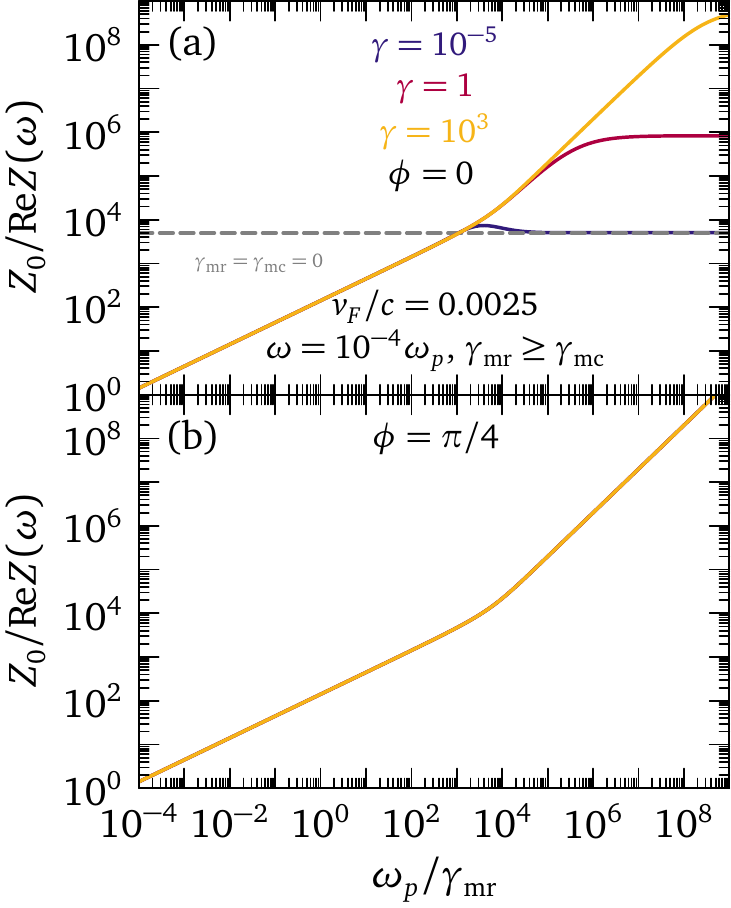} 
\caption{\label{fig:Zs_phi045_taumr_Square} Inverse surface resistance $1/\mathrm{Re}Z(\omega)$ normalized to the vacuum-impedance $Z_0$, as a function inverse momentum-relaxation rate $\omega_p/\gamma_{\rm mr}$, for a square Fermi surface with Fermi-velocity modulus $v_F=2.5 \times 10^{-3} c$, $\omega/\omega_p=10^{-4}$, and $\gamma_{\rm mc}=\mathrm{max}\left\{\gamma_{\rm mr},\bar{\gamma}\right\}$, with $\bar{\gamma}=\left\{10^{-5},1,10^3\right\}$ for the blue, red, and gold curves respectively. Specular boundary conditions are assumed, according to Eq.\@ (\ref{eq:Z_spec}). The dashed gray lines represent the limit $\gamma_{\rm mr}=\gamma_{\rm mc}=0$. (a) Results from Eq.\@ (\ref{eq:sigma_yy_dia_expl}), for tilting angle $\phi=0$ with respect to the surface; see Fig.\@ \ref{fig:Square}(a). The scattering-less limit is given by Eq.\@ (\ref{eq:Z_square_phi0_gamma0}). (b) Results from Eqs.\@ (\ref{eq:G_0_dia}) and (\ref{eq:sigma_yy_hexphi}), for $\phi=\pi/4$; see Fig.\@ \ref{fig:Square}(b). The scattering-less limit is purely imaginary, as it follows from Eq.\@ (\ref{eq:Z_hex_gamma0_highom}).
} 
\end{figure} 

In the scattering-less limit, the frequency dependence in anomalous regime is altered, as demonstrated analytically in Appendix \ref{Square_Z_analys_noscat}: for $\phi=0$ the impedance obeys $\left|Z(\omega)\right|\propto \omega^{1/2}$ as with $\gamma_{\rm mc}>0$, although with a different prefactor. This shows that, similarly to the hexagonal case of Sec.\@ \ref{Z_hex_spec}, scattering (either momentum-conserving or relaxing) qualitatively influences the impedance in anomalous regime. This feature reflects the crossover between regions {\Large \textcircled{\small D$_1$}} and {\Large \textcircled{\small D$_2$}} in Fig.\@ \ref{fig:3D_skin_effect_align}.

The presence of anomalous and viscous skin effect for the square Fermi-surface shape with $\phi=0$ can also be deduced by the saturation of the inverse surface resistance $1/\mathrm{Re}Z(\omega)$ as a function of $\omega_p/\gamma_{\rm mr}$, at fixed frequency $\omega$, as previously shown for the hexagonal case in Fig.\@ \ref{fig:Zs_phi090_taumr_Hex}. 
Our numerical results for $1/\mathrm{Re}Z(\omega)$ are reported in Figs.\@ \ref{fig:Zs_phi045_taumr_Square}(a),(b), for orientations $\phi=0$ and $\phi=\pi/4$ of the square Fermi surface respectively: these configurations correspond to Fig.\@ \ref{fig:Square}(a),(b). For $\phi=0$, the results are very similar for the square and hexagonal shapes, by comparison between Fig.\@ \ref{fig:Zs_phi045_taumr_Square}(a) and Fig.\@ \ref{fig:Zs_phi090_taumr_Hex}(a). The surface resistance saturates to a $\gamma_{\rm mc}$-dependent value in the limit $\gamma_{\rm mr} \rightarrow 0$, which signals either anomalous skin effect (for $\omega \gg \gamma_{\rm mc}$) or viscous skin effect (for $\omega \ll \gamma_{\rm mc}$). A significative difference emerges in Fig.\@ \ref{fig:Zs_phi045_taumr_Square}(b), in that the surface resistance vanishes in the limit $\gamma_{\rm mr}= 0$, and it is independent from $\gamma_{\rm mc}$, due to absence of non-local effects. 

\section{Discussion}\label{Disc}

The results of Sec.\@ \ref{Z_spec} quantitatively describe the ``phase diagrams'' of skin effect, and corroborate the qualitative interpretation of such diagrams in terms of characteristic length scales, as discussed in Sec.\@ \ref{Summ}. This consistency demonstrates that the surface impedance is an ideal probe of spatial non-locality in the electrodynamic response of isotropic and anisotropic systems. Our collision-operator formalism can be readily generalized to include the effects of short- and long-ranged interactions among conducting electrons, and to estimate the influence of different interface boundary conditions. Below we comment on the extent to which such generalizations interplay with the relations among flat Fermi-surface segments, scattering, non-local conductivity, and skin effect. We also offer a method, based on the ``ineffectiveness concept'', to infer the frequency scaling of the skin depth and impedance from how the conductivity scales with $q_x$ and $\omega$ in all regimes discussed in Sec.\@ \ref{Z_spec}. 

\subsection{A generalized ``ineffectiveness concept'' for anisotropic skin effect}\label{Gen_ineffectiveness}

The relations between conductivity, skin depth, and impedance in different anisotropic regimes lend themselves to an intuitive physical interpretation, which is based on the ``ineffectiveness concept'' introduced by Pippard to estimate the crossover between normal and anomalous skin effect in isotropic systems \cite{Pippard-1947a,Pippard-1947b, Dressel-2001}. The argument is summarized as follows: when the skin depth $\delta_s(\omega)$ is much shorter than the mean free path $l_m=v_F \tau_{\rm mr}$, only those electrons that travel approximately parallel to the sample surface are able to react to the incident electric field, and they perceive a constant field. Assuming that those electrons lie within a small angle $\pm \theta_{\rm eff} \approx \beta \delta_s(\omega)/l_m$ with respect to the normal to the surface, the effective density of reacting electrons is $n_{\rm eff}=n \beta \delta_s(\omega)/l_m$, where $n$ is the physical electron density per unit volume. This can be interpreted as an effective Ohmic (Drude-like) conductivity $\sigma_{\rm eff}(\omega)=n_{\rm eff} e^2 \tau_{\rm mr}/m=n \beta e^2 n \delta_s(\omega)/v_F$, which now depends on the skin depth itself. Alternatively, one can think about $\sigma_{\rm eff}(\omega)$ as containing an effective scattering time $\tau_{\rm eff}=\beta \delta_s(\omega)/v_F$ for electrons.  
Inserting this effective conductivity into the expression for normal (Ohmic) skin effect \cite{Reuter-1948,Sondheimer-2001,Dressel-2001}
\begin{equation}\label{eq:skin_depth_normal}
\delta_s(\omega)=\sqrt{2/\left[\mu_0 \omega \delta_s(\omega)\right]},
\end{equation}
we obtain a self-consistent relation for the skin depth that leads to $\delta_s(\omega)=\left[2 l_m/(\mu_0 \beta \sigma_0 \omega)\right]^{1/3}$, with $\sigma_0=n e^2 \tau_{\rm mr}/m$: this is indeed the skin depth in anomalous regime for isotropic systems. 
The assumption on which the above line of thought relies are that the effective conductivity $\sigma_{\rm eff}(\omega) \propto 1/q_x \propto \delta_s(\omega)$. Besides, momentum-conserving collisions $\gamma_{\rm mc}\neq \gamma_{\rm mr}$ are not explicitly considered, which excludes the occurrence of viscous skin effect. Pippard also applied his qualitative arguments to anisotropic systems in 2D and 3D \cite{Pippard-1954} \footnote{Through his argument, Pippard realized in the anisotropic case that the surface resistance is dominated by the portions of the Fermi surface that are parallel to the applied field \cite{Pippard-1954}, in agreement with the present study}, maintaining the above fundamental assumptions, and later Sondheimer quantitatively confirmed these arguments for ellipsoidal Fermi surfaces \cite{Sondheimer-1954}.

Following the ineffectiveness concept, we can generalize the qualitative estimations as follows. Suppose that the conductivity evolves as 
\begin{equation}\label{eq:cond_estim}
\frac{\sigma_{yy}(q_x,\omega)}{\epsilon_0 \omega_p^2}=\frac{\alpha_1}{\omega^\beta q_x^\alpha},
\end{equation}
with $\alpha_1$ a numerical factor independent of frequency and momentum, and $\left\{\eta,\zeta\right\}\in \mathbb{Z}$. From the considerations of Sec.\@ \ref{Skin}, there will eventually be a single polariton branch that is damped less that others at the highest depths into the metal, for which we can employ Eq.\@ (\ref{eq:skin_depth_mode}): hence, we can estimate $\delta_s(\omega) \approx \alpha_2/q_x$, with $\alpha_2$ a numerical factor. This gives 
\begin{equation}\label{eq:cond_estim2}
\frac{\sigma_{yy}[\delta_s(\omega),\omega]}{\epsilon_0 \omega_p^2}=\alpha_3\frac{ \delta_s^{\alpha}(\omega)}{\omega^\beta},
\end{equation}
with $\alpha_3=\alpha_1 \alpha_2^{-\alpha}$. Identifying Eq.\@ (\ref{eq:cond_estim2}) with the effective Ohmic conductivity to utilize for the ineffectiveness argument, we equate
\begin{equation}\label{eq:sigma_eff_ineff}
\sigma_{yy}[\delta_s(\omega),\omega]\equiv \sigma_{\rm eff}(\omega)=\epsilon_0 \omega_p^2 \tau_{\rm eff}(\omega),
\end{equation}
from which we deduce an effective scattering time $\tau_{\rm eff}(\omega)=\alpha_3 \omega^{-\beta} \delta_s^\alpha(\omega)$, or an effective carrier density alternatively. Inserting the effective conductivity (\ref{eq:sigma_eff_ineff}) into the Ohmic skin-depth expression Eq.\@ (\ref{eq:skin_depth_normal}), we obtain a self-consistent relation that we can solve for $\delta_s(\omega)$: 
\begin{equation}\label{eq:skin_depth_gen_ineff}
\delta_s(\omega)=\left(\frac{2}{\alpha_3}\right)^{\frac{1}{2 +\alpha}} \lambda_L^{\frac{2}{2+\alpha}} \omega^{\frac{\beta-1}{\alpha+2}},
\end{equation}
which links the skin depth with the dependence of the conductivity (\ref{eq:cond_estim}) on frequency and momentum. The estimation for the impedance then follows from Eq.\@ (\ref{eq:skin_depth_Z_gen}). The latter relation is derived for specular surface scattering in Appendix \ref{Z_spec_scaling}. 
The qualitative estimation provided by Eq.\@ (\ref{eq:skin_depth_gen_ineff}) can be applied even when the conductivity has various forms of momentum dependence, which occurs for the anomalous and viscous regimes, and to both isotropic and anisotropic cases in different orientation. For instance: in anomalous regime for isotropic systems we have $\beta=0$ and $\alpha=1$, so that $\delta_s(\omega)\propto \omega^{-1/3}$, consistently with Eq.\@ (\ref{eq:Z_iso_anom}); in the hexagonal and square geometries with $\phi=0$ we have $\beta=-1$ and $\alpha=2$, which yields $\delta_s(\omega)\propto \omega^{-1/2}$, in agreement with Eqs.\@ (\ref{eq:aniso_flat2}) and (\ref{eq:aniso_flat4}), and their counterparts for $\gamma_{\rm mc}>0$; in extreme hydrodynamic regime and for isotropic and anisotropic systems alike, we have $\beta=0$ and $\alpha=2$, so that $\delta_s(\omega)\propto \omega^{-1/4}$, which agrees with Figs.\@ \ref{fig:omegacross_iso}, \ref{fig:omegacross_hex} and \ref{fig:omegacross_square}(a),(b). 
Therefore, the generalization of the ineffectiveness concept to momentum-dependent conductivities, as here outlined, can help in guessing the qualitative evolution of the skin depth and the impedance modulus with frequency, from the expression of the conductivity itself. 

\subsection{Quasiparticle interactions}

The kinetic approach employed in this work is ultimately phenomenological, as it does not depend on the microscopic origin of momentum-conserving and relaxing scattering. However, by the same logic, the results here derived can be applied regardless of the specific mechanisms available for electron scattering, i.e.\@, to different origins of spatial non-locality, as long as the kinetic approach remains valid. For example, the awareness that non-local hydrodynamic behavior can not only emerge from electron-electron interactions, but also through the interaction of carriers with phonons \cite{Steinberg-1958,Gurzhi-1972,Levchenko-2020} and impurities \cite{Burmistrov-2019,Hui-2020} has recently been raised. In all these cases, the electrodynamic phenomenology can be investigated through the Boltzmann theory outlined in this paper, in the presence or absence of anisotropy. A microscopic account, on how distinct forms of electron self-energies from different interaction channels produce visco-elastic phenomenologies in electrodynamics, will be reported elsewhere \cite{Valentinis-2021-manybody}. 

A natural extension of the present work involves the inclusion of quasiparticle interactions, both long-ranged (i.e.\@, the Coulomb potential) and short-ranged (e.g.\@, electron-hole excitations) \cite{Lucas-2018}. In the isotropic 3D case, this procedure is at the basis of Landau-Silin theory of charged Fermi liquids \cite{Silin-1958a,Silin-1958b}, where short-ranged interactions are expanded into the basis of spherical harmonics. Likewise, in isotropic 2D systems the appropriate expansion basis is given by angular momentum states \cite{Lucas-2018}; see also Sec.\@ I of the Supplemental Material. 
Schematically, the inclusion of interactions modifies the Boltzmann equation (\ref{eq:BE}) to \cite{Vignale-2005,Lucas-2018}
\begin{multline}\label{eq:BE_Ftot}
\frac{\partial f_{\vec{k}}\left(\vec{r},t\right)}{\partial t}+\vec{F}_{t}(\vec{r},t)\cdot\mathbf{\nabla}_{\vec{k}}f_{\vec{k}}\left(\vec{r},t\right)+\vec{v}_{\vec{k}}\cdot\mathbf{\nabla}_{\vec{r}}f_{\vec{k}}\left(\vec{r},t\right) \\ =-\left({\cal C} f\right)_{\vec{k}},
\end{multline}
where now $\vec{F}_{t}(\vec{r},t)$ contains the contribution of the external electric field, and of the gradient of the interacting quasiparticle dispersion:
\begin{equation}\label{eq:F_tot}
\vec{F}_t(\vec{r},t)=-e \vec{E}(\vec{r},t)-\nabla_{\vec{r}} \mathscr{E}_{\vec{k}}(\vec{r},t). 
\end{equation}
In turn, the interacting dispersion $\mathscr{E}_{\vec{k}}(\vec{r},t)=\varepsilon_{\vec{k}}+\varepsilon_L\left[\delta f_{\vec{k}}(\vec{r},t)\right]+\varepsilon_{\vec{k}}^C(\vec{r},t)$ contains the Landau short-ranged interaction $\varepsilon_L\left[\delta f_{\vec{k}}(\vec{r},t)\right]$ and the long-ranged Coulomb term $\varepsilon_{\vec{k}}^C(\vec{r},t)$ in a self-consistent Vlasov approach. Both interactions have to be expanded in terms of the eigenfunctions of the collision operator (\ref{eq:Coll_op_psi}), which are known to be $\xi_{\vec{k},m}=c_0 e^{i m \theta}$ in the isotropic case, but differ in the presence of anisotropy. 
Then, from Eq.\@ (\ref{eq:F_tot}) it is evident that interactions affect the source term (\ref{eq:s_m_chik}) at the right-hand side of the infinite linear system (\ref{eq:Boltz_chik}), which modifies the form of collective modes \cite{Lucas-2018} and generates an effective renormalized Fermi velocity $v_F^\ast$. We defer the details of this treatment to a subsequent work.
Equally interesting is the case where interactions stem from an external magnetic field, which is able to generate novel forms of anomalous skin effect, such as the \emph{anomaly-induced-non-locality} due to the chiral anomaly in Weyl semimetals \cite{Matus-2021}. 

\subsection{Influence of interface boundary conditions}\label{Bound_cond_disc}

Another crucial aspect of the surface-impedance calculations are the interface boundary conditions \cite{FZVM-2014,Kiselev-2019,Moessner-2019,Valentinis-2021a,Wagner-2015_preprint}. In Sec.\@ \ref{Z_spec}, we assumed specular scattering at the vacuum-metal interface, which is physically similar to a smooth surface with no-stress boundary conditions. Recently, this type of boundary conditions were experimentally realized in transport measurements on GaAs/AlGaAs accumulation-mode heterostructures \cite{Keser-2021}. 
To investigate the influence of a different degree of specularity in interface scattering, one can proceed similarly to Reuter and Sondheimer \cite{Reuter-1948,Sondheimer-1954,Sondheimer-2001}, by considering the extreme cases of specular or diffusive surface scattering for electrons in the metal. An interpolation between these two extremes can be achieved by assuming a specularity factor $p\in\left[0,1\right]$ for both electric field and current density, which means that a portion $p$ of scattering is specular, while a part $1-p$ happens diffusively. In the diffusive-scattering case, the surface electric field $E_y(0^+,\omega)$ can then be linked to the conductivity $\sigma_{yy}(q_x,\omega)$ through a momentum integration over the half-space occupied by the metal, in a similar way as for the specular-scattering case of Eq.\@ (\ref{eq:E0_sigma}). In Sec.\@ IV of the Supplemental Material we checked our calculations for the surface impedance modulus $\left|Z(\omega)\right|$ assuming diffusive scattering ($p=0$), and we found negligible differences with respect to the specular-scattering case of Sec.\@ \ref{Z_spec}. Therefore, the results for the impedance modulus are robust against the effect of boundary conditions. 

However, the surface resistance $\mathrm{Re}Z(\omega)$ is dramatically sensitive to boundary conditions in the anomalous reflection, or extreme anomalous, regime ({\Large \textcircled{\normalsize E}} in Fig.\@ \ref{fig:3D_skin_effect}). This sensitivity appears because the region {\Large \textcircled{\normalsize E}} is least affected by bulk scattering, with both relaxation and momentum-conserving collisions being negligible, so that interface scattering at the vacuum-metal surface assumes a major role for the conduction properties \cite{Holstein-1952, Casimir-1967b,Wooten-1972opt}. We reserve the detailed comparison of boundary conditions for the surface resistance to a follow-up work. 

\subsection{Effect of rounded corners and Fermi-surface curvature}\label{Rounded}

\begin{figure}[ht]
\includegraphics[width=0.8\columnwidth]{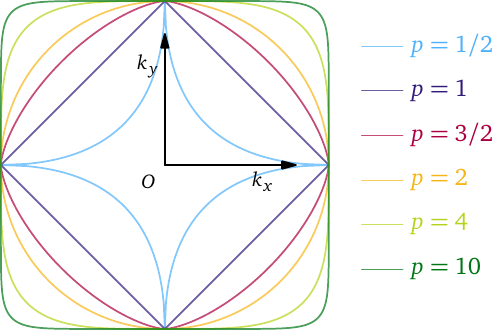}
\caption{\label{fig:supercircle} The ``supercircle'' Fermi-surface geometry, for different values of the parameter $p$ which controls the curvature of individual segments. The ``diamond-shaped'' Fermi surface of Fig.\@ \ref{fig:Square} is retrieved for $p=1$, while for $p=2$ one retrieves a circular Fermi surface.
}
\end{figure}

In Sec.\@ \ref{Z_spec} we obtained the impedance ``phase diagrams'' of Figs.\@ \ref{fig:omegacross_hex} and \ref{fig:omegacross_square} under the assumption that the 2D Fermi surface is a perfect regular polygon. While such geometry allowed us to clearly identify how the different regimes for $Z(\omega)$ are affected by the anisotropy and orientation of the polygonal shape, realistic Fermi surfaces possess more complex shapes. In particular, even when these shapes approximately resemble a polygon, the corresponding faces are not perfectly flat, and the sharp angles of the idealized polygonal cases are replaced by rounded corners. For this reason, it is interesting to investigate the effect of small curvature of Fermi-surface segments and of rounded corners on the shapes studied in Sec.\@ \ref{Z_spec}. Do the effects of anisotropy on the surface impedance persist (at least partially), when we take into account Fermi-surface non-idealities? We will demonstrate in this section that the answer is yes, considering specific examples of non-polygonal geometries. 
\begin{figure*}[ht] \centering
\includegraphics[width=0.8\textwidth]{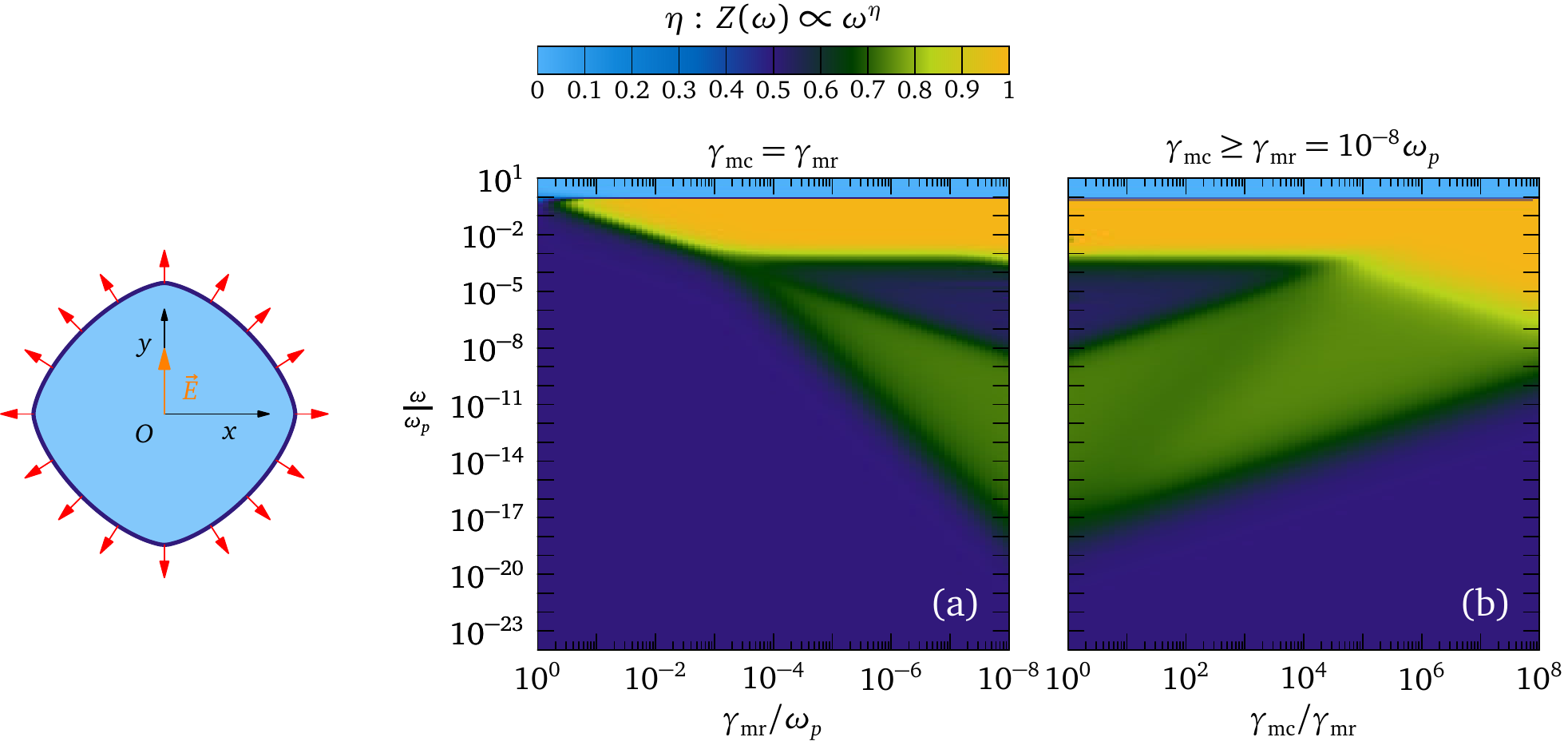}
\caption{\label{fig:omegacross_superc_1p5} Skin effect regimes for a ``supercircle'' Fermi surface (\ref{eq:r_supercircle}) with $p=3/2$, as measured by the surface impedance $Z(\omega)$ as a function of relaxation rate $\gamma_{\rm mr}$, momentum-conserving collision rate $\gamma_{\rm mr}$, and frequency $\omega/\omega_p$, where $\omega_p$ is the plasma frequency (\ref{eq:omega_p_iso}). The Fermi-surface geometry is sketched on the left-hand side of the plot, together with the applied electric field $\vec{E}=E_y \hat{u}_y$ aligned with the $y$ axis, and the local Fermi velocity vectors shown by red arrows. (a) Regimes in the $\left(\gamma_{\rm mr},\omega \right)$ plane, for $\gamma_{\rm mr}=\gamma_{\rm mc}$. (b) Regimes in the $\left(\gamma_{\rm mc},\omega \right)$ plane, for fixed $\gamma_{\rm mr}=10^{-8}\omega_p$. The color palette is the density plot of $\mathrm{Arg}Z(\omega)/(-\pi/2)$, which yields the exponent $\eta$ of $Z(\omega)\propto \omega^\eta$. 
} 
\end{figure*}

To generalize the treatment of Secs.\@ \ref{Cond_poly}-\ref{Z_spec}, let us consider an arbitrarily varying Fermi wave vector $k_F(\theta)$ in the 2D plane, as a function of angle $\theta$. Then, in accordance with the parametrization (\ref{eq:v_k_F}), the components of the unit vector $\hat{n}_{\vec{k}}$ along the $x$ and $y$ axes are
\begin{subequations}\label{eq:polar_Cartesian_kF}
\begin{equation}\label{eq:cart_k_x}
n_{\vec{k},x}=N^{-1}\left[\cos\theta k_F(\theta)+\sin\theta \frac{d k_F(\theta)}{d \theta} \right],
\end{equation}
\begin{equation}\label{eq:cart_k_y}
n_{\vec{k},y}=N^{-1}\left[\sin\theta k_F(\theta)-\cos\theta \frac{d k_F(\theta)}{d \theta}  \right],
\end{equation}
\begin{equation}\label{eq:N_normaliz}
N=n_{\vec{k}}=\sqrt{\left[k_F(\theta)\right]^2+\left[\frac{d k_F(\theta)}{d \theta}\right]^2},
\end{equation}
\end{subequations}
as shown in Appendix \ref{Curvature_2D}. Hence, given a parametrization $k_F(\theta)$, Eqs.\@ (\ref{eq:polar_Cartesian_kF}) allow us to compute the transverse conductivity through Eqs.\@ (\ref{eq:sigma_yy}) and (\ref{eq:G_0_iso}), as well as the surface impedance through Eqs.\@ (\ref{eq:Z_spec_displ}) and (\ref{eq:S_nonloc}). This way, we can compare the corresponding ``phase diagrams'' for $Z(\omega)$ with the ones of the previously considered polygonal shapes.
Notice that the integration line element $d S= n_{\vec{k}}(\theta) d\theta$ varies along the Fermi surface, in accordance with Eq.\@ (\ref{eq:N_normaliz}). This feature holds for polygonal geometries as well, but in the present non-ideal case we cannot split the integration over $d S$ into discrete sums as done in the polygonal case. 

As an example, in Sec.\@ \ref{Supercircle} we consider the ``supercircle'' geometry, in which a square Fermi surface can be morphed into shapes with nonzero curvature of segments and rounding of corners by the means of a single parameter $p$. Our investigation culminates in Sec.\@ \ref{PdCoO2_2D} with the analysis of a more realistic 2D case: the in-plane experimental Fermi surface of PdCoO$_2$, as parametrized through angle-resolved photoemission and quantum oscillations measurements \cite{Hicks-2012,Nandi-2018,Bachmann-2021}. 

\subsubsection{The ``supercircle'' geometry}\label{Supercircle}

The ``supercircle'', a special case of Lamé curve or ``superellipse'' for equal semiaxis, satisfies the equation $\left| x\right|^p+ \left| y\right|^p=r^p$, $p>0$, for radius $r$. This equation allows one to smoothly interpolate an astroid (for $p<1$), a ``diamond-shaped'' geometry (for $p=1$), a circle  (for $p=2$), and a square with rounded corners (for $p>2$), as a function of a single parameter $p$ which controls the shape curvature. In our case, the radius is the Fermi wave vector: $r \equiv k_F$. 
Using polar coordinates $k_x=k_F \cos \theta$ and $k_y=k_F \sin\theta$, the supercircle radius results
\begin{equation}\label{eq:r_supercircle}
k_F=k_F(\theta)=\left(\left|\cos\theta\right|^p+\left|\sin\theta\right|^p\right)^{-1/p}.
\end{equation}
Fig.\@ \ref{fig:supercircle} shows the Fermi surface corresponding to Eq.\@ (\ref{eq:r_supercircle}), for different values of $p>0$. 
In Appendix \ref{Curvature_2D}, we show that the results for the ``diamond-shaped'' and circular geometries, presented in Secs.\@ \ref{Square_dia}, \ref{Z_sq_spec} and in Secs.\@ \ref{Iso_circ}, \ref{App_Z_iso_analysis} respectively, are indeed retrieved as special cases of the parametrization (\ref{eq:r_supercircle}). 
\begin{figure*}[ht] \centering
\includegraphics[width=0.8\textwidth]{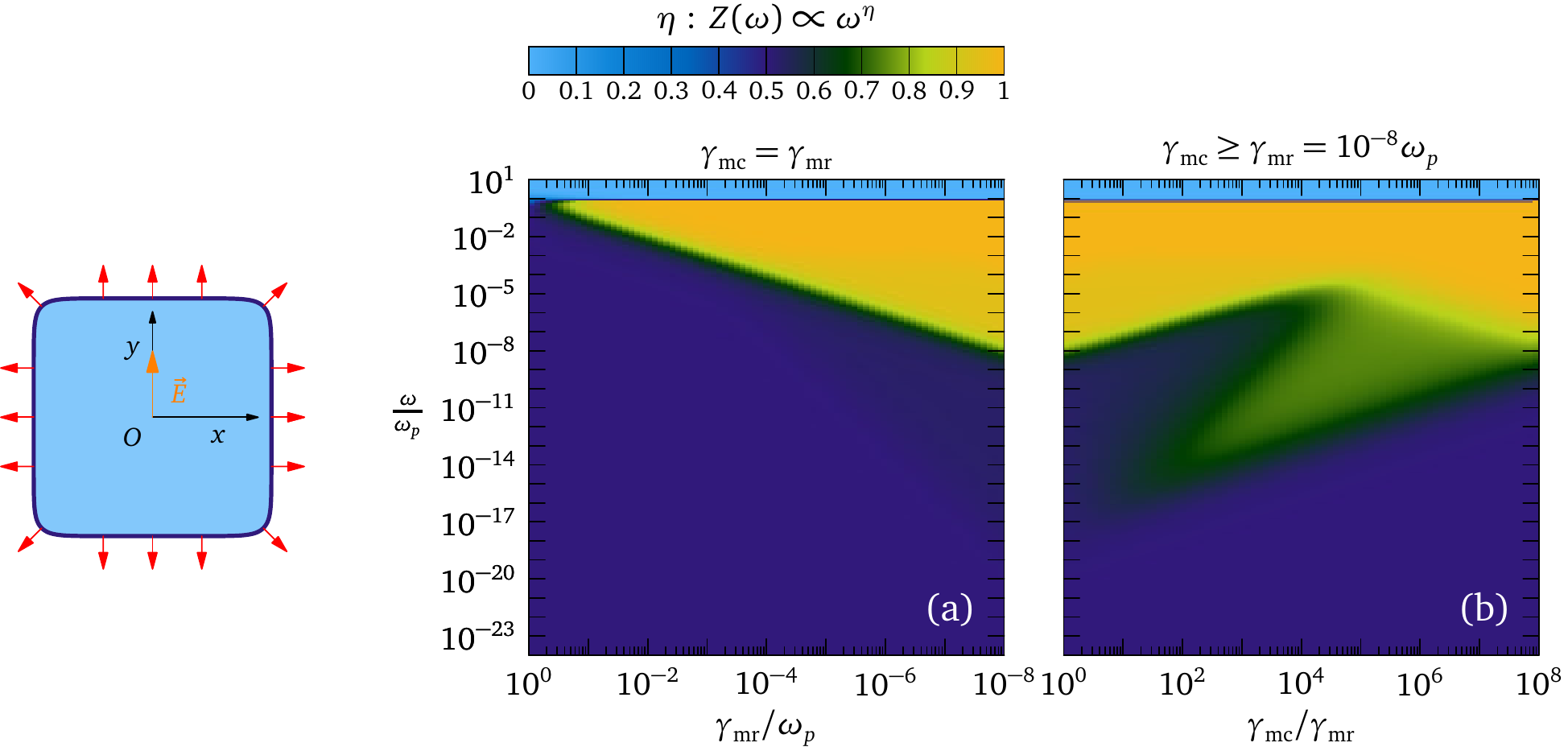}
\caption{\label{fig:omegacross_superc_10} Skin effect regimes for a ``supercircle'' Fermi surface (\ref{eq:r_supercircle}) with $p=10$, as measured by the surface impedance $Z(\omega)$ as a function of relaxation rate $\gamma_{\rm mr}$, momentum-conserving collision rate $\gamma_{\rm mr}$, and frequency $\omega/\omega_p$, where $\omega_p$ is the plasma frequency (\ref{eq:omega_p_iso}). The Fermi-surface geometry is sketched on the left-hand side of the plot, together with the applied electric field $\vec{E}=E_y \hat{u}_y$ aligned with the $y$ axis, and the local Fermi velocity vectors displayed by red arrows. (a) Regimes in the $\left(\gamma_{\rm mr},\omega \right)$ plane, for $\gamma_{\rm mr}=\gamma_{\rm mc}$. (b) Regimes in the $\left(\gamma_{\rm mc},\omega \right)$ plane, for fixed $\gamma_{\rm mr}=10^{-8}\omega_p$. The color palette is the density plot of $\mathrm{Arg}Z(\omega)/(-\pi/2)$, which gives the exponent $\eta$ of $Z(\omega)\propto \omega^\eta$. 
} 
\end{figure*}

Fig.\@ \ref{fig:omegacross_superc_1p5} shows the results for the intermediate case $p=3/2$, which corresponds to a ``diamond'' shape with additional curvature of the four Fermi-surface segments. Such curvature is further enhanced for $p>3/2$, until for $p=2$ the Fermi surface becomes perfectly circular. Hence, the case $1<p<2$ is suitable to analyze the effect of deforming the polygonal geometries by introducing a finite curvature into the flat segments. 
We notice subtle differences with respect to the ``diamond-shaped'' dispersion of Fig.\@ \ref{fig:omegacross_square}(a),(b): the exponent $\eta$ in region {\Large \textcircled{\small D$_1$}} becomes slightly smaller than $\eta=3/4$ (for example, we have $\eta\approx 0.73$ around the center of region {\Large \textcircled{\small D$_1$}} in Fig.\@ \ref{fig:omegacross_superc_1p5}(b)), while in region {\Large \textcircled{\small D$_2$}} $\eta$ is slightly larger than $\eta=1/2$ (specifically, $\eta \approx 0.55$ around the center of region {\Large \textcircled{\small D$_2$}} in Fig.\@ \ref{fig:omegacross_superc_1p5}(b)). This effect is due to the finite curvature of the Fermi-surface segments, which also produces a visually distinguishable difference between regions {\Large \textcircled{\small D$_1$}} and {\Large \textcircled{\normalsize V}} in Fig.\@ \ref{fig:omegacross_superc_1p5}(b). Such differences are gradually enhanced for $3/2<p<2$, until for $p=2$ we retrieve the isotropic ``phase diagram'' of Fig.\@ \ref{fig:omegacross_iso}, where $\eta=2/3$ in both regions {\Large \textcircled{\small D$_1$}} and {\Large \textcircled{\small D$_2$}}. 
Hence, Fig.\@ \ref{fig:omegacross_superc_1p5} provides us a practically relevant information: if Fermi-surface segments are not exactly flat in practice, then the exponents $\eta$ in the anomalous regions {\Large \textcircled{\small D$_1$}} and {\Large \textcircled{\small D$_2$}} quantitatively differ with respect to the ideal flat-segments case. Still, even for a deformation of the ``diamond-shaped'' case as large as $p=3/2$, the ``phase diagrams'' of Figs.\@ \ref{fig:omegacross_superc_1p5} and \ref{fig:omegacross_square}(a),(b) are qualitatively consistent, and the effect of Fermi-surface anisotropy is robust with respect to the curvature of Fermi-surface segments. Such robustness may be rationalized by realizing that a small curvature of segments predominantly affects the low-frequency part of the ``phase diagram'', where the response is anyway diffusive and local; e.g.\@, in region {\Large \textcircled{\normalsize A}} the exponent $\eta=1/2$ is insensitive to the change in Fermi-surface curvature because the response is dominated by momentum relaxation at rate $\gamma_{\rm mr}$. 
Therefore, we conclude that the ``phase diagrams'' derived for the polygonal shapes of Secs.\@ \ref{Z_hex_spec} and \ref{Z_sq_spec} provide qualitative guidance on the effect of dispersion anisotropy on the surface impedance, even in the presence of a small curvature of Fermi-surface segments, for shapes that can still be approximated by an ideal polygonal geometry. Quantitative difference with respect to a perfect polygon emerge at the level of the numerical value of the exponent $\eta$, especially in the anomalous regions {\Large \textcircled{\small D$_1$}} and {\Large \textcircled{\small D$_2$}} which are the most sensitive to Fermi-surface geometry. 

Furthermore, to selectively analyze the effect of rounded corners, we show the results for $p=10$ in Fig.\@ \ref{fig:omegacross_superc_10}: this geometry consist of a square with additional rounding of corners, and as such it can be directly compared with Fig.\@ \ref{fig:omegacross_square}(c),(d). The case of a perfect square is retrieved in the limit $p\rightarrow +\infty$. 
Here we appreciate a qualitative difference in Fig.\@ \ref{fig:omegacross_superc_10}(b) with respect to Fig.\@ \ref{fig:omegacross_square}(d): the rounding of corners has reintroduced the viscous regime {\Large \textcircled{\normalsize V}} with $\eta=3/4$, which was absent in the idealized square shape of Fig.\@ \ref{fig:omegacross_square}(d). Moreover, the crossovers between the viscous, anomalous and perfect-conductor regions are all altered with respect to the square shape. In fact, the ``phase diagram'' of Fig.\@ \ref{fig:omegacross_superc_10} is qualitatively consistent with the hexagonal case of Fig.\@ \ref{fig:omegacross_square}(c),(d): this is because such hexagonal geometry shares all its relevant geometrical features with the ``supercircle'' for $p=10$: two Fermi-surface segments have velocity essentially parallel to the applied field -- see also Secs.\@ \ref{Cond_poly_sigma} and \ref{Cond_poly_skin} -- while the rest of the Fermi surface does not have a velocity identically orthogonal to the field as is the case for a perfect square. Hence, we realize that the perfect square is an idealized special case, for which the non-local character of the response vanishes entirely from the ``phase diagram'' in Fig.\@ \ref{fig:omegacross_square}(c),(d). In reality, rounded corners of an approximately polygonal geometry qualitatively impact the exponent $\eta$ of the surface impedance. Still, the ``phase diagram'' of Fig.\@ \ref{fig:omegacross_superc_10} qualitatively differs from the isotropic case  of Fig.\@ \ref{fig:omegacross_iso}, which demonstrates that rounded corners do not completely suppress the effect of Fermi-surface anisotropy on the electrodynamic response. 

The idealized polygons of Secs.\@ \ref{Hex_FS} and \ref{Square_FS} serve as an illustration of the effect of dispersion anisotropy, as they allow for a clear-cut analysis of the different regimes of anisotropic skin effect. However, in order to test our kinetic theory by comparing it with optical spectroscopy data, it is advantageous to consider more realistic Fermi-surface shapes, which properly model the experimental dispersion of the material under investigation. For this reason, in the next section we consider the exemplary case of the experimental parametrization of the Fermi surface of PdCoO$_2$, which involves both portions with finite curvature and rounded corners. 

\subsubsection{The 2D Fermi surface of PdCoO$_2$}\label{PdCoO2_2D}

\begin{figure*}[ht] \centering
\includegraphics[width=0.8\textwidth]{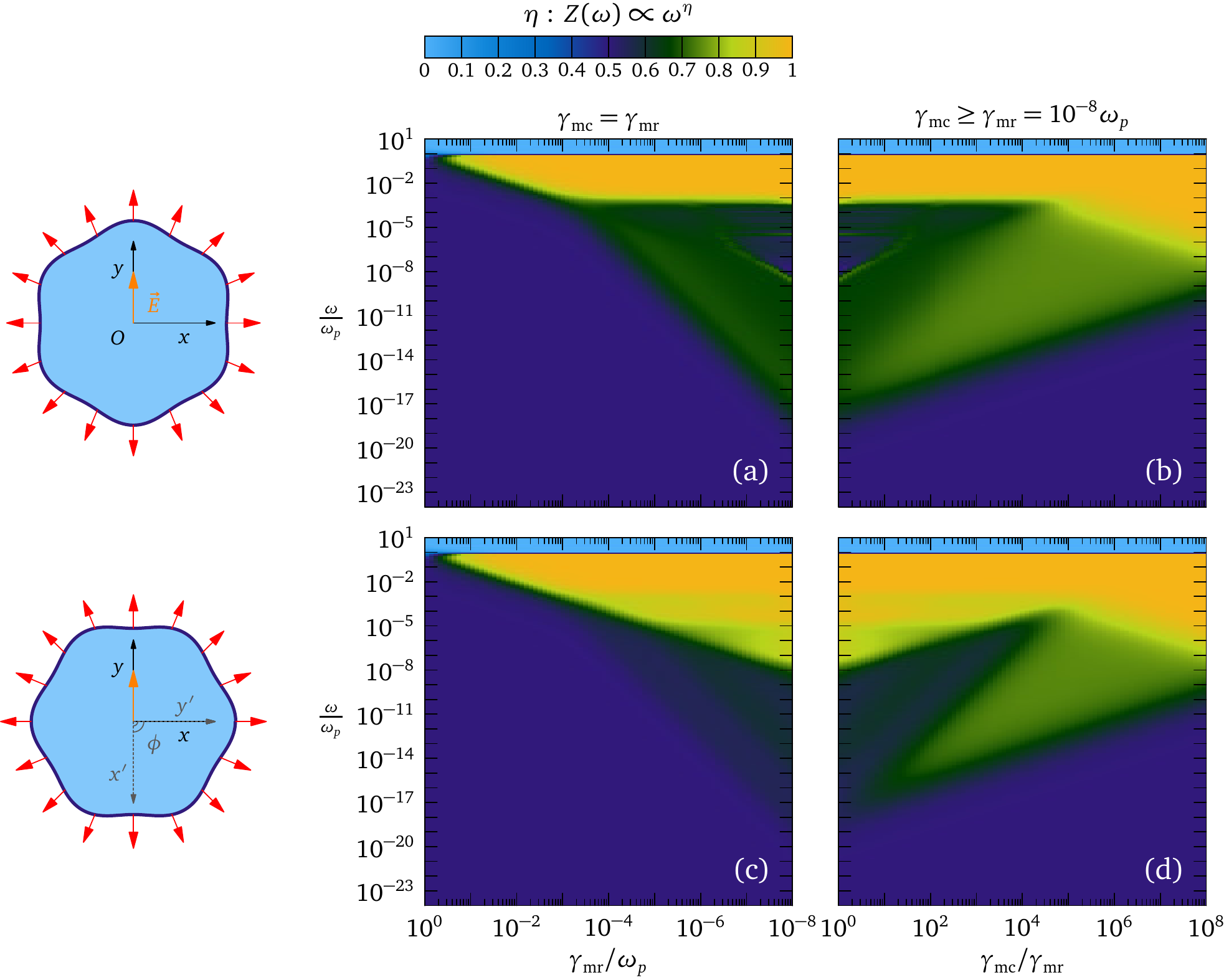}
\caption{\label{fig:omegacross_PCO} Orientational dependence of skin effect regimes for the 2D parametrization (\ref{eq:k_F_PCO_exp}) of the Fermi surface of PdCoO$_2$, as measured by the surface impedance $Z(\omega)$ as a function of relaxation rate $\gamma_{\rm mr}$, momentum-conserving collision rate $\gamma_{\rm mr}$, and frequency $\omega/\omega_p$, where $\omega_p$ is the plasma frequency (\ref{eq:omega_p_iso}). The corresponding orientation is sketched on the left-hand side of each plot, together with the applied electric field $\vec{E}=E_y \hat{u}_y$ aligned with the $y$ axis. Red arrows depict the local Fermi velocity vectors. (a) Configuration with $\theta_0=\pi/2$ in the $\left(\gamma_{\rm mr},\omega \right)$ plane, for $\gamma_{\rm mr}=\gamma_{\rm mc}$. (b) Configuration with $\theta_0=\pi/2$ in the $\left(\gamma_{\rm mc},\omega \right)$ plane, for fixed $\gamma_{\rm mr}=10^{-8}\omega_p$. (c) Geometry with $\theta_0=0$ in the $\left(\gamma_{\rm mr},\omega \right)$ plane, for $\gamma_{\rm mr}=\gamma_{\rm mc}$. (d) Results for $\theta_0=0$ in the $\left(\gamma_{\rm mc},\omega \right)$ plane, for fixed $\gamma_{\rm mr}=10^{-8}\omega_p$. The color palette is the density plot of $\mathrm{Arg}Z(\omega)/(-\pi/2)$, giving the exponent $\eta$ of $Z(\omega)\propto \omega^\eta$. 
} 
\end{figure*}
The experimental parametrization of the Fermi surface of the delafossite compound PdCoO$_2$, according to quantum oscillations experiments, shows a quasi-2D geometry with a slight warping in the direction perpendicular to the Pd planes \cite{Hicks-2012,Baker-2022_preprint}. Here we adopt another common 2D parametrization \cite{Sunko_thesis-2019}:
\begin{subequations}\label{eq:k_F_PCO_exp}
\begin{equation}\label{eq:k_F_PCO_exp_1}
\vec{k}_F(\theta)=\rho(\theta-\theta_0)\left(\cos\theta \hat{u}_x +\sin\theta \hat{u_y} \right),
\end{equation}
\begin{multline}\label{eq:k_F_PCO_exp_2}
\rho(\theta-\theta_0)=k_{0}+k_6\cos\left[6(\theta-\theta_0)\right]\\ +k_{12}\cos\left[12(\theta-\theta_0)\right].
\end{multline}
\end{subequations}
$\theta$ characterizes the angular dependence of $k_F$ in the $\left(k_x,k_y\right)$ plane, while $\theta_0$ takes into account a possible in-plane rotation of the Fermi surface with respect to the $x$ and $y$ axes. The angular harmonics are $k_0=0.9518 {\AA}^{-1}$, $k_6=0.0444 {\AA}^{-1}$, and $k_{12}=0.0048 {\AA}^{-1}$.
Eqs.\@ (\ref{eq:k_F_PCO_exp}) allows us to illustrate the effect of a more realistic 2D parametrization on our impedance ``phase diagrams'', while still taking advantage of the simplification (\ref{eq:mirror_planes}) due to the existence of two mirror symmetry planes in reciprocal space. A more general parametrization is employed in Ref.\@ \onlinecite{Baker-2022_preprint}, where we directly compare our kinetic theory with surface resistance measurements on PdCoO$_2$. In accordance with the fits of bandstructure calculations to angle-resolved photoemission data in Ref.\@ \onlinecite{Nandi-2018}, we still employ Eq.\@ (\ref{eq:v_k_F}), i.e.\@, we assume that the Fermi-velocity modulus has no angular dependence and equals $v_F=7.5 \times 10^5$ m/s \cite{Hicks-2012}.  

Eqs.\@ (\ref{eq:k_F_PCO_exp}) produce the Fermi surfaces on the left of Fig.\@ \ref{fig:omegacross_PCO}: $\theta=\pi/2$ and $\theta=0$ correspond to the sketch at the left of Figs.\@ \ref{fig:omegacross_PCO}(a) and \ref{fig:omegacross_PCO}(c), respectively. The numerical results for the corresponding impedance ``phase diagrams'' are shown in Figs.\@ \ref{fig:omegacross_PCO}(a)-(d), which can be directly compared with the ideal hexagonal geometry of Fig.\@ \ref{fig:omegacross_hex}. As remarked in Sec.\@ \ref{Supercircle} for the rounded-square shape, the rounded corners produced by Eq.\@ (\ref{eq:k_F_PCO_exp}) quantitatively modify the exponent $\eta$ in anomalous regime with respect to the one for a perfect hexagonal Fermi surface. In particular, the differences between the orientations of Figs.\@ \ref{fig:omegacross_PCO}(a),(b) and \ref{fig:omegacross_PCO}(c),(d) are diminished, with respect to the sharper differences between Figs.\@ \ref{fig:omegacross_hex}(a),(b) and \ref{fig:omegacross_hex}(c),(d) for a hexagonal geometry. Therefore, relaxing the approximation of a perfect hexagonal shape quantitatively smoothens some of the features resulting from anisotropy, since the rounded corners and slightly curved segments decrease the overall anisotropy of the Fermi surface. However, there are still discernible differences between Figs.\@ \ref{fig:omegacross_PCO}(a),(b) and \ref{fig:omegacross_PCO}(c),(d), particularly in the anomalous regimes {\Large \textcircled{\small D$_1$}} and {\Large \textcircled{\small D$_2$}}, which qualitatively still resemble the idealized cases of Figs.\@ \ref{fig:omegacross_hex}(a),(b) and \ref{fig:omegacross_hex}(c),(d). Hence, we conclude that anisotropic effects in the skin effect are not completely suppressed, once we model Fermi surfaces that are more realistic than idealized polygonal geometries. Indeed, surface resistance measurements on PdCoO$_2$ using a bolometric microwave spectrometer confirm the anisotropy of skin effect in such compound, in full agreement with our kinetic theory \cite{Baker-2022_preprint}. 
 
\section{Conclusions}\label{Concl}

By expanding the collision integral in the eigenbasis of the collision operator, we solved the Boltzmann equation with an electric-field source term, and obtained the distribution function and the transverse conductivity for arbitrary electronic dispersion, under the minimal assumption of two (in 2D) or three (in 3D) symmetry planes in momentum space, taking into account momentum-preserving collisions at rate $\gamma_{\rm mc}$ as well as momentum relaxation at rate $\gamma_{\rm mr} \leq \gamma_{\rm mc}$. The conductivity (\ref{eq:lonloc_sigma_T}) is valid for both isotropic and anisotropic Fermi surfaces, which allows us to treat spherically symmetric and polygonal geometries within the same formalism. 
After checking the isotropic case of a circular Fermi surface, we specialized to the polygonal geometries of a hexagonal and a square Fermi-surface shape, with isotropic Fermi-velocity modulus. We analyzed the connection between the conductivity and the skin depth for both geometries, and their dependence on Fermi-surface orientation with respect to the direction of the applied field. 
Assuming specular surface scattering, we numerically calculated the surface impedance in all regimes of skin effect, for circular, hexagonal and square geometries, and for different Fermi-surface orientations for the latter two cases. We also analyzed the impedance in each skin-effect regime, and traced quantitative boundaries to characterize the crossovers among the different regimes. Our main results are the impedance ``phase-diagrams'' shown in Figs.\@ \ref{fig:omegacross_iso}, \ref{fig:omegacross_hex}, and \ref{fig:omegacross_square}, where the crossovers between different skin effect regimes are determined, both numerically and analytically, as a function of frequency and ratio between momentum-conserving and relaxing scattering rates. We also interpreted these diagrams, in the presence or absence of anisotropy, in a unified fashion: the qualitative boundaries between adjacent regimes are described in terms of ratios between characteristic lengths for skin effect, collected in Table \ref{tab:skin_lengths} and visualized in Fig.\@ \ref{fig:3D_skin_effect}. This way, we classify the orientational behaviour of anisotropic skin effect in all regimes in the diagrams \ref{fig:3D_skin_effect_misalign}, \ref{fig:3D_skin_effect_align}, and \ref{fig:3D_skin_effect_square}, which highlight that skin effect regimes must be identified within the parameter space set by $\gamma_{\rm mr}$, $\gamma_{\rm mc}$, and $\omega$. 

In classifying the dependence of $\left|Z(\omega)\right|$ on Fermi-surface geometry, a crucial role is played by the presence of Fermi-surface segments with velocity aligned (or nearly parallel) to the direction of the incident electric field. This alignment is capable of altering the frequency scaling exponent $\eta$ of the skin depth and the impedance, and even suppress altogether the non-local character (dependence on $q_x$) of conduction in the anisotropic system. In the light of such phenomenon, one can rationalize the different extension and boundaries among the viscous and anomalous regimes, i.e.\@ regions {\Large \textcircled{\normalsize V}}, {\Large \textcircled{\small D$_1$}} and {\Large \textcircled{\small D$_2$}} in Fig.\@ \ref{fig:3D_skin_effect}, for different orientations. 

Furthermore, we interpreted the frequency dependence of the skin depth and impedance, demonstrated in our theory, in terms of a generalized version of the ``ineffectiveness concept'' introduced by Pippard, which allows one to infer the evolution of skin effect with frequency, from the dependence of the conductivity on frequency and momentum. 

The orientational dependence of skin effect, as well as the modifications of the impedance ``phase diagrams'' due to anisotropy, persist even if we consider more complex non-polygonal Fermi-surface geometries, as we demonstrated by explicitly treating the case of a ``supercircle'' (in Sec.\@ \ref{Supercircle}) and of a simplified 2D experimental parametrization of the Fermi surface of PdCoO$_2$ \cite{Hicks-2012,Nandi-2018,McGuinness-2021} (in Sec.\@ \ref{PdCoO2_2D}). In a companion paper, the full 3D parametrization for PdCoO$_2$ is employed in our kinetic theory and found in agreement with surface-resistance microwave measurements \cite{Baker-2022_preprint}. There, it is shown that the nonzero dispersion perpendicular to the hexagonally coordinated Pd layers has to be taken into account, to provide an accurate description of the non-local electrodynamics of PdCoO$_2$. However, our 2D calculations serve as a useful reference, to estimate to what extent the anisotropic transport in the intralayer dimension affect skin effect, and to locate in which regime the optical data lies in the parameter space of Fig.\@ \ref{fig:3D_skin_effect}. A detailed study of the impact of different dimensionality on non-local optics in the collision-operator formalism represents an interesting future development of our theory. 

Thus, our theory provides a flexible and compact formalism, which makes contact with a basic electrodynamic quantity, the surface impedance, that is a consolidated and accurate method to investigate the spatially non-local current response. Our methods are practically adaptable to many classes of 2D materials with anisotropic Fermi surfaces. 
More generally, our work provides new guidelines for the theoretical and experimental investigation of the ballistic, viscous and Ohmic regimes of conduction in novel 2D and 3D materials, and for the orientational dependence of the optical properties in anisotropic electronic systems. The potential technological impact of such guidelines encompasses many future applications of ultra-high-conductivity materials and integrated circuits operating at GHz frequencies \cite{Sarvari-thesis-2008}. 

\section{Acknowledgments}

We are grateful to G. Ghiringhelli, C. Berthod, A. P. Mackenzie, R. M\"{o}ssner, G. A. Inkof for insightful discussions. 
D. V. acknowledges partial support by the Swiss National Science Foundation (SNSF) through the SNSF Early Postdoc.Mobility Grant P2GEP2$\_$181450. D. V. and J. S. acknowledge partial support by the European Commission’s Horizon 2020 RISE program Hydrotronics (Grant No. 873028). G. B. and D. A. B. acknowledge support from the Max Planck-UBC-UTokyo Center for Quantum Materials and the Canada First Research Excellence Fund, Quantum Materials and Future Technologies Program, as well as the Natural Sciences and Engineering Research Council of Canada (RGPIN-2018-04280).

\appendix

\section{Collision operator and eigenmodes in two-times approximation}\label{App:coll_op_states}

Given the assumptions (\ref{eq:chi_states}) for the eigenfunctions $\chi_{\vec{k}}$ and the eigenvalue equations (\ref{eq:eigen_C}), the collision operator $\hat{C}$ in spatial dimensions $\left\{i \right\}$ can be written as
\begin{multline}\label{eq:coll_op_decomp}
\hat{C}=\sum_m \left| \chi_{\vec{k}, m} \right\rangle \gamma_m  \left\langle \chi_{\vec{k}, m} \right|= \left| \chi_{\vec{k},0} \right\rangle 0  \left\langle \chi_{\vec{k}, 0} \right| \\ + \gamma_{\rm mr} \sum_i \left| \chi_{\vec{k}, i} \right\rangle  \left\langle \chi_{\vec{k}, i}\right| + \gamma_{\rm mc} \sum_{\alpha \neq \left\{0, \left\{i \right\} \right\}} \left| \chi_{\vec{k}, \alpha} \right\rangle  \left\langle \chi_{\vec{k}, \alpha} \right|. 
\end{multline}
From the completeness property (\ref{eq:chi_complete}), we can write
\begin{multline}\label{eq:sumexcl_1}
\sum_{\alpha \neq \left\{0,\left\{i \right\} \right\} }\left| \chi_{\vec{k}, \alpha} \right\rangle  \left\langle \chi_{\vec{k}, \alpha} \right|=1-\left| \chi_{\vec{k},0} \right\rangle  \left\langle \chi_{\vec{k}, 0} \right| \\ - \sum_i \left| \chi_{\vec{k},i} \right\rangle  \left\langle \chi_{\vec{k}, i} \right|,
\end{multline}
which upon insertion in Eq.\@ (\ref{eq:coll_op_decomp}) yields 
\begin{multline}\label{eq:coll_op_decomp2}
\hat{C}= \gamma_{\rm mc} \left(1-\left| \chi_{\vec{k},0} \right\rangle \left\langle \chi_{\vec{k}, 0} \right|\right)-\left(\gamma_{\rm mc}-\gamma_{\rm mr}\right) \times \\ \sum_i  \left| \chi_{\vec{k},i} \right\rangle \left\langle \chi_{\vec{k}, i} \right|.
\end{multline}
Acting with the collision operator (\ref{eq:coll_op_decomp2}) on a generic eigenfunction $\left| \phi_{\vec{k}} \right\rangle$ produces
\begin{multline}\label{eq:C_phi_k}
\hat{C} \left| \phi_{\vec{k}} \right\rangle=\gamma_{\rm mc} \left| \phi_{\vec{k}} \right\rangle -\gamma_{\rm mc} \left| \chi_{\vec{k},0} \right\rangle \left\langle \chi_{\vec{k},0} \right| \left. \phi_{\vec{k}} \right\rangle - \left(\gamma_{\rm mc}-\gamma_{\rm mr}\right) \times \\ \sum_i  \left| \chi_{\vec{k},i} \right\rangle \left\langle \chi_{\vec{k}, i} \right| \left. \phi_{\vec{k}} \right\rangle.
\end{multline}
By virtue of the definition of the scalar product (\ref{eq:scalar_prod}), Eq.\@ (\ref{eq:C_phi_k}) is equivalent to Eq.\@ (\ref{eq:C_op_scal_expl}). 

The normalization of the eigenfunction $\chi_{\vec{k},0}$ imposes 
\begin{multline}\label{eq:c_0_norm}
\left\langle \chi_{\vec{k},0} \right| \left. \chi_{\vec{k},0} \right\rangle=\int_{\vec{k}'}w_{\vec{k}'}\chi_{\vec{k}',0}^{*}\chi_{\vec{k},0} \\ =\frac{2}{(2 \pi)^d \hbar} \int d\epsilon \left[-k_B T \frac{\partial f^{\left(0\right)}\left(\epsilon\right)}{\partial \epsilon} \right] \int_{S(\epsilon)} \frac{d S}{v_{\vec{k}'}} c_0^\ast c_0 \\ \approx \frac{2 k_B T}{ (2 \pi)^d \hbar} \underbrace{\int d \epsilon \frac{-\partial f^{\left(0\right)}\left(\epsilon\right)}{\partial \epsilon}}_{1} \int_{S_F}  \frac{d S}{v_{\vec{k}_F'}}  \left|c_0 \right|^2 \\ = \frac{2 k_B T}{(2 \pi)^d \hbar}\int_{S_F}  \frac{d S}{v_{\vec{k}_F'}} \left|c_0\right|^2 \equiv 1.
\end{multline}
Inverting the last step of Eq.\@ (\ref{eq:c_0_norm}), we find the coefficient (\ref{eq:c_0_vF_var}). 
In the same way, the normalization condition for $\chi_{\vec{k},i}$ demands
\begin{multline}\label{eq:c_i_norm}
\left\langle \chi_{\vec{k},i} \right| \left. \chi_{\vec{k},i} \right\rangle=\int_{\vec{k}'}w_{\vec{k}'}\chi_{\vec{k}',i}^{*}\chi_{\vec{k},i}\\ =\frac{2}{(2 \pi)^d \hbar} \int d\epsilon \left[-k_B T \frac{\partial f^{\left(0\right)}\left(\epsilon\right)}{\partial \epsilon} \right] \\ \times \int_{S(\epsilon)} \frac{d S}{v_{\vec{k}'}} c_0^\ast c_0 c_i^\ast c_i \tilde{v}_{\vec{k}',i} \tilde{v}_{\vec{k},i}  \\ \approx \frac{2 k_B T}{ (2 \pi)^d \hbar} \underbrace{\int d \epsilon \frac{-\partial f^{\left(0\right)}\left(\epsilon\right)}{\partial \epsilon}}_{1} \\ \times \int_{S_F}  \frac{d S}{v_{\vec{k}_F'}}  \left|c_0\right|^2 \left|c_i\right|^2 (\tilde{v}_{\vec{k},i})^2 \equiv 1.
\end{multline}
The last step of Eq.\@ (\ref{eq:c_i_norm}) leads to the coefficient (\ref{eq:c_1_vF_var}).

\section{Analysis of the surface impedance for specular scattering}\label{App:Z_analys}

\subsection{Isotropic (circular) Fermi surface}\label{App_Z_iso_analysis}

\begin{figure}[ht] \centering
\includegraphics[width=0.95\columnwidth]{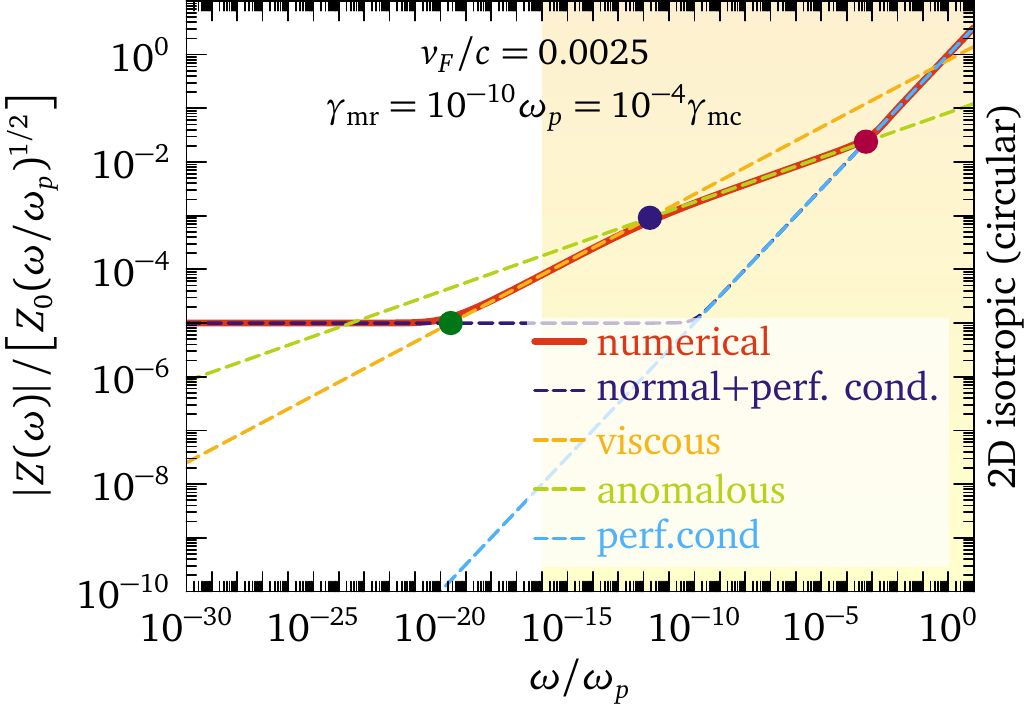}
\caption{\label{fig:AbsZ_iso_crossovers} Absolute value of the surface impedance $\left|Z(\omega)\right|$, divided by the vacuum impedance $Z_0$ and by $(\omega/\omega_p)^{1/2}$, as a function of $\omega/\omega_p$, for a circular Fermi surface (isotropic 2D case). We use the parameters $v_F/c=0.0025$, and $\gamma_{\rm mr}=10^{-4}\gamma_{\rm mc}=10^{-6}\omega_p$. Dashed curves show analytical results valid in each regime, derived in Secs.\@ \ref{Normal_iso}-\ref{Hydro_iso}. The yellow-shaded area gives a qualitative estimation of the parameter space accessible to experiments \cite{Dressel-2001}.
}
\end{figure}

In this appendix, we provide analytical results for the impedance of a 2D isotropic Fermi surface. Each of these results approximates $Z(\omega)$ in a specific regime. The exact $\left|Z(\omega)\right|$ (divided by $Z_0$ and by $(\omega/\omega_p)^{1/2}$) calculated numerically from Eqs.\@ (\ref{eq:S_nonloc}), (\ref{eq:Z_spec}) and (\ref{eq:sigmayy_iso}), is reported as a red solid curve in Fig.\@ \ref{fig:AbsZ_iso_crossovers} as a function of $\omega/\omega_p$, together with the analytical approximations obtained as follows. 

\subsubsection{Normal skin effect and perfect-conductor regime}\label{Normal_iso}

In the limit of large momentum-relaxation rate $\gamma_{\rm mr}\gg \gamma_{\rm mc}$ and small momentum $v_F \left|q_x\right| \ll (\omega+i \gamma_{\rm mc})$, Eq.\@ (\ref{eq:sigmayy_iso}) reduces to the Ohmic result
\begin{equation}\label{eq:sigmayy_iso_lowq}
\frac{\sigma_{yy}(q_x,\omega)}{\epsilon_0 \omega_p^2}=\frac{1}{\gamma_{\rm mr}-i \omega}.
\end{equation}
Inserting Eq.\@ (\ref{eq:sigmayy_iso_lowq}) in Eqs.\@ (\ref{eq:S_nonloc}) and (\ref{eq:Z_spec}), and performing the momentum integrations, we achieve
\begin{equation}\label{eq:Z_iso_lowq}
\frac{Z}{Z_0}=\frac{1-i}{\sqrt{2} \omega_p} \sqrt{\gamma_{\rm mr}-i \omega} \sqrt{\omega}.
\end{equation}
Eq.\@ (\ref{eq:Z_iso_lowq}) gives $\left|Z(\omega)\right|\propto \sqrt{\gamma_{\rm mr} \omega}$ at low frequencies (normal skin effect), while it becomes $\left|Z(\omega)\right|=\mu_0 \lambda_L \omega$ at high frequencies (perfect-conductor regime). Eq.\@ (\ref{eq:Z_iso_lowq}) produces the dashed blue line in Fig.\@ \ref{fig:AbsZ_iso_crossovers}. The high-frequency limit of Eq.\@ (\ref{eq:Z_iso_lowq}) also gives the dashed light-blue line in Fig.\@ \ref{fig:AbsZ_iso_crossovers}, in the perfect-conductivity regime. 

\subsubsection{Anomalous skin effect regime}

For $\omega \gg\left\{\gamma_{\rm mc},\gamma_{\rm mr} \right\}$ (ballistic regime, with negligible scattering) and $v_F \left|q_x\right| \gg (\omega+i \gamma_{\rm mc})$ (large momentum), Eq.\@ (\ref{eq:sigmayy_iso}) becomes 
\begin{equation}\label{eq:sigmayy_iso_highq}
\frac{\sigma_{yy}(q_x,\omega)}{\epsilon_0 \omega_p^2}=\frac{2}{v_F q_x}.
\end{equation}
Using Eq.\@ (\ref{eq:sigmayy_iso_highq}) in performing the integrations over momentum in Eqs.\@ (\ref{eq:S_nonloc}) and (\ref{eq:Z_spec}), we obtain
\begin{equation}\label{eq:Z_iso_anom}
\frac{Z}{Z_0}=\frac{2}{3}\left(\frac{1}{\sqrt{3}}-i\right) \left(\frac{v_F}{2 c}\right)^{\frac{1}{3}}\left(\frac{\omega}{\omega_p}\right)^{\frac{2}{3}}. 
\end{equation}
Hence, Eq.\@ (\ref{eq:Z_iso_anom}) gives $\left|Z(\omega)\right|\propto \omega^{2/3}$, consistently with the literature on ballistic (anomalous) skin effect in isotropic systems \cite{Reuter-1948,Sondheimer-2001,Dressel-2001}. Eq.\@ (\ref{eq:Z_iso_anom}) yields the dashed green line in Fig.\@ \ref{fig:AbsZ_iso_crossovers}. 
\begin{figure*}[ht] \centering
\includegraphics[width=0.8\textwidth]{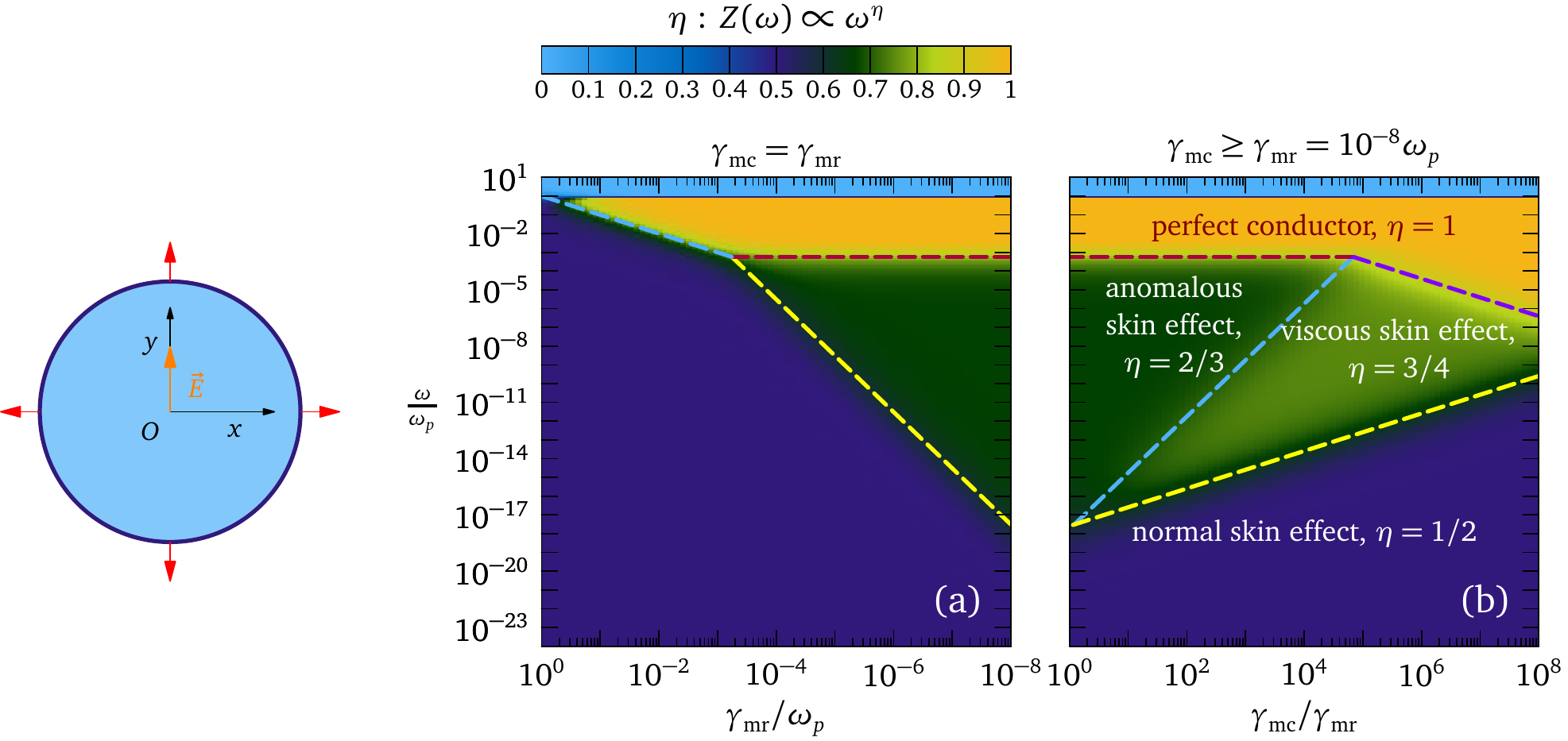}
\caption{\label{fig:omegacross_iso} Skin effect regimes for an isotropic 2D (circular) Fermi surface, as measured by the surface impedance $Z(\omega)$ as a function of relaxation rate $\gamma_{\rm mr}$, momentum-conserving collision rate $\gamma_{\rm mr}$, and frequency $\omega/\omega_p$, where $\omega_p$ is the plasma frequency (\ref{eq:omega_p_iso}). The Fermi-surface geometry is sketched on the left-hand side of the plot, together with the applied electric field $\vec{E}=E_y \hat{u}_y$ aligned with the $y$ axis, and the local Fermi velocity vectors shown by red arrows. (a) Regimes in the $\left(\gamma_{\rm mr},\omega \right)$ plane, for $\gamma_{\rm mr}=\gamma_{\rm mc}$. (b) Regimes in the $\left(\gamma_{\rm mc},\omega \right)$ plane, for fixed $\gamma_{\rm mr}=10^{-8}\omega_p$. The color palette is the density plot of $\mathrm{Arg}Z(\omega)/(-\pi/2)$, giving the exponent $\eta$ of $Z(\omega)\propto \omega^\eta$. Dashed lines are the analytical crossover boundaries derived in Appendix \ref{Iso_crossover_omega}.
}
\end{figure*}

\subsubsection{Hydrodynamic regime}\label{Hydro_iso}

We analyze hydrodynamic skin effect in the isotropic limit by expanding the denominator in the conductivity Eq.\@ (\ref{eq:sigmayy_iso}) for $\gamma_{\rm mc}\rightarrow+\infty$ at first order (assuming negligible momentum relaxation): 
\begin{equation}\label{eq:sigmayy_iso_hydro}
\frac{\sigma_{yy}(q_x,\omega)}{\epsilon_0 \omega_p^2}=\frac{1}{-i \omega+(v_F q_x)^2/(4 \gamma_{\rm mc})}.
\end{equation}
We then perform the integrals over momentum in Eqs.\@ (\ref{eq:S_nonloc}) and (\ref{eq:Z_spec}), using the expansion (\ref{eq:sigmayy_iso_hydro}), and we expand the result in the low-frequency limit $\omega \rightarrow 0^+$, to obtain
\begin{equation}\label{eq:Z_iso_hydro}
\frac{Z}{Z_0}=\frac{(-1+i)(-1)^{\frac{7}{8}} \sqrt{v_F/c}}{2 \sqrt{2 \omega_p} \gamma_{\rm mc}^{\frac{1}{4}}} \omega^{\frac{3}{4}}. 
\end{equation}
Eq.\@ (\ref{eq:Z_iso_hydro}) produces the dashed orange line in Fig.\@ \ref{fig:AbsZ_iso_crossovers}. 

\subsubsection{Crossover frequencies}\label{Iso_crossover_omega}

To estimate characteristic frequencies for the various crossovers between different skin effect regimes, we consider the intersection points between the analytical results for $\left|Z(\omega)\right|$ found in Secs.\@ \ref{Normal_iso}-\ref{Hydro_iso}. 
Let us begin with the crossovers in the $\left(\gamma_{\rm mc},\omega\right)$ plane, for $\gamma_{\rm mc}\geq \gamma_{\rm mr}$; see Fig.\@ \ref{fig:omegacross_iso}(b). 
Using the low-frequency expansion of Eq.\@ (\ref{eq:Z_iso_lowq}) and Eq.\@ (\ref{eq:Z_iso_hydro}), we deduce the frequency $\omega=\omega_{nv}$ for the crossover between normal and viscous skin effect: 
\begin{equation}\label{eq:omegacross_iso_nv}
\omega_{nv}=\frac{16 \gamma_{\rm mc} \gamma_{\rm mr}^2}{\omega_p^2 (v_F/c)^2}. 
\end{equation}
Eq.\@ (\ref{eq:omegacross_iso_nv}) gives the green dot in Fig.\@ \ref{fig:AbsZ_iso_crossovers} and the dashed yellow line in Fig.\@  \ref{fig:omegacross_iso}(b). It is consistent with the qualitative crossover condition {\Large \textcircled{\normalsize A}}--{\Large \textcircled{\normalsize V}} in Table \ref{tab:qual_cross_gammamc}, up to a numerical constant.
Employing Eqs.\@ (\ref{eq:Z_iso_hydro}) and (\ref{eq:Z_iso_anom}), we obtain the crossover frequency $\omega=\omega_{va}$ between the viscous and anomalous regimes:
\begin{equation}\label{eq:omegacross_iso_va}
\omega_{va}= C \gamma_{\rm mc}^3\left(\frac{c}{\omega_p v_F}\right)^2,
\end{equation}
with $C=4294967296/387420489 \approx 11.086$. 
Eq.\@ (\ref{eq:omegacross_iso_va}) produces the blue dot in Fig.\@ \ref{fig:AbsZ_iso_crossovers} and the dashed light-blue line in Fig.\@  \ref{fig:omegacross_iso}(b). It agrees with the qualitative criterion {\Large \textcircled{\small D$_1$}}--{\Large \textcircled{\normalsize V}} in Table \ref{tab:qual_cross_gammamc}.
The crossover frequency $\omega=\omega_{ap}$ between anomalous and perfect-conductor regimes follows from Eq.\@ (\ref{eq:Z_iso_anom}) and $\left|Z(\omega)\right|=Z_0 \omega/\omega_p$: 
\begin{equation}\label{eq:omegacross_iso_ap}
\omega_{ap}=\frac{32}{81 \sqrt{3}}\frac{v_F}{\lambda_L}. 
\end{equation}
The red dot in Fig.\@ \ref{fig:AbsZ_iso_crossovers} and the dashed red line in Fig.\@ \ref{fig:omegacross_iso}(b) are both given by Eq.\@ (\ref{eq:omegacross_iso_ap}). The latter is in qualitative agreement with the length-scale criterion {\Large \textcircled{\small D$_2$}}--{\Large \textcircled{\normalsize E}} in Table \ref{tab:qual_cross_gammamc}.
Finally, the crossover between viscous skin effect and perfect-conductor regime is derived from Eq.\@ (\ref{eq:Z_iso_hydro}) and $\left|Z(\omega)\right|=Z_0 \omega/\omega_p$, giving
\begin{equation}\label{eq:omegacross_iso_vp}
\omega_{vp}=\frac{v_F^2}{16 \gamma_{\rm mc}}\frac{1}{\lambda_L^2}. 
\end{equation}
Eq.\@ (\ref{eq:omegacross_iso_vp}) generates the dashed purple line in Fig.\@ \ref{fig:omegacross_iso}(b), and is consistent with the qualitative condition {\Large \textcircled{\normalsize V}}--{\Large \textcircled{\normalsize F}} in Table \ref{tab:qual_cross_gammamc}.
We now analyze the crossovers in the $\left(\gamma_{\rm mr},\omega\right)$ plane, for $\gamma_{\rm mc}= \gamma_{\rm mr}$; see Fig.\@ \ref{fig:omegacross_iso}(a).
The boundary between normal and anomalous skin effect follows from Eqs.\@ (\ref{eq:Z_iso_lowq}) and (\ref{eq:Z_iso_anom}):
\begin{equation}\label{eq:omegacross_iso_na_gammar}
\omega_{na}'=\frac{19683}{1024} \left( \frac{\lambda_L}{v_F}\right)^2 \gamma_{\rm mr}^3. 
\end{equation}
Eq.\@ (\ref{eq:omegacross_iso_na_gammar}) yields the dashed yellow line in Fig.\@ \ref{fig:omegacross_iso}(a), and is consistent with the qualitative condition {\Large \textcircled{\normalsize A}}--{\Large \textcircled{\small D$_1$}} in Table \ref{tab:qual_cross_gammamr}.
The crossover between anomalous and perfect-conductor regimes is independent from $\gamma_{\rm mr}$, and it follows Eq.\@ (\ref{eq:omegacross_iso_ap}) as previously found from Eq.\@ (\ref{eq:Z_iso_anom}). Therefore, Eq.\@ (\ref{eq:omegacross_iso_ap}) also gives the dashed red line in Fig.\@ \ref{fig:omegacross_iso}(a), in agreement with the condition {\Large \textcircled{\small D$_2$}}--{\Large \textcircled{\normalsize E}} in Table \ref{tab:qual_cross_gammamr}.
Lastly, a boundary between the normal and perfect-conductor regimes stems from the low-frequency expansion of Eq.\@ (\ref{eq:Z_iso_lowq}) and $\left|Z(\omega)\right|=Z_0 \omega/\omega_p$. We simply obtain
\begin{equation}\label{eq:omegacross_iso_np_gammar}
\omega_{np}'=\gamma_{\rm mr}, 
\end{equation}
which is in full agreement with the criterion {\Large \textcircled{\normalsize A}}--{\Large \textcircled{\normalsize B}} in Table \ref{tab:qual_cross_gammamr}, and produces the dashed light-blue line in Fig.\@ \ref{fig:omegacross_iso}(a). 

\subsection{Hexagonal Fermi surface}\label{App_Z_hex_analysis}

This appendix contains the analysis of the surface impedance for a hexagonal Fermi surface and rotation angles $\phi=\left\{0, \pi/2\right\}$. 

\subsubsection{Scattering-less regime}\label{App_Z_hex_analysis_noscat}

\begin{figure}[ht]
\includegraphics[width=0.8\columnwidth]{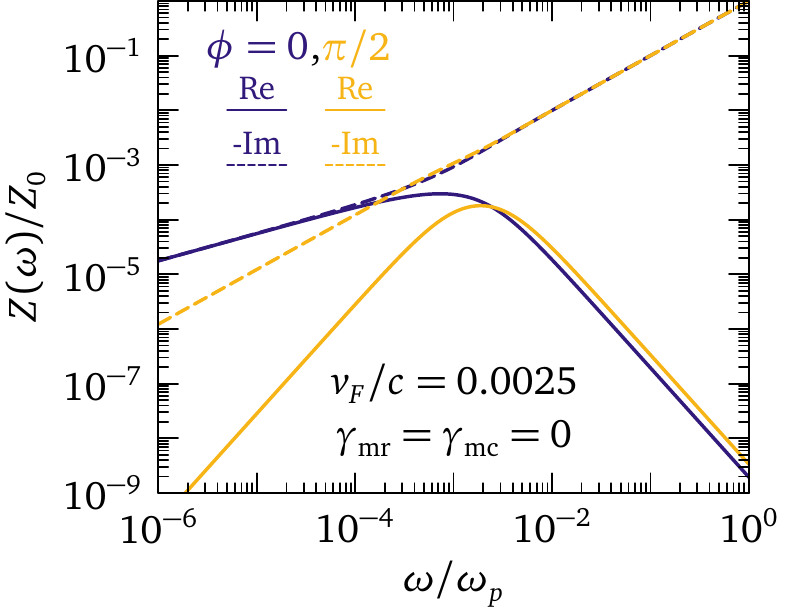}
\caption{\label{fig:Zs_inf_Hex} Surface impedance $Z(\omega)$ normalized to the vacuum-impedance $Z_0$, as a function of dimensionless frequency $\omega/\omega_p$, for a hexagonal Fermi surface with Fermi-velocity modulus $v_F=2.5 \times 10^{-3} c$, in the limit without relaxation and collisions, i.e.\@, $\gamma_{\rm mr}=\gamma_{\rm mc}=0$. Blue (gold) curves show the results for crystals tilted at an angle $\phi=0$ ($\phi=\pi/2$) with respect to the surface, according to Eq.\@ (\ref{eq:Z_noscat_0}) (Eq.\@ (\ref{eq:Z_noscat_90})). Solid and dashed lines refer to $\mathrm{Re} Z(\omega)$ and $-\mathrm{Im} Z(\omega)$, respectively. 
}
\end{figure}

In the absence of scattering, i.e.\@, $\gamma_{\rm mr}=\gamma_{\rm mc}=0$, and for $\phi=0$, the conductivity reduces to Eq.\@ (\ref{eq:sigmayy_hex_par_gamma0}) and the integral over $z$ in Eq.\@ (\ref{eq:Z_spec}) can be evaluated analytically: 
\begin{subequations}\label{eq:Z_noscat_0}
\begin{equation}\label{eq:Z_noscat_0_1}
\frac{Z}{Z_0}=\frac{1}{2 \sqrt{2}}\frac{\omega}{\omega_p}\frac{v_F}{c} \frac{ \left(\alpha_2-i \alpha_3\right)- \omega/\omega_p \left(\alpha_2-i \alpha_3\right)/\alpha_1}{\alpha_2 \alpha_3},
\end{equation}
\begin{equation}
\alpha_1=\sqrt{\left(\frac{\omega}{\omega_p}\right)^2+\left(\frac{v_F}{c}\right)^2},
\end{equation}
\begin{equation}
\alpha_2=\sqrt{\frac{\omega}{\omega_p}\left(\alpha_1-\frac{\omega}{\omega_p}\right)},
\end{equation}
\begin{equation}
\alpha_3=\sqrt{\frac{\omega}{\omega_p}\left(\alpha_1+\frac{\omega}{\omega_p}\right)}.
\end{equation}
\end{subequations}
The surface impedance (\ref{eq:Z_noscat_0}) crosses over from $\mathrm{Re}Z(\omega) \sim \omega^{1/2}$ and $-\mathrm{Im}Z(\omega) \sim \omega^{1/2}$ at low frequency to $\mathrm{Re}Z(\omega) \ll -\mathrm{Im}Z(\omega) \sim \omega$ at high $\omega$, as shown by the solid and dashed blue curves in Fig.\@ \ref{fig:Zs_inf_Hex}, respectively. We can understand this crossover analytically, by combining Eq.\@ (\ref{eq:Z_spec}) with the conductivity (\ref{eq:sigmayy_hex_par_gamma0}) as done in Sec.\@ \ref{Skin_hex}: in the regime $\omega\ll v_{F} \left|q_x\right|$, we directly obtain 
\begin{equation}\label{eq:Z_hex_gamma0_lowom}
Z(\omega)\approx \frac{1-i}{2} Z_0 \sqrt{\frac{v_F}{c}} \sqrt{\frac{\omega}{\omega_p}}=\mu_0 \omega \frac{1-i}{2} \delta_s(\omega),
\end{equation}
with the skin depth $\delta_s(\omega)$ given by Eq.\@ (\ref{eq:aniso_flat2}), while for $\omega\gg v_{F} \left|q_x\right|$ the momentum dependence of the conductivity disappears and 
\begin{equation}\label{eq:Z_hex_gamma0_highom}
Z(\omega)\approx -i Z_0 \frac{\omega}{\omega_p}=-i \mu_0 \omega \lambda_L,
\end{equation}
i.e.\@, with the frequency-independent skin depth $\delta_s \equiv \lambda_L$ associated with a perfect conductivity. In fact, as noticed in Sec.\@ \ref{Skin}, the high-frequency limit (\ref{eq:Z_hex_gamma0_highom}) of the surface impedance is general and independent from the shape of the anisotropic 2D Fermi surface. 

In the limit $\gamma_{\rm mr}=\gamma_{\rm mc}=0$ and $\phi=\pi/2$, the conductivity is given by Eq.\@ (\ref{eq:sigma_yy_hex_90_0}) and the integral over $z$ in Eq.\@ (\ref{eq:Z_spec}) yields the analytical result
\begin{subequations}\label{eq:Z_noscat_90}
\begin{multline}\label{eq:Z_noscat_90_1}
\frac{Z}{Z_0}=i\frac{\sqrt{3}}{2}\frac{\omega}{\omega_p}\frac{v_F}{c} \frac{1}{\beta_1} \left[ \frac{ -2 (\omega/\omega_p)^2-(v_F/c)^2+\beta_1}{-2 (\omega/\omega_p)^2+(v_F/c)^2-\beta_1} \right. \\ \left. - \frac{ 2 (\omega/\omega_p)^2+(v_F/c)^2+\beta_1}{-2 (\omega/\omega_p)^2+(v_F/c)^2+\beta_1}\right],
\end{multline}
\begin{equation}
\beta_1=\sqrt{4\left(\frac{\omega}{\omega_p}\right)^2+8\left(\frac{\omega}{\omega_p}\frac{v_F}{c}\right)^2+\left(\frac{v_F}{c}\right)^4}.
\end{equation}
\end{subequations}
The high-frequency limit of the surface impedance (\ref{eq:Z_noscat_90}) is characteristic of a perfect conductivity with no momentum dependence, as mentioned in the preceding section. However, for $\phi=\pi/2$ we have at low frequencies $\mathrm{Re}Z(\omega) \ll -\mathrm{Im}Z(\omega) \sim \omega$, as shown by the solid and dashed gold curves in Fig.\@ \ref{fig:Zs_inf_Hex}. This behavior is understood by using Eqs.\@ (\ref{eq:Z_spec}) and (\ref{eq:sigma_yy_hex_90_0}) in the limit $\omega\ll v_{F} \left|q_x\right|$, which straightforwardly gives 
\begin{equation}\label{eq:Z_hex_gamma0_90_lowom}
Z(\omega)\approx -i Z_0 \sqrt{\frac{3}{2}} \frac{\omega}{\omega_p}=-i \mu_0 \omega \delta_s,
\end{equation}
with the skin depth $\delta_s=\sqrt{3/2} \lambda_L$ in accordance with Eq.\@ (\ref{eq:aniso_flat3}). 
Therefore, under the assumption $\omega \ll v_F q_x$ valid at low frequencies, we expect a qualitative difference in the skin depth and the surface impedance between the orientations $\phi=0$ and $\phi=\pi/2$. An analogous difference follows from the analysis of $Z(\omega)$ for a square Fermi surface, as performed in Sec.\@ \ref{Square_Z_analys}.

\subsubsection{Normal and perfect-conductor regimes}

For orientation angle $\phi=0$, the low-momentum conductivity follows Eq.\@ (\ref{eq:sigma_yy_phi0_lowq}). To order $o\left[(v_F q_x)^2\right]$, such conductivity is momentum-independent, and equivalent to the one of the Drude model, Eq.\@ (\ref{eq:sigma_yy_sq_pi4_gamma}). Inserting this momentum-independent conductivity in Eqs.\@ (\ref{eq:S_nonloc}) and (\ref{eq:Z_spec}), and performing the momentum integrations, we straightforwardly obtain
\begin{equation}\label{eq:Z_hex_lowq} 
\frac{Z(\omega)}{Z_0}=-i \frac{\omega}{\omega_p} \sqrt{1+\frac{i \gamma_{\rm mr}}{\omega}}.
\end{equation} 
Eq.\@ (\ref{eq:Z_hex_lowq}) gives the dashed blue line in Fig.\@ \ref{fig:AbsZ_hex_crossovers}(a). At low frequencies, it implies $\left|Z(\omega)\right|\propto \omega^{1/2}$, while for $\omega \rightarrow +\infty$ it becomes the perfect-conductor result $Z(\omega)=-i Z_0 \omega/\omega_p=-i \mu_0 \lambda_L \omega$. The latter also gives the dashed light blue line in Fig.\@ \ref{fig:AbsZ_hex_crossovers}(a).

The conductivity for $\phi=\pi/2$ in the low-momentum regime is the same as for $\phi=0$, i.e.\@, Eq.\@ (\ref{eq:sigma_yy_phi0_lowq}), to order $o\left[(v_F q_x)^2\right]$. This is a consequence of the property mentioned in Sec.\@ \ref{Hex_phi}, that the electrodynamics is independent of orientation angle $\phi$ in the $q_x \rightarrow 0^+$ limit for a hexagonal Fermi surface. Hence, Eq.\@ (\ref{eq:Z_hex_lowq}) also gives the dashed blue line in Fig.\@ \ref{fig:AbsZ_hex_crossovers}(b), while the dashed light blue line in the same figure follows $Z(\omega)=-i \mu_0 \lambda_L \omega$.  

\subsubsection{Anomalous regime}

To analyze the high-momentum anomalous regime for angle $\phi=0$, we use the expansion of Eq.\@ (\ref{eq:sigma_yy_hexpar_expl}) to leading order in $q_x \rightarrow +\infty$, which gives Eq.\@ (\ref{eq:sigma_yy_phi0_highq}). Inserting the latter into Eqs.\@ (\ref{eq:S_nonloc}) and (\ref{eq:Z_spec}) and performing the momentum integrations analytically, we obtain 
\begin{equation}\label{eq:Z_hex_par_highq}
\frac{Z(\omega)}{Z_0}=\frac{1-i}{\sqrt{2 \omega_p}}\frac{\sqrt{v_F/c} \omega^{\frac{3}{4}}}{2(\omega+i \gamma_{\rm mc})^{\frac{1}{4}}}. 
\end{equation}
In the regime $\omega \gg \gamma_{\rm mc}$, Eq.\@ (\ref{eq:Z_hex_par_highq}) gives 
\begin{equation}\label{eq:Z_hex_par_anom}
\frac{Z(\omega)}{Z_0}=\frac{1-i}{2 \sqrt{2}}\sqrt{\frac{v_F \omega}{c \omega_p}}. 
\end{equation}
Eq.\@ (\ref{eq:Z_hex_par_anom}) gives the dashed green line in Fig.\@ \ref{fig:AbsZ_hex_crossovers}(a). 

For orientation angle $\phi=\pi/2$, we use the high-momentum expansion (\ref{eq:sigma_yy_hex_90_highq}) of the conductivity in performing the momentum integrals of Eqs.\@ (\ref{eq:S_nonloc})-(\ref{eq:Z_spec}). The result is 
\begin{equation}\label{eq:Z_hex_pi2_highq}
\frac{Z(\omega)}{Z_0}=\frac{-i \omega}{\sqrt{2}}\sqrt{3+\frac{i(\gamma_{\rm mc}+2 \gamma_{\rm mr})}{\omega}}.
\end{equation}
The low-frequency expansion of Eq.\@ (\ref{eq:Z_hex_pi2_highq}) at leading order gives
\begin{equation}\label{eq:Z_hex_pi2_anom}
\frac{Z(\omega)}{Z_0}=\frac{1-i}{2 \omega_p}\sqrt{(\gamma_{\rm mc}+2 \gamma_{\rm mr}) \omega}, 
\end{equation} 
which yields the dashed green line in Fig.\@ \ref{fig:AbsZ_hex_crossovers}(b). We can extract a skin depth from Eq.\@ (\ref{eq:Z_hex_pi2_anom}) in the presence of scattering, by defining $Z(\omega)=(1-i)/2 \mu_0 \omega \delta_s(\omega)$. This gives Eq.\@ (\ref{eq:aniso_flat3_gammamc}) for the skin depth. 

\subsubsection{Hydrodynamic regime}

To obtain the hydrodynamic impedance for $\phi=0$, it is sufficient to consider Eq.\@ (\ref{eq:Z_hex_par_highq}) in the limit $\gamma_{\rm mc} \gg \omega$ (dominant momentum-conserving collisions). This way, we obtain
\begin{equation}\label{eq:Z_hex_par_hydro}
\frac{Z(\omega)}{Z_0}=\frac{1-i}{2 \sqrt{2 \omega_p}}\frac{\sqrt{v_F/c} \omega^{\frac{3}{4}}}{(i \gamma_{\rm mc})^{\frac{1}{4}}}. 
\end{equation}
Eq.\@ (\ref{eq:Z_hex_par_hydro}) produces the dashed orange line in Fig.\@ \ref{fig:AbsZ_hex_crossovers}(a).

For orientation angle $\phi=\pi/2$, the analysis of the hydrodynamic regime requires to push the expansion (\ref{eq:sigma_yy_hex_90_highq}) to the next term of order $(v_F q_x)^{-2}$. Using such term in the momentum integrals in Eqs.\@ (\ref{eq:S_nonloc})-(\ref{eq:Z_spec}), and expanding the result for $\omega \rightarrow 0$ at leading order, produces 
\begin{equation}\label{eq:Z_hex_pi2_hydro}
\frac{Z(\omega)}{Z_0}=-\frac{(-1)^{3/4}}{2 \sqrt{\omega_p}} \left(\frac{\omega}{\gamma_{\rm mc}}\right)^{\frac{3}{4}} \sqrt{ \frac{v_F}{c}(\gamma_{\rm mc}+2 \gamma_{\rm mr}) }.
\end{equation}
Eq.\@ (\ref{eq:Z_hex_pi2_hydro}) gives the dashed orange line in Fig.\@ \ref{fig:AbsZ_hex_crossovers}(b). 

\subsubsection{Crossover frequencies}\label{Hex_crossover_omega}

For $\phi=0$, we begin by finding crossover boundaries in the $\left(\gamma_{\rm mc},\omega\right)$ plane for $\gamma_{\rm mc} \geq \gamma_{\rm mr}$. The crossover frequency between normal and viscous skin effect results from equating $\left|Z(\omega)\right|$ from the low-frequency expansion of Eq.\@ (\ref{eq:Z_hex_lowq}) and from Eq.\@ (\ref{eq:Z_hex_par_hydro}), and solving for $\omega=\omega_{nv}$. We obtain
\begin{equation}\label{eq:omegacr_nv_hex_par}
\omega_{nv}= \frac{16 \gamma_{\rm mc} \gamma_{\rm mr}^2}{\omega_p^2(v_F/c)^2}. 
\end{equation}
Eq.\@ (\ref{eq:omegacr_nv_hex_par}) gives the green dot in Fig.\@ \ref{fig:AbsZ_hex_crossovers}(a) and the dashed yellow curve in Fig.\@ \ref{fig:omegacross_hex}(b). It is also consistent with the qualitative criterion for the crossover {\Large \textcircled{\normalsize A}}--{\Large \textcircled{\normalsize V}} in Table \ref{tab:qual_cross_gammamc}. 
To evaluate the crossover frequency between viscous and anomalous skin effect, we equate $\left|Z(\omega)\right|$ from Eqs.\@ (\ref{eq:Z_hex_par_hydro}) and (\ref{eq:Z_hex_par_anom}), solving for $\omega=\omega_{va}$. This gives simply
\begin{equation}\label{eq:omegacr_va_hex_par}
\omega_{va}=\gamma_{\rm mc}. 
\end{equation}
Eq.\@ (\ref{eq:omegacr_va_hex_par}) yields the blue dot in Fig.\@ \ref{fig:AbsZ_hex_crossovers}(a) and the dashed light-blue curve in Fig.\@ \ref{fig:omegacross_hex}(b). Besides, Eq.\@ (\ref{eq:omegacr_va_hex_par}) fully agrees with the qualitative condition {\Large \textcircled{\small D$_1$}}--{\Large \textcircled{\small D$_2$}} in Table \ref{tab:qual_cross_gammamc}. 
The crossover frequency between anomalous and perfect-conductor regimes results from equating $\left|Z(\omega)\right|$ stemming from Eqs.\@ (\ref{eq:Z_hex_par_anom}) and $\left|Z(\omega)\right|=Z_0 \omega/\omega_p$, solving for $\omega=\omega_{ap}$. We have 
\begin{equation}\label{eq:omegacr_ap_hex_par}
\omega_{ap}=\omega_p \frac{v_F}{4 c}. 
\end{equation}
Eq.\@ (\ref{eq:omegacr_ap_hex_par}) produces the red dot in Fig.\@ \ref{fig:AbsZ_hex_crossovers}(a) and the dashed red curve in Fig.\@ \ref{fig:omegacross_hex}(b). Up to the numerical prefactor $1/4$, it is consistent with the boundary {\Large \textcircled{\small D$_2$}}--{\Large \textcircled{\normalsize E}} in Table \ref{tab:qual_cross_gammamc}.
Finally, there is a crossover between viscous and perfect-conductor regimes, found by equating Eq.\@ (\ref{eq:Z_hex_par_hydro}) and $\left|Z(\omega)\right|=Z_0 \omega/\omega_p$. The corresponding frequency is
\begin{equation}\label{eq:omegacr_vp_hex_par}
\omega_{vp}= \left(\frac{v_F}{\lambda_L} \right)^2 \frac{1}{16 \gamma_{\rm mc}}. 
\end{equation}
Eq.\@ (\ref{eq:omegacr_vp_hex_par}) generates the dashed purple line in Fig.\@ \ref{fig:omegacross_hex}(b), and it is in qualitative agreement with the length-scale criterion {\Large \textcircled{\normalsize V}}--{\Large \textcircled{\normalsize F}} from Table \ref{tab:qual_cross_gammamc}. 

For $\phi=\pi/2$, we evaluate the crossover between normal and viscous skin effect by equating $\left|Z(\omega)\right|$ which result from the low-frequency expansion of Eq.\@ (\ref{eq:Z_hex_lowq}) and from the lengthy analytical expression which produces the dashed orange line in Fig.\@ \ref{fig:AbsZ_hex_crossovers}(b), in hydrodynamic regime. Here we quote the final result: 
\begin{equation}\label{eq:omegacr_nv_hex_pi2}
\omega_{nv}=\frac{16 \gamma_{\rm mc}^3 \gamma_{\rm mr}^2}{(\gamma_{\rm mc}+2 \gamma_{\rm mr})^2(v_F/c)^2 \omega_p^2}.
\end{equation}
Eq.\@ (\ref{eq:omegacr_nv_hex_pi2}) gives the green dot in Fig.\@ \ref{fig:AbsZ_hex_crossovers}(b) and the dashed yellow curve in Fig.\@ \ref{fig:omegacross_hex}(d). It also respects the qualitative criterion {\Large \textcircled{\normalsize A}}--{\Large \textcircled{\normalsize V}} in Table \ref{tab:qual_cross_gammamc}, up to corrections of higher order in $\gamma_{\rm mc}$ and $\delta\gamma$. 
The crossover between viscous and anomalous skin effect results from equating $\left|Z(\omega)\right|$ stemming from the analytical expression in hydrodynamic regime and from Eq.\@ (\ref{eq:Z_hex_pi2_anom}). The final result is
\begin{equation}\label{eq:omegacr_va_hex_pi2}
\omega_{va}=\frac{4 \gamma_{\rm mc}^3}{\omega_p^2 (v_F/c)^2}.
\end{equation}
Eq.\@ (\ref{eq:omegacr_va_hex_pi2}) yields the blue dot in Fig.\@ \ref{fig:AbsZ_hex_crossovers}(b) and the dashed light-blue curve in Fig.\@ \ref{fig:omegacross_hex}(d). It is consistent with the condition {\Large \textcircled{\small D$_1$}}--{\Large \textcircled{\normalsize V}} in Table \ref{tab:qual_cross_gammamc}, up to higher-order corrections in $\gamma_{\rm mc}$ and $\delta\gamma$. 
We obtain the crossover frequency between anomalous and perfect-conductor by equating $\left|Z(\omega)\right|$ stemming from Eqs.\@ (\ref{eq:Z_hex_pi2_anom}) and $\left|Z(\omega)\right|=Z_0 \omega/\omega_p$, solving for $\omega=\omega_{ap}$. We have 
\begin{equation}\label{eq:omegacr_ap_hex_pi2}
\omega_{ap}=\frac{\gamma_{\rm mc}}{2}+\gamma_{\rm mr}. 
\end{equation}
Eq.\@ (\ref{eq:omegacr_ap_hex_pi2}) produces the red dot in Fig.\@ \ref{fig:AbsZ_hex_crossovers}(b) and the dashed red curve in Fig.\@ \ref{fig:omegacross_hex}(d). It qualitatively agrees with the criterion {\Large \textcircled{\small D$_1$}}--{\Large \textcircled{\small D$_2$}} reported in Table \ref{tab:qual_cross_gammamc}, at leading order in $\gamma_{\rm mc}$.
Lastly, the crossover between viscous and perfect-conductor regimes stems from from the analytical expression in hydrodynamic regime and from $\left|Z(\omega)\right|=Z_0 \omega/\omega_p$. We obtain 
\begin{equation}\label{eq:omegacr_vp_hex_pi2}
\omega_{vp}=\frac{(\gamma_{\rm mc}+2 \gamma_{\rm mr})^2 (v_F/c)^2}{16 \gamma_{\rm mc}^3} \omega_p^2. 
\end{equation}
At leading order in $\gamma_{\rm mc}$, Eq.\@ (\ref{eq:omegacr_vp_hex_pi2}) agrees with the qualitative condition {\Large \textcircled{\normalsize V}}--{\Large \textcircled{\normalsize F}} in \ref{tab:qual_cross_gammamc}. The dashed purple line in Fig.\@ \ref{fig:omegacross_hex}(d) stems from Eq.\@ (\ref{eq:omegacr_vp_hex_pi2}). 

We now trace the visible crossover boundaries in the $\left(\gamma_{\rm mr},\omega\right)$ plane for $\gamma_{\rm mc} = \gamma_{\rm mr}$. For $\phi=0$, we find a crossover between normal and viscous skin effect by equating the leading low-frequency expansion of Eq.\@ (\ref{eq:Z_hex_lowq}) and Eq.\@ (\ref{eq:Z_hex_par_hydro}):
\begin{equation}\label{eq:omegacr_nv_hex_par_gammar}
\omega_{nv}'=16\left(\frac{\lambda_L}{v_F} \right)^2\gamma_{\rm mr}^3. 
\end{equation}
Eq.\@ (\ref{eq:omegacr_nv_hex_par_gammar}) yields the dashed yellow line in Fig.\@ \ref{fig:omegacross_hex}(a), and it qualitatively agrees with the length-scale criterion {\Large \textcircled{\normalsize A}}--{\Large \textcircled{\small D$_1$}} in Table \ref{tab:qual_cross_gammamr}. 
The crossover between viscous and anomalous regimes is found through Eqs.\@ (\ref{eq:Z_hex_par_hydro}) and (\ref{eq:Z_hex_par_anom}), with the simple result of Eq.\@ (\ref{eq:omegacross_iso_np_gammar}), as in the isotropic case. It is in perfect agreement with the criterion {\Large \textcircled{\small D$_1$}}--{\Large \textcircled{\small D$_2$}} in Table \ref{tab:qual_cross_gammamr}. Then, Eq.\@ (\ref{eq:omegacross_iso_np_gammar}) produces the dashed light-blue line in Fig.\@ \ref{fig:omegacross_hex}(a). 
The boundary between the anomalous and perfect-conductor regimes can be estimated through Eq.\@ (\ref{eq:Z_hex_par_anom}) and $\left|Z(\omega)\right|=Z_0 \omega/\omega_p$: it is independent from $\gamma_{\rm mr}$ and $\gamma_{\rm mc}$, and gives Eq.\@ (\ref{eq:omegacr_ap_hex_par}). This result gives the dashed red line in Fig.\@ \ref{fig:omegacross_hex}(a), in accordance with the qualitative condition {\Large \textcircled{\small D$_2$}}--{\Large \textcircled{\normalsize E}} in Table \ref{tab:qual_cross_gammamr}.
Finally, we estimate the boundary between the normal and perfect-conductor regimes using Eq.\@ (\ref{eq:Z_hex_lowq}) and $\left|Z(\omega)\right|=Z_0 \omega/\omega_p$. We simply retrieve Eq.\@ (\ref{eq:omegacross_iso_np_gammar}), in the same way as for the crossover between viscous and anomalous skin effect. Hence, Eq.\@ (\ref{eq:omegacross_iso_np_gammar}) corresponds also to the dashed purple line in Fig.\@ \ref{fig:omegacross_hex}(a), in agreement with the condition {\Large \textcircled{\normalsize A}}--{\Large \textcircled{\normalsize B}} in Table \ref{tab:qual_cross_gammamr}.

In the configuration with $\phi=\pi/2$, the only visible crossover with $\gamma_{\rm mr}=\gamma_{\rm mc}$ is the one between anomalous and perfect-conductor regimes. It is found by equating the low-frequency expansion of Eq.\@ (\ref{eq:Z_hex_pi2_anom}) with $\left|Z(\omega)\right|=Z_0 \omega/\omega_p$, and it simply gives
\begin{equation}\label{eq:omegacr_ap_hex_pi2_gammar}
\omega_{ap}'=\frac{3}{2}\gamma_{\rm mr},  
\end{equation}
This equation provides the dashed red line in Fig.\@ \ref{fig:omegacross_hex}(c), and it qualitatively agrees with the criteria {\Large \textcircled{\small D$_1$}}--{\Large \textcircled{\small D$_2$}} and {\Large \textcircled{\normalsize A}}--{\Large \textcircled{\normalsize B}} in Table \ref{tab:qual_cross_gammamr}.

\subsection{Square Fermi surface}\label{Square_Z_analys}

In this appendix, we derive analytical expressions for the surface impedance assuming a square Fermi surface and rotation angles $\phi=\left\{0, \pi/4\right\}$. 

\subsubsection{Scattering-less regime}\label{Square_Z_analys_noscat}

In the same way as in Sec.\@ \ref{Z_hex_spec}, we can analyze the scattering-less limit of the surface impedance for a square Fermi surface. In the geometry with tilting angle $\phi=0$, using Eq.\@ (\ref{eq:Z_spec}) and the conductivity (\ref{eq:sigmayy_square_par_gamma0}), from the integral over $z$ we directly obtain
\begin{subequations}\label{eq:Z_square_phi0_gamma0} 
\begin{equation}\label{eq:Z_square_phi0_gamma0_1}
\frac{Z(\omega)}{Z_0}=i \frac{\omega}{\omega_p}\frac{v_F}{c} \frac{\gamma_1(\gamma_2-\gamma_3)-\omega/\omega_p(\gamma_2+\gamma_3)}{2 \gamma_1 \gamma_2 \gamma_3},
\end{equation}
\begin{equation}
\gamma_1=\sqrt{\left(\frac{\omega}{\omega_p}\right)^2+2 \left(\frac{v_F}{c}\right)^2},
\end{equation}
\begin{equation}
\gamma_2=\sqrt{\frac{\omega}{\omega_p}\left(\gamma_1-\frac{\omega}{\omega_p}\right)},
\end{equation}
\begin{equation}
\gamma_3=\sqrt{-\frac{\omega}{\omega_p}\left(\gamma_1+\frac{\omega}{\omega_p}\right)}
\end{equation}
\end{subequations}
\begin{figure}[ht]
\includegraphics[width=0.8\columnwidth]{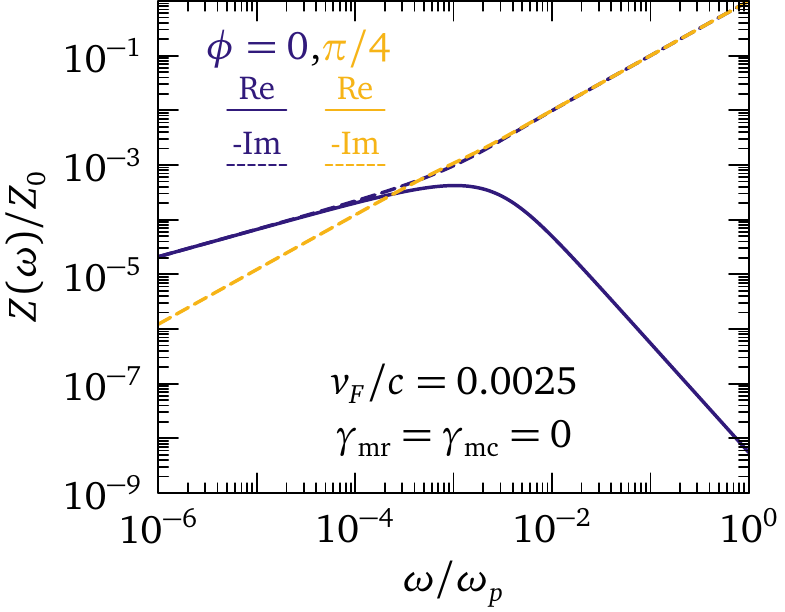}
\caption{\label{fig:Zs_inf_Square} Surface impedance $Z(\omega)$ normalized to the vacuum-impedance $Z_0$, as a function of dimensionless frequency $\omega/\omega_p$, for a square Fermi surface with Fermi-velocity modulus $v_F=2.5 \times 10^{-3} c$, in the limit without relaxation and collisions, i.e.\@, $\gamma_{\rm mr}=\gamma_{\rm mc}=0$. Blue (gold) curves show the results for crystals tilted at an angle $\phi=0$ ($\phi=\pi/4$) with respect to the surface, according to Eq.\@ (\ref{eq:Z_square_phi0_gamma0}) (Eq.\@ (\ref{eq:Z_hex_gamma0_highom})). Solid and dashed lines refer to $\mathrm{Re} Z(\omega)$ and $-\mathrm{Im} Z(\omega)$, respectively. Notice that the impedance is entirely imaginary for the $\phi=\pi/4$ case. 
}
\end{figure}
Fig.\@ \ref{fig:Zs_inf_Square} shows the real and (minus the) imaginary parts of the surface impedance (\ref{eq:Z_square_phi0_gamma0_1}) as a function of $\omega/\omega_p$, as blue solid and dashed curves respectively. The behavior of $Z(\omega)$ is very similar to the one for a hexagonal Fermi surface with $\phi=0$; see Fig.\@ \ref{fig:Zs_inf_Hex}. We can trace back this analogy to the similar form of the conductivities (\ref{eq:sigmayy_hex_par_gamma0}) and (\ref{eq:sigmayy_square_par_gamma0}) for the hexagonal and square case respectively, which also give the same qualitative behavior of the skin depths (\ref{eq:aniso_flat2}) and (\ref{eq:aniso_flat4}). Consequently, in the regime $\omega \ll v_F \left|q_x\right|$ the conductivity (\ref{eq:sigmayy_square_par_gamma0}) and Eq.\@ (\ref{eq:Z_spec}) formally yield Eq.\@ (\ref{eq:Z_hex_gamma0_lowom}), but with Eq.\@ (\ref{eq:aniso_flat4}) for the skin depth in square geometry. In the high-frequency limit $\omega \gg v_F \left|q_x\right|$ we retrieve the smallest possible skin depth $\lambda_L$ and the impedance follows Eq.\@ (\ref{eq:Z_hex_gamma0_highom}). 

As the skin depth analyzed in Sec.\@ \ref{Skin_square}, the surface impedance of the square Fermi surface is qualitatively modified by the presence of two large portions of the Fermi surface parallel to the applied field for $\phi=\pi/4$; see Fig.\@ \ref{fig:Square}. In fact, the disappearance of the momentum dependence of the conductivity generates the London skin depth $\delta\equiv \lambda_L$, so that the surface impedance is simply given by Eq.\@ (\ref{eq:Z_hex_gamma0_highom}): it is purely imaginary and linear in frequency. Such evolution with $\omega$ is shown by the dashed gold line in Fig.\@ \ref{fig:Zs_inf_Square} and is valid at any frequency in the absence of scattering. 
Hence, we expect a qualitative change in $Z(\omega)$ between the orientations $\phi=0$ and $\phi=\pi/4$, with the disappearance of $\mathrm{Re}Z(\omega)$ in the latter case for $\gamma_{\rm mr}=\gamma_{\rm mc}=0$. Notice that such disappearance does not occur for $\gamma_{\rm mc}>0$, as shown by the asymptotic limits for $\gamma_{\rm mr} \rightarrow 0$ shown in Fig.\@ \ref{fig:Zs_phi045_taumr_Square}: in that case, the asymptotic value of the surface resistance is entirely due to momentum-conserving collisions. 

\subsubsection{Normal and perfect-conductor regimes}

Assuming an orientation $\phi=0$, the low-momentum conductivity is given by Eq.\@ (\ref{eq:sigmayy_square_lowq}). To order $o\left[(v_F q_x)^2\right]$, such conductivity does not depend on momentum, and is equivalent to the one of the Drude model. Inserting the momentum-independent term in Eqs.\@ (\ref{eq:S_nonloc}) and (\ref{eq:Z_spec}), and performing the momentum integrations, we retrieve Eq.\@ (\ref{eq:Z_hex_lowq}). 
Hence, Eq.\@ (\ref{eq:Z_hex_lowq}) also gives the dashed blue line in Fig.\@ \ref{fig:AbsZ_sq_crossovers}(a). At low frequencies, it implies $Z(\omega)\propto \omega^{1/2}$, while in the high-frequency limit it turns into the perfect-conductor result $Z(\omega)=-i \mu_0 \lambda_L \omega$, which is also shown by the dashed light blue line in Fig.\@ \ref{fig:AbsZ_sq_crossovers}(a).

For angle $\phi=\pi/4$, the conductivity is (\ref{eq:sigma_yy_sq_pi4_gamma}), identical to the one of the Drude local conductor for any momentum. Therefore, in the regime of normal skin effect we obtain Eq.\@ (\ref{eq:Z_hex_lowq}) for both $\phi=0$ and $\phi=\pi/4$, and this result gives the dashed blue line in Fig.\@ \ref{fig:AbsZ_sq_crossovers}(b) as well. The dashed light-blue line in Fig.\@ \ref{fig:AbsZ_sq_crossovers}(b) shows the perfect-conductor high-frequency limit $Z(\omega)=-i \mu_0 \lambda_L \omega$. 

\subsubsection{Anomalous regime}

For angle $\phi=0$, we use the expansion of Eq.\@ (\ref{eq:sigma_yy_dia_expl}) to leading order in $q_x \rightarrow +\infty$, which gives Eq.\@ (\ref{eq:sigmayy_square_highq}). Using Eqs.\@ (\ref{eq:S_nonloc}) and (\ref{eq:Z_spec}), we obtain 
\begin{equation}\label{eq:Z_square_dia_highq}
\frac{Z(\omega)}{Z_0}=\frac{1-i}{2 \sqrt{\omega_p}}\frac{\sqrt{v_F/c} i \omega }{\left[2 \omega (\omega+i \gamma_{\rm mc})\right]^{\frac{1}{4}}}. 
\end{equation}
In the limit $\omega \gg \gamma_{\rm mc}$, Eq.\@ (\ref{eq:Z_square_dia_highq}) yields 
\begin{equation}\label{eq:Z_square_dia_anom}
\frac{Z(\omega)}{Z_0}=\frac{1-i}{2^{5/4}}\sqrt{\frac{v_F}{c}} \sqrt{\frac{\omega}{\omega_p}}, 
\end{equation}
which gives the dashed green line in Fig.\@ \ref{fig:AbsZ_sq_crossovers}(a). 

Rotating the Fermi surface by an angle $\phi=\pi/4$, the conductivity (\ref{eq:sigma_yy_sq_pi4_gamma}) is momentum-independent, therefore the anomalous regime is absent. 

\subsubsection{Hydrodynamic regime}

The hydrodynamic limit of the impedance for $\phi=0$ follows directly from Eq.\@ (\ref{eq:Z_square_dia_highq}) in the limit $\omega \ll \gamma_{\rm mc}$:
\begin{equation}\label{eq:Z_square_dia_hydro}
\frac{Z(\omega)}{Z_0}=\frac{1-i}{2^{5/4}\sqrt{\omega_p}}\frac{\sqrt{v_F/c} \omega^{\frac{3}{4}}}{(i \gamma_{\rm mc})^{\frac{1}{4}}}. 
\end{equation}
Eq.\@ (\ref{eq:Z_square_dia_hydro}) produces the dashed orange line in Fig.\@ \ref{fig:AbsZ_sq_crossovers}(a). 

For $\phi=\pi/4$ the momentum dependence of the conductivity vanishes, and there is no viscous regime. 

\subsubsection{Crossover frequencies}\label{Sq_crossover_omega}

Let us first define criteria to distinguish the skin effect regimes in the $\left(\gamma_{\rm mc},\omega\right)$ plane for $\gamma_{\rm mc} \geq \gamma_{\rm mr}$.
For $\phi=0$, we obtain the crossover frequency between normal and viscous skin effect from equating $\left|Z(\omega)\right|$ that stem from the low-frequency expansion of Eq.\@ (\ref{eq:Z_hex_lowq}) and from Eq.\@ (\ref{eq:Z_square_dia_hydro}), and we solve for $\omega=\omega_{nv}$. We have
\begin{equation}\label{eq:omegacr_nv_sq_dia}
\omega_{nv}=\frac{8 \gamma_{\rm mc} \gamma_{\rm mr}^2}{\omega_p^2 (v_F/c)^2}. 
\end{equation}
Eq.\@ (\ref{eq:omegacr_nv_sq_dia}) gives the green dot in Fig.\@ \ref{fig:AbsZ_sq_crossovers}(a) and the dashed yellow curve in Fig.\@ \ref{fig:omegacross_square}(b). As for the other Fermi-surface geometries, we can qualitatively interpret Eq.\@ (\ref{eq:omegacr_nv_sq_dia}) as the crossover line {\Large \textcircled{\normalsize A}}--{\Large \textcircled{\normalsize V}} in Table \ref{tab:qual_cross_gammamc}. 
The crossover frequency between viscous and anomalous skin effect stems from equating $\left|Z(\omega)\right|$ from Eqs.\@ (\ref{eq:Z_square_dia_hydro}) and (\ref{eq:Z_square_dia_anom}). The solution is $\omega=\omega_{va}=\gamma_{\rm mc}$, as for the hexagonal case Eq.\@ (\ref{eq:omegacr_va_hex_par}). Therefore, 
Eq.\@ (\ref{eq:omegacr_va_hex_par}) also yields the blue dot in Fig.\@ \ref{fig:AbsZ_sq_crossovers}(a) and the dashed light-blue curve in Fig.\@ \ref{fig:omegacross_square}(b). It coincides with the lenght-scale criterion {\Large \textcircled{\small D$_1$}}--{\Large \textcircled{\small D$_2$}} in Table \ref{tab:qual_cross_gammamc}.
The crossover between anomalous and perfect-conductor regimes results from equating $\left|Z(\omega)\right|$ according to Eqs.\@ (\ref{eq:Z_square_dia_anom}) and $\left|Z(\omega)\right|=Z_0 \omega/\omega_p$, solving for $\omega=\omega_{ap}$. We have 
\begin{equation}\label{eq:omegacr_ap_sq_dia}
\omega_{ap}=\omega_p \frac{v_F}{2 \sqrt{2} c}. 
\end{equation}
Eq.\@ (\ref{eq:omegacr_ap_sq_dia}) generates the red dot in Fig.\@ \ref{fig:AbsZ_sq_crossovers}(a) and the dashed red curve in Fig.\@ \ref{fig:omegacross_square}(b). Up to the prefactor $1/(2 \sqrt{2} )$, it agrees with the condition {\Large \textcircled{\small D$_1$}}--{\Large \textcircled{\normalsize V}} in Table \ref{tab:qual_cross_gammamc}. Finally, we find a boundary between viscous and perfect-conductor regimes by equating Eq.\@ (\ref{eq:Z_square_dia_hydro}) and $\left|Z(\omega)\right|=Z_0 \omega/\omega_p$. We obtain 
\begin{equation}\label{eq:omegacr_vp_sq_dia}
\omega_{vp}=\frac{1}{8} \left(\frac{v_F}{\lambda_L}\right)^2\frac{1}{\gamma_{\rm mc}}. 
\end{equation}
Eq.\@ (\ref{eq:omegacr_vp_sq_dia}) is in qualitative agreement with the criterion {\Large \textcircled{\normalsize V}}--{\Large \textcircled{\normalsize F}} in Table \ref{tab:qual_cross_gammamc}, and it produces the dashed purple line in Fig.\@ \ref{fig:omegacross_square}(b).

For $\phi=\pi/4$, the only crossover occurs between the normal and perfect-conductor regimes. We achieve $\omega_{np}=\gamma_{\rm mr}$, in agreement with Eq.\@ (\ref{eq:omegacross_iso_np_gammar}) and independently from $\gamma_{\rm mc}$. Then, Eq.\@ (\ref{eq:omegacross_iso_np_gammar}) gives the red dot in Fig.\@ \ref{fig:AbsZ_sq_crossovers}(b) and the dashed red line in Fig.\@ \ref{fig:omegacross_square}(d). It is perfectly consistent with the qualitative criteria {\Large \textcircled{\normalsize A}}--{\Large \textcircled{\normalsize B}} and {\Large \textcircled{\small D$_1$}}--{\Large \textcircled{\small D$_2$}} in Table \ref{tab:qual_cross_gammamr}. 

We now define crossover boundaries for the square geometry, in the $\left(\gamma_{\rm mr},\omega\right)$ plane for $\gamma_{\rm mc} = \gamma_{\rm mr}$. For $\phi=0$, there is a crossover between normal and viscous skin effect, found by equating the low-frequency expansion of Eq.\@ (\ref{eq:Z_hex_lowq}) and Eq.\@ (\ref{eq:Z_square_dia_hydro}):
\begin{equation}\label{eq:omegacr_nv_sq_dia_gammar}
\omega_{nv}'=8\left( \frac{\lambda_L}{v_F}\right)^2 \gamma_{\rm mr}^3.
\end{equation}
Eq.\@ (\ref{eq:omegacr_nv_sq_dia_gammar}) is in qualitative agreement with the criterion {\Large \textcircled{\normalsize A}}--{\Large \textcircled{\small D$_1$}} in Table \ref{tab:qual_cross_gammamr}, and it generates the dashed yellow line in Fig.\@ \ref{fig:omegacross_square}(a).
A boundary between the viscous and anomalous regimes is identified by the frequency at which Eqs.\@ (\ref{eq:Z_square_dia_hydro}) and (\ref{eq:Z_square_dia_anom})coincide. As for the hexagonal and circular geometries, this gives Eq.\@ (\ref{eq:omegacross_iso_np_gammar}), which is equivalent to the condition {\Large \textcircled{\small D$_1$}}--{\Large \textcircled{\small D$_2$}} in Table \ref{tab:qual_cross_gammamr}, and it produces the dashed light-blue line in Fig.\@ \ref{fig:omegacross_square}(a).
The crossover between the anomalous and perfect-conductor regimes is estimated by equating Eq.\@ (\ref{eq:Z_square_dia_anom}) and $\left|Z(\omega)\right|=Z_0 \omega/\omega_p$. One retrieves Eq.\@ (\ref{eq:omegacr_ap_sq_dia}), consistently with the previous analysis in the $\left(\gamma_{\rm mc},\omega\right)$ plane. Therefore, Eq.\@ (\ref{eq:omegacr_ap_sq_dia}) is also qualitatively consistent with the condition {\Large \textcircled{\small D$_2$}}--{\Large \textcircled{\normalsize E}} in Table \ref{tab:qual_cross_gammamr}, and it yields the dashed red line in Fig.\@ \ref{fig:omegacross_square}(a). 
Besides, there is a crossover between normal and perfect-conductor regimes, which is identified through the equality between the low-frequency expansion of Eq.\@ (\ref{eq:Z_hex_lowq}) and $\left|Z(\omega)\right|=Z_0 \omega/\omega_p$. In this case, we retrieve Eq.\@ (\ref{eq:omegacross_iso_np_gammar}), in perfect agreement with the length-scale criterion {\Large \textcircled{\normalsize A}}--{\Large \textcircled{\normalsize B}} in Table \ref{tab:qual_cross_gammamr}. Hence, Eq.\@ (\ref{eq:omegacross_iso_np_gammar}) also produces the dashed purple line in Fig.\@ \ref{fig:omegacross_square}(a).

In the configuration with $\phi=\pi/4$, there is but one crossover with $\gamma_{\rm mr}=\gamma_{\rm mc}$: the one between normal and perfect-conductor regimes. It is given by $\omega_{np}=\gamma_{\rm mr}$, which agrees with Eq.\@ (\ref{eq:omegacross_iso_np_gammar}) as in the $\left(\gamma_{\rm mc},\omega\right)$ plane. This equation gives the dashed red line in Fig.\@ \ref{fig:omegacross_hex}(c), and it is equivalent to the qualitative criteria {\Large \textcircled{\small D$_1$}}--{\Large \textcircled{\small D$_2$}} and {\Large \textcircled{\normalsize A}}--{\Large \textcircled{\normalsize B}} in Table \ref{tab:qual_cross_gammamr}.

\section{Frequency scaling of the surface impedance for specular interface scattering}\label{Z_spec_scaling}

In this section, we derive the scaling relation (\ref{eq:skin_depth_Z_gen}) for the frequency dependence of the impedance $Z(\omega)$ analytically for specular interface scattering. We start from Eq.\@ (\ref{eq:Z_spec_displ}) and neglect the displacement current, leading to Eq.\@ (\ref{eq:Z_spec}), that is
\begin{equation}\label{eq:Z_spec_scaling}
Z(\omega) \approx \frac{i \omega \mu_0}{\pi} \int_{-\infty}^{+\infty} d q_x \frac{1}{ \mu_0 i \omega \sigma_{yy}(q_x,\omega)-q_x^2}.
\end{equation}
Assuming the scaling relation (\ref{eq:cond_estim}) for the conductivity, we can write $\mu_0 i \omega \sigma_{yy}(q_x,\omega)=-\alpha_1 (-i \omega)^{-\beta+1}q_x^{-\alpha}$. Inserting the latter equation into Eq.\@ (\ref{eq:Z_spec_scaling}), and performing the momentum integration analytically, we obtain
\begin{multline}\label{eq:Z_scaling_alpha_beta}
Z(\omega)=\frac{2 \mu_0}{\pi} \Gamma\left(\frac{1+\alpha}{2+\alpha}\right)\Gamma\left(\frac{3+\alpha}{2+\alpha}\right)\alpha_1^{-\frac{1}{2+\alpha}} (-i \omega)^{\frac{\beta+\alpha+1}{2+\alpha}} \\ \propto (-i \omega)^{\eta},
\end{multline}
where $\Gamma(z)=\int_0^{+\infty} dt e^{-t} t^{z-1}$ is the Euler Gamma function, and $\eta=(1+\alpha+\beta)/(2+\alpha)$. Using $(-i)^\eta=e^{-i \eta \pi/2}$, we retrieve the scaling relation (\ref{eq:skin_depth_Z_gen}).

\section{Parametrization of arbitrarily shaped 2D Fermi surfaces}\label{Curvature_2D}

This appendix details the methodology employed to calculate the surface-impedance ``phase diagrams'' for non-polygonal 2D Fermi surfaces, including the effects of an arbitrary curvature of Fermi-surface segments. For simplicity, here we retain the assumption of a constant Fermi velocity modulus for any orientation in reciprocal space, in accordance with Eq.\@ (\ref{eq:v_k_F}), however the direction of $\vec{v}_{\vec{k}_F}$ varies with orientation and is locally orthogonal to the Fermi surface by definition \cite{Ashcroft-1976}. A straightforward generalization to cases where even the Fermi-velocity modulus is anisotropic may be performed by referring to the more general parametrization (\ref{eq:vF_anis}). 
In this section, we also retain the hypothesis that two mirror symmetry planes exist in reciprocal space, according to Eq.\@ (\ref{eq:mirror_planes}) and employing $\beta=y$ and $\alpha=x$ for the space directions. 

Consider the 2D parametrization $k_F(\theta)$ for the variation of the Fermi wave vector as a function of angle $\theta$. Then, a vector $\vec{n}_{\vec{k}}$ locally orthogonal to the curve  $k_F(\theta)$ corresponds to
\begin{equation}\label{eq:n_k_vec}
\vec{n}_{\vec{k}}=\frac{d \vec{k}_F(\theta)}{d\theta} \times \hat{u}_z,
\end{equation}
where $\hat{u}_z$ is the unit vector normal to the 2D plane where $k_F(\theta)$ lies. The differential line element $d S$, which enters into Fermi-surface integrations, results
\begin{equation}\label{eq:dS_gen}
d S= n_{\vec{k}} d \theta,
\end{equation}
where 
\begin{equation}\label{eq:n_k_gen_mod}
n_{\vec{k}}=\left|\vec{n}_{\vec{k}}\right|=\sqrt{\left[k_F(\theta)\right]^2+\left[\frac{d k_F(\theta)}{d \theta}\right]^2}. 
\end{equation}
Eq.\@ (\ref{eq:n_k_gen}) corresponds to the rotation by $\pi/2$ of the unit vector locally tangent to the curve $k_F(\theta)$ at a given angle $\theta$. 
The unit vector normal to $k_F(\theta)$ is then
\begin{multline}\label{eq:n_k_gen}
\hat{n}_{\vec{k}}=\frac{\vec{n}_{\vec{k}}}{n_{\vec{k}}}=\frac{1}{\sqrt{\left[k_F(\theta)\right]^2+\left[\frac{d k_F(\theta)}{d \theta}\right]^2}} \\ \times \left[\hat{u}_{k_F} k_F(\theta)-\hat{u}_{\theta} \frac{d k_F(\theta)}{d \theta}\right],
\end{multline}
where $\hat{u}_{k_F}$ and $\hat{u}_{\theta}$ are the unit vectors in the radial and angular directions, respectively.
Then, remembering the transformation of unit vectors from polar to Cartesian coordinates, 
\begin{subequations}\label{eq:polar_Cartesian_kF_decomp}
\begin{equation}
\hat{u}_{k_F}=\cos\theta \hat{u}_{k_x}+\sin\theta \hat{u}_{k_y},
\end{equation}
\begin{equation}
\hat{u}_{\theta}=-\sin\theta \hat{u}_{k_x}+\cos\theta \hat{u}_{k_y},
\end{equation}
\end{subequations}
combining Eqs.\@ (\ref{eq:n_k_gen}) and (\ref{eq:polar_Cartesian_kF_decomp}) we have
\begin{multline}\label{eq:n_k_Cartesian}
\hat{n}_{\vec{k}}=N^{-1}\left\{\left[\cos\theta k_F(\theta)+\sin\theta \frac{d k_F(\theta)}{d \theta} \right] \hat{u}_{k_x}\right. \\ \left.+ \left[\sin\theta k_F(\theta)-\cos\theta \frac{d k_F(\theta)}{d \theta} \right] \hat{u}_{k_y} \right\},
\end{multline}
where $N\equiv n_{\vec{k}}$ obeys Eq.\@ (\ref{eq:n_k_gen_mod}). 
Further decomposing Eq.\@ (\ref{eq:n_k_Cartesian}) into its components along $k_x$ and $k_y$ yields Eq.\@ (\ref{eq:polar_Cartesian_kF}). 

As an exemplary application, in the next section we calculate the components (\ref{eq:polar_Cartesian_kF}) of $\hat{n}_{\vec{k}}$ for the ``supercircle'' geometry (\ref{eq:r_supercircle}) of Sec.\@ \ref{Supercircle}, using selected values of the parameter $p$. 

\subsection{Parametrization of the ``supercircle'' Fermi surface}

For the geometry (\ref{eq:r_supercircle}), the derivatives $d k_F(\theta)/d\theta$ have to be separately evaluated in the four quadrants of the 2D plane. For generic $p$, we have \begin{widetext}
\begin{equation}\label{eq:dr_p_gen}
\frac{d k_F(\theta)}{d \theta}=\left[\left|\cos\theta\right|^{p-1} \mathrm{sign}\left\{\cos \theta\right\} \sin \theta-\left|\sin\theta\right|^{p-1} \mathrm{sign}\left\{\sin\theta\right\}\cos\theta\right] \left[\left|\cos\theta\right|^p+\left|\sin\theta\right|^{p}\right]^{-\frac{1+p}{p}}.
\end{equation}
\end{widetext}
For generic values of $p$, the integrals (\ref{eq:G_0_iso}) which contribute to the transverse conductivity (\ref{eq:sigma_yy}) have to be calculated numerically. Therefore, we do not have a general closed expression for the conductivity for any momentum and frequency. In the following, we separately analyze specific values of $p$: we exactly retrieve the results of Secs.\@ \ref{Square_dia} and \ref{Z_sq_spec} for $p=1$ (``diamond'' shape) and of Secs.\@ \ref{Iso_circ} and \ref{App_Z_iso_analysis} for $p=2$ (circular shape). 

\subsubsection{``Diamond'' shape ($p=1$)}

The derivatives (\ref{eq:dr_p_gen}) for $p=1$ are
\begin{subequations}\label{eq:dr_p_gen_p1}
\begin{equation}\label{eq:dr_q1_p1}
\frac{d k_F(\theta)}{d \theta}=\frac{-\cos \theta+\sin\theta}{(\cos\theta+\sin\theta)^2}, \, 0\leq \theta <\frac{\pi}{2},
\end{equation}
\begin{equation}\label{eq:dr_q2_p1}
\frac{d k_F(\theta)}{d \theta}=\frac{-\cos \theta+\sin\theta}{-1+\sin(2 \theta)}, \, \frac{\pi}{2} \leq \theta <\pi,
\end{equation}
\begin{equation}\label{eq:dr_q3_p1}
\frac{d k_F(\theta)}{d \theta}=\frac{\cos \theta-\sin\theta}{(\cos\theta+\sin\theta)^2}, \, \pi \leq \theta < \frac{3\pi}{2},
\end{equation}
\begin{equation}\label{eq:dr_q4_p1}
\frac{d k_F(\theta)}{d \theta}=\frac{\cos \theta+\sin\theta}{(\cos\theta-\sin\theta)^2}, \, \frac{3\pi}{2} \leq \theta < 2 \pi. 
\end{equation}
\end{subequations}
The normalization factors (\ref{eq:n_k_gen_mod}) are
\begin{subequations}\label{eq:n_k_p1}
\begin{equation}
n_{\vec{k}}=\frac{1}{\sqrt{2}} \left[(\cos\theta + \sin\theta)^4\right]^{-\frac{1}{2}}, \, 0\leq \theta <\frac{\pi}{2} \, v\, \pi \leq \theta < \frac{3\pi}{2},
\end{equation}
\begin{equation}
n_{\vec{k}}=\frac{1}{\sqrt{2}} \left[(\cos\theta - \sin\theta)^4\right]^{-\frac{1}{2}}, \, \frac{\pi}{2} \leq \theta <\pi \, v\, \frac{3\pi}{2} \leq \theta < 2 \pi.
\end{equation}
\end{subequations}
Using Eqs.\@ (\ref{eq:dr_p_gen_p1}) and (\ref{eq:n_k_p1}) into Eqs.\@ (\ref{eq:polar_Cartesian_kF}), we identically retrieve
\begin{subequations}\label{eq:n_k_p1}
\begin{equation}
\hat{n}_{\vec{k}}=\frac{1}{\sqrt{2}}\hat{u}_{k_x}+\frac{1}{\sqrt{2}}\hat{u}_{k_y}, \, 0\leq \theta <\frac{\pi}{2}
\end{equation}
\begin{equation}
\hat{n}_{\vec{k}}=-\frac{1}{\sqrt{2}}\hat{u}_{k_x}+\frac{1}{\sqrt{2}}\hat{u}_{k_y}, \, \frac{\pi}{2} \leq \theta <\pi,
\end{equation}
\begin{equation}
\hat{n}_{\vec{k}}=-\frac{1}{\sqrt{2}}\hat{u}_{k_x}-\frac{1}{\sqrt{2}}\hat{u}_{k_y}, \, \pi \leq \theta < \frac{3\pi}{2},
\end{equation}
\begin{equation}
\hat{n}_{\vec{k}}=\frac{1}{\sqrt{2}}\hat{u}_{k_x}-\frac{1}{\sqrt{2}}\hat{u}_{k_y}, \, \frac{3\pi}{2} \leq \theta < 2 \pi,
\end{equation}  
\end{subequations}
which is fully consistent with the piecewise-constant parametrization of Tab.\@ \ref{tab:v_dia_sample1} in Sec.\@ \ref{Square_dia}. 

\subsubsection{Circular shape ($p=2$)}

For a circular shape, the derivatives (\ref{eq:dr_p_gen}) are null in all quadrants:
\begin{equation}\label{eq:dr_p2}
\frac{d k_F(\theta)}{d \theta}=0.
\end{equation}
Using Eq.\@ (\ref{eq:dr_p2}) into Eq.\@ (\ref{eq:n_k_gen}), we retrieve
\begin{equation}\label{eq:n_k_p2}
\hat{n}_{\vec{k}}\equiv \hat{u}_{k_F},
\end{equation}
as it should be: the direction normal to the Fermi surface is the radial direction for a circular shape. Hence, in this special case we identically retrieve the results of Sec.\@ \ref{Iso_circ}.

\end{document}


\title{Supplemental material for ``Kinetic theory of the non-nocal electrodynamic response in anisotropic metals: skin effect in 2D systems''}

\author{Davide Valentinis}
\affiliation{Institut für Theorie der Kondensierten Materie, Karlsruher Institut
für Technologie, 76131 Karlsruhe, Germany}
\affiliation{Institut für Quantenmaterialien und Technologien, Karlsruher Institut
für Technologie, 76131 Karlsruhe, Germany}
\author{Graham Baker}
\affiliation{Department of Physics and Astronomy, University of British Columbia, Vancouver, BC V6T 1Z1, Canada}
\affiliation{Quantum Matter Institute, University of British Columbia, Vancouver, BC V6T 1Z4, Canada}
\author{Douglas A. Bonn}
\affiliation{Department of Physics and Astronomy, University of British Columbia, Vancouver, BC V6T 1Z1, Canada}
\affiliation{Quantum Matter Institute, University of British Columbia, Vancouver, BC V6T 1Z4, Canada}
\author{J\"{o}rg Schmalian}
\affiliation{Institut für Theorie der Kondensierten Materie, Karlsruher Institut
für Technologie, 76131 Karlsruhe, Germany}
\affiliation{Institut für Quantenmaterialien und Technologien, Karlsruher Institut
für Technologie, 76131 Karlsruhe, Germany}
\date{\today}

\begin{abstract}
In this supplemental material, an alternative derivation of the distribution function for 2D isotropic Fermi surfaces is provided, which gives identical results to the derivation using the collision-operator formalism reported in Sec.\@ IIIa of the main text. Additional graphs for the surface impedance with specular boundary conditions are shown for hexagonal and square Fermi-surface shapes. An analysis of anisotropic hydrodynamics for trigonal systems, based on the Navier-Stokes equations in the absence of rotational invariance, is performed to confirm that the electrodynamic properties are independent of Fermi-surface orientation for D$_{3d}$ symmetry (hexagonal shape). The influence of boundary conditions on the surface impedance, in the isotropic and anisotropic cases, is quantitatively estimated assuming diffusive interface scattering: the differences in the ``phase diagrams'' of the surface impedance modulus with respect to specular surface scattering are found to be negligible, for circular, hexagonal and square Fermi surfaces, and for all analyzed Fermi-surface orientations. 
\end{abstract}

\maketitle

\section{Transverse conductivity of isotropic 2D Fermi Systems: derivation with angular momentum states}

Let us apply the collision-operator formalism to the case of isotropic 2D Fermi systems.
In this case, rotational invariance implies that the eigenfunctions are angular momentum states. If we confine ourselves to energies small compared to the Fermi level $E_{F}$, we can use 
\begin{equation}
\chi_{\vec{k},m}=\chi_{\vec{k},m}(\vec{q},\omega)=c_{0}e^{i m\theta},
\end{equation}
where $\theta$ is the angle formed by the wave vector $\vec{k}=k_{F}\left(\cos\theta,\sin\theta\right)$ in the $x y$ plane. We neglect the dependence of the eigenfunctions on the magnitude of the momentum, as it is small by $\left|k-k_{F}\right|/k_{F}\ll 1$. The coefficient $c_{0}$ follows from the normalization condition (21), which translates in the present case as
\begin{equation}
\left\langle m\mid m'\right\rangle =\int_{\vec{k}}w_{\vec{k}}\chi_{\vec{k},m}^{*}\chi_{\vec{k},m'}=\delta_{m,m'}.
\end{equation}
Using the definition (15) of the scalar product, we have
\begin{multline}
\left\langle m\mid m'\right\rangle =\int_{\vec{k}} \left[-k_B T \frac{\partial f^{0}(\varepsilon_{\vec{k}})}{\partial \varepsilon_{\vec{k}}}\right] \chi_{\vec{k},m}^{*}\chi_{\vec{k},m'} \\ = \int_0^{2 \pi} \frac{d \theta}{2 \pi} \int d \varepsilon N_{\rm el}(\varepsilon) \left[-k_B T \frac{\partial f^{0}(\varepsilon)}{\partial \varepsilon}\right] c_0^\ast e^{-i m \theta} c_0 e^{i m' \theta} \\ \approx N_{\rm el}(0) k_B T \left|c_0\right|^2 \underbrace{\int_{-\infty}^{+\infty} d \epsilon \left[-\frac{\partial f^{0}(\varepsilon)}{\partial \varepsilon}\right]}_{1} \times \\ \underbrace{ \int_0^{2 \pi} \frac{d \theta}{2 \pi} e^{i \theta (m-m')}}_{ \delta_{m,m'}} \equiv \delta_{m,m'}
\end{multline}
such that $c_{0}=1/\sqrt{N_{\rm el}(0) k_B T}$. Here we have used the electronic density of states $N_{\rm el}(\varepsilon)=\sum_{\vec{k}} \delta(\varepsilon-\varepsilon_{\vec{k}})$.
With the above formalism, we only have to determine the matrix elements of the velocity operator 
\begin{multline}\label{eq:v_matr_el}
\left\langle m\left|\vec{v}\right|m'\right\rangle  = \int_{\vec{k}}w_{\vec{k}}\chi_{\vec{k}m}^{*}\boldsymbol{v}_{\vec{k}}\chi_{\vec{k},m'} \\
 = v_{F}\int\frac{d\theta}{2\pi}\left(\begin{array}{c}
\cos\theta\\
\sin\theta
\end{array}\right)e^{-i\theta\left(m-m'\right)} \\
 = \frac{v_{F}}{2}\left[\delta_{m',m+1}\left(\begin{array}{c}
1\\
i
\end{array}\right)+\delta_{m',m-1}\left(\begin{array}{c}
1\\
-i
\end{array}\right)\right]
\end{multline}
and the source term 
\begin{multline}\label{eq:source_matr_el}
s_{m} = -\frac{e}{k_B T}\vec{E}\cdot \int_{\vec{k}}w_{\vec{k}}\chi_{\vec{k},m}^{*}\vec{v}_{\vec{k}} \\
=  -\frac{e}{k_B T}\vec{E} v_{F}\int\frac{d\theta}{2\pi} \int d \varepsilon N_{\rm el}(\varepsilon) \left[-k_B T \frac{\partial f^{0}(\varepsilon)}{\partial \varepsilon}\right] \times \\ c_0^{\ast} e^{-im\theta}\left(\begin{array}{c}
\cos\theta\\
\sin\theta
\end{array}\right)\\
\approx  -\frac{e}{k_B T}\vec{E}N_{\rm el}(0 )k_B T c_{0}^{\ast} v_{F}\int\frac{d\theta}{2\pi}e^{-i m \theta}\left(\begin{array}{c}
\cos\theta\\
\sin\theta
\end{array}\right)\\
 = \sqrt{\frac{N_{\rm el}(0)}{k_B T}}\frac{e v_F}{2} \vec{E}\cdot\left[\begin{array}{c}
\delta_{m,1}+\delta_{m,-1}\\
i\left(\delta_{m,1}-\delta_{m,-1}\right)
\end{array}\right]
\end{multline}
Hence, we obtain the Boltzmann equation with $\vec{q}=q\left(\cos\theta \hat{u}_x+\sin\theta \hat{u}_y \right)$:
\begin{equation}\label{eq:Boltz_iso_theta}
\left(\gamma_{m}-i\omega\right)a_{m}+i\frac{v_F q}{2}\left(e^{i\theta}a_{m+1}+e^{-i\theta}a_{m-1}\right)=s_{m}
\end{equation}
A closely related problem was in fact recently studied in Ref.\@ \onlinecite{Lucas-2018}, where the collective modes of the system were analyzed.
Here we determine the full distribution function, which shall then give us the transverse conductivity where $\vec{q}\parallel\hat{u}_{x}$ and 
$\vec{E}\parallel\hat{u}_{y}$.
The assumption $\vec{q}\parallel\hat{u}_{x}$ implies $\theta=0$. Consistently with Eq.\@ (31), we take $\gamma_0=0$ for charge conservation, $\gamma_1=\gamma_{\rm mr} \geq0$ for slow momentum relaxation, and for higher-order modes $\gamma_m =\gamma_{\rm mc} \geq 0 \, \forall m \geq 2$. The source term $s_m$ is relaxing just the mode with $m=\pm 1$ for electric field parallel to the $y$ axis, hence we write 
\begin{equation}\label{eq:source_term_expl}
s_{\pm 1} =\pm i Y_E,
\end{equation}
where 
\begin{equation}
Y_{E}=\frac{e}{k_B T}\frac{v_{F}}{2c_{0}}E_{y}
\end{equation}
is proportional to the electric field. Now we have the set of equations 
\begin{subequations}
\begin{equation}
-i\omega a_{0}+i\frac{v_F q_x}{2}\left(a_{1}+a_{-1}\right) = 0,
\end{equation}
\begin{equation}
\left(\gamma_{\rm mr}-i\omega\right)a_{\pm 1}+i\frac{v_F q_x}{2}\left(a_{0}+a_{\pm 2}\right) = \pm i Y_{E}, 
\end{equation}
\begin{equation}
\left(\gamma_{\rm mc}-i\omega\right)a_{m}+i\frac{v_F q_x}{2}\left(a_{m+1}+a_{m-1}\right) =0,\, \left|m\right|\geq2.
\end{equation}
\end{subequations}
Next, we divide all equations by $v_F q$ to get 
\begin{subequations}
\begin{equation}
-i s a_{0}+i\frac{1}{2}\left(a_{1}+a_{-1}\right)=0,
\end{equation}
\begin{equation}
\left(\gamma_{\rm mr}-i s\right)a_{\pm 1}+i\frac{1}{2}\left(a_{0}+a_{\pm2}\right) = \pm i\frac{Y_{E}}{v_F q_x}, 
\end{equation}
\begin{equation}
\left(\gamma_{\rm mc}-i s\right)a_{m}+i\frac{1}{2}\left(a_{m+1}+a_{m-1}\right)=0, \, \left|m\right|\geq2,
\end{equation}
\end{subequations}
where $\Gamma_{i}=\gamma_{i}/(v_F q_x)$, $i=\left\{\rm mr, \rm mc\right\}$, and $s=\omega/(v_F q_x)$. We now introduce $a_{n}=b_{n}Y_{E}/(v_F q_x)$, to get the equations 
\begin{eqnarray}
s b_{0}-\frac{1}{2}\left(b_{1}+b_{-1}\right) & = & 0,\nonumber \\
\left(s+i\gamma_{\rm mr}\right)b_{\pm1}-\frac{1}{2}\left(b_{0}+b_{\pm2}\right) & = & \mp1,
\end{eqnarray}
while for $\left|m\right|\geq2$ we have 
\begin{equation}
\left(s+i\Gamma_{\rm mc}\right)b_{m}-\frac{1}{2}\left(b_{m+1}+b_{m-1}\right)=0.
\end{equation}
We start our analysis by assuming that $q$ is real, and we will make the analytic continuation to $q \in \mathbb{C}$ at the end.
The recursion relation for $m\geq2$ has the two solutions 
\[
b_{m}=c_{+}\lambda_{+}^{m}+c_{-}\lambda_{-}^{m},
\]
where $\lambda_{\pm}=z\pm\sqrt{z^{2}-1}$ with $z=s+i\Gamma_{\rm mc}$.
Those are the roots of 
\begin{equation}\label{eq:zquadr}
z=\frac{1}{2}\left(\lambda+\lambda^{-1}\right).
\end{equation}
To achieve convergence it must hold $\left|\lambda_{\pm}\right|\leq1$. Let us assume without restriction that $\omega>0$. Under the assumption of real $q_{x}$, we know that both, the real part and the imaginary part of $z$ are positive.
If we write $\lambda=r e^{i\varphi}$, we can analyze separately the real and imaginary parts of Eq.\@ (\ref{eq:zquadr}):
\begin{eqnarray}
s & = & \frac{\cos\varphi}{2}\left(r+r^{-1}\right)>0\nonumber \\
\Gamma_{\rm mc} & = & \frac{\sin\varphi}{2}\left(r-r^{-1}\right)>0.
\end{eqnarray}
To ensure convergence, it must hold $\left|\lambda\right|=r\leq1$, hence we obtain the conditions $\cos\varphi>0$ and $\sin\varphi<0$, which imply that $\varphi\in\left[\frac{3}{2}\pi,2\pi\right]$, i.e.\@, ${\rm Re}\lambda>0$ and ${\rm Im}\lambda<0$. 
If one analyzes the above solutions  $\lambda_{\pm}$, one then finds that only $\lambda=\lambda_{-}$ is acceptable. Hence, it must hold that $c_{+}=0$. Most important is that one of the two constants must always be zero. This should survive the analytical continuation to complex momenta $q_{x}$.
Now we can always write that $b_{m}=\lambda^{m-1}b_{1}$, which yields in particular $b_{2}=\lambda b_{1}$. 
We can perform the same analysis for negative $m$ and find $b_{m}=\lambda^{-m-1}b_{-1}$, and in particular $b_{-2}=\lambda b_{-1}$, with the same $\lambda$ for positive and negative $m$. 

This yields a closed set of equations with the solution $b_{0}=0$ as well as  
\begin{multline}
b_{\pm1}=\mp\frac{1}{s+i\Gamma_{\rm mr}-\frac{\lambda}{2}} \\= \mp\frac{1}{s+i\Gamma_{\rm mr}-\frac{s+i \Gamma_{\rm mc} -\sqrt{(s+i \Gamma_{\rm mc})^2-1}}{2}}.
\end{multline}
We are now in the position to determine the distribution function $\psi_{\theta}$ from Eq.\@ (22). The linear system to solve is
\begin{eqnarray*}
\psi_{\theta} & = & \sum_{m=-\infty}^{\infty}a_{m}\chi_{\vec{k},m}\\
 & = & c_{0}\sum_{m=-\infty}^{\infty}a_{m}e^{im\theta}=\frac{Y_{E}c_{0}}{v_Fq_{x}}\sum_{m=-\infty}^{\infty}b_{m}e^{im\theta}\\
 & = & -i\frac{2Y_{E}c_{0}}{\omega+i\gamma_{\rm mr}-\frac{\lambda}{2}v_Fq_{x}}\sum_{m=1}^{\infty}\lambda^{m-1}\sin\left(m\theta\right)\\
 & = & \frac{2Y_{E}c_{0}}{i\omega-\gamma_{\rm mr}-i\frac{\lambda}{2}v_Fq_{x}}\frac{\sin\theta}{1+\lambda^{2}-2\lambda\cos\theta}.
\end{eqnarray*}
From Eq.\@ (11) we obtain the distribution function 
\begin{multline}\label{eq:distr_iso_expl}
\delta f_{\vec{k}}\left(q_{x},\omega\right)=e\frac{\partial f^{\left(0\right)}\left(\varepsilon_{\vec{k}}\right)}{\partial\varepsilon_{\vec{k}}}\frac{v_{\vec{k},y}E_{y}}{-i\omega+\gamma_{\rm mr}+i\frac{\lambda}{2}v_F q_{x}} \times \\ \frac{1}{1+\lambda^{2}-2\lambda v_{\vec{k},x}/v_{F}},
\end{multline}
with 
\[
\lambda=\frac{1}{v_F q_{x}}\left(\omega+i\gamma_{\rm mc}-\sqrt{\left(\omega+i\gamma_{\rm mc}\right)^{2}-\left(v_F q_{x}\right)^{2}}\right),
\]
$v_{\vec{k},x}=v_F \cos \theta$, and $v_{\vec{k},y}=v_F \sin \theta$. 
Let us analyze the the denominator in the distribution function (\ref{eq:distr_iso_expl}):
\begin{multline}\label{eq:D_1}
D = \left(-i\omega+\gamma_{\rm mr}+i\frac{\lambda}{2}v_F q_{x}\right)\left(1+\lambda^{2}-2\lambda \frac{v_{\vec{k},x}}{v_{F}}\right)\\
 = \left(-i\omega+\gamma_{\rm mr}+\gamma_{\rm mc}-\gamma_{\rm mc}+i\frac{\lambda}{2}v_F q_{x}\right) \times \\ \left(1+\lambda^{2}-2\lambda \frac{v_{\vec{k},x}}{v_{F}}\right)\\
 = -i\omega+\gamma_{\rm mc}+iv_{\vec{k},x}q_{x} \\ -\left(\gamma_{\rm mc}-\gamma_{\rm mr}\right)2\lambda\left(\frac{\omega+i\gamma_{\rm mc}}{v_F q_{x}}-\frac{v_{\vec{k},x}}{v_{F}}\right).
\end{multline}
In deriving the above result, we used 
\begin{eqnarray}
D_{1} & \equiv & \left(-i\omega+\gamma_{\rm mc}+i\frac{\lambda}{2}v_F q_{x}\right)\left(1+\lambda^{2}-2\lambda \frac{v_{\vec{k},x}}{v_{F}}\right)\nonumber \\
 & = & -iv_F q_{x}\left(z-\frac{\lambda}{2}\right)\left(1+\lambda^{2}-2\lambda \frac{v_{\vec{k},x}}{v_{F}}\right).
\end{eqnarray}
Inserting Eq.\@ (\ref{eq:zquadr}) gives 
\begin{eqnarray}
D_{1} & = & -iv_F q_{x}\left(\frac{\lambda^{-1}+\lambda}{2}-\frac{v_{\vec{k},x}}{v_{F}}\right)\nonumber \\
 & = & -iv_F q_{x}\left(z-\frac{v_{\vec{k},x}}{v_{F}}\right)\nonumber \\
 & = & -i\omega+\gamma_{\rm mc}+i v_{\vec{k}, x}q_{x}.
\end{eqnarray}
This gives the first term in Eq.\@ (\ref{eq:D_1}) for $D$. 

All in all, we obtain the following expression for the distribution function
\begin{multline}\label{eq:distrcircular}
\delta f_{\vec{k}}\left(q_{x},\omega\right)=e\frac{\partial f^{\left(0\right)}\left(\varepsilon_{\vec{k}}\right)}{\partial\varepsilon_{\vec{k}}} \times \\ \frac{v_{\vec{k},y}E_{y}}{-i\omega+\gamma_{\rm mc}+iv_{\vec{k},x}q_{x}+M_{\vec{k}}\left(q_{x},\omega\right)}
\end{multline}
where  
\begin{multline}
M_{\vec{k}}\left(q_{x},\omega\right)=-2\frac{\gamma_{\rm mc}-\gamma_{\rm mr}}{v_F q_{x}} \times \\ \left(\omega+i\gamma_{\rm mc}-\sqrt{\left(\omega+i\gamma_{\rm mc}\right)^{2}-\left(v_F q_{x}\right)^{2}}\right) \times \\ \left(\frac{\omega+i\gamma_{\rm mc}}{v_F q_{x}}-\frac{v_{\vec{k},x}}{v_{F}}\right). 
\end{multline}
The distribution function (\ref{eq:distrcircular}) can be used to describe Ohmic transport, ballistic transport, and viscous or hydrodynamic transport, including the crossover among these distinct regimes for a circular Fermi surface. Eq.\@ (\ref{eq:distrcircular}) coincides with Eq.\@ (83), which was obtained using a completely different strategy, namely as a special case of the collision-operator formalism in the isotropic limit. 

\section{Additional plots for the surface impedance}

\subsection{Surface impedance for equal scattering rates}

In the main text, it is argued that two conditions on the ratio between momentum-relaxation rate $\gamma_{\rm mr}$ and momentum-conserving collision rate $\gamma_{\rm mc}$ are the most interesting for realistic scenarios: $\gamma_{\rm mr} \approx \gamma_{\rm mc}$ at high temperatures, and $\gamma_{\rm mr} \ll \gamma_{\rm mc}$ at low temperatures. The latter case is analyzed in detail in the main text, while in the following we show results for the surface impedance assuming equal scattering rates and specular interface scattering. 

Since the hydrodynamic regime is realized for $\gamma_{\rm mc} \gg \gamma_{\rm mr}$, we can already anticipate that, in all cases considered in this section, viscous skin effect will be absent. Therefore, only normal, anomalous and perfect-conductor skin effects are possible for equal scattering rates. 
\begin{figure}[ht]
\includegraphics[width=0.8\columnwidth]{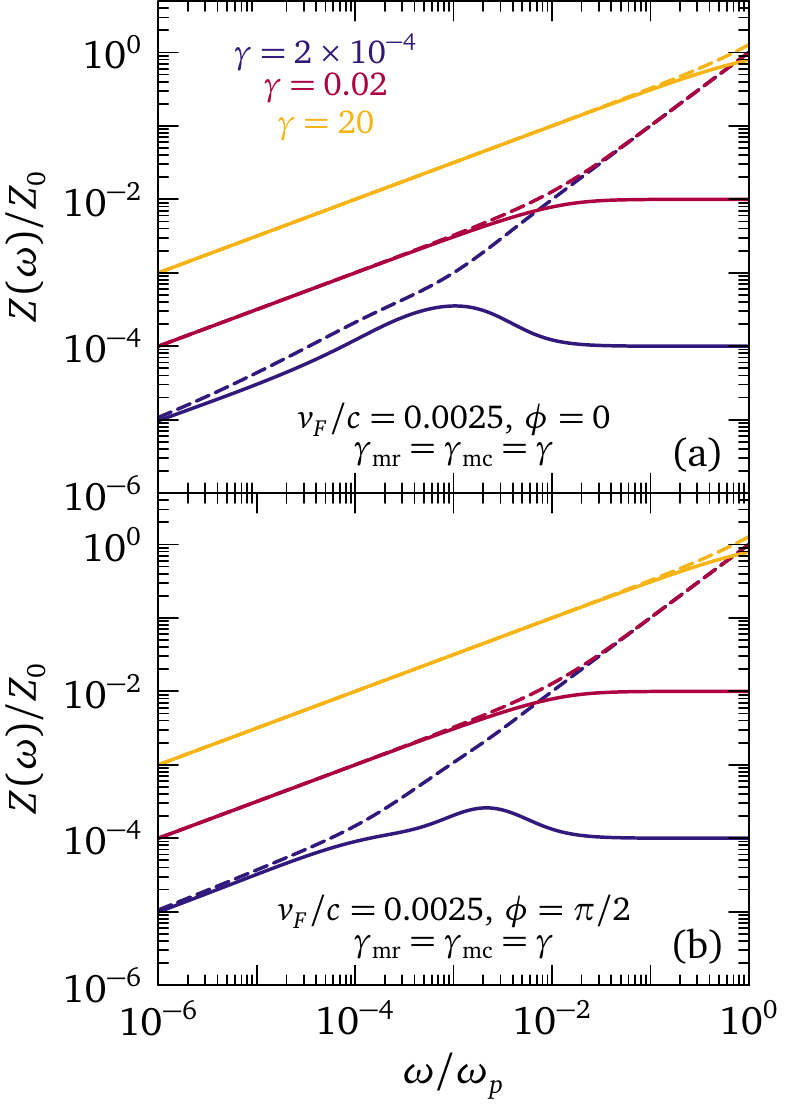} 
\caption{\label{fig:Zs_phi090_om_Hex} Surface impedance $Z(\omega)$ normalized to the vacuum-impedance $Z_0$, as a function of $\omega/\omega_p$, for a hexagonal Fermi surface with Fermi-velocity modulus $v_F=2.5 \times 10^{-3} c$, $\gamma_{\rm mr}=\gamma_{\rm mc}=\gamma$, and specular boundary conditions. Solid (dashed) lines refer to $\mathrm{Re} Z(\omega)$ ($-\mathrm{Im} Z(\omega)$). (a) Results for tilting angle $\phi=0$; see Fig.\@ 7(a). (b) Results for $\phi=\pi/2$; see Fig.\@ 7(b). 
} 
\end{figure}
Fig.\@ \ref{fig:Zs_phi090_om_Hex} shows the real part (solid curves) and imaginary part (dashed curves) of the surface impedance $Z(\omega)$, normalized to the vacuum-impedance $Z_0$, calculated numerically as a function of dimensionless frequency $\omega/\omega_p$ using Eqs.\@ (143) and (144), for a hexagonal Fermi-surface shape. We use the Fermi-velocity modulus $v_F=2.5 \times 10^{-3} c$ and $\gamma_{\rm mr}=\gamma_{\rm mc}=\gamma$. Blue, red and gold curves refer to $\gamma=\left\{2 \times 10^{-4}, 0.02,20\right\}$ respectively. Fig.\@ \ref{fig:Zs_phi090_om_Hex}(a) displays the results in the ``parallel'' configuration for tilting angle $\phi=0$ with respect to the surface (see Fig.\@ 7(a)) using the conductivity (92), while Fig.\@ \ref{fig:Zs_phi090_om_Hex}(b) shows the results for $\phi=\pi/2$ (see Fig.\@ 7(b)), employing the conductivity given by Eqs.\@ (101) and (102). 

Focusing on low frequencies, in all panels of Fig.\@ \ref{fig:Zs_phi090_om_Hex} we retrieve $\left\{\mathrm{Re},\mathrm{Im}\right\}Z(\omega) \propto \omega^{1/2}$: this is the regime of normal skin effect, where $\omega \gamma \ll 1$ and $\omega\ll v_F \left|q_x\right|$, and nonlocal effects are negligible. At high frequency, the impedance is predominantly imaginary and given by the perfect-conductivity result $Z(\omega)=-i \lambda_L \mu_0 \omega$. The first-order correction to the latter behavior for $\omega \rightarrow +\infty$ gives a subdominant real part $\mathrm{Re}Z(\omega)=\gamma/2$, which is constant with frequency and is evident in all panels of Fig.\@ \ref{fig:Zs_phi090_om_Hex}. 
To see this correction, we can analyze the conductivity and the impedance at high frequency. The analysis is analogous for $\phi=0$ and $\phi=\pi/2$, and here we explicitly report only on the $\phi=0$ case for compactness. In the ``parallel'' configuration of Fig.\@ 7(a), the conductivity follows Eq.\@ (93) in the case $\delta \gamma=0$. Series-expanding for $\omega \rightarrow +\infty$, we have
\begin{equation}\label{eq:sigma_hex_par_gamma_highom}
\frac{\sigma_{yy}(q_x,\omega)}{\epsilon_0 \omega_p^2}=\frac{i \omega +\gamma}{\omega}+i\frac{(v_F q_x)^2-4 \gamma^2}{4 \omega^3}. 
\end{equation}
Inserting this result into Eqs.\@ (143) and (144), we obtain the surface impedance
\begin{equation}\label{eq:Z_hex_par_gamma_highom}
\frac{Z(\omega)}{Z_0}=\frac{2 \omega^3}{\omega_p\sqrt{\gamma^2+i \gamma \omega-\omega^2\left[(v_F/c)^2+4\omega^2\right]}}.
\end{equation}
Expanding again Eq.\@ (\ref{eq:Z_hex_par_gamma_highom}) for $\omega \rightarrow +\infty$, we obtain $Z(\omega)/Z_0=-i\omega/\omega_p+ \gamma/(2 \omega_p)+o(1/\omega)$, which implies a frequency-linear imaginary part and a constant real part, exactly as shown in Fig.\@ \@ \ref{fig:Zs_phi090_om_Hex}. 

For the lowest scattering rate $\gamma=2 \times 10^{-4}$ (blue curves in Fig.\@ \ref{fig:Zs_phi090_om_Hex}), we see a bump in the real part of $Z(\omega)$ before its saturation to a constant value: this happens in the crossover between normal and anomalous regimes, when the scattering rate $\gamma$ is low enough that a small frequency window of ballistic (anomalous) skin effect appears between normal and perfect-conductor regimes. A similar non-monotonic evolution occurs in the surface impedance as a function of temperature in the crossover regime between hydrodynamic and anomalous skin effect, as analyzed by Gurzhi for electrons interacting with phonons \cite{Gurzhi-1968}. 

\begin{figure}[ht]
\includegraphics[width=0.8\columnwidth]{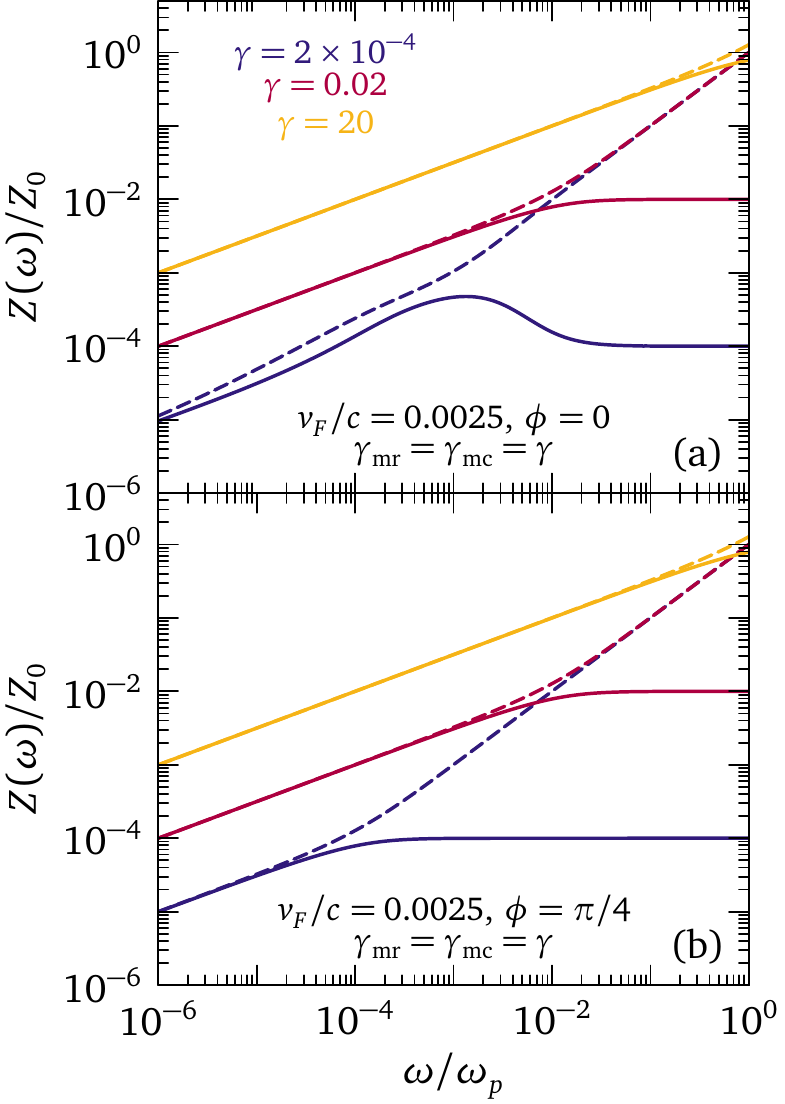} 
\caption{\label{fig:Zs_phi045_om_Square} Surface impedance $Z(\omega)$ normalized to the vacuum-impedance $Z_0$, as a function of dimensionless frequency $\omega/\omega_p$, for a square Fermi surface with Fermi-velocity modulus $v_F=2.5 \times 10^{-3} c$, $\gamma_{\rm mr}=\gamma_{\rm mc}=\gamma$, and specular boundary conditions. Blue, red and gold curves refer to $\gamma=\left\{2 \times 10^{-4}, 0.02,20\right\}$ respectively. Solid (dashed) lines refer to $\mathrm{Re} Z(\omega)$ ($-\mathrm{Im} Z(\omega)$). (a) Results for tilting angle $\phi=0$ with respect to the surface; see Fig.\@ 8(a). (b) Results for $\phi=\pi/4$; see Fig.\@ 8(b). 
} 
\end{figure}

In the case of a square Fermi surface, we obtain the impedance shown in Fig.\@ \ref{fig:Zs_phi045_om_Square}: again, the real part (solid curves) and imaginary part (dashed curves) of the surface impedance $Z(\omega)$, normalized to the vacuum-impedance $Z_0$, are calculated numerically as a function of dimensionless frequency $\omega/\omega_p$ using Eqs.\@ (143)-(144), and we employ the same parameters as for Fig.\@ \ref{fig:Zs_phi090_om_Hex}. 
Fig.\@ \ref{fig:Zs_phi045_om_Square}(a) displays the results in the ``diamond-shaped'' configuration for tilting angle $\phi=0$ (see Fig.\@ 8(a)) using the conductivity (110), while Fig.\@ \ref{fig:Zs_phi090_om_Hex}(b) presents the results for $\phi=\pi/4$ (see Fig.\@ 8(b)) employing the conductivity given by Eqs.\@ (102) and (116). 
The analysis of Fig.\@ \ref{fig:Zs_phi045_om_Square} is analogous to the one of \ref{fig:Zs_phi090_om_Hex}, as is evident by visually comparing the two figures. 
Notice that the bumps in the impedance are absent from Fig.\@ \ref{fig:Zs_phi090_om_Hex}(b), because the conductivity (120) does not allow for anomalous skin effect: the only crossover as a function of frequency is directly from normal skin effect to the high-frequency result $Z(\omega)/Z_0=-i\omega/\omega_p+ \gamma/(2 \omega_p)+o(1/\omega)$. 

We conclude that, if the assumption of equal scattering rates holds at high temperature, we expect an anisotropic surface impedance characterized by normal skin effect at low frequency, which transitions into a highly conducting regime at high frequency, characterized by a linear imaginary part and a constant real part equal to $Z_0 \propto \gamma/\omega_p$. 

\subsection{Imaginary part as a function of relaxation rate}

In this section, we show the imaginary parts of the anisotropic surface impedance associated with the real parts shown in Figs.\@ 11 and 14, as a function of inverse momentum-relaxing scattering rate $1/\gamma_{\rm mr}$, for the same parameters as in Figs.\@ 11 and 14. 
\begin{figure}[ht]
\includegraphics[width=0.8\columnwidth]{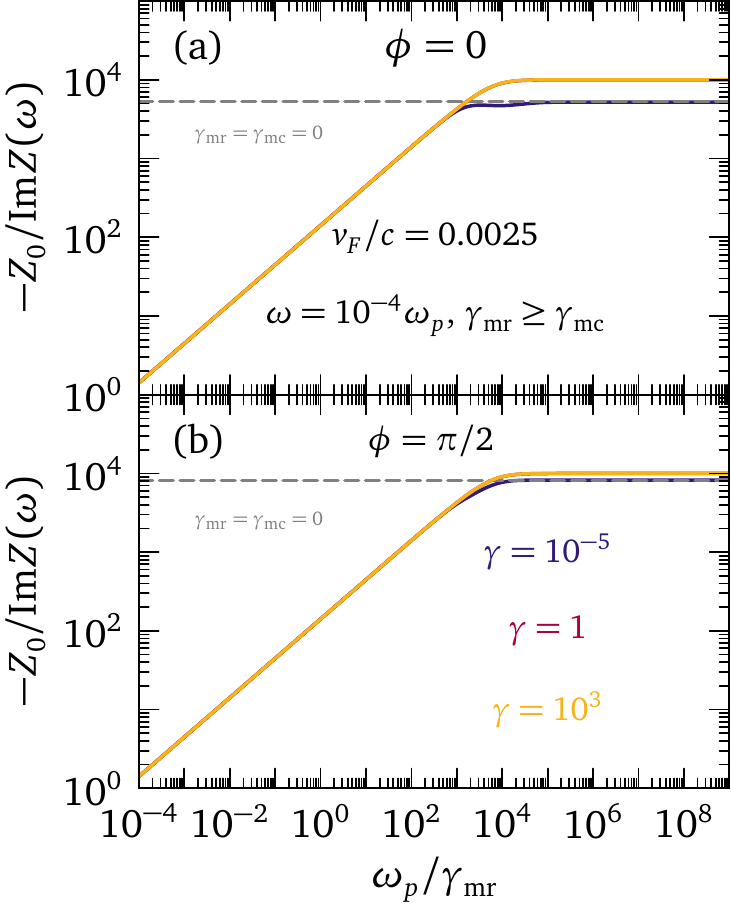} 
\caption{\label{fig:ImZs_phi090_taumr_Hex} Inverse imaginary part $-1/\mathrm{Im}Z(\omega)$ of the surface impedance, normalized to the vacuum-impedance $Z_0$, as a function inverse momentum-relaxation rate $\omega_p/\gamma_{\rm mr}$, for a hexagonal Fermi surface with Fermi-velocity modulus $v_F=2.5 \times 10^{-3} c$, $\omega/\omega_p=10^{-14}$, and $\gamma_{\rm mc}=\mathrm{max}\left\{\gamma_{\rm mr},\bar{\gamma}\right\}$, with $\bar{\gamma}=\left\{10^{-5},1,10^3\right\}$ for the blue, red, and gold curves respectively. Specular boundary conditions are assumed. The dashed gray lines represent the limit $\gamma_{\rm mr}=\gamma_{\rm mc}=0$. (a) Results for tilting angle $\phi=0$ with respect to the surface; see Fig.\@ 7(a). The scattering-less limit is given by Eq.\@ (B13). (b) Results for $\phi=\pi/2$; see Fig.\@ 7(b). The scattering-less limit stems from Eq.\@ (B16).
} 
\end{figure}
Fig.\@ \ref{fig:ImZs_phi090_taumr_Hex}(a) and \ref{fig:ImZs_phi090_taumr_Hex}(b) show the numerical results stemming from Eqs.\@ (143)-(144) for a hexagonal Fermi surface, with orientation angle $\phi=0$ and $\phi=\pi/2$ respectively. 
\begin{figure}[ht]
\includegraphics[width=0.8\columnwidth]{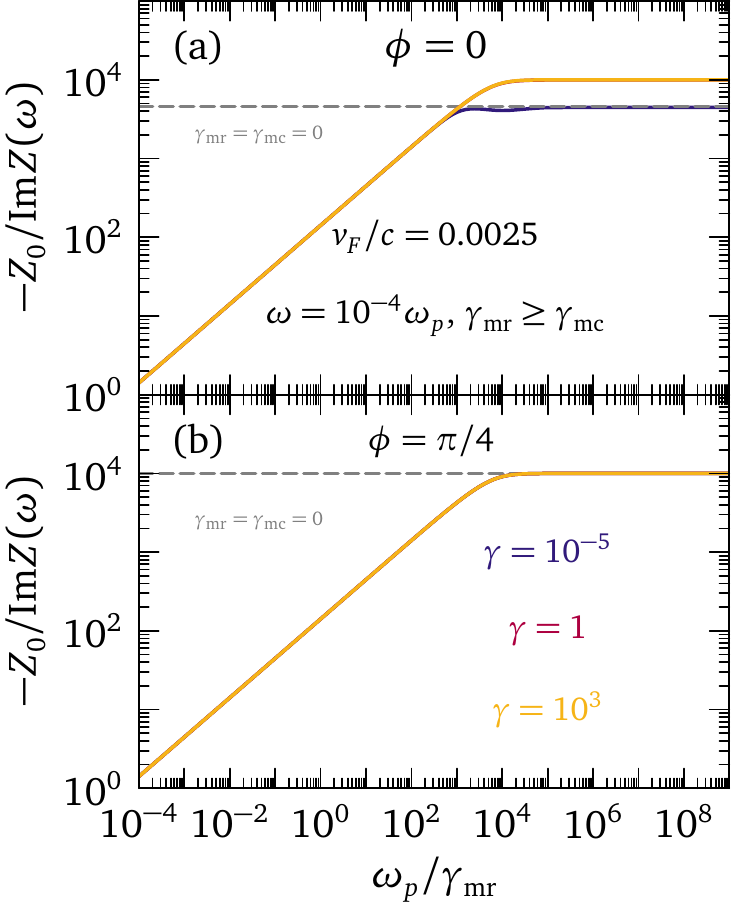} 
\caption{\label{fig:ImZs_phi045_taumr_Square} Inverse imaginary part $-1/\mathrm{Im}Z(\omega)$ of the surface impedance, normalized to the vacuum-impedance $Z_0$, as a function inverse momentum-relaxation rate $\omega_p/\gamma_{\rm mr}$, for a square Fermi surface with Fermi-velocity modulus $v_F=2.5 \times 10^{-3} c$, $\omega/\omega_p=10^{-14}$, and $\gamma_{\rm mc}=\mathrm{max}\left\{\gamma_{\rm mr},\bar{\gamma}\right\}$, with $\bar{\gamma}=\left\{10^{-5},1,10^3\right\}$ for the blue, red, and gold curves respectively. Specular boundary conditions are assumed. The dashed gray lines represent the limit $\gamma_{\rm mr}=\gamma_{\rm mc}=0$. (a) Results for tilting angle $\phi=0$ with respect to the surface; see Fig.\@ 8(a). The scattering-less limit is given by Eq.\@ (B35). (b) Results for $\phi=\pi/4$; see Fig.\@ 8(b). The scattering-less limit stems from Eq.\@ (B15).
} 
\end{figure}
Fig.\@ \ref{fig:ImZs_phi045_taumr_Square}(a) and \ref{fig:ImZs_phi045_taumr_Square}(b) display the numerics for a square Fermi surface, with orientation angle $\phi=0$ and $\phi=\pi/4$ respectively. 

From the comparison between Figs.\@ 11 and \ref{fig:ImZs_phi090_taumr_Hex}, and between Figs.\@ 14 and \ref{fig:ImZs_phi045_taumr_Square}, we see that the crossover from normal to viscous or anomalous skin effect (depending on whether $\omega \ll \gamma_{\rm mc}$ or $\omega \gg \gamma_{\rm mc}$, respectively) is visible both in the real and imaginary parts of the impedance, as a saturation in the limit of vanishing $\gamma_{\rm mr}$. However, the saturation value of the imaginary part is much less sensitive to $\gamma_{\rm mc}$ than the corresponding real part, for the same parameters. Hence, the real part of the impedance allows a more accurate distinction between anomalous and viscous skin effect, in the relaxationless limit. 

\section{Anisotropic hydrodynamics in trigonal systems}

We want to analyze the hydrodynamic flow in anisotropic systems. In particular we are interested in the case of a trigonal system with point group D$_{3d}$. The problem was already discussed in Ref.\@ \onlinecite{Cook-2019} for two-dimensional systems with square and hexagonal symmetry.
In the case of a hexagonal Fermi surface it was found that no dependency on the direction of the flow velocity w.r.t.\@ the crystalline axes exists.
Since Ref.\@ \onlinecite{Cook-2019} made one simplifying assumption for the symmetry of the viscosity tensor (see below), we are double-checking their conclusion for the most general structure of the tensor. We confirm that the flow is indeed independent on the orientation of the sample, in agreement with Ref.\@ \onlinecite{Cook-2019}.

The dissipative contribution to the momentum current is related to the velocity gradient by $\tau_{\alpha\beta}=\eta_{\alpha\beta,\gamma\delta}\partial_{\gamma}u_{\delta}.$ This gives rise to the linearized Navier-Stokes equation of an anisotropic system \cite{Link-2018}
\begin{equation}
m\partial_{t}u_{\beta}-eE_{\beta}=\partial_{\alpha}\eta_{\alpha\beta,\gamma\delta}\partial_{\gamma}u_{\delta}-\gamma_{\beta\alpha}u_{\alpha}.
\end{equation}
Here $\gamma_{\beta\alpha}$ refers to the tensor momentum-relaxing rate.
The point group of PdCoO$_{2}$ is the trigonal group D$_{3d}$.
This implies that the second-rank tensor $\gamma_{\alpha\beta}={\rm diag}\left(\gamma,\gamma,\gamma_{\perp}\right)$ has no in-plane anisotropy.
If we confine the analysis to in-plane transport there are three distinct viscosity elements 
\begin{eqnarray}
\eta_{1} & = & \eta_{xxxx}=\eta_{yyyy},\nonumber \\
\eta_{2} & = & \eta_{xxyy}=\eta_{yyxx},\nonumber \\
\eta_{r} & = & \eta_{xyxy}-\eta_{xyyx},
\end{eqnarray}
where by symmetry $\frac{1}{2}\left(\eta_{1}-\eta_{2}\right)=\eta_{xyxy}+\eta_{xyyx}$. The momentum current in the plane is then given as: 
\begin{equation}
\left(\begin{array}{c}
\tau_{xx}\\
\tau_{xy}\\
\tau_{yx}\\
\tau_{yy}
\end{array}\right)=\left(\begin{array}{cccc}
\eta_{1} & 0 & 0 & \eta_{2}\\
0 & \frac{\eta_{1}-\eta_{2}+\eta_{r}}{2} & \frac{\eta_{1}-\eta_{2}-\eta_{r}}{2} & 0\\
0 & \frac{\eta_{1}-\eta_{2}-\eta_{r}}{2} & \frac{\eta_{1}-\eta_{2}+\eta_{r}}{2} & 0\\
\eta_{2} & 0 & 0 & \eta_{1}
\end{array}\right)\left(\begin{array}{c}
\partial_{x}u_{x}\\
\partial_{x}u_{y}\\
\partial_{y}u_{x}\\
\partial_{y}u_{y}
\end{array}\right).
\end{equation}
It holds by symmetry $\eta_{xyxy}=\eta_{yxyx}$ and $\eta_{xyyx}=\eta_{yxxy}$. If we set the rotational viscosity $\eta_{r}=0$, we recover the usual structure of the elastic constants of a trigonal system, where $\eta_{1}=\eta_{11}$ and $\eta_{2}=\eta_{12}$ within the Voigt notation. However, as pointed out in Ref.\@ \onlinecite{Link-2018}, the broken rotational symmetry of a system implies that the momentum current is no longer a symmetric tensor yielding $\eta_{r}\neq0$.
Then the Voigt notation is no more useful. The name rotational viscosity was coined in Ref.\@ \onlinecite{Cook-2019}.
Here, the form of the viscosity tensor of a hexagonal system was also studied. Notice, in Ref.\@ \onlinecite{Cook-2019} the additional assumption of $\eta_{2}=-\eta_{1}$ was made.

We are now in a position to write down the two Navier-Stokes equations for the two in-plane velocity components.
In full glory of the tensor elements, it holds:
\begin{subequations}
\begin{multline}
m\partial_{t}u_{x}-eE_{x} = \partial_{x}\eta_{xx,xx}\partial_{x}u_{x}+\partial_{x}\eta_{xx,yy}\partial_{y}u_{y} \\ +\partial_{y}\eta_{yx,xy}\partial_{x}u_{y}+\partial_{y}\eta_{yx,yx}\partial_{y}u_{x}-\gamma u_{x} 
\end{multline}
\begin{multline}
m\partial_{t}u_{y}-eE_{y} = \partial_{y}\eta_{yyyy}\partial_{y}u_{y}+\partial_{y}\eta_{yy,xx}\partial_{x}u_{x} \\ +\partial_{x}\eta_{xy,yx}\partial_{y}u_{x}+\partial_{x}\eta_{xy,xy}\partial_{x}u_{y}-\gamma u_{y}.
\end{multline}
\end{subequations}
If we insert our above results for the viscosity tensor, we obtain
\begin{subequations}\label{eq:Navier_Stokes_rot}
\begin{equation}
m\partial_{t}u_{x}-eE_{x} = \eta_{1}\Delta u_{x}+\eta'\partial_{y}\left(\partial_{y}u_{x}-\partial_{x}u_{y}\right)-\gamma u_{x},
\end{equation}
\begin{equation}
m\partial_{t}u_{y}-eE_{y}= \eta_{1}\Delta u_{y}-\eta'\partial_{x}\left(\partial_{y}u_{x}-\partial_{x}u_{y}\right)-\gamma u_{y},
\end{equation}
\end{subequations}
where we introduced $\eta'=-(\eta_{1}+\eta_{2}-\eta_{r})/2$.
Without the terms proportional to $\eta'$, the behavior predicted by Eqs.\@ (\ref{eq:Navier_Stokes_rot}) would be identical to the rotation-invariant case.
 
Let us now look at the viscous skin effect or any other effect where we cut the sample along a given direction with flow parallel to the surface.
Let $x$ and $y$ be along the crystal axes. We want to consider a sample which is cut along the $y'$ direction where $\boldsymbol{x}'=R_{\theta}\boldsymbol{x}$ with 
\begin{equation}
R_{\theta}=\left(\begin{array}{cc}
\cos\theta & -\sin\theta\\
\sin\theta & \cos\theta
\end{array}\right)
\end{equation}
The velocity transforms as $\boldsymbol{u}=R_{\theta}^{-1}\boldsymbol{u'}$. With flow along the $y'$ axis we expect that $u'_{x'}=0$, i.e.\@, no flow orthogonal to the surface and $\partial_{y'}u_{y'}=0$.
Then it follows 
\begin{eqnarray}
\partial_{x}^{2} & \rightarrow & \cos^{2}\theta\partial_{x'}^{2}\nonumber \\
\partial_{y}^{2} & \rightarrow & \sin^{2}\theta\partial_{x'}^{2}\nonumber \\
\partial_{x}\partial_{y} & \rightarrow & -\cos\theta\sin\theta\partial_{x'}^{2}
\end{eqnarray}
as well as $u_{x}\rightarrow\sin\theta u'_{y'}$ and $u_{y}\rightarrow\cos\theta u'_{y'}$. If we insert this and transform the electric field accordingly, it follows:
\begin{equation}
\partial_{t}u_{y'}-\frac{e}{m}E_{y'}=\left(\eta_{1}+\eta'\right)\partial_{x'}^{2}u_{y'}-\gamma u_{y'}.
\end{equation}
The effective viscosity $\eta_{1}+\eta'$ is independent on the angle $\theta$. This result implies that for a trigonal system with flow in the $xy$ plane, there is no dependency of the Navier-Stokes equation w.r.t.\@ the orientation of the surface.

\section{Influence of boundary conditions on the impedance modulus}

\begin{figure*}[ht] \centering
\includegraphics[width=0.8\textwidth]{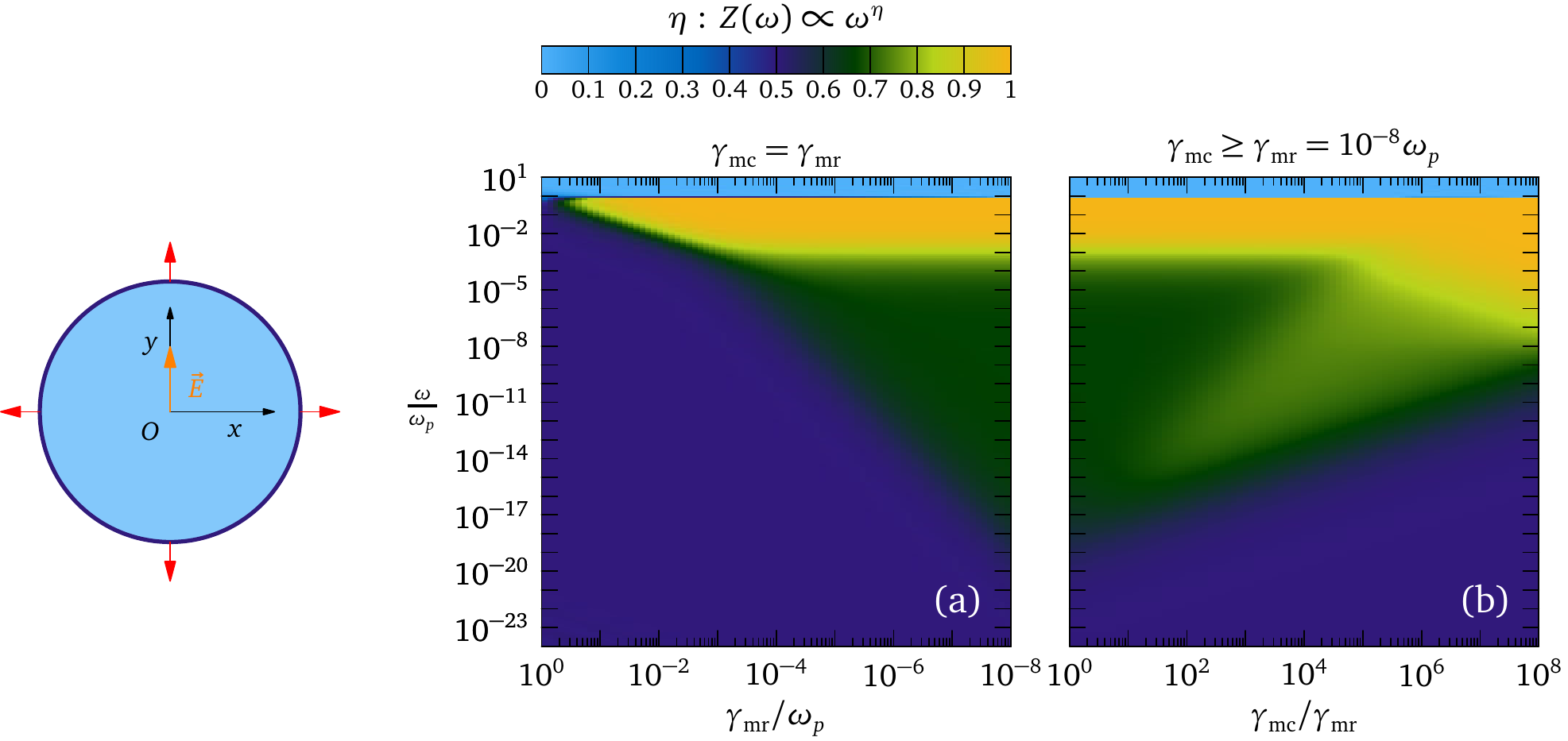} 
\caption{\label{fig:omegacross_iso_diff} Skin effect regimes for an isotropic 2D (circular) Fermi surface, as measured by the surface impedance $Z(\omega)$ as a function of relaxation rate $\gamma_{\rm mr}$, momentum-conserving collision rate $\gamma_{\rm mr}$, and frequency $\omega/\omega_p$, where $\omega_p$ is the plasma frequency (60). Diffusive boundary conditions are assumed, in accordance with Eq.\@ (\ref{eq:Z_diff}). The Fermi-surface geometry is sketched on the left-hand side of the plot, together with the applied electric field $\vec{E}=E_y \hat{u}_y$ aligned with the $y$ axis, and the local Fermi velocity vectors shown by red arrows. (a) Regimes in the $\left(\gamma_{\rm mr},\omega \right)$ plane, for $\gamma_{\rm mr}=\gamma_{\rm mc}$. (b) Regimes in the $\left(\gamma_{\rm mc},\omega \right)$ plane, for fixed $\gamma_{\rm mr}=10^{-8}\omega_p$. The color palette is the density plot of $\mathrm{Arg}Z(\omega)/(-\pi/2)$, giving the exponent $\eta$ of $Z(\omega)\propto \omega^\eta$. 
}
\end{figure*}

\begin{figure*}[ht] \centering
\includegraphics[width=0.8\textwidth]{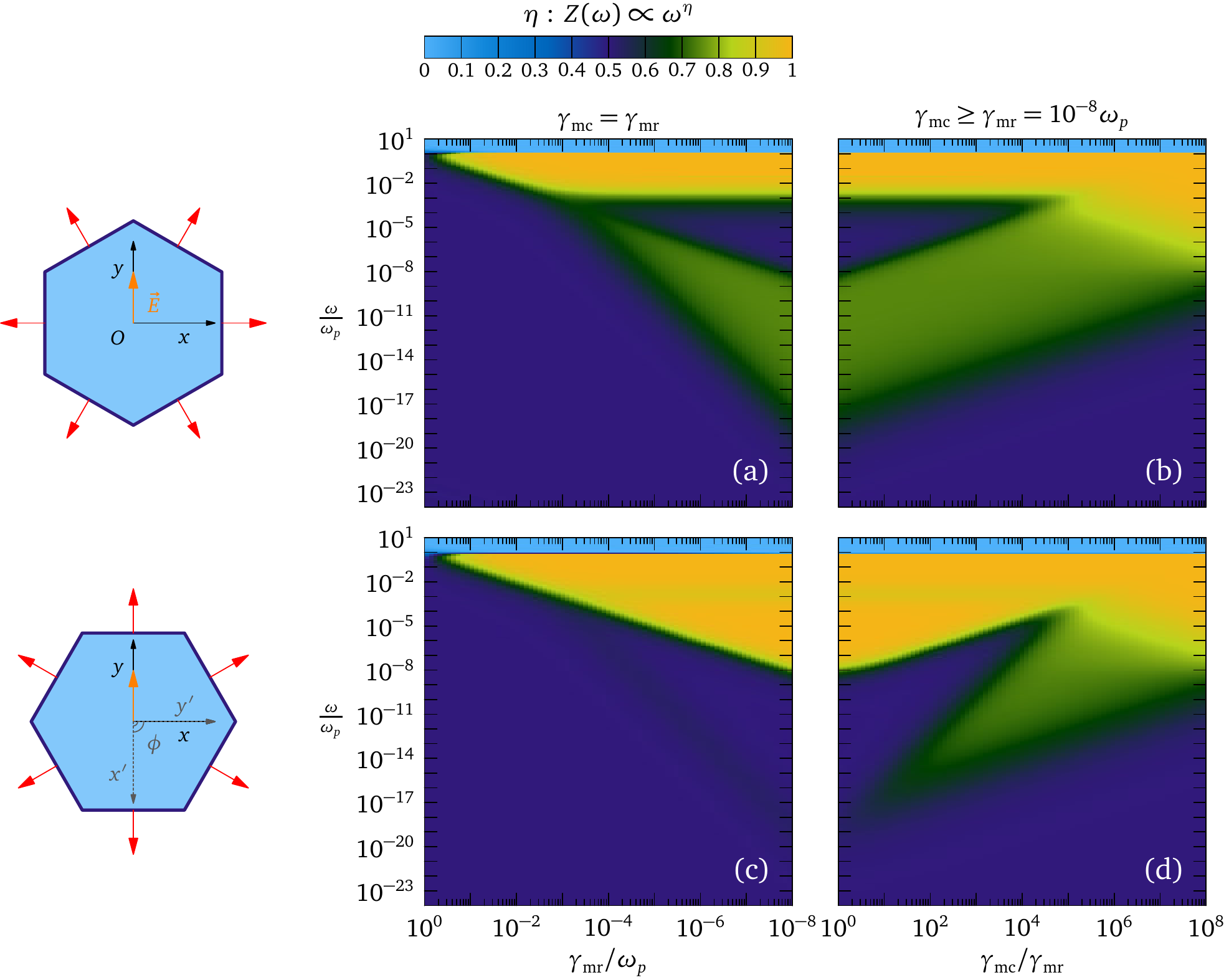}
\caption{\label{fig:omegacross_hex_diff} Orientational dependence of skin effect regimes for a hexagonal Fermi surface, as measured by the surface impedance $Z(\omega)$ as a function of relaxation rate $\gamma_{\rm mr}$, momentum-conserving collision rate $\gamma_{\rm mr}$, and frequency $\omega/\omega_p$, where $\omega_p$ is the plasma frequency (60). Diffusive boundary conditions are assumed, in accordance with Eq.\@ (\ref{eq:Z_diff}). The corresponding orientation is sketched on the left-hand side of each plot, together with the applied electric field $\vec{E}=E_y \hat{u}_y$ aligned with the $y$ axis. Red arrows depict the local Fermi velocity vectors. (a) ``Parallel'' configuration in the $\left(\gamma_{\rm mr},\omega \right)$ plane, for $\gamma_{\rm mr}=\gamma_{\rm mc}$. (b) ``Parallel'' configuration in the $\left(\gamma_{\rm mc},\omega \right)$ plane, for fixed $\gamma_{\rm mr}=10^{-8}\omega_p$. (c) Fermi surface rotated by $\phi=\pi/2$ with respect to panels (a) and (b), in the $\left(\gamma_{\rm mr},\omega \right)$ plane, for $\gamma_{\rm mr}=\gamma_{\rm mc}$. (d) Fermi surface rotated by $\phi=\pi/2$ in the $\left(\gamma_{\rm mc},\omega \right)$ plane, for fixed $\gamma_{\rm mr}=10^{-8}\omega_p$. The color palette is the density plot of $\mathrm{Arg}Z(\omega)/(-\pi/2)$, giving the exponent $\eta$ of $Z(\omega)\propto \omega^\eta$.  
}
\end{figure*}

\begin{figure*}[ht] \centering
\includegraphics[width=0.8\textwidth]{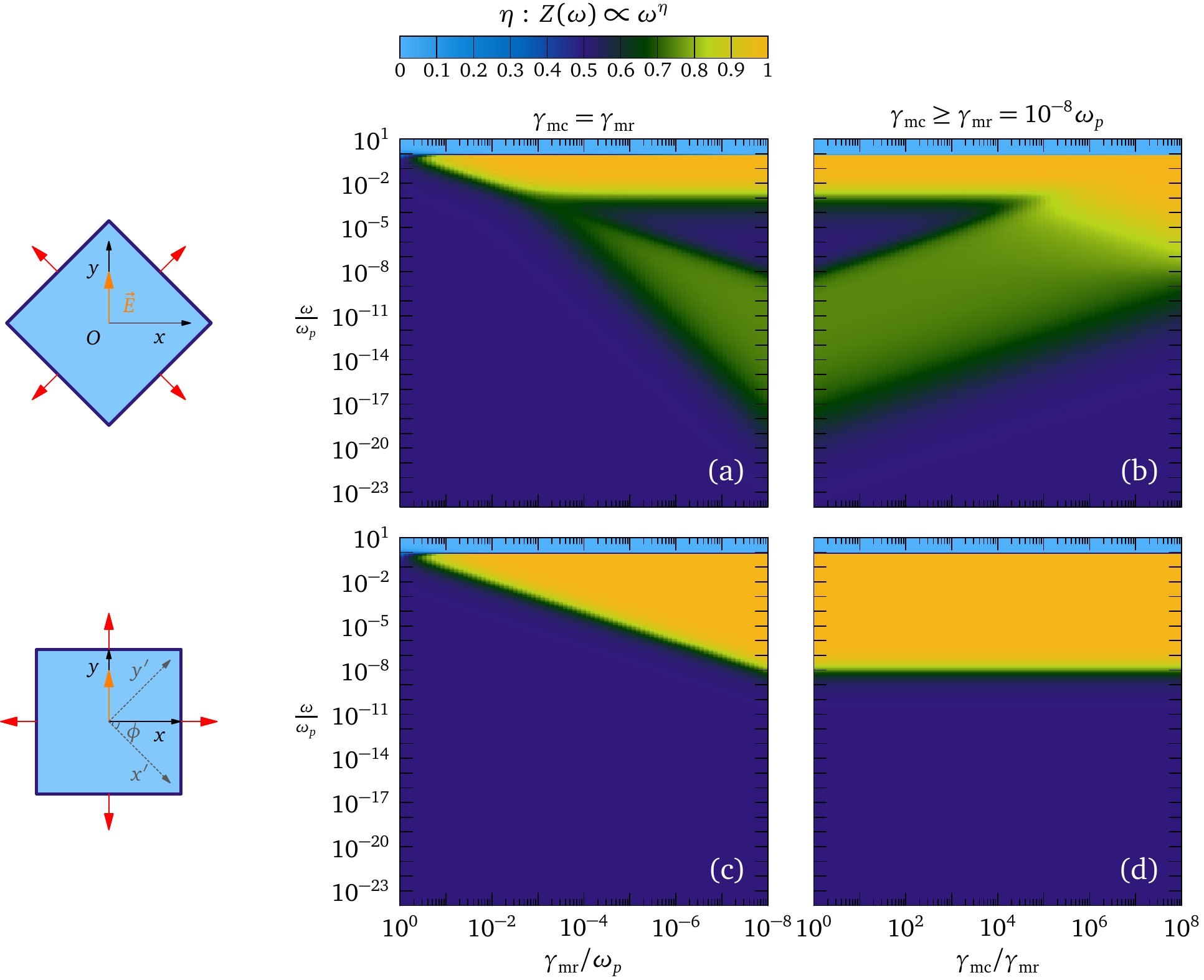}
\caption{\label{fig:omegacross_square_diff} Orientational dependence of skin effect regimes for a square Fermi surface, as measured by the surface impedance $Z(\omega)$ as a function of relaxation rate $\gamma_{\rm mr}$, momentum-conserving collision rate $\gamma_{\rm mr}$, and frequency $\omega/\omega_p$, where $\omega_p$ is the plasma frequency (60). Boundary conditions are diffusive, in accordance with Eq.\@ (\ref{eq:Z_diff}). The corresponding orientation is sketched on the left-hand side of each plot, together with the applied electric field $\vec{E}=E_y \hat{u}_y$ aligned with the $y$ axis. Red arrows depict the local Fermi velocity vectors. (a) ``Diamond-shaped'' configuration in the $\left(\gamma_{\rm mr},\omega \right)$ plane, for $\gamma_{\rm mr}=\gamma_{\rm mc}$. (b) ``Diamond-shaped'' configuration in the $\left(\gamma_{\rm mc},\omega \right)$ plane, for fixed $\gamma_{\rm mr}=10^{-8}\omega_p$. (c) Fermi surface rotated by $\phi=\pi/4$ with respect to panels (a) and (b), in the $\left(\gamma_{\rm mr},\omega \right)$ plane, for $\gamma_{\rm mr}=\gamma_{\rm mc}$. (d) Fermi surface rotated by $\phi=\pi/4$ in the $\left(\gamma_{\rm mc},\omega \right)$ plane, for fixed $\gamma_{\rm mr}=10^{-8}\omega_p$. The color palette is the density plot of $\mathrm{Arg}Z(\omega)/(-\pi/2)$, giving the exponent $\eta$ of $Z(\omega)\propto \omega^\eta$. 
} 
\end{figure*}

In this section, we check the results in Fig.\@ 10, 13, and 16 of the main text against the effect of different boundary conditions at the vacuum-metal interface. While specular interface scattering is assumed in the main text, here we employ diffusive scattering. In this case, the electric field inside the metal satisfies  \cite{Reuter-1948,Sondheimer-2001,Dressel-2001}
\begin{multline}\label{eq:EM_wave_d}
\left(\frac{\partial^2}{\partial x^2}+\frac{\omega^2}{c^2}\right)E_y^d(x,\omega) =-i \mu_0 \omega J_y^d(x,\omega) \\ =-i \mu_0 \omega \int_{0}^{+\infty} d x' \sigma_{yy} (x-x',\omega) E_y^d(x',\omega).
\end{multline}
We assume that $\lim_{x\rightarrow +\infty} \left|E_y^d(x,\omega)\right|=0$, and also that $\lim_{z\rightarrow +\infty} \left|\partial E_y^d(x,\omega)/\partial x\right|=\lim_{x\rightarrow +\infty} \left|\partial^2 E_y^s(x,\omega)/\partial x^2\right|=0$ \cite{Reuter-1948}. Here we have the additional complication that the driving term in Eq.\@ (\ref{eq:EM_wave_d}) only acts in the half-infinite space $x\in \left(0,+\infty\right)$, which in principle prevents us from directly switch to momentum space. We follow Ref.\@ \onlinecite{Reuter-1948} in defining $E_y^d(x,\omega)=0 \, : \, x<0$ and an auxiliary field
\begin{equation}\label{eq:E_tilde_z_om}
\tilde{E}(x,\omega)=\begin{cases} 0, \, x\geq0, \\ i \mu_0 \omega \int_0^{+\infty} d x' \sigma_{yy}(x-x',\omega) E_y^d(x',\omega), \, z<0, 
\end{cases}
\end{equation}
where $\sigma_{yy}(x-x',\omega)$ is the non-local conductivity. 
Then, by construction the wave equation (\ref{eq:EM_wave_d}) is equivalent to the problem
\begin{multline}\label{eq:wave_eq_E_tilde}
\tilde{E}(x,\omega)=\frac{\partial^2 E_y^d(x,\omega)}{\partial x^2} +\frac{\omega^2}{c^2} E_y^d(x,\omega) \\ +i \mu_0 \omega \int_0^{+\infty} d x' \sigma_{yy}(x-x',\omega) E_y^d(x',\omega), \, \forall x \in \mathbb{R}. 
\end{multline}
We see that Eq.\@ (\ref{eq:wave_eq_E_tilde}) is now manageable for all $x$ outside and inside the metal, and this allows one to employ the bilateral Laplace transform $\mathscr{L}$ of the fields \cite{Reuter-1948} $E_y^d(s,\omega)=\mathscr{L}\left\{E_y^d(x,\omega)\right\}$, $\tilde{E}(s,\omega)=\mathscr{L}\left\{\tilde{E}(x,\omega)\right\}$, and $\sigma_{yy}(s,\omega)=\mathscr{L}\left\{\sigma_{yy}(x,\omega)\right\}$. This way, one obtains an equation for $E_y^d(s,\omega)$ in terms of $E_y^d(0^+,\omega)$, and a logarithmic integral over $s$ involving $\sigma_{yy}(s,\omega)$. The integral can then be simplified \cite{Dingle-1953} to give an expression for $\left[E_x^d(0^+,\omega)\right]/\left[\left.\partial E_x^d(z,\omega)/\partial z\right|_{z=0^+}\right]$. The latter is then converted into the surface impedance $Z(\omega)$ through Eq.\@ (120):
\begin{equation}\label{eq:Z_diff}
Z(\omega)=i \mu_0 \omega \frac{2}{\pi} \int_0^{+\infty} d q_x \left[i \mu_0 \omega \sigma_{yy}(q_x,\omega)+\frac{\omega^2}{c^2}-q_x^2\right]^{-1}. 
\end{equation}
Parametrizing the degree of specularity of interface scattering with $p\in\left[0,1\right]$, Eq.\@ (\ref{eq:Z_diff}) corresponds to $p=0$ (completely diffusive scattering), while Eq.\@ (142) of the main text corresponds to $p=1$ (specular scattering). 
Figs.\@ \ref{fig:omegacross_iso_diff}, \ref{fig:omegacross_hex_diff}, and \ref{fig:omegacross_square_diff} show the density-plot ``phase diagrams'' for the surface impedance modulus $\left|Z(\omega)\right|$ of a circular, hexagonal, and square Fermi surface, employing diffusive boundary conditions. The results must be compared with Figs.\@ 16, 10, and 13 of the main text respectively, which refer to the same geometries and are obtained assuming specular interface scattering. 
All other parameters in Figs.\@ \ref{fig:omegacross_iso_diff}, \ref{fig:omegacross_hex_diff}, and \ref{fig:omegacross_square_diff} are the same as in Figs.\@ 16, 10, and 13. Comparing all ``phase diagrams'' of the same Fermi-surface geometry for specular and diffusive scattering, we see that the results for $\left|Z(\omega)\right|$ are in full agreement for the two types of boundary conditions. Hence, we conclude that the results for the impedance modulus, described in the main text, are robust against the effect of non-specular surface scattering. 

